\documentclass[a4paper,openright,twoside,11pt]{book}
\usepackage{amsfonts}
\usepackage{latexsym}
\usepackage{epsfig}
\usepackage{graphicx}
\usepackage{fancyhdr}
\newcommand{\dd}{\partial}

\newcommand{\tr}{{\rm tr \,}}

\newcommand{\be}{\begin{equation}}
\newcommand{\ee}{\end{equation}}
\newcommand{\ben}{\begin{displaymath}}
\newcommand{\een}{\end{displaymath}}
\newcommand{\ba}{\begin{eqnarray}}
\newcommand{\ea}{\end{eqnarray}}

\newcommand{\bean}{\begin{eqnarray*}}
\newcommand{\eean}{\end{eqnarray*}}

\begin{document}
\thispagestyle{empty}

\vspace*{-2cm} \hbox{\hskip 12cm ROM2F-05/06  \hfil} \hbox{\hskip
12cm hep-th/0503204 \hfil}

\vspace{1cm}

\begin{center}

{\LARGE {\bf Issues on tadpoles and vacuum redefinitions in String Theory}}

\vspace{1cm}
{\large Marco Nicolosi} \\
\vspace{0.6cm}
{\large {\it Dipartimento di Fisica, \ Universit{\`a} di Roma \
``Tor Vergata''}} \\
{\large {\it I.N.F.N.\ -\ Sezione di Roma \ ``Tor Vergata''}} \\
{\large {\it Via della Ricerca  Scientifica, 1}} \\
{\large {\it 00133 \ Roma, \ Italy}}
\end{center}

\vspace*{1cm}
{\small
\begin{center} {\bf Abstract} \end{center}
\noindent This Thesis discusses a number of issues related to the
problem of tadpoles and vacuum redefinitions that the breaking of
supersymmetry brings about in String Theory. The idea pursued here
is to try to formulate the theory in a ``wrong'' vacuum (the
vacuum that one naively identifies prior to the redefinitions)
and, gaining some intuition from some simpler field theory
settings, try to set up a calculational scheme for vacuum
redefinitions in String Theory. This requires in general
complicated resummations, but some simpler cases can be
identified. This is true, in principle, for models with fluxes,
where tadpoles can be perturbatively small, and for the one-loop
threshold corrections, that in a large class of models (without
rotated branes) remain finite even in the presence of tadpoles.
The contents of the Thesis elaborate on those of hep-th/0410101,
but include a number of additions, related to the explicit study
of a quartic potential in Field Theory, where some subtleties were
previously overlooked, and to the explicit evaluation of the
one-loop threshold corrections for a number of string models with
broken supersymmetry.
 }

 \vskip 36pt
 \begin{center}
( March , 2005 )
\end{center}
 \pagestyle{empty}
\newpage
\cleardoublepage
\pagestyle{empty}

\newpage

\oddsidemargin +1.5cm \evensidemargin +0.5cm



\thispagestyle{empty}

\vspace*{-1.5truecm}

\begin{center}

\makebox[\textwidth]{
\raisebox{0.68cm}{
\begin{minipage}[h]{13truecm}
        \hspace*{0.2cm}
        {\sc \LARGE UNIVERSIT\`A DEGLI STUDI DI ROMA\\
        \hspace*{3.36cm}
        ``TOR \rule{0pt}{24pt}VERGATA''}\\
    \end{minipage}}}
\end{center}


\begin{center}
    \includegraphics[height=1.45truecm]{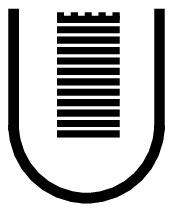}
\end{center}

\vspace{0.5cm}

\begin{center}
FACOLT\`A DI SCIENZE MATEMATICHE, FISICHE E NATURALI \\
Dipartimento di Fisica \\
\end{center}

\vspace{1.3cm}

\begin{center}
{\LARGE {\bf Issues on tadpoles and vacuum redefinitions in String Theory}}
\end{center}

\vspace{1.0cm}

\begin{center}
{\large Tesi di dottorato di ricerca in Fisica} \\
\rule{0pt}{22pt} presentata da \rule{0pt}{22pt} \\
\rule{0pt}{22pt}{\large {\em Marco Nicolosi}} \rule{0pt}{22pt}
\end{center}

\vspace{1.0cm}
\noindent
Candidato \\
\rule{0pt}{20pt}{\large {\em Marco Nicolosi} \\
\rule{0pt}{20pt}} \\
Relatore \\
\rule{0pt}{20pt}{\large Prof. {\em Augusto Sagnotti} \\
\rule{0pt}{20pt}} \\
Coordinatore del dottorato \rule{0pt}{30pt} \\
\rule{0pt}{20pt}{\large Prof. {\em Piergiorgio Picozza}} \\
\vspace{0.8cm}
\begin{center}
\underline{Ciclo XVII} \\
\vspace{0.4 cm}
Anno Accademico 2004-2005
\end{center}
\pagebreak
\newpage
\begin{flushright}
{\it Ai miei genitori}
\end{flushright}
\vfill\eject
\cleardoublepage
\pagenumbering{roman}
\chapter*{\em Acknowledgments \markboth{Acknowledgments}{Acknowledgments}}
\addcontentsline{toc}{chapter}{Acknowledgments} This Thesis is
based on research done at the Physics Department of the {\it
Universit\`a di Roma ``Tor Vergata''} during my Ph.D., from
November 2001 to October 2004, under the supervision of Prof.
Augusto Sagnotti. I would like to thank him for guiding me through
the arguments covered in this Thesis and for his encouragement and
suggestions. I am also grateful to Prof. Gianfranco Pradisi for
his explanations and for many interesting discussions. I wish to
thank Prof. Emilian Dudas for the enjoyable and fruitful
collaboration on the subjects reported in the last two chapters of
this Thesis and for his many suggestions. Let me also thank Marcus
Berg: I benefited greatly from discussions with him at the early
stages of this work. I would also like to acknowledge the
hospitality of {\it CERN}, {\it Scuola Normale Superiore di Pisa}
and CPhT-{\it \'{E}cole Politechnique}, where part of this work
was done. Moreover, I thank the Theoretical Physics group at the
Physics Department of the {\it Universit\`a di Roma ``Tor
Vergata''} for the very stimulating environment I have found
there. In particular I would like to thank Prof. Massimo Bianchi
and Prof. Yassen Stanev, and my referee, Prof. Ignatios
Antoniadis. Finally, I wish to thank Valentina, Enrica, Marco,
Luigi, Guido, Carlo, Dario, Marianna, Enrico, Maurizio, Mauro,
Elisa, Valerio, Oswaldo, Vladimir, that accompanied me during the
years of my Ph.D. studies. A particular thank is to Noemi for her
presence and constant encouragement.
\\
\\
Most of the figures of the first Chapter are taken from the review
``Open Strings'' by Carlo Angelantonj and Augusto Sagnotti.
\newpage
\cleardoublepage
\pagestyle{fancy}\fancyhf{}
\renewcommand{\chaptermark}[1]{\markboth{#1}{}}
\renewcommand{\sectionmark}[1]{\markright{\thesection.\ #1}{}}
\fancyhead[LE,RO]{\bfseries\thepage}
\fancypagestyle{plain}{\fancyhead{} \renewcommand{\headrulewidth}{0pt}}
\oddsidemargin +1.5cm \evensidemargin +0.5cm
\fancyhead[RE]{\bfseries\leftmark}
\fancyhead[LO]{\slshape\rightmark}
\setlength{\headwidth}{14cm}
\pagenumbering{roman}
\tableofcontents
\newpage
\cleardoublepage
\vfill\eject

\pagenumbering{arabic}
\chapter*{Introduction \markboth{Introduction}{Introduction}}
\addcontentsline{toc}{chapter}{Introduction}
\section*{The Standard Model and some of its problems}
Quantum Field Theory is a powerful tool and an extremely
appealing theoretical framework to explain the physics of elementary
particles and their interactions. The Standard Model describes such
interactions in terms of Yang-Mills gauge theories. The gauge group
$SU(3)\times SU(2)\times U(1)$ reflects the presence of three
fundamental forces: electromagnetism, the weak interaction,
and the strong interaction. All these forces are mediated
by spin-one bosons, but they have a very different behavior due to
their abelian or non-abelian nature.

In electromagnetism the gauge bosons are uncharged and thus
a test charge in vacuum can be only affected by the creation and
annihilation of virtual particle-antiparticle pairs around it and
these quantum fluctuations effectively screens its charge.
On the other hand, for the other two interactions there is a further
effect of anti-screening due to radiation of virtual gauge bosons that
now are charged, and this second effect is the one that dominates at short distances
in the strong interactions. Its consequence is the asymptotic
freedom at high energies, well seen in deep inelastic scattering experiments,
and more indirectly the confinement of quarks at low energies, that
explains why there no free colored particles (the particles that feel the
strong interactions) are seen in nature. The weak interactions should have the
same nature (and therefore the same infrared behavior)
as the strong ones, but a mechanism of symmetry breaking
that leaves a residual scalar boson, the Higgs boson, gives mass to
two of the gauge-bosons mediating the interaction, and makes its
intensity effectively weak at energy scales lower than
$M_W\simeq 100 GeV$.

The picture is completed adding the
matter that is given by leptons, that only feel electro-weak interactions, and quarks,
that feel also the strong interactions. Matter is arranged in three
different generations. The peculiar feature of the Standard Model, that makes it consistent
and predictive, is its renormalizability. And indeed the Standard Model
was tested with great precision up to the scale of fractions of a $TeV$.
However, in spite of the agreement with particle experiments and of the number of
successes collected by Standard Model, this theory does not give a fully
satisfactory setting from a conceptual point of view.

The first problem that arises is related to the huge number of free
parameters from which the Standard Model depends, like the gauge
couplings, the Yukawa couplings, the mixing angles in the weak
interactions, to mention some of them. The point is that there is no
theoretical principle to fix their values at a certain scale,
but they have to be tuned from experiments.

The last force to consider in nature is gravity. This force is extremely
weak with respect to the other forces, but contrary to them,
it is purely attractive and hence it dominates at large-scales in the
universe. At low energies, the dynamics of gravity is described in geometrical
terms by General Relativity.

In analogy with the fine-structure constant $\alpha=q^2/\hbar c$
that weights the Coulomb interaction, one can define a dimensionless
coupling for the gravitation interaction of the form $\alpha_G=G_NE^2/\hbar
c^5$, where $G_N$ is the Newton constant. In units of $\hbar=c=1$, one
can see that $\alpha_G\sim 1$ for $E\sim 1/\sqrt{G_N}=M_{Pl}$,
where the Planck mass is $M_{Pl}\sim 10^{19}GeV$. So we see that the
gravitational interaction becomes relevant at the Planck scale, and therefore
one should try to account for quantum corrections.
If the exchange of a graviton between two particles corresponds to an
amplitude proportional to $E^2/M^2_{Pl}$, the exchange of two gravitons
is proportional to
\be
\frac{1}{M_{Pl}^4}\int_0^{\Lambda} E^3 d E \sim \frac{\Lambda^4}{M_{Pl}^4} \ ,
\ee
that is strongly divergent in the ultraviolet. And the situation becomes worse
and worse if one considers the successive orders in perturbation
theory: this is the problem of the short distance divergences in
quantum gravity, that makes the theory non renormalizable.
Of course a solution could be that quantum gravity has a non-trivial
ultraviolet fixed-point, meaning that the divergences are only an
artifact of the perturbative expansion in powers of the coupling and
therefore they cancel if the theory is treated exactly, but to
date it is not known whether this is the case. The other possibility
is that at the Planck scale there is new physics. The situation would then be
like with the Fermi theory of weak interaction, where the divergences
at energy greater then the electro-weak scale, due to the point-like
nature of the interaction in the effective theory, are the signal of new
physics at such scale, and in particular of the existence of an
intermediate gauge boson. In the same way, it is very reasonable and
attractive to think that the theory of gravity be the infrared limit
of a more general theory, and that the divergences of quantum gravity,
actually due to the short distances behavior of the interaction,
could be eliminated smearing the interaction over space-time.

But the problem of the ultraviolet behavior of quantum gravity is
not the only one, when one considers all the forces in nature.
The first strangeness that it is possible to notice is the existence
of numbers that differ by many orders of magnitude. This problem is known
as the {\it hierarchy problem}. For example, between the electro-weak
scale, the typical scale in the Standard Model, and the Planck scale,
whose squared inverse essentially weighs the gravitational
interaction, there are $17$ orders of magnitude. Not only,
the other fundamental scale in gravity, $\Lambda^{1/4}\sim
10^{-13}GeV$, where $\Lambda$ is the cosmological constant, is also very
small if compared with the electro-weak energy $E_W\sim 100GeV$.
Moreover, there are other hierarchy differences in the parameters of
the Standard Model, for instance in the fermion masses.
Differences of many orders of magnitude seem very unnatural, especially
considering that quantum corrections should make such values extremely
unstable. Supersymmetry, introducing bosonic and fermionic particles
degenerate in mass, stabilizes the hierarchy but does not give any explanation
of such differences.

The last problem that we want to mention is the {\it cosmological
constant problem} \cite{cosmological}. One can naturally associate the cosmological
constant to the average curvature of the universe, and of course the
curvature is related to its vacuum energy density. Therefore,
one could try to estimate such a density from the microscopic point of
view $\rho_{micro}$, and compare it with the macroscopic value
$\rho_{macro}$, obtained by astrophysical observations.
The first estimate is provided in Quantum Filed Theory considering
the zero-point energy of the particles in nature. For example at the
Planck scale $\rho_{micro}\sim M_{Pl}^4 c^5/\hbar^3$.
On the other hand, from a simple dimensional analysis, the macroscopic
density can be expressed in terms of the Hubble constant $H$ through the
relation $\rho_{macro}\sim H^2 c^2 /G_N$. The point is that the
theoretical estimate is $120$ orders of magnitude greater than
the observed value. Surely, supersymmetry can improve matters.
Fermions and bosons contribute to the vacuum energy with an opposite
sign and so a supersymmetric theory would give a vanishing result for
$\rho_{micro}$. One can break supersymmetry at the
scale of the Standard Model with $E_{breaking}\sim TeV$ and
considering $\rho_{micro}\sim E_{breaking}^4 /\hbar^3 c^3$, but there is an
improvement of only $30$ orders of magnitudes.
In spite of all the attempts to solve this great mismatch, the
cosmological constant problem up to date remains essentially
unsolved.
\section*{The birth of String Theory and the Dual Models}
In the sixties physicists were facing the problem of the huge zoology
of hadronic resonances that the high energy experiments were
revealing. A fact was that the spin
$J$ and the mass $m^2$ of such resonances appeared to be related linearly trough the
simple relation $m^2=J/\alpha'$, checked up to $J=11/2$,
where $\alpha'\sim 1GeV^{-2}$ became known as the Regge slope.
Another key ingredient of the hadronic scattering amplitudes
was the symmetry under the cyclic permutation of the external
particles. Considering the scattering of two hadrons (1,2) going into
two other hadrons (3,4) and defining the Mandelstam variables as usual
\be
s=-(p_1+p_2)^2 \ , \qquad t=-(p_2+p_3)^2 \ , \qquad u=-(p_1+p_3)^2 \ ,
\ee
the symmetry under the cyclic permutation $(1234)\rightarrow (2341)$
reflects itself in the planar duality under the interchange of $s$ with $t$.
On the other hand, if one attempts to write the interaction due
to the exchange of an hadronic resonance of spin $J>1$ and mass $m^2_J$,
one should obtain a vary bad ultraviolet behavior with increasing
$J$. In fact, the corresponding scattering amplitude in the $t$-channel
at high energy would be proportional to
\be
A_J(s,t) \sim \frac{(-s)^J}{t-m^2_J} \ .
\ee
It was Veneziano \cite{veneziano} that in 1968 wrote a formula for the
scattering amplitude obeying planar duality and with an ultraviolet
behavior far softer then any local quantum field theory amplitude,
\be
A(s,t) \ = \
\frac{\Gamma(-\alpha(s))\Gamma(-\alpha(t))}{\Gamma(-\alpha(s)-\alpha(t))}
\ ,
\ee
where $\Gamma$ is the usual $\Gamma$-function,
and $\alpha(s)=\alpha(0)+\alpha' s$. The Veneziano amplitude has
poles corresponding to the exchange of an infinite number of
resonances of masses $m^2=n-\alpha(0)/\alpha'$, and
it is just the sum over all these exchanges that gives to the dual
amplitude its soft behavior at high energy.

In spite of its beauty and elegance, the Veneziano formula soon
revealed not suitable to describe the hadronic interactions, since
it predicts a decrease of the scattering amplitudes with energy
that is too fast with respect to the indications of the
experimental data. The Veneziano formula and its generalization
due to Shapiro and Virasoro \cite{sv} to a non-planar duality
(symmetry with respect to the exchange of each pair of the
variables $s$, $t$, $u$), were instead associated in a natural way
to the scattering amplitudes respectively of open and closed
strings. In particular, the infinite number of poles of such
amplitudes corresponding to the exchange of particles of higher
spin and mass can be read as a manifestation of the infinite
vibrational modes of a string. One of the peculiar features of the
closed string is that it contains a massless mode of spin 2, that
its low-energy interactions associate naturally to the graviton.
Hence a theory of closed strings seems to be a possible candidate
to describe quantum gravity without the usual pathologies at short
distances, given the soft behavior in the ultraviolet regime of
the interactions in the dual models. The Regge slope in this case
has to be identified with the characteristic scale in quantum
gravity, the Planck scale, $\sqrt{\alpha'} \sim 10^{-33}cm$. A
simple argument to understand why the interaction between two
strings has such a good ultraviolet behavior is to consider that
in the scattering the interaction is spread along a fraction of
the length of the strings. Hence, only a fraction of the total
energy is really involved in the interaction, and the coupling
$\alpha_G$ is effectively replaced by \be \alpha_{eff} \ =
\frac{G_N E^2}{\hbar c^5} \ \left(\frac{\hbar c}{E
  \ell_s}\right)^2 \ ,
\ee
where $\ell_s = \sqrt{\alpha'} $ is the length of the
string. One can observe that the
bad dependence of $E^2$ is thus cancelled in $\alpha_{eff}$.
On the other hand, an open string has in
its spectrum a massless mode of spin one, that
can be associated to a gauge vector. Therefore String Theory seems
also to furnish a way to unify all the forces in nature, giving one and the same origin
for gravity and gauge interactions.

At the beginning in the Veneziano model, String Theory contained
only bosonic degrees of freedom. Moreover it predicted the
existence of a tachyon in its spectrum. It was thanks to the work
of Neveu, Schwarz and Ramond \cite{nsr} that it was understood how
to include fermions in the theory. Moreover, the work of Gliozzi,
Scherk and Olive \cite{gso} was fundamental to understand how to
obtain supersymmetric spectra, projecting away also the tachyon.
Another peculiarity of String Theory is that quantum consistency
requires additional spatial dimensions. The dimensionality of
space-time is $D=26$ for the bosonic String and $D=10$ for the
Superstring. This feature of String Theory of course is very
appealing and elegant from a conceptual point of view (it connects
to the original work of Kaluza and Klein that unified the
description of a graviton, a photon and a massless scalar field in
$D=4$ starting from a theory of pure gravity in $D=5$), but
provides that the additional dimensions be compactified on some
internal manifold, to recover the $3+1$ dimensions to which we are
used. On the one hand, choosing different internal manifolds one
can obtain different four dimensional low-energy effective field
theories. Moreover, one has the possibility of choosing some of
the internal radii large enough, and this is important for trying
to solve the hierarchy problem. By suitable compactifications it
is also possible to break supersymmetry. All these possibilities
are surely key ingredients that compactifications offer to String
Theory, but on the other hand the presence of the additional
dimensions is a major problem for the predictivity of the
parameters of the four-dimensional world. We will come back on
this issue in the following.
\section*{M-theory scenario and dualities}
Today we know that there are five different supersymmetric
ten-dimensional String Models. They are Type IIA, Type IIB,
Type I $SO(32)$, heterotic $SO(32)$ (or HO) and heterotic $E_8\times
E_8$ (or HE). A lot of effort was devoted during the last decade in the
attempt to unify them. It was finally understood that all these models
can be regarded as different limits of a unique theory at
11 dimensions, commonly called M-theory \cite{mtheory}. Moreover, all these theories
are related to one another by some transformations known as dualities.
Surely a string has an infinite number of vibrational modes
corresponding to particles of higher and higher masses. Such masses are
naturally of the order of the Planck scale, but one can consider only
the massless sector. In other words, one can think to make an expansion
in powers of the string length $\ell_s=\sqrt{\alpha'}$, recovering the point-particle
low-energy effective field theory in the $\ell_s \to 0$ limit. At this level what we
find are some supersymmetric generalizations of General Relativity
known as supergravity theories. In ten dimensions exist three
different supersymmetric extensions of gravity: the Type IIA supergravity, that has
supersymmetry ${\mathcal N}=(1,1)$, the Type IIB supergravity, with ${\mathcal
  N}=(2,0)$ and the Type I supergravity, with supersymmetry ${\mathcal N}=(1,0)$,
but all of them have a common sector consisting in a graviton
$G_{\mu\nu}$, a dilaton $\phi$ and an antisymmetric two-tensor,a 2-form,
$B_{\mu\nu}$. The dynamics of such universal sector is
governed by the effective action
\be
\label{ef}
S_{eff} \ = \ \frac{1}{2k_{10}^2}\int d^{10} x \sqrt{-{\rm det}G}e^{-2\phi}\left(
R+4\dd_\mu\phi\dd^\mu\phi-\frac{1}{12}H^2\right) \ ,
\ee
where $R$ is the curvature scalar, $k_{10}^2$ is related to the
ten-dimensional Newton constant, and $H_{\mu\nu\rho}$
is the field strength of the 2-form.

The dilaton plays in String Theory a crucial role, since it weighs
the perturbative expansion. Moreover, its vacuum expectation value
$\langle \phi \rangle$, is a first example of a {\it modulus}
a free dynamical parameter from which the theory depends.
We want to stress that there is no potential to give a vacuum value
to the dilaton, and thus its expectation value remains undetermined.
Hence in ten dimensions one has actually a one-parameter
family of vacua labelled by the arbitrary expectation value of
the dilaton. Notice that the coupling constant $k_{10}^2$ in
the effective action (\ref{ef}) is not really a free parameter of the
theory. In fact, introducing the string coupling constant
$g_s=e^{\langle\phi\rangle}$, one can see that a change of $k_{10}^2$
can be reabsorbed by a shift to the vacuum expectation value of the dilaton.

After this digression on the role of dilaton, we can come back to
the dualities. The Type IIA and IIB superstring theories contain
only oriented closed strings and have as low-energy effective
field theories respectively the ten-dimensional supergravities of
types IIA and IIB. On the other hand, the Type I superstring has
unoriented closed and open strings, and thus we expect that it
describe at same time gravity and gauge interactions. And in fact
its low-energy behavior is governed by the Type I supergravity
together with the supersymmetric generalization of a Yang-Mills
theory with gauge group $SO(32)$. The Heterotic String is a theory
of closed strings. Now a closed string has left and right moving
modes and they are independent, and so one can consider the right
modes of the usual superstring in ten-dimensions together with the
left modes of the bosonic string compactified from $D=26$ to
$D=10$. Notice that the compactification introduces in a natural
way the internal degrees of freedom of a gauge theory without the
need to introduce open strings. The resulting theories are
supersymmetric and free from tachyons. Moreover, string
consistency conditions fix the choice of the internal lattice to
only two possibilities: the first one corresponds to the roots of
the lattice of $E_8\times E_8$, while the second one to the roots
of $SO(32)$. Hence, in the low-energy limit, the two heterotic
strings give the usual Type I supergravity coupled to a Super
Yang-Mills theory with gauge group $E_8\times E_8$ or $SO(32)$.
There exist also other non supersymmetric ten-dimensional
heterotic models corresponding to different projection of the
spectrum. Perhaps the most interesting, not supersymmetric but
free from tachyons, is the $SO(16)\times SO(16)$ model.

At the end of the seventies, Cremmer, Julia and Scherk found the
unique supergravity theory in eleven dimensions. Its bosonic spectrum
contains the metric and a 3-form $A_3$ whose dynamics is given by the action
\be
S_{11} \ = \ \frac{1}{2k^2_{11}}\int d^{11}x \sqrt{-{\rm det}G} \
\left(R-\frac{1}{24}F_{IJKL}F^{IJKL}\right) \ - \ \frac{\sqrt{2}}{k^2_{11}}\int A_3 \wedge F_4 \wedge F_4 \ ,
\ee
where $k_{11}$ is related to the eleven dimensional Newton constant,
and $F_4 = 6 d A_3$ is the field strength of the 3-form. Notice that, in net
contrast with the ten-dimensional supergravities, here the spectrum
does not contain any 2-form.

However compactifying the eleventh dimension on a circle one
recovers the ten-dimensional Type IIA supergravity. And this is
not all. If one compactifies the eleven-dimensional supergravity
on a segment $S^1/{\mathbb Z}_2$, one recovers the low-energy
theory of the heterotic $E_8\times E_8$ string \cite{hwi}. At this
point it is quite natural to think that, just like all
ten-dimensional supergravities are low-energy limits of the
corresponding superstring theories, so even the eleven-dimensional
supergravity can be regarded as the low-energy limit of a more
fundamental theory, that is commonly called M-theory
\cite{mtheory}. What is M-theory up to date is not known. In
particular, we do not know what are its fundamental degrees of
freedom. Surely what we can say is that it is not a theory of
strings. In fact, as we already said, all ten-dimensional
supergravity theories contain in their spectrum a 2-form. Now a
two-form has just the right tensorial structure to describe the
potential for a one-dimensional object (a string), just like in
the usual case the potential for a point charge is a vector.
Therefore the presence of the 2-form in the spectrum is a clear
signal that the dynamics is described by strings, while its
absence in the spectrum of the eleven-dimensional supergravity
reveals that M-theory is not related to strings.

Up to now we discussed how to recover the Type IIA
and the Heterotic $E_8\times E_8$ models from the mysterious
M-theory. But the surprises are not finished. In fact, the other
ten-dimensional models are also related to one another through some
transformations known as {\it dualities}. In general a duality is an
invertible map that connects the states of a theory to the ones of
another theory (or of the same theory) preserving interactions and symmetries. The importance and
utility of a duality can be appreciated already in
Quantum Field Theory, where generally one makes the perturbative
expansion in powers of $\hbar$. The point is that not all
quantities can be described in terms of a perturbative series, and a
duality can help because it allows to see the same phenomenon in
another description. A case of particular interest is provided by a duality
that maps the perturbative region of a theory into the non-perturbative
region of the same theory. This is the case of the $S$-duality \cite{Sduality} in
String Theory.
\begin{figure}
\begin{center}
\includegraphics[width=7cm,height=5cm]{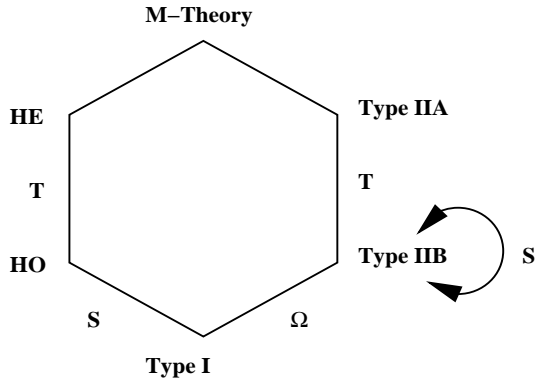}
\end{center}
\caption{Dualities between different ten-dimensional String Theories.}
\label{esagono}
\end{figure}
Such duality inverts the string coupling constant
\be
S \ : \ g_s \ \longleftrightarrow \ \frac{1}{g_s} \ ,
\ee
and, as can be seen from the figure\ref{esagono}, maps the $SO(32)$ Heterotic
String to the Type I $SO(32)$ String. More precisely, $S$-duality
identifies the weak coupling limit of one theory with the strong limit
of the other. Moreover, the Type IIB model is self-dual.
The weak-strong coupling $S$-duality is manifest in
the low-energy effective field theories, but really the duality,
being non-perturbative, remains only a conjecture, as the strong
coupling limit of String Theory is not fully under control. However,
up to date, all non perturbative tests revealed no
discrepancy with the conjecture of duality.

A crucial step in the understanding of the $S$-duality is the
existence in the spectra of the various string models of
extended objects with $p$ spatial dimensions, whose presence is
fundamental for the right counting and the matching of the degrees of
freedom after a non perturbative duality is performed. These objects corresponds to
solitonic configurations with tension proportional to the inverse of
the string coupling constant (in net contrast with the usual case in
Field Theory where the tension is proportional to the squared inverse of
the coupling), and are mapped by an $S$-duality in the
usual perturbative string states.
They are known as $p$-branes and if originally they appeared as classical
supergravity solutions, then it was realized that some of these
objects, known as D$p$-branes \cite{pol95}, can be thought as
topological defects where open strings terminate, with Dirichlet
boundary conditions in the directions orthogonal to them and Neumann
ones in the parallel directions. A D-brane is
characterized by its tension and by a charge that is defined by the coupling of the brane
to a corresponding tensor potential. Together with D-branes, one can also define
antibranes, $\bar{\rm D}$-branes, that are characterized by the same
tension but by a reversed charge.

Another important duality, that in contrast with the previous one
is perturbative, is the $T$-duality between the Type
IIA and Type IIB theories or between the Heterotic $E_8\times E_8$ and
$SO(32)$ theories. In particular, a $T$-duality
identifies one theory compactified on a circle of radius $R$ with
the other theory compactified on a circle of radius $1/R$.
A peculiar feature of $T$-duality is that it interchanges Neumann and
Dirichlet boundary conditions \cite{dlp,horava2,green}, and thus it changes the
dimensionality of a D-brane. And indeed the content of D$p$-branes of
Type IIA and Type IIB models are just the right one to respect the
T-duality relating them (D$p$-branes with $p$ odd for Type IIB and $p$ even for Type IIA).

The last link we need to unify all the five ten-dimensional superstring
models is the orientifold projection $\Omega$ \cite{cargese}
that connects the Type IIB String with the Type I String.
$\Omega$ exchanges the left and right modes of a closed string,
and the Type I String is obtained identifying the left and right modes
of Type IIB. The fixed points of such projection correspond to
some extended non dynamical space-time objects known as the
orientifold-planes or briefly O-planes. In contrast with a D-brane,
whose tension is always positive, an O-plane can also have a negative
tension. Moreover, like a D-brane, an O-plane carries a charge with respect
to some tensor potential.

Now the hexagon of dualities is closed (see figure\ref{esagono}) and what we learn is
that in spite of their apparent differences, all the ten-dimensional
superstring theories can be thought of really as different limits in a certain
parameter space of a unique underlying theory that is identified with the
M-theory. Notice that in this appealing picture the fact that the
eleven-dimensional supergravity is unique is very compelling from the
unification point of view.

A crucial matter that we have to stress before closing this
discussion is the consistency of all these ten-dimensional Superstring
Theories, and in particular the absence of anomalies in their spectra.
Anomalies arise already in Field Theory, and are quantum violations of
classical symmetries. The violation of a global
symmetry is not dangerous and often can be useful from a
phenomenological point of view. For example, in the theory of the strong
interaction with massless quarks, the quantum violation of the
classical scale invariance is a mechanism that gives mass to the
hadrons. On the other hand a violation of a local symmetry,
like the gauge symmetry in a Yang-Mills theory or the invariance under
diffeomorphisms in General Relativity, is a real problem since
the unphysical longitudinal degrees of freedom no longer decouple,
and as a consequence the theory loses its unitarity. Therefore, the
cancellation of all (gauge, gravitational, mixed)-anomalies is a
fundamental property to verify in String Theory.
The first type of cancellation of anomalies in String Theory is
achieved imposing the tadpole condition in the Ramond-Ramond (R-R)
sector. We will come back to the issue of tadpoles in the following, but
for the moment what we really need to know is that such condition
from the space-time point of view corresponds to imposing that the
Faraday-lines emitted by the branes present in the model be absorbed
by the O-planes, or in other words that the compactified space-time be globally
neutral. Such a condition fix also the gauge group for the Type I models.
The other anomalies arising in String Theory from the so called
non-planar diagrams are cancelled thanks to a mechanism due
to Green and Schwarz \cite{gs}. The anomaly of the one-loop hexagon-diagram, the analog of the
triangle-diagram in four dimensions that one meets in gauge theory, is exactly
cancelled by a tree-level diagram in which the 2-form propagates. This
mechanism works in all the ten-dimensional theories we saw\footnote{Really also the heterotic
$SO(16)\times SO(16)$ model we cited is anomaly-free.} (the Type IIA is not
anomalous because is not chiral). The mechanism of Green and Schwarz can be generalized to the case of
several 2-forms \cite{as92} and is at the heart of the consistency of
string models.
\section*{Compactifications and supersymmetry breaking}
Up to now we presented some arguments why String Theory should be
consider a good candidate for quantum gravity. Moreover, we saw
that all the consistent supersymmetric ten-dimensional models are
really dual one to the other and that all of them can be linked to
a unique eleven-dimensional theory. Finally, String Theory
describes together gravity and gauge interactions, giving a
concrete setting for unifying in a consistent fashion all the
forces of the Standard Model with Quantum Gravity. The following
step we need to recover our four dimensional world is a closer
look at the compactification of the six additional dimensions on
an internal manifold. A single string state gives an infinite
tower of massive excitations with masses that are related to the
inverse of the internal dimensions, but from the low-energy point
of view one can effectively think that all the massive recurrences
disentangle if the typical size of the internal volume is small
enough. The other key ingredient in order to get a realistic
four-dimensional physics is supersymmetry breaking, that really
can be also related to the issue of compactification. We will
therefore review these two arguments together, showing also how
the presence of D-branes can provide some new natural settings to
break supersymmetry.

The simplest way to realize compactifications in String Theory is to choose as
internal manifold a torus. This follows the lines traced by
Kaluza and Klein, but a closed string offers more possibilities with
respect to a point particle, since a string can also wrap around a
compact dimension. Another interesting setting is provided by orbifold \cite{dhvw}
compactifications, obtained identifying points of a certain internal
manifold under the action of a discrete group defined on it.
Such identifications in general leave a number of fixed points
where the Field Theory would be singular but String Theory is well defined on it.
A further interesting and elegant setting for compactification is
furnished by Calabi-Yau spaces, that in contrast with the orbifold
compactifications are smooth manifolds and in suitable limits reduce to
orbifolds that are exactly solvable in String Theory. A Calabi-Yau $n$-fold is a complex manifold
on which a Ricci-flat K\"{a}hler
metric can be defined. As a consequence a non trivial $SU(n)$
holonomy group emerges that in turn is responsible for
supersymmetry breaking on such spaces.
For instance, a six-dimensional Calabi-Yau with holonomy group $SU(3)$ preserves only
${\mathcal N}=1$ supersymmetry. Different $SU(3)$ Calabi-Yau manifolds can be recovered
blowing up in different ways the fixed points of the orbifold $T^6/\mathbb{Z}_3$. Another interesting example
of a Calabi-Yau manifold
is provided by the space $K3$ \cite{aspin}, with holonomy $SU(2)$, that gives a
four-dimensional ${\cal N}=2$ supersymmetry and in a suitable limit reduces to the orbifold $T^4/{\mathbb Z}_2$.

Supersymmetry breaking can be obtained in standard toroidal or orbifold
compactifications. The important thing to notice is that in the
first case supersymmetry is broken at a scale fixed by the radius of
the internal manifold. In fact, one can generalize the Scherk-Schwarz \cite{schschwstr,scherkschwarz}
mechanism to String Theory, for instance giving periodic boundary condition on a
circle to bosons and antiperiodic conditions to fermions. In this way
the masses of the Kaluza-Klein excitations are proportional to $n/R$
for bosons and $(n+1/2)/R$ for fermions, and thus the gauginos or
gravitinos are lifted in mass and supersymmetry is broken at the scale
$1/R$. This way to break supersymmetry
is like a spontaneous breaking and in the limit of
decompactification one recovers all the original supersymmetry. This fact is in net
contrast with the case of breaking through orbifold compactification,
where the breaking is obtained projecting away some states,
and after the orbifolding no trace of the original supersymmetry remains.
On the other hand, it would be interesting to break supersymmetry at a scale which is
independent from $R$, for example at the string scale, that in some
recent models requiring large extra-dimensions, can even be of the order of
$TeV$.

A new interesting phenomenon happens when one breaks supersymmetry
by toroidal compactification in the presence of D-branes
\cite{ads1,bsb}. So let us consider the case of
some branes parallel to the direction of breaking. This case is
called usually ``Scherk-Schwarz breaking'', and supersymmetry is
broken both in the bulk (closed sector) and on the branes (open
sector). The spectrum is a deformation of a supersymmetric one
that can be recovered in the decompactification limit. Something
new happens if the direction of breaking is orthogonal to the
branes (really we are thinking in a T-dual picture). In this case,
at least at the massless level, supersymmetry is preserved at
tree-level on the branes. This phenomenon is commonly called
``brane supersymmetry'', and indeed at tree level the gaugino does
not take any mass. Really supersymmetry breaking on the branes is
mediated by radiative corrections due to the gravitational
interactions, and so also the gaugino eventually gets a mass that
with respect to the one of gravitino is suppressed by the Planck
mass, being a quantum effect. This phenomenon however has not been
fully studied to date. In contrast with the previous case one can
break supersymmetry on the branes, and there are essentially two
ways to do that. The first one is provided by models that require
configurations with the simultaneous presence of branes and
antibranes of different types \cite{bsb}. In this case the closed
sector is supersymmetric, but is generally different from the
standard supersymmetric one, while supersymmetry is broken at the
string scale on the open sector, as branes and antibranes preserve
different halves of the original supersymmetry. This kind of
configurations, however is classically stable and free of
tachyons. The second one is obtained deforming a supersymmetric
open spectrum with a system of separated brane-antibrane pairs of
the same type. Such a configuration is unstable due the attractive
force between branes and antibranes and a tachyon develops if
their distance if small. The last known way to break supersymmetry
in String Theory is with intersecting D-branes
\cite{Uranga:2003pz}, that in a T-dual picture corresponds to
turning on constant magnetic fields on the internal manifold
\cite{bachasmag}. A very appealing feature of this kind of
breaking is that  chiral fermions in different generations can
live at the intersections of the branes.

Directly related to the issue of dimensional reduction and
supersymmetry breaking are two of the greatest problems of
String Theory. A common problem one has to face in String Theory when one performs a
reduction from the ten-dimensional world to the four-dimensional one
is the emergence of quantities that remain arbitrary. These are
the moduli, an example of which we already met in the vacuum
expectation value of the dilaton. At the beginning of this
Introduction we mentioned the problem of the large number of free parameters present in
the Standard Model. Of course it would be great to get some
predictions on them from String Theory, but this is not what happens.
The four-dimensional low-energy parameters after the reduction from
ten-dimensions depend on some moduli, like for example the size and
shape of the extra dimensions, and their predictivity is related to the
predictivity of such moduli. Now in gravity there is
no global minimum principle to select a certain configuration
energetically more stable than another one, and so really the moduli
remain arbitrary. This problem is known as the {\it moduli problem},
and up to date represents the greatest obstacle for the predictivity
of the Standard Model parameters from String Theory. Therefore, in
spite of a unique (thanks to dualities) ten-dimensional theory, in
four dimensions we have typically a continuum of vacua labelled by the
expectation values of such moduli.
Really the moduli problem is always associated to supersymmetric
configurations. And in fact supersymmetry if on the one hand stabilizes
the space-time geometry, on the other hand makes the moduli
arbitrary in perturbation theory (this is not true non perturbatively for ${\mathcal N}=1$).
Of course, one has to break supersymmetry at a certain scale
to recover the observed physics, but in that case the arising quantum
fluctuations are not under control, and moreover some infrared
divergences due to the propagation of massless states going
into the vacuum can affect the string computations. These
divergences typically arise when one breaks supersymmetry, and
are related to the presence of uncancelled Neveu-Schwarz
Neveu-Schwarz (NS-NS) tadpoles.  Physically, their emergence
means that the Minkowski background around which one
quantizes the theory after supersymmetry breaking is no more a real vacuum,
and has to be corrected in order to define reliable quantities.
This is the problem of the vacuum redefinition in String Theory
with broken supersymmetry and we will deal at length with this issue in
the following, since it is the main theme of this Thesis. For the moment, let us
return to the moduli problem.

An important attempt to stabilize the moduli is provided by
compactifications with non trivial internal fluxes \cite{fluxes}.
After supersymmetry breaking a modulus takes some vacuum
expectation value but in general, for the technical reasons we
already stressed, it is not possible to compute it. In contrast
with this fact, if one turns on some background potentials on the
internal manifold, a low-energy effective potential arises and
many of the moduli are frozen. The nice thing to underline in the
presence of fluxes is the possibility to compute the effective
potential. Really what one can do for general Calabi-Yau
compactifications is to build an approximate expansion of the
potential around its extrema, but there are also simple orbifold
compactifications in which the potential for some moduli is known
globally. Moreover, playing with the internal fluxes, one can also
stabilize most of the moduli without breaking supersymmetry.
\section*{Brane-worlds}
A very interesting scenario that has been developed in the last
years is provided by the so called brane-worlds
\cite{Antoniadis:1998ig,Arkani-Hamed:1998rs}. As we already said,
the tension of a brane scales as the inverse of the string
coupling constant, and so it seems to be like a rigid wall at low
energy, but of course its dynamics can be described by the open
strings terminating on it. In fact, the low-energy open string
fluctuations orthogonal to the branes correspond to the
oscillations of the brane from its equilibrium position and from
the brane point of view such modes are effectively seen as scalar
fields arising as Goldstone bosons of the translational symmetry
of the vacuum broken by the presence of the brane.  In the same
way, the fermions living on the branes can be seen as goldstinos
arising after the breaking of supersymmetry introduced by the
presence of the brane. On the other hand, the parallel
fluctuations of a string terminating on the brane describe at low
energy a $U(1)$ gauge boson. A supersymmetric D$p$-brane is really
a BPS state, meaning that it preserves only half of the total
supersymmetry of Type II vacua, and that its tension and charge
are equal. This implies that between two parallel branes the
gravitational attraction is compensated by the Coulomb repulsion
and no net force remains. So one has the possibility to superpose
some branes, let say $N$, with the consequence of enhancing the
gauge symmetry to $U(N)$ due to the fact that now an open string
has  $N\times N$ ways to start and end on the $N$ D-branes. In
other words a stack of $N$ coincident D-branes gives the
possibility to realize non abelian interactions for the open
strings, while their displacement can be seen as a sort of
spontaneous symmetry breaking preserving the total rank. A nice
thing to notice is that while the low-energy dynamics of the gauge
filed living on the branes is the usual one described by
Yang-Mills theory, the higher energy string corrections remove the
usual divergence at $r \to 0$ of the Coulomb interaction between
point charges. And in fact  the low-energy effective action for
the open string modes, at least in the abelian case, is given by
the Born-Infeld action \cite{bidirac}. For example, in the case of
a static electric field, the usual power law of the Coulomb
interaction $1/r^2$ is replaced by $1/\sqrt{r^4+(2\pi\alpha')^2}$
and so we see that the string once more time regulates the
short-distance divergence.

At this point we saw that stacks of D-branes describe non abelian gauge
groups and that the displacement of D-branes is responsible for gauge
symmetry breaking. But with branes one can obtain more. And in fact, as
already stressed, the intersection of D-branes can lead to chiral fermions
coming from the open strings stretched between them (really,
considering the low lying modes, these chiral fermions live in the
intersection volume of the branes). The interesting possibility
provided by this setting is that chiral fermions are obtained in a number
of replicas giving a realistic set up in which one can try to reproduce the
matter fields (and the gauge group) of the Standard Model \cite{standardModel}.
A simple example to understand the origin of the
matter replication is given by a configuration with two stacks of D6-branes that
intersect in a four-dimensional volume. Now we have to think the
other dimensions of the branes as wrapped around some 3-cycles of an
internal compact six-dimensional space. Two such 3-cycles
can intersect several times in the internal manifold, thus leading to
replicas of the chiral matter living at the four-dimensional intersection.

One of the most important issues related to the brane-world
scenario is the geometrical explanation that one can give in this
context to the hierarchy between the electro-weak scale of the
Standard Model and the Planck scale \cite{Antoniadis:1990ew,Antoniadis:1998ig,Arkani-Hamed:1998rs}. In other words, brane-worlds
can provide a simple argument to explain the weakness of gravity
with respect to the other forces. And in fact, while the forces
mediating gauge interactions are constrained to  the branes,
gravity spreads on the whole space-time so that only a part of its
Faraday lines are effectively felt by the brane-world.
A simple way to see how this argument works is to consider that
the four-dimensional Newton force, in the case of $n$ additional
transverse compact dimensions of radius $R$, at short distances is
\be
\frac{1}{\left(M^{Pl}_{4+n}\right)^{2+n}} \ \frac{1}{r^{2+n}} \ ,
\ee
where $M^{Pl}_{4+n}$ is the Planck mass in $4+n$ dimensions. On the
other hand for distances greater then the scale of compactification one
should observe the usual power law of the Newton force
\be
\frac{1}{\left(M^{Pl}_{4}\right)^2} \ \frac{1}{r^2} \ ,
\ee
where $M^{Pl}_{4}=10^{19}GeV$ is the four-dimensional Planck
mass. Continuity at $r=R$ gives
\be
M^{Pl}_{4} \ = \ \left(M^{Pl}_{4+n}\right)^{2+n} \ R^n
 \ ,
\ee
so that one can fix the String scale $M_s$, or the $4+n$-dimensional
Planck scale, at the $TeV$ scale, $M_s=M^{Pl}_{4+n}\sim TeV$, and
obtain the usual value of the four-dimensional Planck scale provided
the size of the transverse dimensions is given by
\be
R \sim 10^{32/n} 10^{-17} cm \ .
\ee
For example, the case of only one transverse extra-dimension
is excluded, since it would give $R\sim 10^{10}Km$, but $n=2$ is interesting
because it gives already $R\sim mm$ \cite{Antoniadis:1990ew}. The case with $n>2$ would give dimensions
that should be too small, completely inaccessible for Newton low-energy measurements
(the case $n=6$ for example corresponds to $R\sim fm$). Up to date
the limit on the size of the transverse dimensions is at the
sub-millimeter scale $\sim 200 \mu m$, at which no deviations from the
power law of Newton force has been discovered. On the other hand,
surely there can be some extra dimensions parallel to the branes and
these ones have to be microscopic, at least of the order of
$10^{-16}cm$ in order to not have other physics in the
well explored region of the Standard Model.
Therefore, if on the one hand it seems that the brane-world scenario could solve
the hierarchy problem, giving the possibility of choosing a
string scale of the order of the electro-weak scale, on the other
hand a geometrical hierarchy emerges between the macroscopic
transverse directions and the microscopic parallel ones.
\section*{Tadpoles in String Theory}
As we already stressed, when one breaks supersymmetry in String Theory
some bosonic one-point functions going into the vacuum usually
emerge. These functions are commonly called {\it tadpoles} and are
associated to the NS-NS sector, to distinguish them from tadpoles in
the R-R sector. In the presence of open strings, the latter identify from the space-time point of
view a configuration of D-branes and O-planes with a non-vanishing
total charge. Such tadpoles typically signal an inconsistency of the theory,
the presence of quantum anomalies, and therefore R-R tadpoles in all cases where
the charge cannot escape should be cancelled. On the other hand NS-NS tadpoles correspond from the
space-time point of view to configurations of branes
with a non-vanishing tension that gives rise to a net gravitational
attraction between them. Hence, a
redefinition of the background is necessary. Let us try ti be more concrete.
Up to now one is able to do string computations essentially only
around the flat Minkowski background, a case
that is allowed and protected by supersymmetry.
Supersymmetry breaking then destabilizes the space-time, producing a
potential for the dilaton
\be
V_{\phi} \sim  T \int d^{10} x \sqrt{-{\rm det} G} \ e^{-\phi}
\ee
that in turn acts as a source in the equation of
motion of the graviton. Thus the flat Minkowski background is no more
a solution,  and a vacuum redefinition is necessary.
And in fact the emergence of NS-NS tadpoles is
always accompanied by the emergence of infrared divergences
in string amplitudes due to the propagation of NS-NS massless
 states that are absorbed by tadpoles at vanishing momentum.
The tadpole problem was faced for the first time in the eighties by
Fishler and Susskind \cite{fs} for the bosonic closed string where a non-vanishing dilaton tadpole emerges
at one-loop. In particular, they showed that the
 one-loop conformal anomaly from the small handle divergences in the bosonic closed string
can be cancelled by a shift of the background.

The dilaton tadpole problem is today one of the most important issues to
understand in order to have a clearer understanding of supersymmetry breaking in String Theory.
On the other hand, this last step is fundamental if one wants to construct
realistic low-energy scenarios to compare with Standard Model.
Up to date it has proved impossible to carry out
background redefinitions {\it \`{a} la } Fischler and Susskind
in a systematic way.
In this context, our proposal is to insist on quantizing the string in a Minkowski
background, correcting the quantities so obtained with
suitable counterterms that reabsorb the infrared divergences and
lead these quantities to their proper values. This way to proceed of course seems
unnatural if one thinks about the usual saddle-point perturbative expansion in field Theory.
And in fact this means that we are building the perturbation theory not around a
saddle-point, sice the Minkowski background is no more the real vacuum. Nevertheless we
think that this approach can be a possible way to face the problem in String Theory
where one is basically able to perform string computations only in a Minkowski background. What
should happen is that quantities computed in a ``wrong'' vacuum recover their right values after
suitable tadpole resummations are performed while the corresponding infrared divergences are at the same time
cancelled. The typical problems one has to overcome when one faces the problem with our approach is that
in most of the models that realize supersymmetry breaking in String Theory, tadpoles arise already at
the disk level. Hence, even in a perturbative region of small string coupling constant, the first
tadpole correction can be large. Therefore the power series expansion in tadpoles becomes out of control
already at the first orders, and any perturbative treatment typically looses its meaning.
On the other hand, the higher order corrections due to tadpoles correspond to
Riemann surfaces of increasing genus, and the computation becomes more and more involved,
and essentially impossible to perform beyond genus $3/2$.

At this point, and with the previous premise we seem to have only
two possible ways to follow in String Theory. The first is to
search for quantities that are protected against tadpole
corrections. An example of such quantities is provided by the
one-loop threshold corrections (string corrections to gauge
couplings) for models with parallel branes, but in general for all
model with supersymmetry breaking without a closed tachyon
propagating in the bulk. As we will see, one-loop threshold
corrections are ultraviolet (infrared in the transverse tree-level
closed channel) finite in spite of the presence of NS-NS tadpoles.

The second issue to investigate is the possibility of models with
perturbative tadpoles. Turning on suitable fluxes it is possible
to have ``small'' tadpoles. In this case, in addition to the usual
expansion in powers of the string coupling, one can consider
also another perturbative expansion in tadpole insertions. In
these kinds of models we do not need the resummation, but the first
few tadpole corrections should be sufficient to recover a reliable
result in a perturbative sense.

This Thesis is organized in the following way. The first chapter
is devoted to basic issues in String Theory, with particular
attention to the orientifold construction. Simple examples of
toroidal and orbifold compactifications together with their
orientifolds are discussed. In the second chapter we review some
models in which supersymmetry breaking is realized: Type 0 models,
compactifications with Scherk-Schwarz deformations, the orbifold
$T^4/{\mathbb Z}_2$ with brane supersymmetry breaking, models with
internal magnetic fields. In the third chapter we discuss how one
can carry out our program in a number of field Theory toy models.
In particular we try to recover the right answer at the classical
level starting from a ``wrong'' vacuum.  The cases of cubic and
quartic potentials are simple, but also very interesting, and
provide us some general features related to tadpole resummations
and convergence domains around inflection points of the potential,
where the tadpole expansion breaks down. Moreover, some explicit
tadpole resummation are explicitly performed in a string inspired
model with tadpoles localized on D-branes or O-planes. The
inclusion of gravity should give further complications, but really
resummation works without any particular attention also in this
case. In the last chapter we begin to face the problem at the
string theory level. In particular, we analyze an example of
string model where the vacuum redefinition can be understood
explicitly not only at the level of the low-energy effective field
theory, but even at the full string theory level and where the
vacuum of a Type II orientifold with a compact dimension and local
tadpoles is given by a Type-0 orientifold with no compact
dimensions. These results are contained in a paper to appear in
Nuclear Physics B \cite{Dudas:2004nd}. Then we pass to compute
threshold corrections in a number of examples with supersymmetry
breaking, including models with brane supersymmetry breaking,
models with brane-antibrane pairs and the Type 0$^{\prime}$B
string, finding that the one-loop results are ultraviolet finite
and insensitive to the tadpoles. Such computations, that we
performed independently, will be contained in a future work
\cite{bdnps}.
\chapter{Superstring theory}
\section{Classical action and light-cone quantization}
\subsection{The action}
Let us take as our starting point the action for a free point particle of mass $m$ moving
in a $D$ dimensional space-time of metric $G_{\mu\nu}$\footnote{We use the convention of a
mostly positive definite metric.}. This action is well known to be
proportional to the length of the particle's world-line
\be
\label{A.a.1}
S \ = \ -m \ \int\sqrt{-G_{\mu\nu} \ \dot{X}^\mu \dot{X}^\nu} \ dt \ ,
\ee
and extremizing it with respect to the coordinates $X^\mu$ gives the geodesic equation.
This action has two drawbacks: it contains a square root and is valid only for
massive particles. One can solve these problems
introducing a Lagrange multiplier, a 1-bein $e(t)$ for the world-line, that has no dynamics,
and whose equation of motion is a constraint. The new action, classically equivalent to (\ref{A.a.1}), is
\be
\label{A.a.2}
S \ = \ \frac{1}{2} \ \int dt \ e  \ (e^{-2} \ \dot{X}^\mu\dot{X}_\mu \ - \ m^2) \ ,
\ee
and the mass-shell condition is provided by the constraint.

We now pass to describe the dynamics of an extended object, a
string whose coordinates are $X^\mu(\sigma,\tau)$, where $0\leq
\sigma\leq\pi$ runs over the length of the string and $\tau$ is the
proper time of the string that in its motion sweeps a world sheet.
Hence, the coordinates of the string map a two-dimensional variety with
metric $h_{\alpha\beta}(\sigma,\tau)$ into a $D$-dimensional
target space with metric $G_{\mu\nu}(X)$. In the following, we
will use $\xi^0$ for $\tau$ and $\xi^1$ for $\sigma$.

In analogy with the point particle case, we should write
an action proportional to the surface of the world-sheet
swept by the string (Nambu-Goto 1970) but, in order to have an action quadratic in the coordinates,
that from a two-dimensional point of view are fields with an internal symmetry index,
we introduce a Lagrangian multiplier, the metric of the world-sheet,
and  we write the classically equivalent action \cite{pol1}
\be
\label{A.a.3}
S \ = \ -\frac{T}{2} \ \int \ d\sigma d\tau \ \sqrt{-h} \ h^{\alpha\beta} \ \dd_\alpha X^{\mu}
\dd_\beta X^{\nu} \ \eta_{\mu\nu} \ ,
\ee
where $T=1/2\pi\alpha'$ is the tension of the string and $\alpha'=l_s^2$
is the squared string length. The signature of the world sheet
metric is $(-,+)$. Note that in (\ref{A.a.3}) we are just considering
a flat target metric, but one can generalize the construction to a curved
target space-time replacing $\eta_{\mu\nu}$ with $G_{\mu\nu}$.

The action (\ref{A.a.3}) is invariant under two-dimensional general coordinate transformation
(the coordinates $X^\mu(\xi)$ behave like  two-dimensional
scalars). Using such transformations it is always possible, at least
locally, to fix the metric to the form $h_{\alpha\beta}
= e^\phi\eta_{\alpha\beta}$, where $\eta_{\alpha\beta}$ is
the flat world sheet metric. This choice of gauge is known as the
conformal gauge. In two dimensions the conformal factor
$e^\phi$ then disappears from the classical action, that in the conformal gauge
reads
\be
\label{A.a.6}
S \ = \ -\frac{T}{2} \ \int \ d\sigma d\tau \ \eta^{\alpha\beta} \ \dd_\alpha X^{\mu}
\dd_\beta X^{\nu} \ \eta_{\mu\nu} \ ,
\ee
but not from the functional measure of the path integral, unless the
dimension of the target space is fixed to the critical one,
$D=26$ for the bosonic string \footnote{$D=26$ can be
recovered also by imposing that the squared BRST charge vanish). }.
The action is also invariant under Weyl rescaling
\ba
\label{A.a.4}
h_{\alpha\beta}(\xi) \ \rightarrow \ h'_{\alpha\beta}(\xi) \ &=& \
e^{\phi(\xi)} \ h_{\alpha\beta}(\xi) \nonumber\\
X^\mu(\xi)\rightarrow X'^\mu(\xi) \ &=& \ X^\mu(\xi) \ . \ea Gauge
fixing leaves still a residual infinite gauge symmetry that, after
a Wick rotation, is parameterized by analytic and anti-analytic
transformations. This is the conformal invariance of the
two-dimensional theory \cite{Ginsp,bpz}. In the light-cone
quantization, we will use such a residual symmetry to eliminate
the non physical longitudinal degrees of freedom.

The equation of motion for $X^\mu(\xi)$ in the conformal gauge is
simply the wave equation
\be
\label{A.a.7}
(\dd^2_\tau-\dd^2_\sigma) \ X^\mu(\tau,\sigma) \ = \ 0 \ ,
\ee
while the equation for $h_{\alpha\beta}(\xi)$ is a constraint corresponding to
the vanishing of the energy-momentum tensor
\be
\label{A.a.5}
T_{\alpha\beta}  = \ \dd_\alpha X^\mu \dd_\beta X_{\mu} \ - \ \frac{1}{2} \
h_{\alpha\beta} \ (h^{\gamma\delta} \ \dd_\gamma X^\mu \dd_\delta X_\mu) \ = \ 0 \ .
\ee
As a consequence of Weyl invariance, $T_{\alpha\beta}$ is traceless.

We can now generalize the bosonic action (\ref{A.a.6}) to
the supersymmetric case. To this end, let us introduce some fermionic coordinates
$\psi^\mu(\xi)$. This D-plet is a vector from the point of view of
the target Lorentz group and its components are Majorana spinors.
One can generalize (\ref{A.a.6}) simply adding to the kinetic term of D
two-dimensional free bosons the kinetic term of D two-dimensional free fermions,
\be
\label{A.a.8}
S \ = \ -\frac{T}{2} \ \int \ d\sigma d\tau \ \eta^{\alpha\beta} \
\left(
\dd_\alpha X^{\mu}
\dd_\beta X^{\nu} \ -i \ \bar\psi^\mu \gamma_\alpha \dd_\beta \psi^\nu
\right) \ \eta_{\mu\nu} \ .
\ee
The action (\ref{A.a.8}) has a global supersymmetry that is the
residual of a gauge fixing of the more general action \cite{pol2,bdhdz}
\ba
\label{A.a.9}
S & = & -\frac{T}{2}  \int d\sigma d\tau \ \sqrt{-h} \ h^{\alpha\beta}
\bigg[ \dd_\alpha X^{\mu} \dd_\beta X^{\nu} -i  \bar\psi^\mu
  \gamma_\alpha \dd_\beta \psi^\nu\nonumber\\
  &&- i \bar\chi_\alpha \gamma_\sigma\gamma_\beta\psi^\mu
  \left( \dd_\rho X^\nu -\frac{i}{4}\bar\chi_\rho \psi^\nu \right)
  h^{\sigma\rho}\bigg]\eta_{\mu\nu} \ ,
\ea
where $\chi_\alpha$ is a Majorana gravitino. Note that neither the
graviton nor the gravitino have a kinetic term. The reason is that the
kinetic term for the metric in two dimensions is a topological
invariant, so that it does not give dynamics, but is of crucial
importance in the loop Polyakov expansion and we will come back to
this point in the second section of this chapter. The kinetic term for
gravitino is the Rarita-Schwinger action and contains a totally
antisymmetric tensor with three indices,
$\gamma_{\alpha\beta\delta}$,
but in two dimensions such a tensor vanishes.
Our conventions for the two-dimensional $\gamma$-matrices are:
$\gamma^0=\sigma^2 \ , \ \gamma^1=i\sigma^1$,
where $\sigma^\alpha$ are the Pauli matrices. With this representation
of the Clifford algebra, the Majorana spinors $\psi^\mu$ are real.

The action (\ref{A.a.9}) has a local supersymmetry. Just as
the local reparameterization invariance of the theory can be used to fix the conformal
gauge for the metric, the local supersymmetry can be used to put the
gravitino in the form $\chi_\alpha=\gamma_\alpha\chi$, with
$\chi$ a Majorana fermion. In this particular gauge, the gravitino term
in (\ref{A.a.9}) becomes proportional to $\bar\chi \ \gamma_\alpha\gamma^\beta\gamma^\alpha
 \ \psi^\mu = (D-2) \ \bar\chi\gamma^\beta\psi^\mu$, that is zero in two
dimensions, and the action (\ref{A.a.9}) reduces to the form
(\ref{A.a.8}). The conformal factors for the metric and the field
$\chi$ disentangle from the functional measure of the path
integral only in the critical dimension $D=10$ \cite{pol2}. On the
other hand, one can recover the critical dimension also imposing
that $Q_{BRST}^2$ vanish. After gauge fixing, the theory is still
invariant under a residual infinite symmetry, the superconformal
symmetry \cite{bpz,fms}.

Like in the point particle case and in the bosonic string,
in order to describe the dynamics of the superstring, the action (\ref{A.a.8}) has
to be taken together with the constraints
\ba
\label{constraint1}
T_{\alpha\beta}  &=& \ \dd_\alpha X^\mu \dd_\beta X_{\mu} \
+ \ \frac{i}{4} \ \bar\psi^\mu \ (\gamma_\alpha\dd_\beta \ + \
\gamma_\beta\dd_\alpha) \ \psi_\mu \nonumber\\
&& - \ \frac{1}{2} \ \eta_{\alpha\beta} \ (\dd^\rho X^\mu
\dd_\rho X_\mu \ + \ \frac{1}{2} \ \bar\psi^\mu\gamma\cdot\dd \
\psi_\mu) \ = \ 0 \ ,
\ea
and
\be
\label{constraint2}
J^\alpha \ = \ \frac{1}{2} \ \gamma^\beta\gamma^\alpha \ \psi^\mu \
\dd_\beta X_\mu \ = \ 0 \ ,
\ee
where $J^\alpha$ is the supercurrent.

The equation of motion for the bosons, in the conformal gauge, is simply the wave equation
(\ref{A.a.7}). The surface term that comes from the variation of the
action vanishes both for periodic boundary conditions $X^\mu(\tau,\sigma=0)
= X^\mu(\tau,\sigma=\pi)$, that correspond to a closed string, with
the solution \cite{GSW}
\be
\label{modes_exp_bos}
X^\mu \ = \ x^\mu \ + \ 2\alpha'p^\mu\tau \ + \
i\frac{\sqrt{2\alpha'}}{2} \ \sum_{n\neq
  0}\left(\frac{\alpha^\mu_n}{n} \ e^{-2in(\tau-\sigma)}+
\frac{\tilde\alpha^\mu_n}{n} \ e^{-2in(\tau+\sigma)}\right) \ ,
\ee
and for Neumann\footnote{For an open string there is also the possibility to impose
Dirichlet boundary conditions, that means to fix the ends of the
string on hyperplanes. his possibility will be discussed further in the section on toroidal compactifications.} boundary conditions
$X'^\mu=0$ at $\sigma=0,\pi$, that
correspond to an open string, with the solution  \cite{GSW}
\be
X^\mu \ = \ x^\mu \ + \ 2\alpha'p^\mu\tau \ + \
i\sqrt{2\alpha'} \ \sum_{n\neq
  0}\frac{\alpha^\mu_n}{n} \ e^{-in\tau}\cos{n\sigma} \ .
\ee
The zero mode in the expansion describes the center of mass motion, while
$\tilde\alpha^\mu_n$, $\alpha^\mu_n$ are oscillators corresponding to
left and right moving modes.

For the fermionic coordinates, the Dirac equation
splits into
\be
(\dd_\tau-\dd_\sigma)\psi_2^\mu \ = \ 0 \ , \qquad
(\dd_\tau+\dd_\sigma)\psi_1^\mu \ = \ 0 \ ,
\ee
where $\psi_{1,2}^\mu$, the two components of
$\psi^\mu$, are Majorana-Weyl spinors. From their equations of motion
we see that $\psi_1$ depends only on $\tau-\sigma$, so that we prefer to
call it $\psi_-$, while $\psi_2$ depends only on $\tau+\sigma$, and we
call it $\psi_+$. For a closed string, the surface term vanishes both for periodic
(Ramond (R)) and for antiperiodic (Neveu-Schwarz (N-S)) boundary conditions separately
for each component. In the first case (R), the decomposition is
\ba
\psi_+ \ &=& \ \sum_{n \ \in \ \mathbb{Z}} \ \tilde{d}_n^\mu \
e^{-2in(\tau+\sigma)}\nonumber\\
\psi_- \ &=& \ \sum_{n \ \in \ \mathbb{Z}}
\ d_n^\mu \ e^{-2in(\tau-\sigma)} \ ,
\ea
while in the second one (N-S)
\ba
\psi_+ \ &=& \ \sum_{r \ \in \ \mathbb{Z}+\frac{1}{2}}
\ \tilde b_r^\mu  \ e^{-2ir(\tau+\sigma)}\nonumber\\
\psi_- \ &=& \ \sum_{r \ \in \ \mathbb{Z}+\frac{1}{2}}
\ b_r^\mu  \ e^{-2ir(\tau-\sigma)} \ ,
\ea
where we used the convention $ 2 \alpha'=1$.
Hence, for a closed string we have four sectors: R-R, R-NS, NS-R, NS-NS.
Note the presence of zero mode in the R sectors.
For the open strings, the boundary conditions
are $\psi_+=\psi_-$ at $\sigma=0,\pi$ in the Ramond sector, and
$\psi_+=\psi_-$ at $\sigma=0$, $\psi_+=-\psi_-$ at $\sigma=\pi$
in the Neveu-Schwarz sector, meaning that the left and right oscillators have to be identified.
The mode expansions differ from those one for closed strings in the frequency of oscillators,
that has to be halved, and in the overall factor, that now is $\frac{1}{\sqrt{2}}\ $
(for more details see \cite{GSW}).

It is very useful at this point to pass to the coordinates
$\xi^\pm = \tau\pm\sigma \ $. In such a coordinate system, the
metric becomes off-diagonal, $\eta_{+-}=\eta_{-+}=-\frac{1}{2} \ ,
\quad \eta_{\pm\pm}=0$, and the energy momentum tensor decomposes
in a holomorphic part $T_{++}$ depending only on $\xi^+$ and in an
anti-holomorphic part $T_{--}$ depending only on $\xi_- \ $ \be
\label{A.a.10} T_{\pm\pm} \ = \ \dd_\pm X^\mu\dd_\pm X_\mu \ + \
\frac{i}{2} \ \psi_\pm\dd_\pm\psi_\pm \ . \ee The energy-momentum
tensor is traceless, due to the conformal invariance, and
therefore $T_{+-}=T_{-+}=0 \ $. The supercurrent also decomposes
in a holomorphic and an anti-holomorphic part, according to \be
J_\pm(\xi^\pm) \ = \ \psi_\pm^\mu\dd_\pm X_\mu \ . \ee As a result
the two-dimensional superconformal theory we are dealing with
actually splits into two identical one-dimensional superconformal
theories \cite{Ginsp,bpz}.
\subsection{Light-cone quantization}
We now discuss the quantization of the string. Imposing
the usual commutation relation,
\ba
[\dot{X}^\mu(\sigma,\tau),X^\nu(\sigma',\tau)] \ &=& \ -i\pi\delta(\sigma-\sigma')\eta^{\mu\nu}
\ , \nonumber\\
\{\psi^\mu_\pm(\sigma,\tau),\psi^\nu_\pm(\sigma',\tau)\} \ &=& \
\pi\delta(\sigma-\sigma')\eta^{\mu\nu} \ , \ea for the holomorphic
oscillators one gets (and identically is for the anti-holomorphic
ones): \ba \label{algebra_oscillator}
[\alpha^\mu_m,\alpha^\nu_n] \ &=& \ m\delta_{m+n}\eta^{\mu\nu} \ , \nonumber\\
\{d_m^\mu,d_n^\nu\} \ &=& \ \eta^{\mu\nu}\delta_{m+n} \quad (R) \ , \nonumber\\
\{b_r^\mu,b_s^\nu\} \ &=& \ \eta^{\mu\nu}\delta_{r+s}  \quad (NS) \ ,
\ea
where $r,s$ are half integers.
After a suitable rescaling, the oscillators satisfy the usual
bosonic and fermionic harmonic oscillator algebra, where
the oscillators with negative frequency correspond to creation operators.
Note that because of the signature of the target-space metric, the resulting spectrum apparently
contains ghosts. The quantization has to be carried out
implementing the constraints
\be
T_{++} \ = \ T_{--} \ = \ J_+ \ = \ J_- \ = \ 0 \ ,
\ee
or equivalently in terms of their normal mode,
\be
\label{mode_constraints}
L_n \ = \ {\tilde L}_n \ = \ 0 \ , \qquad
F_n \ = \ {\tilde F}_n \ = \ 0 \quad (R) \ , \qquad
G_r \ = \ {\tilde G}_r \ = \ 0 \quad (NS) \ ,
\ee
where
\ba
L_n \ &=& \ \frac{T}{2} \ \int_0^\pi d\sigma \ T_{--} \ e^{-2in\sigma} \nonumber\\
F_n \ &=& \ T \ \int_0^\pi d\sigma J_-^{(R)} \ \ e^{-2in\sigma} \nonumber\\
G_r \ &=& \ T \ \int_0^\pi d\sigma J_-^{(NS)} \ \ e^{-2ir\sigma} \ ,
\ea
and similar expansions hold for the holomorphic part.
Imposing these constraints \textit{\`{a} la} Gupta-Bleuler, one can see that
the negative norm states disappear from the spectrum (no ghost theorem).

Actually, there is another way to quantize the theory that gives directly a spectrum free of ghosts.
We can use the superconformal symmetry to choice a particular gauge that
does not change the form of the world sheet metric and of the gravitino, the light-cone gauge, in which
the constraints are solved in terms of the transverse physical states. The disadvantage of this procedure
is that the Lorentz covariance of the theory is not manifest. However, it can be seen that the
Lorentz invariance holds if the dimension is the critical one $D=10$ ($D=26$ for the bosonic string),
and if the vacuum energy is fixed to a particular value,  as we shall see shortly.

We define $X^\pm = (X^0\pm X^{D-1})/\sqrt{2}$ and similarly for $\psi_A$, where $A=+,-$.
Then we fix to zero the oscillators in the $+$ direction
\ba
X^+ \ &=& \ x^+ +2\alpha'p^+\tau\nonumber\\
\psi_A^+ \ &=& \ 0 \ ,
\ea
and solve the constraints (\ref{mode_constraints}) for $\alpha^-_n$, $ \ b_r^-$, and $d_n^-$
\ba
\label{solve_constr}
\alpha^-_n \ &=& \frac{2}{\sqrt{2\alpha'} \ p^+} \ \Bigg[\sum_{i=1}^{D-2}
\ \frac{1}{2}\sum_{m \ \in \ \mathbb{Z}}:\alpha^i_{n-m}\alpha^i_m:
\nonumber\\
&&+ \ \sum_{i=1}^{D-2} \ \frac{1}{2}\sum_{r \ \in \ \mathbb{Z}+\frac{1}{2}}(r-\frac{n}{2}):b^i_{n-r}b^i_r: +a^{NS}
\ \delta_n\Bigg] \ , \nonumber\\
b_r^- \ &=& \ \frac{2}{\sqrt{2\alpha'} \ p^+} \ \sum_{i=1}^{D-2}
\sum_{s \ \in \ \mathbb{Z}+\frac{1}{2}} \ \alpha^i_{r-s}b^i_s \
\qquad (NS) \ \ , \ea where we are using the notation
$2\alpha^\pm_0 = \sqrt{2\alpha'} \ p^\pm$ and $2\alpha^i_0 =
\sqrt{2\alpha'} \ p^i$. The corresponding solution for the R
sector is obtained with $b^i_r\rightarrow d^i_k, \;\; k \ \in \
\mathbb{Z}$ and $a^{NS}\rightarrow a^{R}$.

The constants $a^{NS}$ and $a^{R}$ have the meaning of vacuum
energy in the respective sectors and come from the normal
ordering, after a suitable regularization. For example for a
single bosonic oscillator the infinite quantity to regularize is
$\frac{1}{2}\sum_{k=1} k$. We regularize it computing the limit
for $x\rightarrow 0^+$ of $\frac{1}{2}\sum_{k=1} k \ e^{-kx}$ and
taking only the finite part. The last sum defines the Riemann
$\zeta(x)$ function. Making the limit of the more general function
\be \zeta_\alpha(-1,x) \ = \ \sum_{n=1}^\infty(n+\alpha) \
e^{-(n+\alpha)x} \ , \ee that for $x\rightarrow 0^+$ goes to
$\zeta_\alpha(-1,0^+)=-\frac{1}{12}[6\alpha(\alpha-1)+1]$ plus a
divergent term, it is possible to fix also the zero point energy
for a fermionic oscillator both in the R and NS sector \cite{bn}.
The result is that every boson contributes to the zero-point
energy with $a=-\frac{1}{24}$, every periodic fermion with
$a=+\frac{1}{24}$ and every antiperiodic fermion with
$a=-\frac{1}{48}$. Therefore, for the bosonic string, where
$D=26$, the shift of the vacuum energy is $a=-1$, as we have $24$
physical bosonic degree of freedom. On the other hand, for the
superstring, where we have $8$ physical bosonic oscillators and
$8$ physical fermionic oscillators, the shift due to the normal
ordering is $a^{R}=0$ and $a^{NS}=-\frac{1}{2}$. The important
thing to notice here is that this regularization is compatible
with the closure of the Lorentz algebra in the light-cone gauge in
the critical dimension.

The transverse Virasoro operators defined through the relation
\be
L_n^\bot \ = \ \sum_{i=1}^{D-2}
\ \left(\frac{1}{2}\sum_{m \ \in \ \mathbb{Z}}:\alpha^i_{n-m}\alpha^i_m:
\ + \ \frac{1}{2}\sum_{r \ \in \ \mathbb{Z}+\frac{1}{2}}(r-\frac{n}{2}):b^i_{n-r}b^i_r: \right) \ , (NS)
\ee
and their analogs in the Ramond sector satisfy the transverse Virasoro algebra
\be
\label{Virasoro}
[L_m^\bot \ , \ L_n^\bot] \ = \ (m-n) \ L_{m+n}^\bot \ + \ A(m)\delta_{m+n} \ ,
\ee
where the central term is $A(m)=\frac{1}{8} (D-2) \ m(m^2-1)$ in the NS sector
and $A(m)=\frac{1}{8}(D-2) \ m^3$ in the R sector.
The number $\frac{1}{8}(D-2)$ is equal to $\frac{1}{12} \ c_{tot} \ $ where $ \ c_{tot}=(D-2)(1+\frac{1}{2}) \ $
is the total central charge of the left (or right) theory \footnote{The central charge is $c=1$ for a boson,
while for a fermion is $c=\frac{1}{2}$.}.
The Fourier modes of the supercurrent together with the Virasoro operators build together the
superconformal (super-Virasoro) algebra \cite{GSW, Kiritsis, Polchinski}.

The relation for $\alpha^-_0$ is of particular importance, because
it gives the mass-shell condition. Recalling that $2\alpha^-_0 =
\sqrt{2\alpha'} \ p^-$ and $2\alpha^i_0 = \sqrt{2\alpha'} \ p^i$
and introducing the number operators \be \
N_B=\sum_{i=1}^{D-2}\sum_{n=1}^\infty\alpha_{-n}^i\alpha_n^i \ ,
\quad \ N_F^{(NS)}=\sum_{i=1}^{D-2}\sum_{r=\frac{1}{2}}^\infty r \
b_{-r}^i b_r^i \ , \quad \
N_F^{(R)}=\sum_{i=1}^{D-2}\sum_{n=1}^\infty n  \ d_{-n}^i d_n^i \
, \ee gives \be \label{mass-shell} M^2 \ = \ \frac{4}{\alpha'} \
\left[N_B+N_F+a \right] \ , \ee where $N_F$ and $a$ are either for
the NS sector or for the R one, and an analog condition  follows
from the anti-holomorphic part. Putting together left and right
sectors gives the mass-shell condition \be \label{mass-shell2} M^2
\ = \ \frac{2}{\alpha'} \ \left[N_B + \bar N_B + N_F + \bar N_F +
a + \bar a \right] \ , \ee together with the level matching
condition for the physical states \be \label{level-matching}
N_B+N_F+a \ = \ \bar N_B+ \bar N_F+ \bar a \ . \ee The mass
formula for the bosonic string is simply obtained removing $N_F$
and $\bar N_F$ and fixing $a=\bar a =-1$.

We can now describe the spectrum, and in particular the low-energy
states of the string. In the NS sector the ground state $\mid
0\rangle_{NS}$ is a scalar tachyon due to the negative shift
$a^{NS}$. The first excited state is a massless vector
$b_{-\frac{1}{2}}^i\mid 0\rangle_{NS}$. Here there is a
peculiarity due to the fact that an anticommuting operator acts on
a boson and gives a boson. The tachyon is eliminated projecting
the spectrum with the Gliozzi-Scherk-Olive (GSO) projector
$P_{GSO}^{(NS)}=\frac{1-(-)^F}{2}$ \cite{gso}, where $F$ counts
the number of fermionic operators, so that after the projection
the ground state is $b_{-\frac{1}{2}}^i\mid 0\rangle_{NS}$.
Moreover, the higher states that remain are only the ones obtained
acting with an even number of fermionic operators on the new
ground state. This prescription also removes the difficulty we
mentioned. On the other hand, in the R sector the states have to
be fermions. In fact, the operators $d_0^i$ satisfy the algebra
$\{d_0^i,d_0^j\}=\delta^{ij}$, that after a rescaling is the
Clifford algebra, and commute with $N_F^{(NS)}$. Therefore, the
mass eigenstates have to be representations of the Clifford
algebra, and in particular the ground state is a massless Majorana
fermion. We project also this sector with $ \
P^{(R)}_{GSO}=\frac{1+(-1)^F\Gamma_9}{2}$, where $\ \Gamma_9$ is
the chirality matrix in the transverse space. The ground state in
the resulting spectrum is a Majorana-Weyl fermion with positive
chirality (the chirality of the ground state is a matter of
convention) and the excited states have alternatively negative and
positive chirality.

The (GSO) truncation gives at low energy the spectrum of
${\mathcal N}=2$ supergravity in $D=10$. If the left and right
ground states in the R sector, $\mid 0 \rangle_{R}$ and $\mid \bar
0 \rangle_{R}$, have the same chirality, then the supergravity is
of Type $IIB$, otherwise, if the chiralities are the opposite, the
supergravity is of Type $IIA$. The massless states, after
decomposing the direct product of left and right sectors in
representations of $SO(8)$, are in the NS-NS sector a symmetric
traceless tensor $G_{ij}$ identified with the graviton, an
antisymmetric $2$-tensor $B_{ij}$, and a scalar $\phi$ called the
dilaton. In the R-R sector they are a scalar, a $2$-form and a
$4$-form with self-dual (antiself-dual) field-strength for Type
$IIB$ or a vector and a $3$-form for Type $IIA$. The mixed sectors
contain two gravitinos and two spinors (called dilatinos). In Type
$IIA$  the two gravitinos are of opposite chirality as the two
dilatinos, while in Type $IIB$ the two gravitinos have the same
chirality, that is opposite to the chirality of the two dilatinos.
It is important to note that although the Type $IIB$ spectrum is
chiral, it is free of gravitational anomalies \cite{alvgaum}.

For an open string there are only left or right oscillators, and
the mass-shell condition is given  by \be M^2 \ = \
\frac{1}{\alpha'} \ \left[N_B+N_F+a \right] \ . \ee The difference
of the Regge slope $\frac{1}{\alpha'}$ with respect to the closed
string $\frac{4}{\alpha'}$ is understood recalling that
$p^{\mu}_{\it open}$ is only half of the total momentum of a
closed string $p^{\mu}_{\it closed}$. Thus, the result for the
open mass formula can be recovered simply substituting
$\frac{1}{2}p^\mu_{closed}$ with $p^\mu_{open}$, or
$\frac{1}{4}M^2_{closed}$ with $M^2_{open}$. The low-energy GSO
projected spectrum has a massless vector in the NS sector and a
Majorana-Weyl fermion in the R sector, that together give the
super Yang-Mills multiplet in $D=10$. After this discussion, the
spectrum for the bosonic string can be extracted without any
difficulty. At the massless level it contains the graviton, the
2-form and the dilaton, in the closed sector, and the vector in
the open sector. The bosonic string contains also open and closed
tachyons, due to the shift of the vacuum energy \cite{tachyon}.
\section{One-loop vacuum amplitudes}
In this section we introduce the Riemann surfaces corresponding to the world-sheets swept by strings at one-loop,
following \cite{open_string}.
As we will see in the next sections, where we will face the orientifold construction, in String Theory it is possible
to construct consistent models with unoriented closed and open strings, starting from a theory of only oriented closed strings.
The important thing to stress is that while for the oriented closed string there is only a contribution for each
order of perturbation theory, corresponding to a closed orientable Riemann surface with a certain number of handles $h$,
for the unoriented closed and open strings there are more amplitudes that contribute to the same order.
For example, at one loop an oriented closed string sweeps a torus that is a closed orientable Riemann
surface with one handle. The next order is given by a double torus, with two handles, and successive orders of
perturbation theory correspond to increasing numbers of handles.
\begin{figure}
\begin{center}
\epsfbox{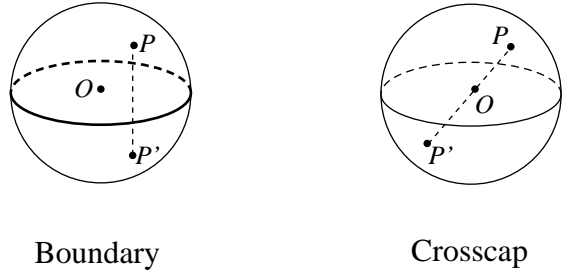}
\end{center}
\caption{Boundary and crosscap.}
\label{fig1}
\end{figure}
In order to elucidate what happens with unoriented closed and open strings, we have to introduce two new important
objects: the boundary $b$ and the crosscap $c$. The meaning of a boundary is understood taking a sphere and identifying
points like in figure \ref{fig1}. The resulting surface is the disk. The line of fixed points of this involution (the equator of the sphere)
is the boundary of the disk. Also the crosscap is better understood considering the simplest surface in which it appears:
the real projective plane. This is obtained identifying antipodal points in a sphere. The crosscap is any equator
of the sphere with the opposite points identified. Let us note that the presence of a crosscap causes the
loss of orientability of a surface, due to the antipodal identification.

The perturbation expansion in String Theory is weighted by $g_s^{-\chi}$, where $g_s=e^{\langle\varphi\rangle}$ is the
string coupling constant, determined by the vacuum expectation value of the dilaton $\langle\varphi\rangle$,
and $\chi$ is the Euler character of the surface corresponding to a certain string amplitude.
A surface with $b$ boundaries, $c$ crosscaps and $h$ handles has
\be
\chi \ = \ 2-2h-b-c \ .
\ee
Of particular importance are the surfaces of genus $g=h+\frac{b}{2}+\frac{c}{2}=1$, or equivalently $\chi=0$,
that describe the one loop vacuum amplitudes. From their partition functions in fact it is possible
to read the free spectrum of the string and to extract consistency conditions that make the theory finite
and free of anomalies. The Riemann surfaces of genus $g=1$ are the torus ($h=1,b=c=0$), the Klein-bottle
($h=0,b=0,c=2$), the annulus ($h=0,b=2,c=0$) and the M\"{o}bius strip ($h=0,b=c=1$).

\begin{figure}
\begin{center}
\epsfbox{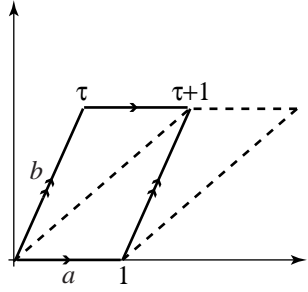}
\end{center}
\caption{The fundamental cell for the torus lattice.}
\label{fig2}
\end{figure}

The torus represents an oriented closed string that propagates in
a loop. With two cuts the torus can be mapped into a parallelogram
whose opposite sides are identified. Rescaling the horizontal side
to length one, we get the fundamental cell for the torus (see figure \ref{fig2}) that
is characterized by a single complex number $\tau=\tau_1+i\tau_2,
\ \tau_2>0$, the ratio between the oblique side and the horizontal
one, known as the Teichm\"{u}ller parameter or modulus.
Actually this cell defines a lattice, and we can choose as fundamental cell
also the one with the oblique side translated by one horizontal
length or the one with the horizontal and the oblique sides
exchanged. These two operations are given respectively by
\be
T \ : \ \tau \rightarrow \tau+1 \ , \qquad \qquad S \ : \ \tau
\rightarrow -\frac{1}{\tau} \ .
\ee
The transformations $T$ and $S$ generate the modular group
$PSL(2,\mathbb{Z})=SL(2,\mathbb{Z})/\mathbb{Z}_{2}$ whose
action on $\tau$ is given by
\be
\tau\rightarrow\frac{a\tau+b}{c\tau+d} \ ,
\ee
with $ad-bc=1$ and $a,b,c,d \in \mathbb{Z}$. All
the cells obtained acting on $\tau$ with the modular group define
equivalent tori. As a result, the values of $\tau$ in the upper half plane that
define inequivalent tori can be chosen for example to belong to the region
$\mathcal{F}=\{-\frac{1}{2}<\tau_1\leq\frac{1}{2} , \
\mid\tau\mid \ \geq 1\}$ (see figure \ref{fig3}).This can be foreseen with a $T$ transformation we can
map all the values of $\tau$ in the strip
$\{-\frac{1}{2}<\tau_1\leq\frac{1}{2}\}$ and with an $S$-modular
transformation we can map $\{\tau: \ \mid\tau\mid \ \leq1\}$ to
$\{\tau: \ \mid\tau\mid \ \geq1\}$.
\begin{figure}
\begin{center}
\epsfbox{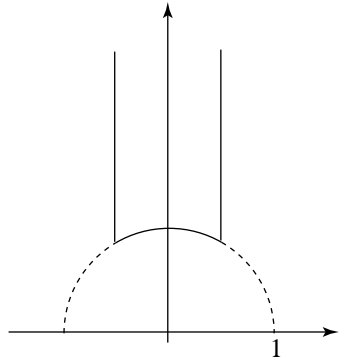}
\end{center}
\caption{Fundamental domain for the torus.}
\label{fig3}
\end{figure}
The Teichm\"{u}ller parameter $\tau$ has the physical meaning of the proper time elapsed
while a closed string sweeps the torus, and the modular invariance of the
torus means that we have an infinity of equivalent choices for it.
Let us note that modular invariance is a peculiar characteristic
only of the torus, and is of fundamental importance since it introduces a natural ultraviolet
cut-off on $\tau$.
For the other surfaces of genus $g=1$ there is no symmetry that protects
from divergences, but all ultraviolet divergences can be related to infrared ones.
\begin{figure}
\begin{center}
\includegraphics[width=4cm,height=6cm]{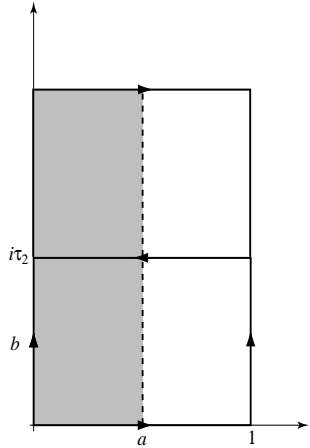}
\end{center}
\caption{Fundamental polygons for the Klein-bottle.}
\label{fig4}
\end{figure}

The Klein-bottle projects the states that propagate in the torus
to give the propagation of an unoriented closed string at one
loop. The modulus of the Klein-bottle is purely imaginary,
$\tau=i\tau_2$, and the fundamental polygon for it, obtained cutting
its surface, is a rectangle with the horizontal side rescaled to
one, the vertical side equal to $i\tau_2$, and the opposite sides
identified after a change of the relative orientation for the
horizontal ones (see figure \ref{fig4}). A vertical doubling of the fundamental polygon of
the Klein-bottle gives the doubly-covering torus with
Teichm\"{u}ller parameter equal to $it_2=i2\tau_2$. The modulus
$\tau_2$ is interpreted as the proper time that a closed string
needs to sweep the Klein-bottle. But there is another choice for
the fundamental polygon, obtained halving the horizontal side and
doubling the vertical one, so that the area remains unchanged. The
vertical sides of the new polygon are actually two crosscaps and
the horizontal ones are identified now with the same orientation.
This polygon corresponds to a tube ending at two crosscaps, so
that the Klein-bottle can also be interpreted as describing a
closed string that propagates between two crosscaps in a proper
time represented by the horizontal side of the second
polygon.

\begin{figure}
\begin{center}
\includegraphics[width=5cm,height=5cm]{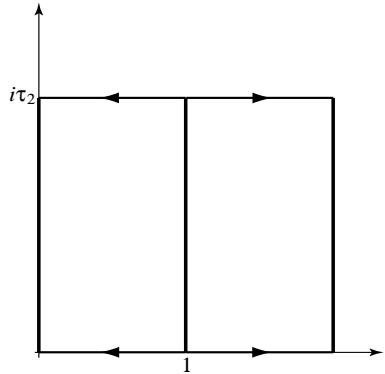}
\end{center}
\caption{Fundamental polygon for the annulus.}
\label{fig5}
\end{figure}
The propagation at one loop of an oriented open string is
described by the annulus. After a cut it is mapped into a
rectangle with the horizontal sides identified (see figure \ref{fig5}). The vertical ones
are the two boundaries of the annulus. The modulus of the annulus
is once more purely imaginary, $\tau=i\tau_2$, and represents the
proper time elapsed while an open string sweeps it. The
doubly-covering torus is obtained doubling the horizontal sides
and its Teichm\"{u}ller parameter is $it_2=i\frac{\tau_2}{2}$. But
there is another equivalent representation of the annulus as a
tube that ends in two boundaries. In this new picture, we can see a
closed string that propagates between the two boundaries, and the
modulus of the tube is just the proper time elapsed in this
propagation.
\begin{figure}
\begin{center}
\includegraphics[width=5cm,height=6cm]{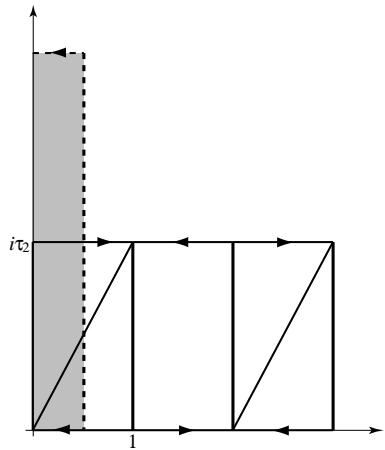}
\end{center}
\caption{Fundamental polygons for the M\"{o}bius strip.}
\label{fig6}
\end{figure}

Finally, at genus $g=1$ we have also the M\"{o}bius strip, that
projects the annulus amplitude to describe the one loop
propagation of an unoriented open string. The  fundamental polygon
is a rectangle where the horizontal sides are identified with the
opposite orientation (see figure \ref{fig6}). The two vertical sides together form the
boundary of the M\"{o}bius strip. The modulus is $i\tau_2$ and has
the meaning of the time elapsed while the open string sweeps the M\"{o}bius.
The doubly covering torus in this case is not obtained simply
horizontally or vertically doubling the polygon but horizontally doubling two times
it (see figure \ref{fig6}). The result is that the Teichm\"{u}ller parameter of the
torus now is not purely imaginary but has a real part: $t=\frac{1}{2}+i\frac{\tau_2}{2}$.
Also in the case of the M\"{o}bius strip it is possible to give an equivalent representation
of the surface, halving the horizontal side while doubling the vertical one.
The vertical sides are now a boundary and a crosscap and
the horizontal side is the proper time elapsed while a closed string propagates
from the boundary to the crosscap trough a tube.
\begin{figure}
\begin{center}
\includegraphics[width=10cm,height=9cm]{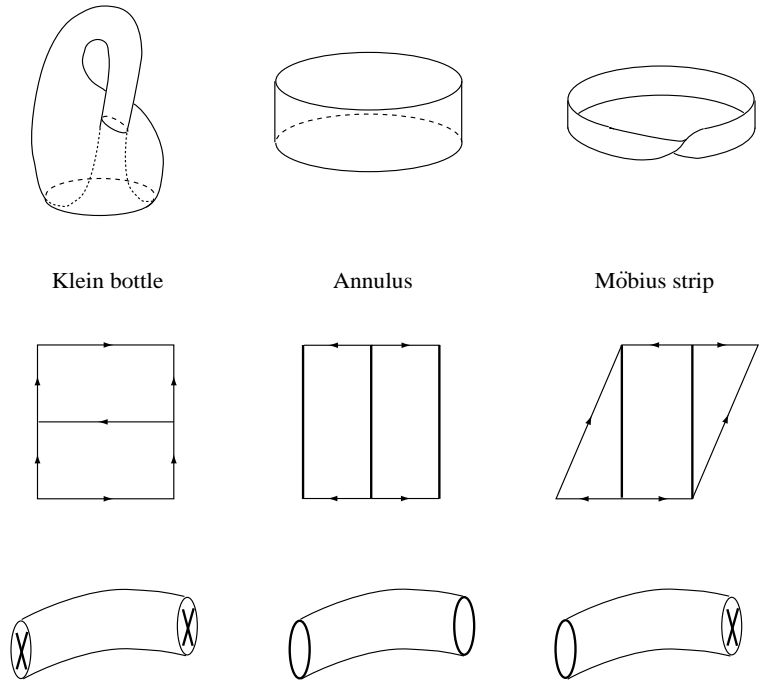}
\end{center}
\caption{Klein-bottle, annulus and M\"{o}bius strip.}
\label{figsurfs}
\end{figure}

In the following, we will refer to the amplitudes corresponding
to the first fundamental polygon as the amplitudes in the direct
channel, while the other choice describes the amplitudes in the
transverse channel. The direct and transverse channels are related through
an $S$-modular transformation that maps the ``vertical time'' of the direct channel
in to the ``horizontal time'' of the transverse one. A subtlety
for the M\"{o}bius strip is due to the fact that the modulus of the
doubly covering torus is not purely imaginary and we will come back
to this point in the following.
\\
\\
\\
\section{The torus partition function}
After having introduced the Riemann surfaces with vanishing Euler characteristic,
corresponding in String Theory to the one-loop vacuum amplitudes, we can begin to write their
partition functions. We start at first from the simplest case of field theory,
deriving the one loop vacuum energy for a massive scalar field in $D$ dimensions
\be
S^{(E)} \ = \ \int \ d^Dx \ \frac{1}{2} \ \left(\dd_\mu\phi \ \dd^\mu\phi+M^2\phi^2 \right) \ .
\ee
The one loop vacuum energy $\Gamma$ is expressed trough the relation
\be
e^{-\Gamma} \ = \ \int [\mathcal{D}\phi] \ e^{-S^{(E)}}\sim \left[\mathrm{det}(-\Box+M^2)
\right]^{-\frac{1}{2}} \ ,
\ee
that means
\be
\Gamma \ = \ \frac{1}{2} \mathrm{tr}\left[\ln(-\Box+M^2)\right] \ .
\ee
Using the formula for the trace of the logarithm of a matrix $A$
\be
\mathrm{tr}(\ln{A}) \ = \ -\int_{\epsilon}^{\infty}\frac{dt}{t} \
\mathrm{tr}\left(e^{-tA}\right) \ ,
\ee
where $\epsilon$ is an ultraviolet cut-off, and inserting a complete set of eigenstates
of the kinetic operator, we get:
\be
\label{imp}
\Gamma \ = \ -\frac{V}{2}\int_{\epsilon}^{\infty} \frac{dt}{t} \
\int\frac{d^Dp}{(2\pi)^D} \ e^{-tp^2} \ \mathrm{tr}\left(e^{-tM^2}\right) \ ,
\ee
where $V$ is the space-time volume. The integral on $p$ is gaussian and can be computed. The result
is that the one loop vacuum energy for a bosonic degree of freedom is
\be
\label{Gamma_function}
\Gamma \ = \ -\frac{V}{2(4\pi)^{\frac{D}{2}}} \
\int_{\epsilon}^{\infty} \frac{dt}{t^{\frac{D}{2}+1}} \ \mathrm{tr}\left(e^{-tM^2}\right) \ .
\ee
For a Dirac fermion there is only a change of sign due to the anticommuting nature
of the integration variables, and we have also to multiply for the number of degrees
of freedom of a Dirac fermion that in $D$ dimension is $2^{\frac{D}{2}}$.
The end result, for a theory with bosons and fermions is
\be
\label{Gamma}
\Gamma \ = \ -\frac{V}{2(4\pi)^{\frac{D}{2}}} \
\int_{\epsilon}^{\infty} \frac{dt}{t^{\frac{D}{2}+1}} \ \mathrm{Str}\left(e^{-tM^2}\right) \ ,
\ee
where the supertrace $\mathrm{Str} \ $ is
\be
\mathrm{Str} \ = \ \sum_{bosons} \ - \ 2^{\frac{D}{2}}\sum_{fermions} \ .
\ee
We now use the expression (\ref{Gamma}) in the case of superstring theory,
recalling the mass formula (\ref{mass-shell2}), that here we report in terms
of $L_0^\bot=N_B+N_F$ and $\bar L_0^\bot=\bar N_B+\bar N_F$
\be
M^2 \ = \ \frac{2}{\alpha'} \ \left[L_0^\bot+\bar L_0^\bot + a + \bar a \right] \ .
\ee
In order to take properly into account the level-matching condition for
the physical states, we have to introduce a delta-function in the integral (\ref{Gamma}).
Setting the dimension to the critical value $D=10$ gives
\be
\Gamma \ = \ -\frac{V}{2(4\pi)^5} \ \int_{-\frac{1}{2}}^{\frac{1}{2}}d s
\int_{\epsilon}^{\infty} \frac{dt}{t^6} \ \mathrm{Str}
\left(e^{ -\frac{2}{\alpha'} \left[L_0^\bot \ + \ \bar L_0^\bot \ + \ a  \
+ \ \bar a \right]t} \ e^{ 2\pi i  \left[L_0^\bot \ + \ a \ - \ \bar L_0^\bot
 \ - \ \bar a \right]s} \right) \ ,
\ee
that, defining  $\tau=s+i\frac{t}{\pi\alpha'} \ $ and
$q \ = \ e^{2\pi i \tau} \ $, $ \ \bar q \ = \ e^{-2\pi i \bar\tau} \ $, becomes
\be
\Gamma \ = \ -\frac{V}{2(4\alpha'\pi^2)^5} \ \int_{-\frac{1}{2}}^{\frac{1}{2}}d \tau_1
\int_{\epsilon}^{\infty} \frac{d\tau_2}{\tau_2^6} \ \mathrm{Str}
\left(q^{ \ L_0^\bot+ a} \ \bar q^{ \ \bar L_0^\bot + \bar a} \right) \ .
\ee
This is the partition function for a closed string that propagates in a loop,
so that it is the torus amplitude with $\tau$ its Teichm\"{u}ller parameter.
Recalling that all inequivalent tori correspond to values of $\tau$ in the
fundamental region $\mathcal{F}$, and apart from an overall normalization constant,
we can write the torus amplitude in the form
\be
\label{torus}
\mathcal{T} \ = \ \int_{\mathcal{F}}\frac{d^2\tau}{\tau_2^2} \
\frac{1}{\tau_2^4} \ \mathrm{Str}\left( q^{ \ L_0^\bot+ a} \
\bar q^{ \ \bar L_0^\bot + \bar a}\right) \ .
\ee
Let us remark again that the modular invariance of $\mathcal{T}$ (that we will
check in a while) allows one to exclude from the integration region the
ultraviolet point $\tau=0$, where the integrand would diverge.

At this point we have to compute the traces. In the NS sector
$a=-\frac{1}{2}$ and the trace is
\be
\mathrm{tr} \ q^{ \ N_B+N_F-\frac{1}{2}} \ = \
\frac{1}{\sqrt{q}} \
\mathrm{tr}\left(q^{ \ \sum_{n=1}^\infty \alpha_{-n}^i\alpha_n^i}\right)
\ \mathrm{tr}\left(q^{ \ \sum_{r=\frac{1}{2}}^\infty r \ b_{-r}^i b_r^i}\right) \ ,
\ee
where a sum over $i=1\ldots 8$ is understood. The bosonic trace is like
the partition function for a Bose gas. Using the algebra
(\ref{algebra_oscillator}), a state with $k$ oscillators of frequency
$n$ gives a contribution $q^{nk}$, so that we have to compute
\be
\prod_{i=1}^8\prod_{n=1}^\infty \sum_k q^{nk}
\ee
and the result is
\be
\frac{1}{\prod_{n=1}(1-q^n)^8} \ .
\ee
The fermionic trace is instead like the
partition function of a Fermi-Dirac gas.
By the Pauli exclusion principle, any oscillator can
have occupation number only equal to $0$ or $1$, and thus the fermionic
trace is simply
\be
\prod_{i=1}^8\prod_{r=1/2}(1+q^r) \ .
\ee
In the R sector $a=0$, and we have to multiply
by $2^{\frac{D-2}{2}}=16$ to take into account the degeneracy of the
Ramond vacuum. Putting the bosonic and the fermionic contributions together
gives
\ba
\mathrm{tr}  \ q^{ \ N_B+N_F-\frac{1}{2}} \ &=& \
\frac{\prod_{r=1/2}(1+q^r)^8}
{\sqrt{q} \ \prod_{n=1}(1-q^n)^8} \quad (NS), \nonumber\\
\mathrm{tr}  \ q^{ \ N_B+N_F} \ &=& \
16 \ \frac{\prod_{n=1}(1+q^n)^8}
{\prod_{n=1}(1-q^n)^8} \quad (R) \ .
\ea
The previous quantities can be expressed in terms of the Jacobi
$\vartheta$-functions of argument $z=0$
\be
\vartheta \left[{\textstyle {\alpha \atop \beta}} \right] (z|\tau) = \sum_n \
q^{\frac{1}{2} (n+ \alpha)^2} \ e^{2 \pi i (n + \alpha)(z+\beta)} \ ,
\ee
where $ \ \alpha, \beta = 0, \ \frac{1}{2} \ $. Equivalently
the Jacobi $\vartheta$-functions can be defined by the infinite products
\ba
\label{theta}
\vartheta \left[ {\textstyle {\alpha \atop \beta}} \right] (z|\tau) &=&
e^{2 i \pi \alpha (z+\beta)} \ q^{\alpha^2/2} \prod_{n=1}^\infty \
( 1 - q^n)\prod_{n=1}^\infty \ (1 + q^{n + \alpha - 1/2} e^{2 i \pi (z+\beta)} )\nonumber\\
&& \times \prod_{n=1}^\infty (1 + q^{n - \alpha - 1/2} e^{-2 i \pi (z+\beta)} ) \ .
\ea
Moreover, we have to introduce the Dedekind $\eta$-function
\be
\label{eta}
\eta(\tau) \ = \ q^{\frac{1}{24}} \prod_{n=1}^\infty \ (1-q^n) \ .
\ee
Using the definition (\ref{eta}) and (\ref{theta}), it is then possible to
write the following quantities:
\ba
& &\frac{\vartheta^4 \left[ {\small 1/2} \atop {\small 1/2} \right] (0|\tau)}
{\eta^{12}(\tau)} = \frac{\vartheta_1^4(0|\tau)}{\eta^{12}(\tau)} = 0 \, , \\
& &\frac{\vartheta^4 \left[ {\small 1/2} \atop {\small 0} \right] (0|\tau)}
{\eta^{12}(\tau)} = \frac{\vartheta_2^4(0|\tau)}{\eta^{12}(\tau)} =
16 \frac{\prod_{n=1}^{\infty} (1 + q^{n})^8}
{\prod_{n=1}^{\infty} (1 - q^n)^8} \, , \\
& &\frac{\vartheta^4 \left[ {\small 0} \atop {\small 0} \right] (0|\tau)}
{\eta^{12}(\tau)} = \frac{\vartheta_3^4(0|\tau)}{\eta^{12}(\tau)} =
\frac{\prod_{n=1}^{\infty} (1 + q^{n-1/2})^8}{
q^{1/2} \prod_{n=1}^{\infty} (1 - q^n)^8} \, , \\
& &\frac{\vartheta^4 \left[ {\small 0} \atop {\small 1/2} \right] (0|\tau)}
{\eta^{12}(\tau)} = \frac{\vartheta_4^4(0|\tau)}{\eta^{12}(\tau)} =
\frac{\prod_{n=1}^{\infty} (1 - q^{n-1/2})^8}{
q^{1/2} \prod_{n=1}^{\infty} (1 - q^n)^8} \ , \label{vacuumd=10}
\ea
that are directly related to the ones that appear in the string
amplitudes after the GSO projection
\ba
\mathrm{tr} \ \left[q^{ \ N_B+N_F-\frac{1}{2}} \
    \frac{1-(-1)^F}{2}\right] \ &=& \ \frac{1}{\eta^8} \
\frac{\vartheta_3^4(0|\tau)
-\vartheta_4^4(0|\tau)}{2\eta^4(\tau) } \quad (NS) \nonumber\\
\mathrm{tr} \ \left[q^{ \ N_B+N_F} \
    \frac{1\pm \Gamma_9(-1)^F}{2}\right] \ &=& \ \frac{1}{\eta^8} \
\frac{\vartheta_2^4(0|\tau)
\pm\vartheta_1^4(0|\tau)}{2\eta^4(\tau) } \quad (R) \ ,
\ea
where the sign in the Ramond sector selects the chirality of the
vacuum. The four $\vartheta$-functions are related to the four spin
structures of the fermionic determinant on the torus, and $\vartheta_1$
vanishes because it is the contribution of the periodic-periodic structure
that is the only one which containing a zero-mode.
In order to have a modular invariant partition function, we need
all the four GSO-projected sectors NS-NS, NS-R, R-NS, R-R. Let us note
again the importance of the projection, that is instrumental to
reconstruct the modular invariant.

At this point, it is useful to introduce the
characters of the affine extension $(k=1)$ of the algebra $so(8)$,
decomposing into two orthogonal subspaces both the NS and the R sectors
\ba
\label{characters}
O_{8} \ &=& \ \frac{\vartheta_3^4+\vartheta_4^4}{2\eta^4} \qquad\qquad
V_{8} \ = \ \frac{\vartheta_3^4-\vartheta_4^4}{2\eta^4}\qquad\qquad (NS)\nonumber\\
S_{8} \ &=& \ \frac{\vartheta_2^4+\vartheta_1^4}{2\eta^4}\qquad\qquad
C_{8} \ = \ \frac{\vartheta_2^4-\vartheta_1^4}{2\eta^4} \qquad\qquad
(R) \ .
\ea
Each of these characters in the language of CFT is a trace over
the corresponding Verma module (or conformal family) that consists in an infinite tower
of descendants of increasing mass and spin whose state with the lowest
conformal weight $L_0^\bot=h$ is called the primary field. The general
form of a character is
\be
\label{form}
\chi(q) \ = \ q^{h-c/24}\sum_k d_k q^k ,
\ee
where $c$ is the central charge, that is $c=12$ if we take into account
also the bosonic degrees of freedom dividing the characters of $so(8)$
by $\eta^8$, $h$ is the conformal weight of the primary field, $h+k$
are the conformal weights of the descendants and the $d_k$ are integers.
For example $O_8/\eta^8$ starts with a scalar tachyon of conformal
weight $ \ h=0 \ $
and squared mass proportional to $-\frac{1}{2}$.
$ \ V_8/\eta^8, \ S_8/\eta^8, \ $ and $ \ C_8/\eta^8 \ $
start respectively with a massless vector, a massless
left Majorana-Weyl spinor and a massless right Majorana-Weyl
spinor, that have the same conformal weight $ \ h = \frac{1}{2} \ $.

The partition function for the torus can now be written rather
simply. Then,
if the relative chirality of the left and right R sectors is the same, we get the torus partition function for
the Type IIB string,
\be
\label{IIB}
\mathcal{T}_{IIB} \ = \ \int \frac{d^2\tau}{\tau_2^2} \
\frac{1}{\tau_2^4 \ (\eta \ \bar\eta)^8} \ |V_8-S_8|^2 \ ,
\ee
otherwise we get the partition function for the Type IIA string,
\be
\label{IIA}
\mathcal{T}_{IIA} \ = \ \int \frac{d^2\tau}{\tau_2^2} \
\frac{1}{\tau_2^4 \ (\eta \ \bar\eta)^8} \ (\bar V_8-\bar S_8) \ ( V_8-
C_8) \ .
\ee
Let us note the minus sign in front of $S_8$ and $C_8$, due to their fermionic
nature. The factor $1/(\eta\bar\eta)^8$ is the contribution of the
$8$ world-sheet bosonic degrees of freedom.

We can write rather simply also the torus partition function for the bosonic string, where we do not
have any fermionic oscillators and there are $D-2=24$ bosonic degrees of freedom, each of which contributes a
factor $1/\sqrt{\tau_2} \ |\eta(\tau)|^2$, giving a torus amplitude equal to
\be
\label{bosonic}
\mathcal{T}_{\it bosonic} \ = \ \int \frac{d^2\tau}{\tau_2^2} \
\frac{1}{\tau_2^{12} \ (\eta \ \bar\eta)^{24}} \ .
\ee

Coming back to the superstring, and after what we said about the
characters, it is simple to read off the low-energy spectra of the
Type IIA and IIB theories from their partition functions and to
convince oneself that they coincide with the ones we already
recovered at the end of the first section. For example, $V_8\bar
V_8$ contains the graviton, the dilaton and the antisymmetric
2-tensor, both in the type IIA and type IIB supergravity.
Moreover, the type IIB superstring contains another scalar another
2-form and a 4-form with self-dual field-strength that come from
$S_8\bar S_8$, two gravitinos with the same chirality and two
dilatinos also with the same chirality that come from $V_8\bar
S_8+S_8\bar V_8$. For the type IIA supergravity $S_8\bar C_8$
gives an abelian vector and a 3-form, while the mixed term gives
two gravitinos and two dilatinos with
 opposite chiralities.

An important thing to observe is that the relation
\be
\vartheta_3^4-\vartheta_4^4-\vartheta_2^4 \ = \ 0 \ ,
\ee
known as the \textit{aequatio identica satis abstrusa} of Jacobi, implies
$ \ V_8-S_8=0 \ $, which means that at each level the spectrum of Type
IIB superstring contains the same number of fermionic and bosonic
degrees of freedom, and again this happens thanks to the GSO-projection.

We can now discuss the modular invariance of the
amplitudes we wrote. First of all, the integration measure is invariant.
The transformation laws for the Dedekind function
\be
T \ : \ \eta(\tau)\rightarrow\eta(\tau+1) \ = \ e^{\frac{i\pi}{12}} \
\eta(\tau) \ , \qquad
S \ : \ \eta(\tau)\rightarrow \eta\left(-\frac{1}{\tau}\right) \ = \
\sqrt{-i\tau} \ \eta(\tau) \ ,
\ee
then imply that also the denominator $\tau_2^4 \ (\eta\bar\eta)^8$ is clearly invariant.
The modular invariance of the torus is then demonstrated if we consider that
on the basis $ \ O_8 \ , \ V_8 \ , \ S_8 \ , \ C_8 \ $ a $T$-modular transformation acts like
\be
T = e^{ -i \pi/3} \ {\rm diag} \left( 1,-1,-1,-1 \right) \ ,
\ee
that can be proved recalling that the form of the
characters is of the type given in (\ref{form}) with $c=4$ (because
the $so(8)$ characters contain only the contribution of the fermionic oscillators),
while an $S$-modular transformation is described by
\ba
S = \frac{1}{2} \ \left( \begin{array}{rrrr}
1 & 1 & 1 & 1 \\
1 & 1 & -1 & -1 \\
1 & -1 & 1 & -1 \\
1 & -1 & -1 & 1
\end{array} \right) \ ,
\ea
that can be determined from the $S$-modular transformation law
for $\vartheta$-functions
\be
\vartheta \left[ {\textstyle {\alpha \atop \beta}} \right]
\left(\frac{z}{\tau}\right|\left.-\frac{1}{\tau}\right) =
(-i \tau)^{1/2} \ e^{2 i \pi \alpha \beta + i \pi z^2/\tau} \ \vartheta
\left[{\textstyle {\beta \atop -\alpha}}\right] (z|\tau) \ .
\ee
For completeness, here we give also the $T$-modular transformation
\be
\vartheta \left[ {\textstyle {\alpha \atop \beta}}
 \right] (z|\tau+1)
= e^{-i \pi \alpha
(\alpha -1)} \vartheta \left[ {\textstyle {\alpha \atop \beta +\alpha - 1/2}}
\right]
(z|\tau) \ .
\ee

Before closing this part, we would like to reconsider the torus partition function
from the point of view of CFT \cite{Ginsp,bpz}.
In fact, apart from the integral with its measure, the general form of a torus partition function is
\be
\label{general torus}
{\cal T} \ = \ \sum_{i,j}\bar \chi_i X_{ij} \chi_j \ ,
\ee
where $X_{ij}$ is a matrix of non negative integer numbers, that for
the rational models is finite-dimensional.
Modular invariance restricts the general form of $X_{ij}$, imposing the constraints
\be
\label{modular_constraints}
S^\dag  \, X  \, S = X \,, \qquad T^\dag  \, X  \, T = X \ .
\ee
This observation is very important and will allow us to write in the
next chapter the torus partition functions for the type 0 models.
\section{The orientifold projection}
This section is devoted to review
the orientifold projection for the Superstring Theory \cite{cargese,ps,orientifolds,bs,bs2,bps,revDudas,open_string}.
This is an algorithm to build a consistent theory of
unoriented closed and open strings. The starting point is a torus partition function
for the Type IIB string that is left-right symmetric. The resulting model is known as
the Type I string.
In order to make the string unoriented,
we have to identify the left and right modes of the string, projecting the spectrum to
left-right symmetric or antisymmetric states.
The projection has to be consistent with the interacting theory, and this means that
only one of the two possible choices can be implemented.
Actually, what we have to do is insert in the trace of (\ref{torus}) the projector
$P=\frac{1+\Omega}{2}$, where $\Omega$ is the world-sheet parity,
whose action is to exchange the left $(L)$ oscillators with the right
$(R)$ ones. The result is that $\mathcal{T}\rightarrow \mathcal{T}/2+\mathcal{K}$,
where $\mathcal{K}$ is the
partition function for the Klein-bottle
\be
{\cal K} \ = \ \frac{1}{2} \int_0^{\infty} \ \frac{d^2\tau}{\tau_2^6} \ \mathrm{Str} \ \left( q^{L_0^\bot-a}
\bar q^{\bar L_0^\bot-\bar a} \Omega\right) \ .
\ee
The trace to perform is
\be
\sum_{L,R}\langle L,R\mid q^{L_0^\bot-a}\bar
q^{\bar L_0^\bot-\bar a}\Omega\mid L,R\rangle
\ = \ \sum_{L,R}\langle L,R\mid q^{L_0^\bot-a}\bar
q^{\bar L_0^\bot-\bar a}\mid R,L\rangle \ ,
\ee
and using the orthogonality of the states, we can reduce this to the sum
over only the diagonal sub-space, identifying effectively $L_0^\bot$ with
$\bar L_0^\bot$,
\be
\sum_{L}\langle L\mid (q\bar q)^{L_0^\bot-a}\mid L\rangle \ .
\ee
The last expression depends naturally from $2i\tau_2$, which we recognize as the
modulus of the doubly covering torus for the Klein-bottle.
Moreover, we understand that the mixed sectors in the torus have
to be simply halved because their contribution after the action of
$\Omega$ would be zero for orthogonality.
Performing the last sum, we get
\be
\label{KI}
{\cal K} = \frac{1}{2} \int_0^{\infty} \ \frac{ d \tau_2}{\tau_2^6} \
\frac{ (V_8 - S_8 ) ( 2 i \tau_2)}{\eta^8(2 i \tau_2)} \ .
\ee
The projection symmetrizes the NS-NS sector and antisymmetrizes the R-R sector, while the
mixed sectors in $\mathcal{T}$ are simply halved, as we said.
This is consistent with the fusion rules
between the conformal families corresponding to the characters $V_8, \  O_8, \  -S_8, \  -C_8$.
These rules, encoded in the Verlinde formula \cite{verlinde}, say that $V_8$ behaves like the identity
in the fusion with the other characters, while $-S_8$ fuses with
$-C_8$ to give $O_8$.

The resulting massless spectrum is obtained from the Type IIB one eliminating the $2$-form from the NS-NS
sector, the scalar and the $4$-form from the R-R sector, one gravitino and one dilatino from the
mixed sector. What survives the projection is a graviton, a dilaton and a RR $2$-form for the bosons,
and a gravitino and a dilatino of opposite chirality for the fermions. This spectrum corresponds
to the $\mathcal{N}=1$ minimal supergravity in $D=10$ dimensions.

In order to go in the transverse channel, the
first thing to do is to express (\ref{KI}) in terms of the modulus of the doubly covering torus, and
the change of variable $2\tau_2=t_2$, (\ref{KI}) uncovers the
important factor $2^5$. Then we have to perform an $S$ modular
transformation: $t_2\rightarrow 1/\ell$, where $\ell$ is the time in the
transverse channel. The ``supercharacter'' $(V_8-S_8)$ is invariant, while
the factor $t_2^4\eta(it_2)^8\rightarrow\eta(i\ell)^8$, and thus the transverse
Klein-bottle is
\be
\label{tildeKI}
\tilde{\cal K} = \frac{2^5}{2} \int_0^{\infty} \ d \ell  \
\frac{ (V_8 - S_8 ) (i \ell )}{\eta^8(i \ell)} \, .
\ee
Let us stress that (\ref{tildeKI}) is a sum over the characters $V_8$
and $-S_8$ flowing in the transverse channel, each of
which is multiplied by a positive coefficient that can be
interpreted as the square of the one-point function of the character
in front of the crosscap.
The Klein-bottle projection has must be consistent with the positivity of
these coefficients in the transverse channel, as in fact is the case in (\ref{tildeKI}).

At this point we introduce the open strings.
As we already saw in the first section, an open string allows the propagation of the
vector and therefore it describes naturally the Yang-Mills interaction.
The important thing to stress is that a scattering amplitude with
only open strings is planar and, as was observed by Chan and Paton \cite{cp},
its cyclic symmetry remains preserved also
multiplying the amplitude for a factor that is the trace over the product of
matrices of a certain gauge group.
So, it is possible to associate to an open
string a Chan-Paton matrix or equivalently to attach at the ends of the string
some Chan-Paton charges. Marcus and Sagnotti showed
that the Chan-Paton group can be only one of the classical groups \cite{Marcus_Sagnotti},
thus excluding the exceptional groups.

Let us begin to write the one-loop partition function for an oriented
open string. Its general form is
\be
\label{open}
N^2 \ \int_0^\infty \ \frac{d
\tau_2}{\tau_2^6} \ {\rm Str} \ q^{\frac{1}{2}(L_0^\bot - a)}  \ ,
\ee
where $N$ is the multiplicity of the Chan-Paton charges. Let us note
the factor $\frac{1}{2}$ in the exponent, reflecting the
different mass-shell condition for an open string. This factor will
make the annulus depend on the modulus of the doubly covering torus.

The amplitudes for unoriented
open strings are obtained inserting in the trace the projector
$P=(1+\epsilon \Omega)/2$, where the world-sheet parity $\Omega$
flips the ends of the string, $\Omega:\sigma\rightarrow\pi-\sigma$,
while $\epsilon$ is a sign that will be
fixed in order to have a consistent model.
After the projection, the
trace (\ref{open}) splits in the amplitude for the annulus
\be
{\cal A} = \frac{N^2}{2} \int_0^{\infty} \ \frac{ d \tau_2}{\tau_2^6} \
\frac{ (V_8 - S_8 ) ( \frac{1}{2}
i \tau_2 )}{\eta^8( {\textstyle{1\over 2}} i \tau_2)} \ ,
\ee
and the amplitude for the M\"{o}bius strip
\be
{\cal M} = \frac{\epsilon N}{2} \int_0^{\infty} \
\frac{ d \tau_2}{\tau_2^6} \
\frac{ (\hat{V}_8 - \hat{S}_8 ) ( {\textstyle{1\over 2}}
i \tau_2 + {\textstyle{1\over 2}})}{\hat{\eta}^8(
{\textstyle{1\over 2}} i \tau_2 + {\textstyle{1\over 2}})} \ .
\ee
The last amplitude is understood if we consider
that $\Omega$ acts on a state $\alpha_{-n}\mid ij\rangle$ (where $i,j$
run on the multiplicity of the Chan-Paton charges at the end of the
string) giving $(-)^n\alpha_{-n}\mid ji\rangle$. Thus, when we compute
the trace, the sum over $i,j$ is restricted only to the diagonal states
$\mid i\rangle$, giving a factor $N$, while $q^{\frac{n}{2}}$ is shifted
to $q^{\frac{n}{2}}(-)^n$, that means $e^{2\pi i n(1/2+i\tau_2/2)}$,
in which we recognize the modulus of the doubly covering torus.
In order to have a real integrand for the
M\"{o}bius strip, we introduced the real hatted characters whose generic form
is
\be
\hat{\chi}(i \tau_2 +{\textstyle{1\over 2}}) =
q^{h - c/24} \sum_k (-1)^k d_k q^k \ , \qquad q=e^{-2\pi\tau_2} \ ,
\label{chireal}
\ee
that differ from the standard $\chi(i\tau_2+1/2)$ in the overall phase $e^{-i\pi(h-c/24)}$.

The transverse channel for the Annulus amplitude is simply obtained making the
change of variable $\frac{\tau_2}{2}=t_2$, that gives an overall
factor $2^{-5}$, and then performing an $S$ transformation
$t_2\rightarrow\frac{1}{\ell}$:
\be
\tilde{\cal A} = \frac{2^{-5} \ N^2}{2} \int_0^{\infty} \
d \ell \
\frac{ (V_8 - S_8 ) ( i \ell )}{\eta^8( i \ell )} \ .
\ee
Note that also for the transverse annulus the characters $V_8$
and $-S_8$ are multiplied by positive numbers whose square roots are
the one-point functions of those characters in front of the boundaries.

Writing the transverse channel for the M\"{o}bius presents some
subtleties due to the real part of $\tau$, and as a result,
the transformation $\tau_2\rightarrow 1/t$ is achieved performing a $P=T^{1/2}ST^2ST^{1/2}$ transformation \cite{gpp}.
On the basis
\be
\frac{O_8}{\tau_2^4 \eta^8} \, , \quad
\frac{V_8}{\tau_2^4 \eta^8} \, , \quad \frac{S_8}{\tau_2^4
\eta^8} \, , \quad \frac{C_8}{\tau_2^4 \eta^8}
\ee
the action of $T$ is simply $T = \ {\rm diag}(-1,1,1,1) \ $,
so that, considering that $S$ squares to the conjugation matrix, that
is the identity for $so(8)$ because its representations are
self-conjugate, the $P$ matrix takes the very simple form
\be
P \ = \ T \ = {\rm diag}(-1,1,1,1) \ ,
\ee
apart from an overall power of $\tau_2$ that disappears in the transverse channel.
In terms of $\ell=t/2$ the transverse M\"{o}bius amplitude then reads
\be
\label{MtildeI}
\tilde{\cal M} = 2 \ \frac{\epsilon N}{2} \int_0^{\infty} \ d \ell  \
\frac{ (\hat{V}_8 - \hat{S}_8 ) ( i \ell + {\textstyle{1\over 2}})}
{\hat{\eta}^8(i \ell + {\textstyle{1\over 2}})} \ .
\ee
Na\"{i}vely  the transverse M\"{o}bius amplitude, that is a tube between a crosscap
and a boundary, can be seen as the geometric mean $\sqrt{\tilde{\cal K}\tilde{\cal A}}$,
so that the factor $2$ in (\ref{MtildeI}) is a combinatoric factor,
while the sign $\epsilon$ reflects the sign ambiguity of
the one-point functions in front of the crosscap and of the
boundary, that we know only squared from $\tilde{\cal A}$ and $\tilde{\cal K}$.

The massless spectrum, that we read from the direct channel, gives
the $N=1$ super Yang-Mills multiplet in $D=10$ and consists of a
vector and a Majorana-Weyl fermion with multiplicity
$N(N-\epsilon)/2$. Therefore, $\epsilon=1$ corresponds to the
adjoint representation of the gauge group $SO(N)$, while
$\epsilon=-1$ corresponds to the adjoint representation of
$USp(N)$.

At this point it is useful to summarize the orientifold construction. To this end, let us consider
the general form of the torus partition function
(\ref{general torus})
\be
{\cal T} \ = \ \sum_{i,j}\bar \chi_i X_{ij} \chi_j \ ,
\ee
with its constraints (\ref{modular_constraints}).
Moreover let us suppose that $X_{ij}=0,1$. A value of $X_{ij}$ different from $0,1$
means that we have more conformal families starting with the same conformal weight,
and thus we have to resolve an ambiguity.

In the Klein-bottle only the left-right symmetric
sectors can propagate, the ones with $X_{ii}\neq 0$
\be
{\cal K} \ = \ \frac{1}{2}\sum_i{\cal K}^i \chi_i \ , \qquad \qquad
{\cal K}^i = \pm X_{ii} \ .
\ee
While in the transverse channel we read
\be
\tilde{\cal K} =  \frac{1}{2} \sum_i  ({\Gamma}^i)^2 \,
\chi_i \ .
\ee
The coefficients $\Gamma_i$ are the one-point functions (or the reflection coefficients)
of $\chi_i$ in front of a crosscap.
The signs of ${\cal K}^i$ have to be chosen in order to have a consistent interacting theory
with positive coefficients $(\Gamma^i)^2$ in the transverse channel.

The transverse annulus has the form
\be
\tilde{\cal A} = \frac{1}{2} \sum_{i} \, \chi_i
\left( \sum_a B^i_a \, n^a \right)^2 \ ,
\ee
where $B_a^i$ is the one-point function of $\chi_i$ in front of a hole with boundary condition labelled
by the index $a$ and where $n^a$ is the corresponding Chan-Paton multiplicity.
As a bulk sector reflecting on a boundary turns into its conjugate,
the transverse annulus propagates only the characters $\chi_i$ that appear in the torus
in the form $\bar \chi_i^C \chi_i $, where $\chi_i^C$
is the conjugate of $\chi_i$. In the case of $X=\mathcal{C}$, where $\mathcal{C}$
is the conjugation matrix, all bulk sectors can
reflect on a hole and the number of different boundary conditions is equal to the number of bulk sectors.
One can turn an important observation of Cardy \cite{cardy2}
on the annulus amplitude into an ansatz to write the orientifold of a theory with $X=\mathcal{C}$.
But let us to proceed with the general case.
After an $S$-modular transformation we get the direct channel for the annulus
\be
{\cal A} = \frac{1}{2} \sum_{i,a,b} \, {\cal A}^i_{ab}
\, n^a\, n^b\,
\chi_i \, .
\ee

Finally, the transverse M\"{o}bius, apart from a combinatoric factor $2$, is the square root of $\tilde{\cal K} \times \tilde{\cal A}$
\be
\tilde{\cal M} = \frac{1}{2} \sum_{i}\,
\hat{\chi}_i \, {\Gamma}^i  \left( \sum_a B^i_a \, n^a \right) \ ,
\ee
while in the direct channel it reads
\be
{\cal M} = {\textstyle \frac{1}{2}}\sum_{i,a} \, {\cal M}^i{}_{a} \, n^a\,
\hat{\chi}_i \ .
\ee
The construction then is consistent if ${\cal M}$ is the projection of ${\cal A}$,
meaning that $M^i_a=\pm A^i_{aa}$, supposing $A_{aa}^i=0,1 \ $.

For completeness, now we report also the amplitudes for the bosonic string.
The orientifold projection proceeds along the same steps we traced for the superstring,
but it is much easier because there are only bosonic oscillators.
Apart from the torus partition function, the other amplitudes are \cite{marcus86,bs88}
\ba
{\cal K} &=& \frac{1}{2}
\ \frac{1}{\tau_2^{12} \ \eta(2 i \tau_2)^{24}} \ , \nonumber\\
{\cal A} &=& \frac{N^2}{2} \ \frac{1}{\tau_2^{12} \
\eta({\textstyle{1\over 2}} i \tau_2)^{24}} \ ,\nonumber\\
{\cal M} &=& \frac{\epsilon \, N}{2} \ \frac{1}{\tau_2^{12} \
\hat{\eta}(\frac{1}{2} i \tau_2 + \frac{1}{2})^{24}} \, ,
\ea
while in the transverse channel they read
\ba
\tilde{\cal K} &=& \frac{2^{13}}{2} \ \frac{1}{\eta(i \ell)^{24}} \, ,
\nonumber \\
\tilde{\cal A} &=& \frac{2^{-13} \ N^2}{2} \ \frac{1}{\eta(i \ell)^{24}}
\, , \nonumber \\
\tilde{\cal M} &=& 2 \ \frac{\epsilon \, N}{2} \
\frac{1}{\hat{\eta} (i \ell+{\textstyle{1\over 2}})^{24}} \, , \ea
where the integration is understood. The Klein-bottle amplitude
symmetrizes the torus, and the resulting unoriented closed
spectrum at the massless level contains only the graviton and the
dilaton, while the massless unoriented open spectrum has a vector
in the adjoint representation of $SO(N)$ or $USp(N)$, depending on
the choice of the sign $\epsilon$.

All those amplitudes have an ultraviolet divergence for
$\tau_2\rightarrow 0$. The origin of this divergence is well
understood if we pass in the transverse channel and we consider
that a closed string state of squared mass equal to $M^2$ and with
degeneracy $c_M$ enters in those amplitudes with a term
proportional to \be \sum_M c_M\int_0^\infty d\ell e^{-\ell M^2} \
= \ \sum_M c_M \left.\frac{1}{p^2+M^2}\right|_{p^2=0} \ . \ee
Thus, in the infrared limit $\ell \rightarrow\infty$, the only
states that survive are the massless ones, and the propagation of
such states at vanishing momentum gives the infrared divergence we
said.
\begin{figure}
\begin{center}
\epsfbox{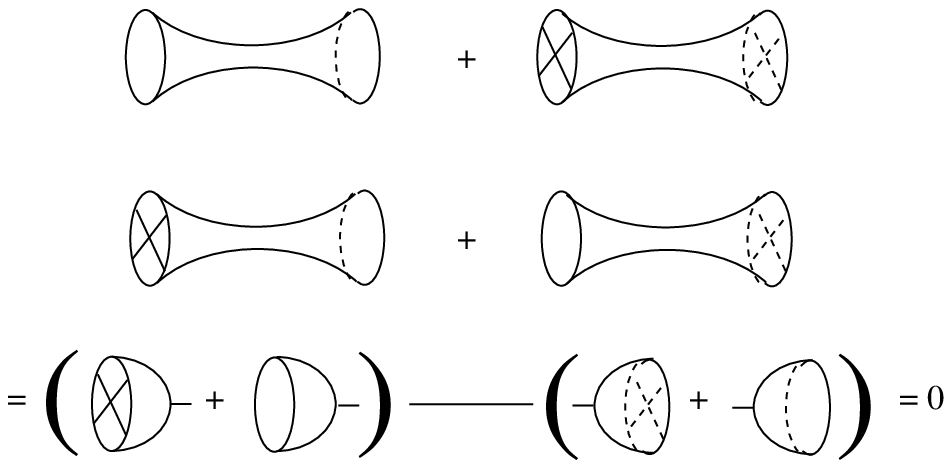}
\end{center}
\caption{Tadpole condition.}
\label{figtadpoles}
\end{figure}
Nevertheless, it is possible to cancel this divergence noting that
in the limit $\ell \rightarrow\infty$ the transverse amplitudes
factorize in a one-point function of a massless closed string
state in front of a boundary or a crosscap times the propagator of
such a state at vanishing momentum times another one-point
function. Therefore, eliminating the pole one is imposing the
vanishing of the residue. As these one-point functions, that are
commonly called tadpoles, depend on $N$ and $\epsilon$, the
conditions one imposes, known as tadpole conditions, will fix also
the gauge group. For the bosonic string, all the (transverse)
infrared divergences cancel by imposing the tadpole condition \be
 2^{13} + 2^{-13} \ N^2 - 2 \, \epsilon \, N  = 2^{-13} \
(N - \epsilon \, 2^{13})^2 \, ,
\ee
that fixes $\epsilon=+1$ and the gauge group $SO(8192)$ \cite{douglas1,weinberg,marcus86,bs88}.

For the Type I superstring \cite{gsop}, thanks to the
supersymmetry, the tadpole condition is the same for both the
NS-NS and the R-R sectors \be \frac{2^5}{2} + \frac{2^{-5} \
N^2}{2} + 2 \ \frac{\epsilon N}{2} = \frac{2^{-5}}{2} \ ( N + 32
\epsilon )^2 = 0 \ , \label{gsso32} \ee and fixes $N=32$ and
$\epsilon=-1$. Therefore the gauge group is $SO(32)$.

Let us note that the tadpole cancellation is possible in general
only if the model contains both unoriented closed and open strings
and that it does not depend on the torus partition function that
is free of divergencies due to the modular invariance. The Type I
model is free of anomaly thanks to the Green Schwarz mechanism
\cite{gs}.

Although the R-R and NS-NS tadpoles are cancelled at the same time for the Type I superstring,
conceptually the tadpole condition in the two sectors has very different meanings.
To explain this point, we have to introduce new space-time objects, the D$p$-branes
and the O$p$ planes. A D$p$-brane is a dynamical extended object with $p+1$ dimensions. The ends of an open string can
attach  to it with Dirichlet boundary conditions in the $9-p$ directions  orthogonal to the brane and
Neumann boundary conditions in the $p+1$ directions parallel to it. These hyperplanes
have a tension that is always positive and are charged with respect to a potential described by a $(p+1)$-form \cite{pol95}.
Moreover, there are also $\overline{{\rm D}p}$-branes that have the same tension but the opposite charge. By convention
we consider positive the charge of a D$p$-brane and negative the one of a $\overline{{\rm D}p}$-brane.

The O$p$ planes, or orientifold planes, are the fixed points of
the world sheet parity, invade $p+1$ dimensions and have no
dynamics. They carry a charge with respect to a $(p+1)$-potential
and a tension that now can be both positive and negative. Thus
there are two types of these hyperplanes together with their
conjugates (that have an opposite charge): the O$_+$ planes with
negative tension and charge, the O$_-$ planes with positive
tension and charge, the $\overline{\rm O}_+$ planes with negative
tension and positive charge and the $\overline{\rm O}_-$ with
positive tension and negative charge.

Actually, the space-time D$p$-branes correspond to the world-sheet
boundaries, while the orientifold planes are the space-time
counterparts of the crosscaps.

At this point, it should be clear that from the NS-NS and the R-R
sectors of the transverse annulus one can read the squared tension
of a brane (the tension is the charge for the gravitational field
that flows in the NS-NS sector) and its charge (the potentials
with respect to the branes are charged are described by the forms
flowing in the R-R sector), while from the NS-NS and R-R sectors
of the transverse Klein-bottle one can read the tension and the charge of
an O-plane.

For example, the Type I superstring contains $N=32$ D$9$-branes
and a O$9_+$ plane. Coming back to the tadpole condition, now we
understood that the R-R tadpole cancellation is related to the
charge neutrality of the space-time, and its absence would give
quantum anomalies. On the other hand, the NS-NS tadpole condition
can be relaxed breaking supersymmetry and giving a dilaton
dependent correction to the low-energy effective field theory
proportional to \be \label{tad} \int d^{10} x
\sqrt{-g}e^{-\varphi} \ , \ee where the coefficient in front of
the dilaton $\varphi$ is related to the Euler characteristic of
the disk.

A term like (\ref{tad}) is a source for the gravitational field
equation, and hence curves the space-time. The maximally symmetric
Minkowski space is no more a solution, and this reflects itself in
the infrared divergences that we met in the transverse channel.
This crucial point is the heart of this Thesis, and we will deal
with it in the last two chapters, where we will show in a number
of field theory toy models how to recover the proper finite
results starting from a ``wrong'' vacuum, after suitable
resummations are carried up, and where some examples of string
models where tadpoles seem not to affect the results, at least at
the lowest order, are discussed.

Before closing this section, we would like to describe a first
example of a model with a NS-NS tadpole. Starting from the
transverse amplitudes $\tilde{\cal A}$ and $\tilde{\cal M}$ of the
Type I $SO(32)$ superstring, one can actually generalize them,
allowing different signs for the NS-NS and R-R sectors \ba
\tilde{\cal A} &=& \frac{2^{-5}}{2} \int d \ell \
\frac{ (n_+ + n_-)^2 \, V_8 - (n_+-n_-)^2 \, S_8 }{\eta^8} \, , \\
\tilde{\cal M} &=& \frac{2}{2} \int d \ell  \ \frac{ \epsilon_{NS}
\, (n_+ + n_- ) \, \hat{V}_8 - \epsilon_R \, (n_+ - n_-) \,
\hat{S}_8 }{\hat{\eta}^8} \, . \ea From $\tilde{\cal A}$ we see
that $n_+$ counts the number of D-branes while $n_-$ counts the
number of $\overline{\rm{D}}$-branes, while the type of O-planes
involved depends on the choice of the two signs $\epsilon_R$ and
$\epsilon_{NS}$. The tadpole conditions in the NS sector
\be \frac{2^5}{2} + \frac{2^{-5} \ (n_++n_-)^2}{2} + 2 \
\frac{\epsilon_{NS} (n_++n_-)}{2} = 0  \qquad NS \ , \ee and in
the R sector \be \frac{2^5}{2} + \frac{2^{-5} \ (n_+-n_-)^2}{2} +
2 \ \frac{\epsilon_{R} (n_+-n_-)}{2} = 0  \qquad R \ , \ee
are clearly solved by the $SO(32)$ superstring, but one can relax
the condition in the NS-NS sector. In that case, the R tadpole
condition fixes $n_+-n_-=32$ and $\epsilon_R=-1$. The sign
$\epsilon_{NS}$ remains still undetermined, and its value
specifies the kind of orientifold planes present in the model. In
particular, $\epsilon_{NS}=+1$ means that $\overline{\rm O}_-$
planes are present, while $\epsilon_{NS}=-1$ means that O$_+$
planes are present. Let us stress that the simultaneous presence
of branes and antibranes brings about a tachyonic instability that
reflect their mutual attraction (see, in the direct channel, the
tachyonic term $n_+n_-O_8 \ $)
\ba \label{Sugimoto_amplitudes} {\cal A} \ &=& \ \frac{1}{2}
\int_0^\infty \, \frac{d t}{t^6 \eta^8} \left[ (n_+^2 + n_-^2) \,
( V_8 - S_8) + 2 n_+ n_- \,
(O_8 - C_8) \right]  \, , \\
{\cal M} \ &=& \ \frac{1}{2} \int_0^\infty \,  \frac{d t}{t^6
\hat{\eta}^8} \left[ \epsilon_{NS} \, (n_+ + n_- ) \, \hat{V}_8 -
\epsilon_R \, (n_+ - n_-) \, \hat{S}_8 \right] \ . \ea
The spectrum is no more supersymmetric, and a ``minimal'' choice
is the Sugimoto model \cite{ sugimoto}, that corresponds to $n_+=0
\ $, $n_-=32$ and $\epsilon_{R}=\epsilon_{NS}=+1$. The resulting
spectrum is not supersymmetric, with a massless vector and a
massless Majorana-Weyl spinor respectively in the adjoint and in
the antisymmetric representation of the gauge group $USp(32)$, but
the tachyon instability is no more present.
\section{Toroidal compactification}
In this section we study the compactification of one dimension,
say $X$, on a circle $S^1$. Certainly the momentum in the
compactified dimension is quantized in inverse units of the
compactification radius $R \ $: $p=m/R \ $, but a closed string
has also the possibility to wrap itself around the circle $n$
times. The integer $n$ is called the winding number. What we have
to do is to identify \be X(\tau,\sigma) \sim
X(\tau,\sigma+\pi)+2\pi Rn  \ , \ee obtaining in this way the
expansion \be X \ = \ x \ + \ 2\alpha'\frac{m}{R}\tau+2nR\sigma+
i\frac{\sqrt{2\alpha'}}{2} \ \sum_{n\neq
  0}\left(\frac{\alpha_n}{n} \ e^{-2in(\tau-\sigma)}+
\frac{\tilde\alpha_n}{n} \ e^{-2in(\tau+\sigma)}\right) \ . \ee
Then, in order to separate also the zero mode in a left and in a
right part, it is convenient to introduce \be
p_L=\frac{m}{R}+\frac{n R}{\alpha'} \ , \qquad \qquad
p_R=\frac{m}{R}-\frac{n R}{\alpha'} \ . \ee In terms of these
momenta $X$ decomposes in $X_L+X_R$, with \be X_{L,R} \ = \
\frac{1}{2}x+\alpha'p_{L,R}(\tau\mp\sigma)+oscillators \ , \ee and
the mass shell condition becomes \be m^2 \ = \
\frac{2}{\alpha'}\left[ \frac{\alpha'}{4}p_L^2+
\frac{\alpha'}{4}p_R^2+L_0^\bot+\bar{L}_0^\bot+a+\bar a\right] \ ,
\ee together with the level-matching condition \be
\frac{\alpha'}{4}p_R^2+L_0^\bot+a \ = \
\frac{\alpha'}{4}p_L^2+\bar{L}_0^\bot +\bar a \ . \ee In computing
the vacuum amplitudes, one has to replace \be
\mathrm{tr}(q^{L_0^\bot+a} \ \bar q^{\bar L_0^\bot + \bar a}) \
\rightarrow \
\mathrm{tr}\left(q^{L_0^\bot+a+\frac{\alpha'}{4}p_R^2} \ \bar
q^{\bar L_0^\bot + \bar a+\frac{\alpha'}{4}p_L^2}\right) \ , \ee
and since in equation (\ref{imp}) the gaussian integral actually
is over the $D-1$ non-compact momenta, a different power of the
modulus $\tau_2$ \be \frac{1}{\tau_2^{D/2+1}} \ \rightarrow \
\frac{1}{\tau_2^{(D-1)/2+1}} \  \ee is obtained. Therefore, in
order to take into account the compactification, what one has to
do is to replace in the vacuum amplitudes \be \frac{1}{\eta\bar
\eta \ \sqrt{\tau_2}} \ \rightarrow \
\sum_{m,n}\frac{q^{\frac{\alpha'}{4}p_R^2} \ \bar
  q^{\frac{\alpha'}{4}p_L^2}}{\eta \bar \eta}
\ee for each compact dimension.

The quantity on the right-hand side of the previous equation is
clearly invariant under a $T$-modular transformation. Proving its
invariance under an $S$-modular transformation requires the use of
the Poisson transformation \be \sum_{\{n_i\} \in \mathbb{Z}} \,
e^{-\pi \, n^{\rm T}\, A \,n \ + \ 2 \, i \, \pi \, b^{\rm T} \,
n} \ = \ \frac{1}{\sqrt{{\rm det}(A)}} \, \sum_{\{m_i\} \in
\mathbb{Z}}
 \, e^{-\pi \, (m - b)^{\rm T} \, A^{-1} \, (m - b)} \, , \label{poissonsum}
\ee
where, denoting with $d$ the number of compact dimensions,
 $A$ is a $d \times d$ square matrix and $m$ and $n$
are $d$-dimensional vectors of integers.

For example, the sum in the case of one compact direction, after
an $S$ transformation, is in terms of $m$ and $n$ \be
\sum_{m,n}e^{-\pi m^2\left(\frac{|\tau|^2
      R^2}{\alpha'\tau_2}\right)^{-1}} \ e^{2\pi i
    m\left(\frac{n\tau_1}{|\tau|^2}\right)} \
    e^{\frac{-\pi\tau_2 n^2 R^2}{\alpha'|\tau|^2}} \ ,
\ee from which we recognize $A^{-1}=|\tau|^2 R^2/\alpha'\tau_2$
and $b=n\tau_1/|\tau|^2$. Performing a Poisson resummation with
respect to $m$ gives \be
\frac{|\tau|R}{\sqrt{\alpha'\tau_2}}\sum_{m',n}e^{-\frac{\pi|\tau|^2
    R^2}{\alpha'\tau_2}\left(m'-\frac{n\tau_1}{|\tau|^2}\right)^2}
e^{\frac{-\pi\tau_2 n^2 R^2}{\alpha'|\tau|^2}} \ ,
\ee
that can be put in the form
\be
\frac{|\tau|R}{\sqrt{\alpha'\tau_2}}\sum_{m',n}e^{-\pi
  n^2\left(\frac{\alpha'\tau_2}{R^2}\right)^{-1}}
 \ e^{-2\pi i m'\frac{iR^2 n \tau_1}{\alpha'\tau_2}} \ e^{-\pi
   m'^2\frac{|\tau|^2R^2}{\alpha'\tau_2}} \ .
\ee

At this point one has to perform another Poisson resummation with
respect to $n$, with $A^{-1}=\alpha'\tau_2/R^2$ and $b=-iR^2 m'
\tau_1/\alpha'\tau_2$ \be
|\tau|\sum_{m',n'}e^{-\pi\frac{\alpha'\tau_2}{R^2}\left(n'+i\frac{R^2
m'\tau_1}{\alpha'\tau_2}\right)^2} \ e^{-\pi\frac{|\tau|^2 R^2
m'^2}{\alpha'\tau_2}} \ . \ee The last expression is just the
original sum, with windings and momenta interchanged and with an
overall factor of $|\tau|$. Such a factor is then absorbed by
$S$-transforming $1/\eta\bar\eta$. The Poisson resummation is
fundamental in the following in order to connect the direct and
transverse channels.

An important comment concerns the special value $R=\sqrt{\alpha'}$
of the internal radius. To understand this point, let us consider
for example the case of a bosonic string, whose critical dimension
is $D=26$ and whose spectrum involves a shift of the vacuum energy
$a=-1$ both for the left and right sectors, with $X^{25}$
compactified on a circle. For a generic value of the radius, the
massless spectrum is solely given acting with the oscillators on
the state $|m=0,n=0\rangle$ as follows:
\be \alpha^i_{-1}\tilde\alpha^j_{-1}|0,0\rangle \ , \qquad
\alpha^i_{-1}\tilde\alpha^{25}_{-1}|0,0\rangle \ , \qquad
\alpha^{25}_{-1}\tilde\alpha^j_{-1}|0,0\rangle \ , \qquad
\alpha^{25}_{-1}\tilde\alpha^{25}_{-1}|0,0\rangle \ . \ee
The first state decomposes in the graviton, the 2-form and the
dilaton, the second and the third states are $U(1)$ vectors, while
the last state is a scalar. This is just the conventional
Kaluza-Klein spectrum familiar from circle compactification in
Field Theory. Actually, the vacuum value of this scalar can not be
fixed and remains undetermined: it is a modulus of the theory, and
determines the effective size of the internal manifold. Now this
scalar is emitted by the operator $\dd X^{25} \bar \dd X^{25}$,
that can be seen as a perturbation of the internal metric, so that
having only one scalar means that one can deform the internal
manifold only in one way.

But for the special value $R=\sqrt{\alpha'}$ there are more
massless states, and the result is a stringy enhancement of the
gauge symmetry. Taking into account the level matching condition
$\bar{N_B}-N_B=-mn$, eight additional scalars  \be
\tilde\alpha_{-1}^{25}|\pm1,\mp1\rangle \ , \qquad
\alpha_{-1}^{25}|\pm1,\pm1\rangle \ , \qquad |\pm2,0\rangle \ ,
\qquad |0,\mp2\rangle \  \ee and four additional vectors \be
\tilde\alpha_{-1}^i|\pm1,\mp1\rangle \ , \qquad
\alpha_{-1}^i|\pm1,\pm1\rangle \  \ee emerge. As a result, there
are now 9 massless scalars, and more importantly 3 vectors from
the left modes, 3 vectors from the right modes, so that the gauge
group is enhanced to $SU(2)\times SU(2)$, while the scalars fill
the $(3,3)$. Moving away from the special value of the radius, a
Higgs mechanism takes place, and the group is broken to
$U(1)\times U(1)$, four scalars are reabsorbed to give mass to
four of the six vectors and four become massive.

A related key ingredient in toroidal compactification is T-duality
\cite{cs}. Such a transformation maps $R\leftrightarrow \alpha'/R$
and interchanges windings with momenta. The net effect of a
T-duality is that $p_L\leftrightarrow p_L$ and $p_R\leftrightarrow
-p_R$. Therefore, defining $ \ T:\tilde\alpha_n\leftrightarrow
\tilde\alpha_n \ $ and $ \ T:\alpha_n\leftrightarrow -\alpha_n \
$, we see that a T-duality acts like a parity only on the right
sector. The special value of the radius $R=\sqrt{\alpha'}$ can be
characterized as the unique self-dual point.

For an open string a T-duality is even more interesting, since it
affects the boundary conditions at its ends. Let us consider the
coordinate of an open string along the compact dimension, with
Neumann boundary conditions $\dd_\sigma X(0)=\dd_\sigma X(\pi) =0$
\ba X=X_L+X_R &=& \frac{x}{2}+\alpha'\frac{m}{R}(\tau+\sigma)+
i\frac{\sqrt{2\alpha'}}{2}\sum_n\frac{\alpha_n}{n}e^{-in(\tau+\sigma)}+\nonumber\\
&+& \frac{x}{2}+\alpha'\frac{m}{R}(\tau-\sigma)+
i\frac{\sqrt{2\alpha'}}{2}\sum_n\frac{\alpha_n}{n}e^{-in(\tau-\sigma)}
\ . \ea Notice that there are no windings, because one can not
interlace an open string. After a T-duality, that is a parity only
on $X_R$, we have a T-dual coordinate \be
X^T=X_L-X_R=2nR^T\sigma+\textit{oscillators} \ee that satisfies
Dirichlet boundary conditions along the compact direction \be
\label{dir} X^T(0)=X^T(\pi)=0 \quad \textrm{mod} \quad  2\pi n R^T
\ . \ee Here, $R^T=\alpha'/R$ is the dual radius, while $n$ has
now the interpretation of a winding number. Eq. (\ref{dir}) means
that in the T-dual picture the ends of the open string are
attached to a hyperplane, actually a D-brane, placed at the origin
of the compact direction and orthogonal to it.

More generally, a D$p$-brane is mapped to a D$(p-1)$-brane, if one
performs a T-duality along a direction tangent to the brane, while
it is mapped to a D$(p+1)$-brane if the T-duality is along a
direction transverse to it. This is consistent with the fact that
the Type IIA and the Type IIB theories compactified on a circle
are one the T-dual version of the other, where the T-duality,
acting as a parity only on the right sector, affects only the
chirality of the space-time spinors emerging from the right
sector. The mentioned consistency can be appreciated if one
considers that the Type IIA theory contains in its spectrum only
D$p$-branes with $p$ even, while the Type IIB theory contains only
D$p$-branes with $p$ odd \footnote{In general a
  $p$-brane couples to a RR-$(p+1)$-form and its magnetic dual in $D$
  dimensions is a $(D-p-4)$-brane. So, for example, the Type IIB
theory that has in the RR sector a scalar a 2-form and a self-dual
4-form contains $p=(-1,1,3)$-branes together with their dual
$p=(5,7)$-branes. The $3$-brane is self-dual. Moreover there are
$9$-branes, but they couple to a 10-form that has no dynamics.
After the orientifold projection, only $p=(1,5,9)$-branes survive
\cite{pol95}.}, so that a T-duality, changing the dimensionality
of a $p$-brane, maps branes with $p$ odd in branes with $p$ even
and vice versa.

The last ingredient that we want to introduce in theories with
compact dimensions is the possibility to break the Chan-Paton
group while preserving its rank, introducing Wilson lines
\cite{bps} on the boundary. For simplicity, let us consider again
the bosonic string with $X^{25}$ compactified on a circle, and let
us turn on a Wilson line via the minimal coupling \be
S=-\frac{T}{2}\int d\tau d\sigma (\dot{X}^2-X'^2)-q\int d \tau
\left.A_{25}\dot{X}^{25}\right|_{\sigma=0} \ , \ee where $A_{25}$
is a constant abelian gauge field. Thus, the mechanical momentum
undergoes a shift \be \dot{X}^{25}=p^{25}+qA^{25}=\frac{m+a}{R} \
, \qquad \textrm{where} \qquad a=qRA^{25}=\textrm{constant} \ .
\ee In a T-dual picture, where momenta become windings, the shift
becomes a displacement of the branes where open strings are
attached. The result is the same, a breaking of the gauge group,
but the T-dual picture gives a more conventional interpretation of
this phenomenon. A displacement of branes induces a stretching of
the strings terminating on them, and consequently some of their
massless states become massive \cite{pol95}.

More in detail, let us consider a stack of $N$ branes orthogonal
to the compact direction, placed at positions given by $0<a_i<1$,
where $i=1,\ldots N$. The internal coordinate of an open string
whose ends are attached to different D-branes is \be X^T=2\pi
R^Ta_i+2R^T(a_j-a_i+n)\sigma+\textit{oscillators} \ee and
satisfies the following boundary conditions \ba
X^T(0)&=& 2\pi R^T a_i \ ,\nonumber\\
X^T(\pi)&=& 2\pi R^T a_j +2\pi R^T n \ . \ea The spectrum is given
(in the bosonic case) by \be
m^2=\frac{1}{\alpha'}(N_B-1)+(a_j-a_i+n)^2\frac{1}{R^2} \ . \ee
Hence, for $N_B=1$ and $n=0$ massless states are present only if
$i=j$. Such states are scalars $\alpha_{-1}^{25}|i,i\rangle$ and
$U(1)$ vectors $\alpha_{-1}^i|i,i\rangle$. One can see that for
each brane there is a gauge group $U(1)$ corresponding to an open
string whose ends are both attached to the same brane.

Now let us consider an oriented open string. If $r$ of the branes
are coincident, there are $r^2$ massless vectors (plus the same
number of scalars) given by $\alpha_{-1}^i|i,j\rangle$ for $i,j\in
1,\ldots r$. These states correspond to the open strings
terminating with both the ends on one of the $r$ coincident
branes. In this case the gauge group is $U(r)\times U(1)^{N-r}$.
In the case of an unoriented open string, one has to consider the
action of the orientifold projection on $X^T$ \be
\Omega:X^T=(X_L-X_R) \ \rightarrow \ (X_R-X_L)=-X^T \ . \ee Thus,
$\Omega$ acts on $X^T$ like a $\mathbb{Z}_2$ orbifold and the
circle is mapped from the projection in a segment whose extrema
$X^T=0,\pi R^T$ are the fixed points of the involution and
correspond to the O-planes. Now, only half of the $N$ branes are
in the interval between the two O-planes, while the others are
their images. If the branes are all separated, the gauge group is
$U(1)^{N/2}$. A stack of $r$ coincident branes gives a factor
$U(r)$ but here there is a new possibility: $r$ branes can
coincide with an orientifold plane and so also with their images.
The result is a factor $SO(2r)$ (or a symplectic group) to the
gauge group. The maximal gauge group is recovered if all the
branes are coincident with one of the two O-planes and in this
case it is $SO(N)$. As we said, what is preserved in the breaking
of gauge group through a displacement of branes is its rank.

At this point we have all the elements to write the partition
functions for the Type I superstring compactified on a circle. We
begin as usual with the torus partition function \be {\cal T} =
|V_8 - S_8 |^2 \sum_{m,n} \frac{ q^{\alpha' p_ R^2/4} \;
\bar{q}^{\alpha' p_L^2/4}}{\eta(\tau) \eta(\bar{\tau})} \ , \ee
where the modular invariant factor $1/(\tau_2^{4-1/2}(\eta\bar
\eta)^7)$ and the integral with the modular invariant measure
$d^2\tau/\tau_2^2$ are left implicit.

The Klein-bottle  amplitude propagates only the left-right
symmetric states, that are the states with vanishing windings (for
which $p_L=p_R$) \be {\cal K} = \frac{1}{2} \, (V_8 - S_8)
(2i\tau_2) \sum_{m} \frac{ \left(e^{-2\pi\tau_2}\right)^{\alpha'
m^2/2 R^2}}{\eta(2 i \tau_2)} \ . \label{k1s} \ee The transverse
channel amplitude is obtained expressing the Klein-bottle in terms
of the doubly covering torus modulus $t=2\tau_2$, then performing
an $S$-modular transformation, and finally a Poisson resummation
\be \tilde{\cal K} = \frac{2^5}{2} \frac{R}{\sqrt{\alpha'}} \ (V_8
- S_8) (i \ell) \sum_{n} \frac{ \left(e^{-2\pi\ell}\right)^{(2n
R)^2/4\alpha'}}{\eta(i \ell)} \ , \ee where the powers of $2$ can
be recovered considering the total power of $\tau_2$. We would
like to notice the overall factor of $R/\sqrt{\alpha'}$, that
takes into account the volume of the compact dimension. Moreover,
we see that in the transverse channel momenta are turned into
windings. In order to compare this expression with the one for the
transverse annulus, we have normalized the windings with the
factor $1/4\alpha'$, so that it is manifest that only even
windings contribute.

Actually, there is another way to project consistently the torus
amplitude in the Klein-bottle amplitude, giving to $\Omega$ an eigenvalue
$+1$ or $-1$ respectively for the even or odd momenta \be {\cal
K'} = \frac{1}{2} \, (V_8 - S_8) (2i\tau_2) \sum_{m}(-)^m \frac{
\left(e^{-2\pi\tau_2}\right)^{\alpha' m^2/2 R^2}}{\eta(2 i
\tau_2)} \ . \ee The phase $(-)^m$ after the Poisson resummation
gives a shift to the windings in the transverse channel \be
\tilde{\cal K'} = \frac{2^5}{2} \frac{R}{\sqrt{\alpha'}} \ (V_8 -
S_8) (i \ell) \sum_{n} \frac{ \left(e^{-2\pi\ell}\right)^{(2n+1)^2
R^2/4\alpha'}}{\eta(i \ell)} \ . \ee These amplitude does not
describe the propagation of massless states because in the sum
there is no a term starting with $q^0$. Therefore, there is not a
RR-tadpole and it is not necessary to introduce the open strings.
This model contains only unoriented closed strings.

Coming back to the first projection, one can write immediately the
transverse annulus, that propagates only states that in the torus
appear in the form $\chi^c \bar \chi$. Such states have
$p_L=-p_R$, so that they correspond to states with vanishing
momenta
\be \tilde{\cal A} = \frac{2^{-5}}{2}\, N^2 \,
\frac{R}{\sqrt{\alpha'}} \ (V_8 - S_8) \ (i \ell) \sum_{n} \
\frac{ \left(e^{-2\pi\ell}\right)^{n^2 R^2/ 4 \alpha'}}{\eta(i
\ell)} \ . \ee The direct channel is easily obtained with an
$S$-modular transformation and a Poisson resummation \be {\cal A}
= \frac{1}{2} \, N^2 \, (V_8 - S_8) ({\textstyle{1\over 2}}
i\tau_2) \sum_{m} \ \frac{ \left(e^{-2\pi\tau_2}\right)^{\alpha'
m^2/2 R^2}}{ \eta({\textstyle{1\over 2}} i \tau_2)} \ . \ee
Finally, we write the transverse M\"{o}bius amplitude \be
\tilde{\cal M} = - \frac{2}{2} \, N \, \frac{R}{\sqrt{\alpha'}} \
(\hat{V}_8 - \hat{S}_8) (i \ell+ {\textstyle\frac{1}{2}}) \sum_{n}
\ \frac{ \left(e^{-2\pi\ell}\right)^{(2n
R)^2/4\alpha'}}{\hat{\eta}(i \ell+ {\textstyle\frac{1}{2}} )} \ ,
\ee
that can propagate only even windings, since it describes a tube
that terminates at a boundary (that allows all windings) and at a
crosscap (that allows only even ones), and the corresponding
direct channel
\be {\cal M} = - \frac{1}{2} \, N \, (\hat{V}_8 - \hat{S}_8)
({\textstyle{1\over 2}} i \tau_2 + {\textstyle{1\over 2}})
\sum_{m} \ \frac{ \left(e^{-2\pi\tau_2}\right)^{\alpha' m^2/2
R^2}}{\hat{\eta} ({\textstyle{1\over 2}} i
\tau_2+{\textstyle{1\over 2}})} \ . \ee
Both the tadpoles of NS-NS and of R-R are cancelled fixing the
Chan-Paton group to be $SO(32)$, that corresponds to absorb the
charge of $N=32$ D$9$-branes with an O$9_+$-plane.

Now, as we already said, it is possible to break the gauge group,
introducing a Wilson line on the boundary \cite{bps}. The
consequent shift of momenta translates in the transverse channel
into different reflection coefficients for the different sectors
labelled by windings \be \tilde{\cal A} = \frac{2^{-5}}{2} \
\frac{R}{\sqrt{\alpha'}} \ (V_8 - S_8) \ (i \ell) \sum_{n} \
\frac{ \left(\rm{tr}W^n\right)^2\left(e^{-2\pi\ell}\right)^{n^2
R^2/ 4 \alpha'}}{\eta(i \ell)} \ . \ee Here $W$ is a diagonal
constant matrix of the form \be W=\rm{diag}\left(e^{2\pi i a_1},
e^{2\pi i a_2}, \  \ldots \ , e^{2\pi i a_{32}}\right) \ , \ee
with $a_2=-a_1, \ a_4=-a_3, \ \ldots a_{32}=-a_{31}$, $ \ 0 <
|a_{i}| < 1$, and its squared trace is \be (\rm{tr}W^n)^2 \ = \
\sum_{i,j}e^{2\pi i n (a_i+a_j)} \ . \ee The transverse M\"{o}bius
amplitude is also modified \be \tilde{\cal M} = - \frac{2}{2} \
\frac{R}{\sqrt{\alpha'}} \ (\hat{V}_8 - \hat{S}_8) (i \ell+
{\textstyle\frac{1}{2}}) \sum_{n} \ \frac{ \rm{tr}W^{2n}
\left(e^{-2\pi\ell}\right)^{(2n
    R)^2/4\alpha'}}{\hat{\eta}(i \ell+ {\textstyle\frac{1}{2}}
)} \ . \ee In the direct channel the $a_i$ are turned into shifts
of the momenta \ba {\cal A} &=& \frac{1}{2} \ (V_8 - S_8)
({\textstyle{1\over 2}} i\tau_2) \sum_{m,i,j} \ \frac{
\left(e^{-2\pi\tau_2}\right)^{\alpha' (m+a_i+a_j)^2/2 R^2}}{
\eta({\textstyle{1\over 2}} i \tau_2)} \ , \nonumber\\
{\cal M} &=& - \frac{1}{2} \ (\hat{V}_8 - \hat{S}_8)
({\textstyle{1\over 2}} i \tau_2 + {\textstyle{1\over 2}})
\sum_{m,i} \ \frac{ \left(e^{-2\pi\tau_2}\right)^{\alpha'
(m+2a_i)^2/2 R^2}}{\hat{\eta} ({\textstyle{1\over 2}} i
\tau_2+{\textstyle{1\over 2}})} \ , \ea where we note that ${\cal
M}$ propagates only the sectors with $a_i=a_j$, consistently with
the lack of orientability of the M\"{o}bius strip.

The gauge group $SO(32)$ is recovered if $a_i=0$ for each $i$. If
$a_i\neq 0$ for each $i$ and  $a_i\neq a_j $ for $i\neq j$, the
massless states are the ones satisfying $m+a_i+a_j=0$ that means
$m=0$ and $a_i=-a_j$. Therefore, from the annulus one reads $16$
massless vectors (the M\"{o}bius has no massless contribution)
giving a gauge group $U(1)^{16}$. This corresponds in the T-dual
picture to separating all the branes.

The case with $a_1\ldots a_{2M}$ all different and non vanishing
and all the others $a_i$ equal to zero gives a gauge group
$SO(32-2M)\times U(1)^M$, corresponding in the T-dual picture to
$M$ separated D$8$-branes (together with their images) and
$(32-2M)$ D$8$-branes coincident with the O-plane at the origin of
the compact dimension.

A case of particular interest is $a_1=a_3=\ldots a_{2M-1}=A$,
$a_2=a_4=\ldots a_{2M}=-A$ and $a_{2M+1}=\ldots a_{32}=0$. Letting
$N=32-2M$, the partition functions read
\ba {\cal A} &=&  (V_8 - S_8) ({\textstyle\frac{1}{2}}i \tau_2)
\sum_{m} \ \left\{ \left( M \bar{M} + \frac{1}{2} \, N^2 \right)
\frac{q^{\alpha'
m^2/2 R^2}}{\eta({\textstyle\frac{1}{2}} i \tau_2)} \right. \nonumber\\
& & \left.+ M N \
\frac{q^{\alpha' (m+A)^2/2 R^2}}{\eta(
{\textstyle\frac{1}{2}} i \tau_2)} +  \bar{M} N \
\frac{q^{\alpha' (m-A)^2/2 R^2}}{\eta({\textstyle\frac{1}{2}}i \tau_2)}
\right.\nonumber \\
& &\left.+
\frac{1}{2} \, M^2 \ \frac{q^{\alpha' (m+2A)^2/2
R^2}}{\eta({\textstyle\frac{1}{2}} i \tau_2)} +
\frac{1}{2}\, \bar{M}^2
\ \frac{q^{\alpha'
(m-2A)^2/2 R^2}}{\eta({\textstyle\frac{1}{2}} i \tau_2)} \right\} \, ,
\ea
\ba
{\cal M} &=& - (\hat{V}_8 - \hat{S}_8)
({\textstyle\frac{1}{2}} i \tau_2 + {\textstyle\frac{1}{2}})
\sum_{m} \ \left\{
 \frac{1}{2} \ N  \frac{q^{\alpha' m^2/2
R^2}}{\hat{\eta}(
{\textstyle\frac{1}{2}} i \tau_2+ {\textstyle\frac{1}{2}})}
\right.\nonumber \\
& &\left.+
\frac{1}{2} \ M \ \frac{q^{\alpha' (m+2A)^2/2
R^2}}{\hat{\eta}
({\textstyle\frac{1}{2}}i \tau_2+{\textstyle\frac{1}{2}})}
+ \frac{1}{2} \, \bar{M} \
\frac{q^{\alpha' (m-2A)^2/2 R^2}}{\hat{\eta}
({\textstyle\frac{1}{2}} i \tau_2+{\textstyle\frac{1}{2}})}
\right\} \, .
\ea
Here $M$ and $\bar M$ are numerically equal but the notation is
meant to stress that the two multiplicities refer to conjugate
charges. In the massless sector we read $N^2/2$ and $M\bar M$
states from the annulus amplitude, and $-N/2$ states from the
M\"{o}bius, giving the adjoint representation of the gauge group
$SO(N)\times U(M)$. Then, from the transverse channel \ba
\tilde{\cal A} &=& \frac{2^{-5}}{2} \ (V_8 - S_8) (i \ell) \
\frac{R}{\sqrt{\alpha'}} \sum_{n} \ \frac{ q^{n^2 R^2/ 4
\alpha'}}{\eta(i \ell)} \ \left( N + M \, e^{2 i \pi A n} +
\bar{M} \, e^{- 2 i \pi A n}    \right)^2 \ ,\nonumber\\
\tilde{\cal M} &=& - \frac{2}{2} \
(\hat{V}_8 - \hat{S}_8)  (i \ell + {\textstyle\frac{1}{2}}) \
\frac{R}{\sqrt{\alpha'}}
 \sum_{n} \
\frac{ q^{(2n R)^2/ 4\alpha'}}{\eta(i \ell+ {\textstyle\frac{1}{2}})}
\ \left( N + M\,
e^{4 i \pi A n} +
\bar{M}\, e^{- 4 i \pi A n} \right) ,\nonumber\\
\ea we can extract the tadpole conditions, corresponding to fixing
$N=32-2M$. In the T-dual picture, the gauge group $SO(N)\times
U(M)$ is understood to emerge from a stack of $32-2M$ D$8$-branes
on the top of the O-plane at the origin of the compact dimension
and a stack of $M$ branes at $X^T\sim \pi R^T A$, where $X^T$ is
the T-dual compact coordinate.

If $A=1/2$, the $U(M)$ factor is enhanced to $SO(2M)$. In fact, in
this case, $M$ and $\bar M$ have the same reflection coefficients
in the transverse channel, so that the direct amplitudes have to
depend only from their sum. In the T-dual picture this means that
two stacks of branes coincide with the two O-planes placed at
$X^T=0$ and $X^T=\pi R^T$.

A $d$ dimensional torus can be defined as the result of the
involution $T^d=\mathbb{R}^d/\Lambda$, where
$\Lambda=\{\sum_{a=1}^d n^a\vec{e}_a: \ n^a\in\mathbb{Z}\}$ is the
torus lattice. One can also define a dual lattice
$\Lambda^*=\{\sum_{a=1}^d m_a\vec{e}^{*a}: \ m_a\in\mathbb{Z}\}$,
where $\vec{e}_a\cdot\vec{e}^{*b}=\delta_a^b$. We have to think of
the windings $w^I$ in the lattice $\Lambda$ and of the momenta
$p_I$ in the dual lattice $\Lambda^*$. Defining
$\Pi_{L,R}=\sqrt{\alpha'}p_{L,R}/\sqrt{2}$, the Lorentzian lattice
$\Gamma_{dd}=(\Pi_L,\Pi_R)$ has to be even and self-dual, in order
to have respectively the invariance under a $T$ and an $S$-modular
transformation.

We would like to stress that, together with the internal metric
$G_{IJ}$, one can turn on an antisymmetric 2-tensor $B_{IJ}$ that
modifies the definition of $p_{L,R}$ \be
p_{I}^{L,R}=m_I\pm\frac{1}{\alpha'}(G_{IJ}-B_{IJ})n^J \ . \ee The
presence of such a tensor affects only the annulus and the
M\"{o}bius, but in order to have consistency with the orientifold
projection, one has to impose a quantization condition for it:
$2B_{IJ}/\alpha'\in \mathbb {Z}$. The fact is that $B_{IJ}$ allows
one to break the gauge group reducing the rank \cite{bps,
bianchitor,wittor,kuej}, in contrast to the mechanism of
displacement of branes we already saw. For example, in absence of
Wilson lines, the gauge group $SO(32)$ of Type I superstring is
reduced to $SO(2^{5-r/2})$, if $r$ is the rank of $B_{IJ}$.
%
\section{Orbifold compactification: $T^4/\mathbb{Z}^2$ orbifold}
\subsection{Orbifolds in general and an example: $S^1/\mathbb{Z}_2$}
Let us consider a manifold $\mathcal{M}$ on which a
discrete symmetry group $G$ acts. An orbifold \cite{dhvw} is the quotient space $\mathcal{M}/G$
obtained identifying points $X$ of $\mathcal{M}$ under the equivalence
relation $X\sim g X$ for all $g \in G$ . The action of the group in
general leaves a number of fixed points $x_0$, such that $x_0=gx_0$ for some
$g\in G$, in which the quotient space has
conical singularities. In spite of them, the spectrum of the resulting string
theory is well defined. Moreover it is
always possible to remove such singularities ``blowing up'' some
moduli, obtaining in such a way a smooth manifold.

A very simple example is the case of $\mathcal{M}=S^1$ and
$G=\mathbb{Z}_2$ \cite{Ginsp}. The coordinate on the circle is
$X\sim X+2\pi R$, while the action of $\mathbb{Z}_2$ is defined by
the generator $g \ : \ X \ \rightarrow \ -X$. The orbifold
$S^1/\mathbb{Z}^2$ is a segment whose extrema $X=0, \pi R$ are the
fixed points of the involution. The issue is how to project
the Hilbert space of a modular invariant theory in a subspace
invariant under the group action, consistently with the interaction of
states, and preserving modular invariance. Let
$Z_{\mathcal{T}}$ be the partition function of a modular invariant
theory, for example the torus amplitude of a given closed-string
model. The projection is obtained inserting in the trace over the
states the operator \be P=\frac{1}{|G|}\sum_{g\in G} g \ , \ee
where $|G|$ is the number of elements in the group $G$. The new
boundary conditions along the time direction of the torus, for a
generic field $X(z)$, are $X(z+\tau)=gX(z)$ for each $g\in G$,
where $z$ is the analytic coordinate on the torus. If we denote
with $\mathcal{T}_{\mathbf{1},g}$ the torus partition function
with g-twisted boundary condition along the time direction, the
projected partition function is
\be
\frac{1}{|G|}\sum_{g\in G} \mathcal{T}_{\mathbf{1},g} \ .
\ee
The resulting theory is clearly not modular invariant because, for
example, an $S$-modular transformation interchanges the
``horizontal'' and the ``vertical'' sides of the torus, thus
mapping $\mathcal{T}_{\mathbf{1},g}$ in
$\mathcal{T}_{g,\mathbf{1}}$. In order to recover modular
invariance, one has to add to the previous untwisted sector
twisted sectors corresponding to twisted boundary conditions along
the spatial direction, so that $X(z+1)=hX(z)$. The resulting modular
invariant partition function of the orbifold $\mathcal{T}/G$ is then
\be
Z_{\mathcal{T}/G}=\frac{1}{|G|}\sum_{h,g\in
  G}\mathcal{T}_{h,g} \ .
\label{abelian}
\ee
The Hilbert space decomposes into a set of
twisted sectors labelled by $h$, and each twisted sector is
projected onto $G$-invariant states. Actually,
(\ref{abelian}) applies to an abelian group $G$, for which the
total number of sectors is equal to the number of elements in the
group. For a non-abelian group, matters are a bit different. In
fact, the $gh$ and $hg$ twisted boundary conditions, corresponding
respectively to $ghX(z)=X(z+\tau+1)$ and $hgX(z)=X(z+\tau+1)$, are
ambiguous unless $gh=hg$. Thus, for a non-abelian group $G$ the
summation in (\ref{abelian}) has to be restricted to those $g,h\in
G$ such that $gh=hg$. In this way, also the number of sectors
changes, and in particular it is equal to the number of conjugacy
classes. This is because, given a certain twisted sector defined
by $hX(z)=X(z+1)$, and acting on it with an operator $g$
satisfying $gh=hg$, one obtains an equivalent sector,
$g^{-1}hgX(z)=X(z+1)$, and the statement is proved since $h$ and
$g^{-1}hg$ belong to the same conjugacy class.

Let us begin to study the simplest orbifold,
$S^1/\mathbb{Z}_2$, beginning from a free boson $X$ on a circle,
whose partition function is
\be
 Z \ = \ (q\bar q)^{-\frac{1}{24}} \
{\rm
  tr}\left(q^{N_B+\frac{\alpha'}{4}p_R^2}
 \ \bar{q}^{\bar N_B +\frac{\alpha'}{4} p_L^2} \right) \ .
\ee
In order to build the untwisted sector of the orbifold,
we have to introduce in the trace the projector operator $P=(1+g)/2$,
where $g : X \rightarrow -X$. This means that the operator $g$ acts on a
general state of momentum $m$ and winding $n$ as
\be
 g\prod_{i=1}^{N}\alpha_{-n_i}\prod_{j=1}^{\bar N} \bar \alpha_{-n_j}
\ |m,n\rangle\ = \ (-)^{N+\bar N}\prod_{i=1}^{N}\alpha_{-n_i} \ \prod_{j=1}^{\bar N} \bar \alpha_{-n_j}
\ |-m,-n\rangle \ .
\ee
Therefore, after the action of $g$, only the states
with $m=n=0$ survive. Moreover, if we denote with $N_n$ the number of oscillators  with
frequency $n$, each factor $q^n$ acquires a sign, and the resulting trace becomes
\be
{\rm tr} \left(q^{N_B+\frac{\alpha'}{4}p_R^2}g\right) =
\prod_{n=1} \sum_{N_n=0} (-q^n)^{N_n}=\prod_{n=1}\frac{1}{1+q^n} \ .
\ee
The partition function for the untwisted sector is then
\ba
Z_{untwisted} &=& \frac{1}{2}Z_{circle}(R)+\frac{1}{2}\frac{(q\bar
  q)^{-\frac{1}{24}}}{\prod_{n=1}(1+q^n)(1+\bar q^n)}\nonumber\\
& & = \frac{1}{2}Z_{circle}(R)+\left|\frac{\eta}{\vartheta_2}\right| \ ,
\ea
where
\be
Z_{circle}=\frac{\sum_{m,n} q^{\frac{\alpha'}{4}p_R^2}
 \ \bar{q}^{\frac{\alpha'}{4} p_L^2} }{\eta \bar \eta} \ .
\ee
In practice, we decomposed the Hilbert space into two subspaces
with eigenvalues $g=\pm 1$
\ba
H^{\pm} &=& \left\{ \alpha_{-n_1}\ldots\alpha_{-n_l}\bar{
\alpha}_{-n_{l+1}}\ldots \bar{\alpha}_{-n_{2k}} \  ( \ |m,n \rangle \pm
|-m,-n \rangle  \ ) \right\} \nonumber\\
& &
+\left\{ \alpha_{-n_1}\ldots\alpha_{-n_l}\bar{
\alpha}_{-n_{l+1}}\ldots \bar{\alpha}_{-n_{2k+1}} \  ( \ |m,n \rangle \mp
|-m,-n \rangle  \ ) \right\} , \nonumber\\
\ea and we projected away the $g=-1$ eigenstates.

Now, modular invariance is recovered adding the twisted
sector. $Z_{circle}$ is already a modular invariant, while the
term $\left|\eta/\vartheta_2\right|$ under an $S$ transformation
gives \be S: \left|\frac{\eta}{\vartheta_2}\right| \
\longrightarrow \left|\frac{\eta}{\vartheta_4}\right| \ . \ee Then
this term has to be $\mathbb{Z}_2$ projected \be g:
\left|\frac{\eta}{\vartheta_4}\right| \ \longrightarrow
\left|\frac{\eta}{\vartheta_3}\right| \ , \ee and the partition
function of the orbifold is finally given by \be Z_{S^1/\mathbb{Z}_2} \ =
\ \frac{1}{2}\left(
Z_{circle}(R)+2\left|\frac{\eta}{\vartheta_2}\right|
+2\left|\frac{\eta}{\vartheta_4}\right|
+2\left|\frac{\eta}{\vartheta_3}\right|\right) \ , \ee that, using
the identity $\vartheta_2\vartheta_3\vartheta_4=2\eta^3$, can also
be written \be Z_{S^1/\mathbb{Z}_2} \ = \ \frac{1}{2}\left(
Z_{circle}(R)+\frac{\left|\vartheta_3\vartheta_4\right|}{\eta\bar\eta}
+\frac{\left|\vartheta_2\vartheta_3\right|}{\eta\bar\eta}
+\frac{\left|\vartheta_2\vartheta_4\right|}{\eta\bar\eta} \right)
\ . \ee
Modular invariance, that can be checked directly, can be justified
because the four terms in $Z_{S^1/\mathbb{Z}_2}$
correspond to the bosonic determinant on the torus with all
possible boundary conditions. In fact, if ``$+$'' and ``$-$''
stand respectively for periodic and antiperiodic boundary
conditions, $Z_{S^1/\mathbb{Z}_2}$ is the sum
$\mathcal{T}_{(++)}+\mathcal{T}_{(+-)}+\mathcal{T}_{(-+)}+\mathcal{T}_{(--)}$,
where $\mathcal{T}_{(++)}=Z_{circle}$ is the only possible
structure for a boson before the orbifold projection. Now, under
an $S$-modular transformation $\mathcal{T}_{(\pm\pm)}
\leftrightarrow \mathcal{T}_{(\pm\pm)}$, and $\mathcal{T}_{(+-)}
\leftrightarrow \mathcal{T}_{(-+)}$, while under a $T$-modular
transformation $\mathcal{T}_{(\pm +)} \rightarrow
\mathcal{T}_{(\pm +)}$, and $\mathcal{T}_{(+-)} \rightarrow
\mathcal{T}_{(--)}$. Let us stress that in the twisted sector the
boson has an antiperiodic boundary condition along the spatial
direction of the torus, now allowed thanks to the
orbifold identification $X\sim -X$.

The mode expansion of $X$ in the twisted sector is \be X \ = \ x_0
\ + \ i\frac{\sqrt{2\alpha'}}{2} \ \sum_{n\in \mathbb{Z}+1/2}
\left(\frac{\alpha_n}{n} \ e^{-2in(\tau-\sigma)}+
\frac{\tilde\alpha_n}{n} \ e^{-2in(\tau+\sigma)}\right) \ , \ee
where $n$ is half-integer, and windings and momenta are forced to
vanish by the antiperiodic boundary condition. Here the zero mode
$x_0$ can be only one of the two fixed points $x_0=0,\pi R$. The
multiplicity of $2$ in the twisted sector is just due to the fact
that one has only two possible choices for the point, actually a
fixed point, around which the filed $X$ can be expanded, with a consequent doubling
of twisted sectors. Thus, in
general, such a multiplicity is equal to the total number
of fixed points.
Moreover, expanding the twisted sector in powers of $q \bar q$, we
see that it starts with $(q\bar q)^{1/48}$, where $1/48$ is just
the shift to the vacuum energy introduced by an antiperiodic boson.
Such a value can be recovered acting with the $\mathbb{Z}_2$-twist
operator, that interchanges periodic and  antiperiodic boundary
conditions, whose conformal weight is $h=1/16$, on the vacuum
state of the untwisted sector, whose shift to the vacuum energy,
due to the holomorphic part of a periodic boson, is $-1/24$. Thus,
the vacuum of the twisted sector is lifted in energy with respect
to the vacuum of the untwisted one.
\subsection{Orbifold $T^4/\mathbb{Z}_2$}
In this subsection we want to study an interesting example of
orbifold compactification of the type IIB superstring theory and
its orientifold. Let us thus consider the space-time
$\mathcal{M}_{10}=\mathcal{M}_6\times T^4/\mathbb{Z}_2$.

The starting point is the torus partition function for
$T^4$ compactification,
\be
\mathcal{T}_{++} \ = \ |V_8-S_8|^2 \Sigma_{m,n}
\ee
where the contribution of the four transverse bosons in
$\mathcal{M}_6$ is left implicit, and $\Sigma_{m,n}$ is the sum over
the lattice of the internal manifold, whose metric is denoted by $g$
\be \Sigma_{m,n}= \sum_{m,n} {q^{{\alpha' \over 4} p_{L}^{{\rm T}}
g^{-1} p_{ L}} \bar q^{{\alpha' \over 4} p_{R}^{{\rm T}} g^{-1}
p_{R}} \over \eta^4 \bar \eta ^4 } \ . \ee
It is convenient
to decompose the characters of $SO(8)$ in representations of
$SO(4)\times SO(4)$ where the first $SO(4)$ is the light-cone of
$\mathcal{M}_6$
\ba V_8 &= V_4 O_4 + O_4 V_4 \, , \qquad O_8
&= O_4 O_4 + V_4 V_4 \, ,
\nonumber \\
S_8 &= C_4 C_4 + S_4 S_4 \, , \qquad C_8 &= S_4 C_4 + C_4 S_4 \, .
\ea
Moreover, we introduce the four combinations of characters
\ba
\label{so4}
& Q_o = V_4 O_4 - C_4 C_4 \, , \qquad \quad
& Q_v = O_4 V_4 - S_4 S_4 \, ,
\nonumber \\
& Q_s = O_4 C_4 - S_4 O_4 \, , \qquad \quad
& Q_c = V_4 S_4 - C_4 V_4 \, ,
\ea
that are the eigenvectors of $\mathbb{Z}_2$. In fact, in order to have consistency
between the action of $\mathbb{Z}_2$ and world-sheet
supersymmetry \cite{dhvw,4d}, the internal $O_4$ and $C_4$ have to be even under
$\mathbb{Z}_2$, while the internal $V_4$ and $S_4$ have to be odd.
In this way, $Q_o$ and $Q_s$ are the positive eigenvectors, while $Q_v$
and $Q_c$ are the negative ones.

At this point we proceed with the orbifold projection, and to this end we recall
that in the untwisted sector only the zero mode of the sum can be
projected, while all the contributions of all other states have to be simply halved.
Thus we separate the contribution of the zero mode in the sum,
denoting with $\Sigma'$ the sum without the zero mode \be
\Sigma_{m,n}=\Sigma_{m,n}'+\frac{1}{(\eta\bar \eta)^4} \ . \ee
Then, letting \be \lambda_{++}=V_8-S_8=Q_o+Q_v \ , \qquad
\Lambda_{++}=\frac{1}{\eta^4} \ , \qquad
\rho_{++}=\lambda_{++}\Lambda_{++} \ , \ee we can write ${\cal
T}_{++}$ as \be {\cal T}_{++} \ = \
|\rho_{++}|^2+|\lambda_{++}|^2\Sigma' \ , \ee where the projection
involves only $|\rho_{++}|^2$.

Let us continue as we learnt from the previous subsection \ba
\mathbb{Z}_2 : \lambda_{++} &=& Q_0+Q_v \ \longrightarrow \ \lambda_{+-}=Q_o-Q_v \nonumber\\
S : \lambda_{+-} &=& Q_0-Q_v \ \longrightarrow \ \lambda_{-+}=Q_s+Q_c \nonumber\\
\mathbb{Z}_2 : \lambda_{-+} &=& Q_s+Q_c \ \longrightarrow \
\lambda_{--}=Q_s-Q_c \ ,
\ea
with the action of the $S$-modular transformation on
$O_4$, $V_4$, $S_4$, $C_4$ defined by
\ba
S = \frac{1}{2} \ \left( \begin{array}{rrrr}
1 & 1 & 1 & 1 \\
1 & 1 & -1 & -1 \\
1 & -1 & -1 & 1 \\
1 & -1 & 1 & -1
\end{array} \right) \, .
\ea
On the other hand, we already know that
\ba
\mathbb{Z}_2 &:& \Lambda_{++} = \frac{1}{\eta^4} \ \longrightarrow \
\left(\frac{2\eta}{\vartheta_2}\right)^2=\Lambda_{+-}\nonumber\\
S &:& \Lambda_{+-}
\ \longrightarrow \ 4\Lambda_{-+}=4\left(\frac{\eta}{\vartheta_4}\right)^2\nonumber\\
\mathbb{Z}_2 &:& 4\Lambda_{-+} \ \longrightarrow \
4\Lambda_{--}=4\left(\frac{\eta}{\vartheta_3}\right)^2 \ .
\ea
Therefore, defining in analogy to $\rho_{++}$ also
$\rho_{+-}=\lambda_{+-}\Lambda_{+-} \ $,
$\rho_{-+}=\lambda_{-+}\Lambda_{-+} \ $ and
$\rho_{--}=\lambda_{--}\Lambda_{--} \ $, the torus partition function can
be written
\be
{\cal T} \ = \
\frac{1}{2}\left(|\rho_{++}|^2+|\rho_{+-}|^2+|\lambda_{++}|^2\Sigma_{m,n}'\right)+
\frac{16}{2}\left(|\rho_{-+}|^2|\rho_{--}|^2\right) \ ,
\ee
or better in the form
\be
{\cal T} = \frac{1}{2} \Biggl[ |Q_o + Q_v|^2
\Sigma_{m,n}
 + |Q_o - Q_v |^2 \left| {2 \eta \over
\vartheta_2} \right|^4
+ 16 |Q_s + Q_c |^2 \left| {\eta \over \vartheta_4}\right|^4
+ 16 |Q_s - Q_c |^2 \left| {\eta \over \vartheta_3} \right|^4 \Biggr]\, ,
\nonumber\\
\ee
where the factor $16=2^4$ in front of the twisted sector takes into account
the total number of fixed points for $T^4/\mathbb{Z}_2$.

In order to read the massless spectrum, it is convenient to introduce
the following characters
\ba
\chi_{++}&=&\frac{\rho_{++}+\rho_{+-}}{2} \ = \ Q_o \
\frac{\Lambda_{++}+\Lambda_{+-}}{2} \ + \
Q_v \ \frac{\Lambda_{++}-\Lambda_{+-}}{2} \nonumber\\
\chi_{+-}&=&\frac{\rho_{++}-\rho_{+-}}{2} \ = \ Q_o \
\frac{\Lambda_{++}-\Lambda_{+-}}{2} \ + \
Q_v \ \frac{\Lambda_{++}+\Lambda_{+-}}{2}\nonumber\\
\chi_{-+}&=&\frac{\rho_{-+}+\rho_{--}}{2} \ = \ Q_s \
\frac{\Lambda_{-+}+\Lambda_{--}}{2} \ + \
Q_c \ \frac{\Lambda_{-+}-\Lambda_{--}}{2}\nonumber\\
\chi_{--}&=&\frac{\rho_{-+}+\rho_{--}}{2} \ = \ Q_s
\frac{\Lambda_{-+}-\Lambda_{--}}{2} \ + \
Q_c \ \frac{\Lambda_{-+}+\Lambda_{--}}{2} \ ,\nonumber\\
\ea in terms of which the torus amplitude becomes \be
\label{torus_chi} {\cal T} \ = \ \frac{1}{2}
|\lambda_{++}|^2\Sigma_{m,n}'+|\chi_{++}|^2+|\chi_{+-}|^2+16|\chi_{-+}|^2
+16|\chi_{--}|^2 \ . \ee The sum $\Sigma_{m,n}'$ gives an infinite
tower of massive states.
Then, let us recall that the conformal weight of $S_{2n}$ and
$C_{2n}$ for a generic $n$ is $h=n/8$, so that for $n=2$ the
weight of the two Weyl spinors is $h=1/4$. Moreover, the weights
of $V_4$ and $O_4$ are respectively $h=1/2$ and $h=0$.
The last ingredient needed for studying the massless
spectrum is the lifting of the vacuum energy in the twisted
sector. The NS vacuum state in the untwisted sector has squared
mass proportional to $-1/2$, as usual. Then one has to lift such a
value by a quantity equal to $1/4$ to obtain a shift to the vacuum
energy in the twisted sector equal to $-1/2+1/4=-1/4$. This is
because in general the vacuum of the $k^{th}$ twisted sector for a
$\mathbb{Z}_N$ orbifold is shifted of the quantity $k/N-(k/N)^2$.

After the previous discussion, is now easy to understand what are
the massless contributions of the characters $\chi$. For example,
$\chi_{++}$ has the first addend that starts with
$q^{h_{Q_o}-1/2}=q^0$, where $h_{Q_o}=1/2$ and $-1/2$ is the shift
to the vacuum energy in the untwisted sector. Therefore, this term
is massless. On the other hand, the second addend of $\chi_{++}$
is massive because in the difference $\Lambda_{++}-\Lambda_{+-}$
the power $q^{-1/2}$ cancels. Thus, at the massless level
$\chi_{++}$ gives $V_4-2C_4$, where the multiplicity $2$ is
provided by the internal $C_4$. The same argument can be applied
to $\chi_{+-}$, that at the massless level gives $4O_4-2S_4$ where
here the factor $4$ is given by the internal $V_4$. In the twisted
sector the shift of the vacuum energy is $-1/4$, so that only the
first addend of $\chi_{-+}$, starting with $q^{h_{Q_s}-1/4} \ $,
$h_{Q_s}=1/4$, contains a massless contribution, that is $2O_4-S_4$,
while the character $\chi_{--}$ has only massive contributions.

At this point we can write the massless spectrum. We collect it in
the table \ref{tabe} where, anticipating the Klein-bottle projection, we
prefer to indicate also which states are symmetric and which
are antisymmetric under the interchange of left and right modes. The
resulting massless oriented closed states organize themselves in
multiplets of $\mathcal{N}=(2,0)$ in $D=6$ and are: 1
gravitational multiplet, that contains the graviton, five
self-dual 2-forms and two left-handed gravitinos, and 21 tensor
multiplets, 16 of which from the twisted sectors, each
containing an antiself-dual 2-form, five scalars and two
right-handed spinors. This is the unique anomaly-free spectrum in
$D=6$ with
this supersymmetry.
\begin{table}
\begin{center}
\begin{tabular}{|c|c|c|c|}
\hline
 & & & \\
 $m^2=0$ & NS-NS & R-R & R-NS $+$ NS-R \\
 & & & \\
\hline
 & & & \\
$|\chi_{++}|^2$ & $g_{\mu\nu}$, $\phi$ $\quad$ (s) &
3$B_{\mu\nu}^+$, $\beta$ $\quad$ (s) &  2$\psi_L^\mu$, 2$\psi_R$\\
 & $B_{\mu\nu}=B_{\mu\nu}^++B_{\mu\nu}^-$ $\quad$ (a) & $B_{\mu\nu}^+$, 3$\beta$ $\quad$ (a) & \\
 & & & \\
\hline
 & & & \\
$|\chi_{+-}|^2$ & 10$\phi$ $\quad$ (s) & 3$B_{\mu\nu}^-$, $\beta$
 $\quad$ (s)
  & 8$\psi_R$\\
 & 6$\phi$ $\quad$ (a) & $B_{\mu\nu}^-$, 3$\beta$ $\quad$ (a) & \\
 & & & \\
\hline
 & & & \\
 16 $|\chi_{-+}|^2$ & $16\times 3$ $\phi$ $\quad$ (s) &
 16$B_{\mu\nu}^-$ $\quad$ (s) & 32$\psi_R$ \\
 & 16$\phi$ $\quad$ (a) & 16$\beta$ $\quad$ (a)&  \\
 & & & \\
\hline
\end{tabular}
\end{center}
\caption{Massless spectrum for the oriented closed sector. Here (s)
  and (a) indicate respectively states that are symmetric or antisymmetric with
  respect to the interchange of left and right modes. $\beta$ stands for a
  scalar in the RR sector. }
\label{tabe}
\end{table}
We stress that here the orbifold compactification broke half of the
supersymmetries. In fact, the simple toroidal compactification from
$\mathcal{N}=(2,0)$ in $D=10$ to $D=6$ would give $\mathcal{N}=(4,0)$,
while here we have $\mathcal{N}=(2,0)$ in $D=6$. This can be understood
thinking that the orbifold $T^4/\mathbb{Z}_2$ is the singular limit
of a smooth manifold K3 with holonomy group $SU(2)$, so that
only half of the supersymmetries are preserved after the
compactification \cite{aspin}.

The Klein-bottle projection is simple to write starting from
(\ref{torus_chi})
\be
{\cal K} \ = \ \frac{1}{4}(Q_o+Q_v)(P_m'+W_n')+
\frac{1}{2}\left[\chi_{++}+\chi_{+-}+16(\chi_{-+}+\chi_{--})\right] \ ,
\ee
where $P_m$ and $W_n$ are sums respectively over momenta and windings
\be
P_m(q) = \sum_m \; {q^{{\alpha'\over 2} m^{{\rm T}} g^{-1} m} \over \eta(q)^4} \, , \qquad
W_n(q) = \sum_n \; {q^{{1\over 2\alpha'} n^{{\rm T}} g n} \over \eta(q)^4} \, ,
\ee
while the corresponding ``primed'' sums are defined removing the
lattice zero mode from them.
In the Klein-bottle amplitude now not only momenta, but also
windings can flow. The reason is that on an orbifold $X$ is identified with
$-X$, so that  $p_{L,R} \sim -p_{L,R}$. This means that the
states with $p_L=-p_R$, corresponding to winding states, survive
to the orientifold projection and propagate on the Klein-bottle.
At this point, recalling that $\chi_{++}+\chi_{+-}=(Q_o+Q_v)/\eta^4$,
and that $\chi_{-+}+\chi_{--}=(Q_s+Q_c)(\eta/\vartheta_4)^2$, we can
write
\be
{\cal K} = \frac{1}{4} \Biggl[ (Q_o + Q_v) \left(
P_m + W_n  \right) + 2 \times 16 \  (Q_s + Q_c) \left(
\frac{\eta}{\vartheta_4} \right)^2 \Biggr] \, ,
\label{Klein_orb}
\ee
where the factor 16 reflects the fact that all 16 twisted sectors are
projected in the same way.

The massless spectrum is simply obtained from the one of the torus
projecting away the left-right antisymmetric states in the NS-NS
sector, and the left-right symmetric states in the R-R, while the  mixed NS-R
sectors have to be halved. The resulting massless states accommodate in
multiplets of $\mathcal{N}=(1,0)$ in $D=6$, and precisely  give one
gravitational multiplet, containing the graviton a self-dual 2-form
and a left-handed gravitino, one tensor multiplet, containing an antiself-dual
2-form a scalar and a right-handed spinor, 20 hyper multiplets, of
which 16 from the twisted sector, each of which containing four scalars and a
right-handed spinor.

The transverse channel for the Klein-bottle amplitude is \be \tilde{\cal
K} = \frac{2^5}{4} \left[ (Q_o + Q_v) \left( v_4 W_n^{(e)} +
\frac{1}{v_4} P_m^{(e)} \right) + 2 (Q_o - Q_v)  \left( \frac{2
\eta}{\vartheta_2} \right)^2 \right] \, , \ee where
$v_4=\sqrt{{\rm det} g /(\alpha')^4}$ is proportional to the
internal volume, and the sums are restricted over the even
windings and momenta.

We want to underline that, despite we use the same notations for
the sums in the direct and transverse channels, the sums in
the transverse channel are defined in a slightly different way \be
P_m = \sum_m \; {q^{{\alpha'\over 4} m^{{\rm T}} g^{-1} m} \over
\eta(i\ell)^4} \, , \qquad W_n = \sum_n \; {q^{{1\over 4\alpha'}
n^{{\rm T}} g n} \over \eta(i\ell)^4} \, , \ee
with $q=e^{-2\pi \ell}$.

At the massless level the transverse Klein-bottle amplitude is
\be
\tilde{\cal K}_0 = \frac{2^5}{4} \left[ Q_o \left(
\sqrt{v_4} + \frac{1}{\sqrt{v_4}} \right)^2  +
Q_v \left(
\sqrt{v_4} - \frac{1}{\sqrt{v_4}} \right)^2
 \right] \, ,
\ee from which one can extract the content of O-planes, whose
tension and R-R charge can be read from $Q_o=V_4O_4-C_4C_4$. The
term proportional to $v_4$ corresponds to the propagation of
states between the usual O9-planes. Now, four T-dualities along
the internal directions map O9-planes into O5-planes and $v_4$ in
$1/v_4$. Thus the term proportional to the inverse of the volume
corresponds to the propagation between O5-O5. Finally there is a
term proportional to $2\sqrt{v_4}\times 1/\sqrt{v_4}$
corresponding to the exchange between O5-O9 and O9-O5.
The presence of both O5 and O9-planes requires the simultaneous
presence of D5 and D9-branes in order to saturate their tensions
and charges.

Before continuing with the open sector, we want to mention that there
are other consistent Klein-bottle projections. For example, one can project
even and odd windings and momenta, along one or more internal
dimensions, in different ways. Moreover, one can symmetrize half of the
16 identical contributions in the twisted sector and antisymmetrize
the other part
\ba
{\cal K} &=& \frac{1}{4} \Biggl[ (Q_o + Q_v) \left(
\sum_m (-1)^m \, {q^{{\alpha'\over 2} m^{{\rm T}} g^{-1} m} \over \eta^4}
+ \sum_n (-1)^n \,
{q^{{1\over 2\alpha'} n^{{\rm T}} g n} \over \eta^4}
\right)
\nonumber \\
& & + 2 \times (8-8) (Q_s + Q_c) \left( \frac{\eta}{\vartheta_4}
\right)^2 \Biggr] \, . \ea
Such a projection does not require the introduction of
open strings, because there are no tadpoles to cancel. In
fact, the signs shift windings and momenta in the transverse
channel, and so the corresponding $\tilde{\cal K}$ starts with
massive states.

Actually, there is a third consistent projection that differs from
the first one for a sign in front of the twisted sector. The
corresponding orientifold breaks supersymmetry in the open sector, and
so we will discuss it in the next chapter.

At this point we proceed with the orientifold construction, writing the
annulus amplitude. From $\tilde{\cal K}_0$ it is clear that
the transverse annulus, at the massless level, must contain the
following untwisted contribution,
\be
\tilde{\cal A}^{(u)}_0=\frac{2^{-5}}{4} \left[ Q_o \left(
N \sqrt{v_4} + \frac{D}{\sqrt{v_4}} \right)^2 +
Q_v \left( N \sqrt{v_4} - \frac{D}{\sqrt{v_4}} \right)^2\right] \ ,
\ee
where $N$ and $D$ count the multiplicities of the open string extrema
respectively with Neumann and Dirichlet boundary conditions.
One can then complete the lattice sums
\be
\tilde{\cal A}^{(u)}=
\frac{2^{-5}}{4} \left[ (Q_o+Q_v) \left(
N^2 v_4 W_n + D^2 P_m\frac{1}{v_4} \right) +
2ND(Q_o-Q_v) \ \left(
\frac{2 \eta}{\vartheta_2} \right)^2\right] \ ,
\ee
and perform an $S$-modular transformation, to obtain
\be
S : \tilde{\cal A}^{(u)} \longrightarrow
\frac{1}{4} \left[ (Q_o + Q_v) \left(
N^2 P_m + D^2 W_n\right) +
2 N D \, (Q_s + Q_c ) \left( {\eta \over \vartheta_4}\right)^2
\right] \, .
\ee
Finally, one has to project this quantity under $\mathbb{Z}_2$,
taking into account that the orbifold involution acts also on
the Chan-Paton charges, thus projecting $N$ in $R_N$ and $D$ in $R_D$ \cite{ps}.
Therefore, the complete expression for the annulus amplitude reads
\ba
\label{anellosuperorbifold}
{\cal A} &=& \frac{1}{4} \Biggl[ (Q_o + Q_v) \left(
N^2 P_m + D^2 W_n\right)  +
\left(R_N^2 + R_D^2 \right) (Q_o - Q_v) \left( {2\eta \over
\vartheta_2}\right)^2
\nonumber\\
& & + 2 N D \, (Q_s + Q_c ) \left( {\eta \over \vartheta_4}\right)^2
+ 2 R_N R_D \, (Q_s - Q_c ) \left( {\eta \over \vartheta_3}\right)^2
\Biggr] \, ,
\ea
while in the transverse channel
\ba
\tilde {\cal A} &=& \frac{2^{-5}}{4} \Biggl[ (Q_o + Q_v) \left(
N^2 v_4 W_n + \frac{D^2}{v_4}P_m \right)  +
2 N D \, (Q_o - Q_v ) \left( {2 \eta \over \vartheta_2}\right)^2
\nonumber \\
& & + 16 \left(R_N^2 + R_D^2 \right) (Q_s + Q_c) \left( {\eta \over
\vartheta_4}\right)^2 - 2 \times 4 R_N R_D \, (Q_s - Q_c )
\left( {\eta \over \vartheta_3}\right)^2
\Biggr] \, . \nonumber\\
\label{Annulus_orb}
\ea
Let us stress that the twisted sector corresponds to mixed
Neumann-Dirichlet boundary conditions.
From the massless level of $\tilde {\cal A}$
\ba
\tilde {\cal A}_0 &=& \frac{2^{-5}}{4} \Biggl\{ Q_o \left(
N \sqrt{v_4} + \frac{D}{\sqrt{v_4}} \right)^2 +
Q_v \left( N \sqrt{v_4} - \frac{D}{\sqrt{v_4}} \right)^2
\nonumber \\
& + & Q_s \left[ 15 R_N^2 + \left(R_N - 4 R_D \right)^2 \right]+
Q_c \left[ 15 R_N^2 + \left(R_N + 4 R_D \right)^2 \right] \Biggr\}
\, , \ea one can distinguish clearly the contributions of open
strings with Neumann-Neumann, Neumann-Dirichlet and
Dirichlet-Dirichlet boundary conditions, corresponding
respectively to D9-D9, D9-D5 and D5-D9,
D5-D5 brane configurations. The particular form of the coefficients in front of $Q_s$
and $Q_c$ takes into account the fact that all D5-branes are
at the same fixed point. The factor 15 in front of $R_N^2$ counts
the number of free fixed points, while the factor 1 in front of
$(R_N\pm 4R_D)^2$ counts the number of fixed points coincident
with some of the D5-branes.

The orientifold projection is then completed if one writes the
M\"{o}bius amplitude. First of all, starting from $\tilde{\cal
A}_0$ and $\tilde{\cal K}_0$ we can write the massless part of the
transverse M\"{o}bius amplitude \ba \label{M_0} \tilde{\cal M}_0
&=& -\frac{2}{4}\left[\hat{Q}_o \left( \sqrt{v_4} +
\frac{1}{\sqrt{v_4}} \right) \left(
N\sqrt{v_4} + \frac{D}{\sqrt{v_4}} \right)\right.\nonumber\\
&&\left.+\hat{Q}_v \left( \sqrt{v_4} - \frac{1}{\sqrt{v_4}}
\right)\left( N\sqrt{v_4} - \frac{D}{\sqrt{v_4}} \right) \right] \
, \ea where we chose the same minus sign for both $\hat Q_o$ and
$\hat Q_v$, in order to cancel all R-R tadpoles.

We note that, since in only the untwisted sector
flows $\tilde{\cal K}$, the complete expression for $\tilde{\cal M}$ is simply
obtained introducing in $\tilde{\cal M}_0$ the lattice sums, \be
\label{MOebiusSuperOrbifoldtran} \tilde{\cal M} = -{2\over 4}
\left[ (\hat Q_o + \hat Q_v ) \left( N v_4 W_n^{(e)}+
D\frac{P_m^{(e)}}{v_4}\right)+ \left( N + D\right) (\hat Q_o -
\hat Q_v ) \left( {2\hat \eta \over \hat \vartheta_2}\right)^2
\right] \, . \ee Finally, we can write the direct channel acting
with a $P$ transformation on $\tilde{\cal M}$. The action of $P$
on the $SO(4)$ characters $O_4$, $V_4$, $S_4$, $C_4$ is given by
the $4\times 4$ block-diagonal matrix $P={\rm
diag}(\sigma_1,\sigma_1)$. The net result is the interchange of
$\hat{Q}_v$ with $\hat{Q}_o$ and of $\hat{Q}_s$ with $\hat{Q}_c$.
Moreover, under $P$ the factors $\Lambda_{+\pm}$ remain unchanged,
while $\Lambda_{-+}$ is interchanged with $i\Lambda_{--}$, so that the
direct channel for the M\"{o}bius strip is \be
\label{MOebiusSuperOrbifold} {\cal M} = -{1\over 4} \left[ (\hat Q
_o + \hat Q _v ) \left( N P_m + D W_n \right) - \left( N +
D\right) (\hat Q _o - \hat Q _v ) \left( {2\hat\eta\over
\hat\vartheta_2} \right)^2 \right] \, . \ee

The tadpole
cancellation conditions can be read from the expressions for
$\tilde{\cal A}_0$, $\tilde{\cal K}_0$, $\tilde{\cal M}_0$. First
of all, since the terms with the projected Chan-Paton
multiplicities appear only in $\tilde{\cal A}_0$, in order to
cancel their tadpoles, one has to impose \be R_N=R_D=0 \
. \ee On the other hand, the tadpole cancellations of $\hat{Q}_o$
and $\hat{Q}_v$ require \be \left(N\sqrt{v_4}\pm
D\frac{1}{\sqrt{v_4}}\right)=32\left(\sqrt{v_4}\pm\frac{1}{\sqrt{v_4}}\right)
\ , \ee that is solved separately for $\sqrt{v_4}$ and
$1/\sqrt{v_4}$ giving \be N=32 \ , \qquad \qquad D=32 \ . \ee
Concerning the open spectrum, it is clear that at the massless
level $Q_0$, that contains the vector, flows only in the annulus
amplitude and not in the M\"{o}bius one. This means that the gauge
group is unitary and that the Chan-Paton charges at the ends of
the string have to be complex conjugate. Therefore, we can
parameterize the Chan-Paton multiplicities with \ba
N &=& n+\bar n \ , \qquad \qquad n=\bar n = 16 \ , \nonumber\\
D &=& d+\bar d \ , \qquad \qquad d=\bar d  =16 \ , \ea and,
consistently with the conditions $R_N=R_D=0$, \be R_N=i(n-\bar n)
\ , \qquad\qquad R_D=i(d-\bar d) \ . \ee With the given
parametrization, at the massless level the open amplitudes are \ba
{\cal A}_0 &=& ( n \bar{n} + d \bar{d} ) Q_0 + {\textstyle{1\over
2}} ( n^2 + \bar{n}^2 + d^2 + \bar{d}^2 ) Q_v
+ (n \bar{d} + \bar{n} d) Q_s \nonumber \\
{\cal M}_0 &=& -  {\textstyle{1\over 2}} (n + \bar{n} + d + \bar{d})
\hat{Q}_v \, ,
\ea
from which one can read the anomaly-free spectrum \cite{bs2,gp}:
$Q_o$ gives a gauge multiplet of $\mathcal{N}=(1,0)$, containing
a vector and a left-handed spinor in the adjoint
representation of $U(16)_{D9}\times U(16)_{D5}$, $Q_v$ gives hyper
multiplets in the $((16\times 16)_a,1)$ and $(1,(16\times 16)_a)$, together with
their complex conjugate representations. Finally $Q_s$ describes only
one half of a hyper multiplet, but in the $(16,\overline {16})$ together with its
conjugate $(\overline{ 16},16)$. Thus, in the end, these two representations
give a single complete hyper multiplet.
\chapter{Supersymmetry breaking}
\section{The 0A and 0B models}
In this section we will discuss other ten-dimensional models that
are not supersymmetric \cite{dhsw} and have a tachyon instability but whose open
descendants can be free of it. Starting
from the general form of the torus partition function (\ref{general torus}),
one can construct (apart from the Type IIA and IIB torus partition functions)
the modular invariants \cite{dhsw}
\ba
\label{torus0}
{\cal T}_{0A} &=& |O_8|^2 + |V_8|^2 + \bar{S}_8 C_8 +
\bar{C}_8 S_8 \, , \nonumber \\
{\cal T}_{0B} &=& |O_8|^2 + |V_8|^2 + |S_8|^2 + |C_8|^2 \ ,
\label{0AB}
\ea
where the contribution of the $8$ transverse bosons and
the integral over the modulus of the torus with its integration measure are
left implicit. These two models do not contain mixed sectors, so that
they have no fermions in their spectra. Both of them have a tachyon, a graviton, an
antisymmetric $2$-form and a scalar in the NS-NS sector. In the R-R sector the Type
0A theory contains a pair of abelian vectors and a pair of $3$-forms, while the Type 0B theory
has two scalars, a pair of $2$-forms and a $4$-form. These spectra are not chiral
and are thus free of anomalies.

Now we start with the orientifold projection \cite{bs} for the
Type 0A theory. First of all we have to write the Klein-bottle, that
propagates only the left-right symmetric sectors \be {\cal K} \ =
\ \frac{1}{2}(O_8+V_8) \ . \ee The Klein-bottle projects away the 2-form
from the NS-NS sector, while halving the R-R ones to leave only one
vector and one 3-form. After an S transformation, that leaves
$O_8+V_8$ unchanged, one gets \be \tilde{\cal{K}} \ = \
\frac{2^5}{2}(O_8+V_8) \ . \ee Since all the $so(8)$ characters
are self-conjugate, only $O_8$
and $V_8$, that enter diagonally in the torus partition function,
propagate in the transverse annulus. We
write them with two independent reflection coefficients \be
\tilde{\cal{A}} \ = \
\frac{2^{-5}}{2}\left[(n_b+n_f)^2V_8+(n_b-n_f)^2O_8\right] \ , \ee
and an $S$-modular transformation gives the direct channel amplitude
\be {\cal
A} \ = \
\frac{1}{2}\left[(n_b^2+n_f^2)(O_8+V_8)-2n_bn_f(S_8+C_8)\right] \
. \ee Finally, we can write the transverse M\"{o}bius amplitude
that is the square root of the product of $\tilde{\cal{K}}$ and
$\tilde{\cal{A}}$ times a combinatoric factor $2$ \be \tilde{\cal
M} \ = \ \epsilon
\frac{2}{2}\left[(n_b+n_f)\hat{V}_8+(n_b-n_f)\hat{O}_8\right] \ .
\ee
The sign $\epsilon$ could be determined imposing the NS-NS
tadpole condition. Let us notice that a relative sign between
$\hat{V}_8$ and $\hat{O}_8$ can be reabsorbed simply interchanging
the roles of $n_b$ and $n_f$. A $P$-modular transformation together
with a rescaling of the modulus gives \be {\cal M} \ = \ \epsilon
\frac{1}{2}\left[(n_b+n_f)\hat{V}_8-(n_b-n_f)\hat{O}_8\right] \ ,
\ee
that is consistent with the projection of the annulus whose R-R
sector can not be projected. Since there are no R-R states flowing
in the transverse channel, one can choose whether or not imposing the
NS-NS tadpole condition. In the first case the sign in the
M\"{o}bius amplitude is $\epsilon=-1$ and $n_b+n_f=32$. The gauge
group is $SO(n_b)\times SO(n_f)$ and the open spectrum contains a
vector in the adjoint, a Majorana fermion $S_8+C_8$ in the
bi-fundamental $(n_b,n_f)$, and tachyons  in the
$(\frac{n_b^2+n_b}{2},1)$ and in the $(1,\frac{n_f^2-n_f}{2})$.
Otherwise, relaxing the tadpole condition means choosing
$\epsilon=+1$, and this gives a gauge group $USp(n_b)\times
USp(n_f)$.

The case of Type 0B  string is particularly interesting,
because it actually allows three different orientifold constructions \cite{susy95},
a freedom suggested by the results in \cite{fps,pss,pss2}.
There are indeed three different Klein-bottle projections compatible with the fusion-rules
\ba
\label{klein0B}
{\cal K}_1 &=& \frac{1}{2}(O_8+V_8-S_8-C_8) \ ,\nonumber\\
{\cal K}_2 &=& \frac{1}{2}(O_8+V_8+S_8+C_8) \ ,\nonumber\\
{\cal K}_3 &=& \frac{1}{2}(-O_8+V_8+S_8-C_8) \ . \ea
We recall that the proper $so(8)$ characters are $V_8$, $O_8$, $-S_8$, $-C_8$, and
that $V_8$ is the identity of the algebra, so that each character
fusing with itself gives $V_8$, while $-S_8$ and $-C_8$ fuse
together in $O_8$. The first Klein-bottle projection symmetrizes
all the characters, and the resulting massless spectrum thus
contains a tachyon, the graviton and the dilaton in the NS-NS
sector, and a pair of 2-forms in the R-R sector. This projection was introduced in \cite{bs}.

The second projection symmetrizes the NS-NS sector, giving the
same spectrum as the previous case, and antisymmetrizes the R-R
sector, giving a complex scalar and an unconstrained 4-form.

The third projection, introduced in \cite{susy95} projects away the tachyon, leaving the graviton and
the dilaton in the NS-NS sector, an R-R scalar, an R-R self-dual
4-form from the antisymmetrization of $|-S_8|^2$ and an R-R 2-form
from the symmetrization of $|-C_8|^2$. Due to the different
projections of $-S_8$ and $-C_8$, the resulting closed spectrum is clearly chiral.

After a rescaling of the modulus and an S-modular transformation,
the three transverse channel amplitudes are
\be
\tilde{\cal{K}}_1 = \frac{2^6}{2}V_8 \ , \qquad\qquad \tilde{\cal{K}}_2 = \frac{2^6}{2}O_8 \ , \qquad\qquad
\tilde{\cal{K}}_3 = \frac{2^6}{2}(-C_8) \ .
\ee

We now add the open sector. In the first case, all characters can flow
in the transverse annulus, since
the matrix $X$ in the torus coincides with the conjugation matrix $\mathcal{C}$. Each of them can have an independent
reflection coefficient, that we parameterize in the following way
\ba
\tilde{\cal A}_1 &=& \frac{2^{-6}}{2} \left[
(n_o + n_v + n_s + n_c)^2 V_8 + (n_o + n_v - n_s - n_c)^2 O_8 \right.\nonumber \\
& &\left. -  (-n_o + n_v + n_s - n_c)^2 S_8 - (-n_o + n_v - n_s + n_c)^2 C_8
\right] \ ,
\ea
and in the direct channel the annulus amplitude then reads
\ba
\label{A1}
{\cal A}_1 &=& {\textstyle\frac{1}{2}} \left[( n_o^2 + n_v^2 + n_s^2 + n_c^2) V_8 + 2(n_o n_v + n_s n_c)O_8\right.\nonumber \\
& & \left. - 2(n_v n_s + n_o n_c) S_8 - 2(n_v n_c + n_o n_s) C_8\right] \ .
\ea
This is the Cardy case \cite{cardy2}.
As $X=\mathcal{C}$, all the closed sectors can reflect on a hole,
and thus there are as many independent boundary conditions,
or different reflection coefficients, as
bulk sectors. Therefore, the matrix giving the fusion
rules $\mathcal{N}_{ij}^k$, that counts how many
times a component of the $k$-th conformal family is contained
in the fusion of the $i$-th and $j$-th families,
has just the right structure to give the content of the $k$-th
state with boundary conditions labelled by $i$ and $j$.
The Cardy ansatz \cite{cardy2} leads in general to
\be
{\cal A} \ = \ \frac{1}{2}\sum_{i,j,k}\mathcal{N}_{ij}^k n^i n^j \chi_k \ ,
\ee
where the fusion coefficients are related to the $S$ matrix through
the Verlinde formula \cite{verlinde}
\be
{\cal N}_{ij}{}^k = \sum_l \, \frac{S_i^l \, S_j^l \, {S^\dag}{}_l^k}{S_1^l} \ .
\label{Verlinde}
\ee
For example, each character fusing with itself gives $V_8$, and thus in (\ref{A1}) $V_8$ is multiplied by
$n_o^2+n_v^2+n_s^2+n_c^2$ . $O_8$ is obtained from the fusion of itself with $V_8$, giving the term $n_on_v O_8 \ $ and
from the fusion of $-S_8$ with $-C_8$, giving the term $n_sn_c O_8$, and so on.

Finally, from $\tilde{\cal K}_1$ and $\tilde{\cal A}_1$ one can write the
transverse M\"{o}bius amplitude
\be
\tilde{\cal M}_1 \ = \ -\frac{2}{2}(n_o+n_v+n_s+n_c)\hat{V}_8 \ ,
\ee
and the corresponding direct-channel amplitude
\be
\label{M1}
{\cal M}_1 \ = \ -\frac{1}{2}(n_o+n_v+n_s+n_c)\hat{V}_8 \ ,
\ee
that is compatible with the projection of (\ref{A1}). In fact, as usual the states with different charges
in the direct annulus can not flow in the M\"{o}bius strip.
The sign of the M\"{o}bius amplitude is chosen to impose the NS-NS tadpole condition, but it could also be reversed.
Retaining NS-NS tadpoles, the tadpole conditions for the three sectors containing massless modes are
\ba
n_o + n_v + n_s + n_c &=& 64 \, ,\nonumber \\
n_o - n_v - n_s + n_c &=& 0 \, ,\nonumber \\
n_o - n_v + n_s - n_c &=& 0 \, , \ \ea
giving $n_0=n_v$ and
$n_s=n_c$ and gauge group $SO(n_o)\times SO(n_v)\times
SO(n_s)\times SO(n_c) \ $. The low-energy open spectrum has gauge
vectors in the adjoint, tachyons in different bi-fundamental
representations $(n_o,n_v,1,1) \ $, $(1,1,n_s,n_c)$, left fermions
in $(1,n_v,n_s,1) \ $ and in $(n_o,1,1,n_c) \ $, and right
fermions in $(1,n_v,1,n_c)$ and $(n_o,1,n_s,1) \ $. The spectrum
is chiral due to the different gauge representation of the left
and right fermions, but the RR tadpole conditions eliminate all
gauge anomalies.

The second choice of Klein-bottle projection is compatible with the following open sector
\ba
{\cal A}_2 &=& {\textstyle\frac{1}{2}} \left[( n_o^2 + n_v^2 +
n_s^2 + n_c^2) O_8 + 2(n_o n_v + n_s n_c)
V_8\right.
\nonumber \\
& &\left.- 2(n_v n_s + n_o n_c) C_8 - 2(n_v n_c + n_o n_s) S_8\right] \ ,
\ea
and
\be
{\cal M}_2 = \mp \frac{1}{2}
(n_o + n_v - n_s - n_c) \hat{O}_8 \ ,
\ee
that in the transverse channel leads to
\ba
\tilde{\cal A}_2 &=& \frac{2^{-6}}{2} \left[
(n_o + n_v + n_s + n_c)^2 V_8 + (n_o + n_v - n_s - n_c)^2 O_8 \right.\nonumber \\
& & \left.+ (n_o - n_v + n_s - n_c)^2 C_8 + (n_o - n_v - n_s +
n_c)^2 S_8 \right] \ea and \be \tilde{\cal M}_2 = \pm \frac{2}{2}
(n_o + n_v - n_s - n_c) \hat{O}_8 \ . \ee
First of all, we would like to
stress that the annulus here can be simply obtained from the
previous one fusing the characters flowing in ${\cal A}_1$ with
$O_8$, that corresponds to using the Cardy ansatz with $S_1^l$
replaced by $S_2^l$ in (\ref{Verlinde}).
Moreover, the sign of the M\"{o}bius strip remains
undetermined because no tadpole condition
involves that amplitude.

Since the vector does not flow in the M\"{o}bius amplitude, this means
that the gauge group is unitary. In terms of $n_b=n_o\ $,
$\bar n_b = n_v\ $, $n_f = n_s\ $ and $\bar n_f = n_c\ $, the R-R tadpole conditions, that one can read only from
$\tilde{\cal A}_2$, give $n_b=\bar n_b$ and $n_f=\bar n_f$. Notice that in the transverse annulus the characters
$-S_8$ and $-C_8$ appear with a negative (unphysical) reflection coefficient, but the previous conditions cancel their contributions.
The gauge group is $U(n_b)\times U(n_f)$ but its total dimension remains undetermined.
The low-energy spectrum contains vectors in the adjoint, left Majorana-Weyl fermions in the $(1,\bar n_b,1,\bar n_f)$ and
in the $(n_b,1,n_f,1)$, right Majorana-Weyl fermions in the $(1,\bar n_b,n_f,1)$ and in the $(n_b,1,1,\bar n_f)$, and tachyons in different
symmetric and antisymmetric representations. The open sector is chiral but free of anomalies.

Finally, we discuss the third orientifold model
commonly known as the 0$^\prime$B model \cite{susy95,c0b,bfl1}, that is the most
interesting one because there is a choice that makes also the open sector
free of tachyons.
As in the previous case, the annulus amplitude can be
determined through the Cardy ansatz with $S_3^l$ in (\ref{Verlinde}), that corresponds to fusing the characters of
${\cal A}_1$ with $-C_8$. The open amplitudes are
\ba
\label{A3}
{\cal A}_3 &=& - \frac{1}{2} \left[
(n_o^2 + n_v^2 + n_s^2 + n_c^2 ) C_8 -
2(n_o n_v + n_s n_c) S_8\right.
\nonumber \\
& &\left. +2(n_v n_s + n_o n_c) V_8 + 2(n_v n_c + n_o n_s) O_8\right] \ ,
\ea
whose transverse channel is
\ba
\label{tildeA3}
\tilde{\cal A}_3 &=& \frac{2^{-6}}{2} \left[
(n_o + n_v + n_s + n_c)^2 V_8 - (n_o + n_v - n_s - n_c)^2 O_8\right. \nonumber \\
& &\left.- (n_o - n_v - n_s + n_c)^2 C_8 + (n_o - n_v + n_s - n_c)^2 S_8\right] \ ,
\ea
and
\be
\label{M3}
{\cal M}_3 = \frac{1}{2} (n_o - n_v - n_s + n_c) \hat{C}_8 \ ,
\ee
that in the transverse channel reads
\be
\label{tildeM3}
\tilde{\cal M}_3 = \frac{2}{2}
(n_o - n_v - n_s + n_c) \hat{C}_8 \ .
\ee
Also in this case the vector does not appear in the M\"{o}bius strip and therefore the
gauge group is unitary. Letting $n_v=n \ $, $n_s=\bar n \ $, $n_o=m \
$ and $n_c=\bar m$, the R-R $S_8$ tadpole condition fixes $m=\bar m$ and
$n=\bar n$ while the R-R $C_8$ tadpole condition fixes $m-n=32$, giving the gauge
group $U(m)\times U(n) \ $. The choice $n=0$ eliminates also the tachyons
from the open spectrum and gives the gauge group $U(32)$, but
actually an $U(1)$ vector acquires mass as a consequence of anomalies reducing the effective gauge group
to $SU(32)$ \cite{wito32,dsw}. The massless open spectrum contains a vector in the adjoint
and right fermions in the $\frac{m(m-1)}{2}$ and in the $\frac{\bar
  m(\bar m -1)}{2} \ $.

\begin{table}
\begin{center}
\begin{tabular}{c c c c | c}
$O_8$ & $V_8$ & $-S_8$ & $-C_8$ \\
\hline
+ & + & + & + & D9$^{(1)}$ \\
+ & + & $-$ & $-$ & $\overline{\rm{D9}}^{(1)}$\\
$-$ & + & + & $-$ & D9$^{(2)}$ \\
$-$ & + & $-$ & + & $\overline{\rm{D9}}^{(2)}$
\end{tabular}
\hspace {2 cm}
\begin{tabular}{c c c c | c}
$O_8$ & $V_8$ & $-S_8$ & $-C_8$ \\
\hline
$\mp$ & $\mp$ & $\mp$ & $\mp$ & O9$_{\pm}^{(1)}$ \\
$\mp$ & $\mp$ & $\pm$ & $\pm$ & $\overline{\rm{O9}}_{\pm}^{(1)}$\\
$\pm$ & $\mp$ & $\mp$ & $\pm$ & O9$_{\pm}^{(2)}$ \\
$\pm$ & $\mp$ & $\pm$ & $\mp$ & $\overline{\rm{O9}}_{\pm}^{(2)}$
\end{tabular}
\end{center}
\caption{D-branes and O-planes for the orientifolds of the 0B model.}
\end{table}
It is possible to read from the transverse channels of the three
projections the D-brane and O-plane content of the orientifolds
of the 0B model. Actually, since there are two different R-R charges, we have two
types of D-branes and O-planes with the corresponding
$\overline{\rm{D}}$-branes and
$\overline{\rm{O}}$-planes that have the same tension but both
the R-R charges reversed (see Tables 2.1). In particular one can see that $\tilde{\cal K}_1$
contains the following combination of O-planes $\rm{O9}^{(1)}_\pm \oplus \rm{O9}^{(2)}_\pm \oplus
\overline{\rm{O9}}^{(1)}_\pm \oplus \overline{\rm{O9}}^{(2)}_\pm \ $,
$\tilde{\cal K}_2$ contains $\rm{O9}^{(1)}_\mp \oplus \rm{O9}^{(2)}_\pm \oplus
\overline{\rm{O9}}^{(1)}_\mp \oplus \overline{\rm{O9}}^{(2)}_\pm \ $,
and finally $\tilde{\cal K}_3$ gives
$\rm{O9}^{(1)}_\mp \oplus \rm{O9}^{(2)}_\pm \oplus
\overline{\rm{O9}}^{(1)}_\pm \oplus \overline{\rm{O9}}^{(2)}_\mp \ $,
where the double choice of sign is due to the possibility of reversing
the M\"{o}bius projection leaving the R-R tadpole conditions unchanged.
Concerning the D-branes content, it is easy to read from
$\tilde{\cal{A}}_1$ that $n_0$ counts the number of
$\overline{\rm{D9}}^{(1)}$, $n_v$ the number of D9$^{(1)}$,
$n_s$ the number of D9$^{(2)}$, and $n_c$ gives the number of
$\overline{\rm{D9}}^{(2)}$-branes.

The second and third projections are instead more complicated,
because their branes are actually complex superpositions of
those of the previous model. For example in $\tilde{\cal{A}}_2$,
in order to have positive coefficients in front of $-S_8$ and $-C_8$,
we have to absorb a minus sign in the squares of the charges. This is as though
$n_b$ counted objects with charges
$(1,1,e^{-i\pi/2},e^{-i\pi/2}) \ $ while $n_f$ corresponded to objects
with charges $(-1,1,e^{i\pi/2},e^{-i\pi/2}) \ $, together with $\bar n_b$ and $\bar
n_f$ that have R-R complex conjugate charges. This is the case of the $+$
sign determination in $\tilde{\cal M}_2$.
These properties are recovered if one combines with complex coefficients
$n_o \ $ $ \ \overline{\rm{D9}}^{(1)}$ and $n_v \ $ D9$^{(1)}$
$ \ (n_o=n_v) \ $ to give
\be
n_b \ = \ \frac{n_o e^{i\pi/4}+n_ve^{-i\pi/4}}{\sqrt{2}} \qquad {\rm and}\qquad
\bar{n}_b \ = \ \frac{n_o e^{-i\pi/4}+n_ve^{+i\pi/4}}{\sqrt{2}} \ ,
\ee
and $n_s \ $ D9$^{(2)}$ with $n_c \ $ $ \ \overline{\rm{D9}}^{(2)} \ $
$(n_s=n_c)$ to give
\be
n_f \ = \ \frac{n_s e^{i\pi/4}+n_c e^{-i\pi/4}}{\sqrt{2}} \qquad {\rm and}\qquad
\bar{n}_f \ = \ \frac{n_s e^{-i\pi/4}+n_c e^{+i\pi/4}}{\sqrt{2}} \ .
\ee
In the same way it is possible to show that for the third model, with
the M\"{o}bius sign of (\ref{tildeM3}), the
right combinations are
\be
n \ = \ \frac{n_v e^{i\pi/4}+n_c e^{-i\pi/4}}{\sqrt{2}} \qquad \rm{and} \qquad
m \ = \ \frac{n_o e^{i\pi/4}+n_s e^{-i\pi/4}}{\sqrt{2}} \ ,
\ee
together with their conjugates.
\section{Scherk-Schwarz deformations}
In this section we want to implement in String Theory the
Scherk-Schwarz mechanism \cite{schschwstr,scherkschwarz}, that
provides a simple and interesting setting in which one can realize
the breaking of supersymmetry \cite{ads1}. While in Field Theory
one has only the option shifting the Kaluza-Klein momenta, lifting
in different ways the masses of bosonic and fermionic fields, in
String Theory one has the further possibility of shifting
windings. A T-duality can turn winding shifts into momentum
shifts, relating these two types of models (at least at the closed
oriented sector), but the orientifold projection gives origin to
very different features in the two cases.

Shifting momenta, the supersymmetry is broken both in the closed
and in the open sectors. We will refer to this model as
``\textit{Scherk-Schwarz supersymmetry breaking}'' \cite{ads1}. On
the other hand, shifting windings corresponds in the T-dual
picture to shifting momenta in the direction orthogonal to
D$8$-branes (we are thinking about the Type I model with only a
compact dimension). Hence, it should be intuitively clear that
such a shift should not affect the low-energy excitations of the
branes, and indeed in this case the bulk is non supersymmetric but
the open sector, at least at the massless level, remains
supersymmetric. Actually, the breaking of supersymmetry in the
bulk sector and in the massive sector of the branes will cause the
breaking of supersymmetry also in the massless open sector via
radiative corrections. We will refer to this phenomenon as
``\textit{brane supersymmetry}'', or also as ``\textit{M-theory
breaking}'', since the winding shifts can be related via a
T-duality to momentum shifts along the $11^{th}$ dimension of
M-theory \cite{hwi}.

\subsection{Momentum shifts}
Let us consider the Type IIB string theory compactified on a
circle. The points on the circle are identified according $X\sim
X+2\pi R n$. Let us now build a shift-orbifold, introducing an
operator $\delta$ whose action is $\delta:X\rightarrow X+\pi R$.
The orbifold operation with respect to $\delta$ identifies the
points $X\sim X+\pi R$. Thus, the shift-orbifold acts halving
effectively the radius of the circle, a fact that has as a
consequence the survival of only even momenta. Therefore, if we
denote with $\Lambda_{m+a,n+b}$ the lattice sum for the circle
\be \Lambda_{m+a,n+b} = {\sum_{m,n} q^{{\alpha ' \over 4} \left(
{(m+a)\over R} + {(n+b)R \over \alpha '}\right)^2}\; \bar q
^{{\alpha ' \over 4} \left( {(m+a)\over R} - {(n+b)R \over \alpha
'} \right)^2} \over \eta (q) \, \eta (\bar q)} \, , \ee
the untwisted sector of the shift-orbifold is simply given,
leaving the summation symbol implicit, by
$\left(\Lambda_{m,n}+(-)^m\Lambda_{m,n}\right)/2$, and as usual
one has to complete the modular-invariant partition function
adding the twisted sector. The result is
\be \Lambda_{m,n}\rightarrow \frac{1}{2}\left(\Lambda_{m,n}+(-)^m
\Lambda_{m,n}+\Lambda_{m,n+\frac{1}{2}}+(-)^m\Lambda_{m,n+\frac{1}{2}}
\right) \ . \label{lattice} \ee
This orbifold is really freely acting, since there are no fixed
points, and the result (\ref{lattice}) is equivalent to the
lattice $\Lambda_{2m,n/2}(R)=\Lambda_{m,n}(R/2)$, meaning that the
action of a shift-orbifold amounts only to a rescaling of the
radius, as anticipated.

In order to have a non-trivial orbifold, we consider a new
$\mathbb{Z}_2$ orbifold whose involution is given by
$(-)^F\delta$, where $F=F_L+F_R$ counts space-time fermions. This
new operator projects $|V_8-S_8|^2\Lambda_{m,n}$ according to
\be |V_8-S_8|^2\Lambda_{m,n} \ \rightarrow \
\frac{1}{2}\left(|V_8-S_8|^2\Lambda_{m,n} +
|V_8+S_8|^2(-)^m\Lambda_{m,n}\right) \ , \ee
and then completing the modular-invariant, gives
\ba {\cal T}_{{\rm KK}} &=& \frac{1}{2} \biggl[ |V_8 - S_8 |^2 \;
\Lambda_{m,n}
+ |V_8 + S_8 |^2 \; (-1)^m \Lambda_{m,n}\nonumber\\
& &  + |O_8 - C_8 |^2 \; \Lambda_{m,n+{1\over 2}} +
|O_8 + C_8 |^2 \; (-1)^m \Lambda_{m,n+{1\over 2}} \biggr] \, ,
\ea
that can be written in the more natural form
\ba {\cal T}_{{\rm KK}} &=& (V_8 \bar V_8 + S_8 \bar S_8 )
\Lambda_{2m,n}
+ (O_8 \bar O_8 + C_8 \bar C_8) \Lambda_{2m,n+{1\over 2}} \nonumber \\
& & - (V_8 \bar S_8 + S_8 \bar V_8 ) \Lambda_{2m+1,n}
- (O_8 \bar C_8 + C_8 \bar O_8 ) \Lambda_{2m+1,n+{1\over 2}} \, .
\label{torus_KK}
\ea
In the decompactification limit $R\rightarrow \infty$ momenta
become a continuum, while windings have to be set to zero, and the
last expression reduces to the torus partition function of Type
IIB string. On the other hand, from $O_8\bar O_8 \
\Lambda_{2m,n+1/2}$ we see that for a special value of the radius
a tachyon develops. In fact $O_8$ starts with $h=-1/2$ while from
the lattice at $m=n=0$ we have a further power of $q$ equal to
$q^{\frac{\alpha'}{4}\left(\frac{R}{2\alpha'}\right)^2}$. Thus,
one has a tachyon for all values of the radius such that
$-\frac{1}{2}+\frac{R^2}{16\alpha'}< 0 $, or $R<2\sqrt{2\alpha'}$.
In the following we will assume that the value of the radius lies
in the region free of the tachyonic instability.

Let us proceed with the orientifold construction. The Klein-bottle
amplitude propagates only states with vanishing windings
\be {\cal K}_{{\rm KK}} = \frac{1}{2} (V_8 - S_8) \; P_{2m} \ ,
\ee
and  the corresponding transverse-channel amplitude reads
\be \tilde{\cal K}_{{\rm KK}} = {2^5 \over 4} \, v  \, (V_8 -
S_8)\; W_{n} \, , \ee
where $v=R/\sqrt{\alpha '}$ is proportional the volume of the
internal manifold. On the other hand, the transverse annulus
propagates only windings, giving an amplitude of the form
\ba \tilde{\cal A}_{{\rm KK}} &=& {2^{-5}\over 4} \, v \,
 \Bigl\{
\bigl[ (n_1 + n_2 + n_3 + n_4 )^2 V_8 \ - \ (n_1 + n_2 - n_3 - n_4
)^2 S_8 \bigr] \, W_n
\nonumber\\
& & + \bigl[ (n_1 - n_2 + n_3 - n_4 )^2 O_8 \ - \ (n_1 - n_2 - n_3
+ n_4 )^2 C_8 \bigr] \, W_{n+{1\over 2}}
\Bigr\} \, , \nonumber\\
\ea
where we have introduced a different reflection coefficient for
each character. As usual, from the relative signs of the $n_i$,
one can see that $n_1$, $n_2$ on the one hand, and $n_3$, $n_4$ on
the other hand, count respectively the numbers of D9-branes and
$\overline{\rm{D}9}$-branes. An S-modular transformation and a
Poisson resummation turn $\tilde{\cal A}_{{\rm KK}}$ into the
direct amplitude
\ba {\cal A}_{{\rm KK}} &=& \frac{1}{2} (n_1^2 + n_2^2 + n_3^2 +
n_4^2 )
\left[ V_8\, P_{2m} - S_8 \, P_{2m+1} \right]+\nonumber\\
& & + (n_1 n_2 + n_3 n_4 ) \left[ V_8 \, P_{2m+1} - S_8 \, P_{2m}
  \right]\nonumber\\
& & + (n_1 n_3 + n_2 n_4 ) \left[ O_8 \, P_{2m} - C_8 \, P_{2m+1} \right]\nonumber\\
& & + (n_1 n_4 + n_2 n_3 ) \left[ O_8 \, P_{2m+1} - C_8 P_{2m}
  \right] \ .
\ea
Finally, the transverse M\"{o}bius amplitude is determined by the
product of $\tilde{\cal K}_{{\rm KK}}$ and $\tilde{\cal A}_{{\rm
KK}}$, and
 \be \tilde{\cal M}_{{\rm KK}} = - \, \frac{v}{2} \,
\Bigl[ (n_1 + n_2 + n_3 + n_4 ) \, \hat V _8 \; W_{n}
 - (n_1 + n_2 - n_3 - n_4 ) \, \hat S _8 \; (-1)^n W_{n}
\Bigr] \, , \ee
where the signs $(-)^n$ for the windings multiplying $\hat S _8 $
are chosen in order to have in the direct channel a term of type
$P_{2m+1}\hat S _8 $
\be {\cal M}_{{\rm KK}} = - \frac{1}{2} (n_1 + n_2 + n_3 + n_4 )
\, \hat V _8 \, P_{2m} + \frac{1}{2}(n_1 + n_2 - n_3 - n_4 )\,
\hat S _8 \, P_{2m+1} \ee
consistently with the projection of the annulus where only the
first line can be projected, the other terms (mixed in charges)
corresponding to non-diagonal contributions. The tadpole
conditions are
\ba \hbox{{\rm NS-NS:}} & \quad n_1 + n_2 + n_3 + n_4 =& 32 \  ,
\nonumber \\
\hbox{{\rm R-R:}} & \quad n_1 + n_2 - n_3 - n_4 =& 32\, ,
\label{KKtad} \ea
and are satisfied by the choice $n_2=n_4=0$ and $n_1+n_2=32$,
fixing the total number of the branes.

The open spectrum does not contain the tachyon instability,
because there are no anti-branes (in this case the NS-NS tadpole
condition is enforced). For what concerns supersymmetry, it is
broken in the closed sector because the mixed (NS-R) terms, from
which one reads the mass of gravitino, are now multiplied by the
lattice $\Lambda_{2m+1,n}$, and so the momentum-shift lifts its
mass. Supersymmetry is also broken in the open sector, where we
have vectors in the adjoint representation of the gauge group
$SO(n_1)\times SO(32-n_1)$, but chiral fermions in the
bi-fundamental representation.

The last point we want to stress here is the possibility to write
the previous amplitudes in a nicer form that allows a direct
comparison of the Scherk-Schwarz deformation in String Theory with
the analogous mechanism in Field Theory, where after
compactification one expects to have periodic bosons (integer
momenta) and antiperiodic fermions (half-integer momenta). One can
recover an amplitude with bosons and fermions with the right
momenta simply rescaling the radius
\be R \ \rightarrow \ R^{\rm SS}=\frac{R}{2} \ , \ee and
consequently windings and momenta \be m \ \rightarrow \
\frac{m}{2} \ , \qquad \qquad n \ \rightarrow \ 2n \ . \ee
For example in this new basis the torus partition function reads
\ba {\cal T}_{{\rm SS}} &=& (V_8 \bar V_8 + S_8 \bar S_8 )
\Lambda_{m,2n} (R_{\rm SS})
+ (O_8 \bar O_8 + C_8 \bar C_8) \Lambda_{m,2n+1}(R_{\rm SS})  \\
& & - (V_8 \bar S_8 + S_8 \bar V_8 ) \Lambda_{m +{1\over 2},2n}(R_{\rm SS})
- (O_8 \bar C_8 + C_8 \bar O_8 ) \Lambda_{m +{1\over 2},2n+1}(R_{\rm
SS}) \, . \nonumber
\ea
\subsection{Winding shifts}
The torus amplitude for the Type IIB superstring with winding
shifts along the compact dimension is simply obtained from
(\ref{torus_KK}) interchanging windings and momenta
\ba {\cal T}_{{\rm W}} &=& ( V_8 \bar V_8 + S_8 \bar S_8 )
\Lambda_{m,2n} + ( O_8 \bar O_8 + C_8 \bar C_8 )
\Lambda_{m+{1\over 2},2n} \nonumber \\
& & - (V_8 \bar S_8 + S_8 \bar V_8)
\Lambda_{m,2n+1} - (O_8 \bar C_8 + C_8 \bar O_8 )
\Lambda_{m+{1\over 2},2n+1} \, .
\ea
Here the supersymmetric case is recovered for $R \rightarrow 0$, a
limit for which momenta disappear and windings become a continuum.
The model is stable and free of a closed tachyon for values of the
radius such that $R < \frac{1}{2}\sqrt{\frac{\alpha'}{2}}$.

The Klein-bottle amplitude can be read from those sectors of the
torus with vanishing windings
\be {\cal K}_{{\rm W}} = \frac{1}{2} (V_8 - S_8 ) \; P_{m} +
\frac{1}{2} (O_8 - C_8 ) \; P_{m+\frac{1}{2}} \, . \ee
In the transverse channel the amplitude is
\be \tilde{\cal
K}_{{\rm W}} = \frac{2^5}{2}\, 2 \,  v \, \left( V_8\; W_{4n} -
S_8 \; W_{4n+2} \right) \, , \ee
from which one can see that at the massless level there is a
contribution only from the NS-NS sector thus meaning that the
total R-R charge has to vanish. Therefore, the model contains
O9-planes and $\overline{\rm O9}$-planes.

The transverse annulus propagates only zero-momentum states. Now
$V_8$ and $-S_8$ are accompanied by $W_{2n}=W_{4n}+W_{4n+2}$,
where the last splitting of the windings is necessary to compare
$\tilde{\cal A}_{{\rm W}}$ with $\tilde{\cal K}_{{\rm W}}$. One
can then let the resulting four sectors flow in the transverse
annulus with different reflection coefficients
\ba \tilde{\cal A}_{{\rm W}} &=& \frac{2^{-5}}{2} \, 2 \, v \,
\Bigl\{ \Bigl[ (n_1+n_2+n_3+n_4)^2 \; V_8  - \;
(n_1+n_2-n_3-n_4)^2 \; S_8 \Bigl]
W_{4n} \nonumber\\
& & + \Bigl[ (n_1-n_2+n_3-n_4)^2 \; V_8 - \; (n_1-n_2-n_3+n_4)^2 \; S_8
\Bigr]  W_{4n+2} \Bigr\} \, .\nonumber\\
\ea
As usual, $n_1$ and $n_2$ count the numbers of D9-branes while
$n_3$ and $n_4$ count the number of $\overline{{\rm D}9}$-branes.

Finally, the transverse amplitude for the M\"{o}bius strip is
obtained from $\tilde{\cal K}_{{\rm W}}$ and $\tilde{\cal
  A}_{{\rm W}}$ in the usual way
\be \tilde{\cal M}_{{\rm W}} = - 2 \, v \,  \Bigl[
(n_1+n_2+n_3+n_4) \, \hat V_8\, W_{4n}
 - (n_1-n_2-n_3+n_4) \, \hat S_8\,
W_{4n+2} \Bigr] \ , \ee
and the tadpole conditions are
\be \hbox{{\rm NS-NS:}} \quad n_1+n_2+n_3+n_4 = 32 \, , \qquad
\qquad \hbox{{\rm R-R:}} \quad n_1+n_2=n_3+n_4 \ . \ee
Moreover, in the limit $R \rightarrow 0$ (that is inside the
stability region) there are additional tadpoles to be cancelled,
due to the fact that in this limit $W_{4n+2}\rightarrow 1$,
\be \hbox{{\rm NS-NS:}} \quad n_1-n_2+n_3-n_4 = 0 \, , \qquad
\qquad \hbox{{\rm R-R:}} \quad n_1-n_2-n_3+n_4=32 \ . \ee
The solution to all four tadpole conditions is $n_1=n_4=16$,
$n_2=n_3=0$, corresponding in a T-dual picture to 16 D8-branes on
the top of an ${\rm O8}_+$-plane placed at the origin of the
compact dimension, and 16 $\overline{\rm D8}$-branes on the top of
an $\overline{\rm O8}_+$-plane placed at $\pi R^T$.

The resulting gauge group is $SO(16)\times SO(16)$. In fact, the
open amplitudes in the direct channel are
\be
{\cal A}_{{\rm W}} = \frac{1}{2} (n_1^2 + n_4^2)\,
(V_8 - S_8 )\;
(P_{m} +  P_{m+\frac{1}{2}}) + n_1 n_4 \, (O_8 - C_8) \;
(P_{m+ \frac{1}{4}} + P_{m+ \frac{3}{4}})
\ee
and
\be
{\cal M}_{{\rm W}} = - \frac{1}{2}
 (n_1+n_4) \left[ (\hat V _8  - \hat S_8)\,  P_{m} +
(\hat V _8  + \hat S_8)\, P_{m+ \frac{1}{2}} \right]  \, , \ee
from which one can read that the massless spectrum contains a
vector and a fermion, both in the adjoint representation of the
gauge group. Thus, as anticipated, supersymmetry is preserved, at
least at the massless level, on the branes, while supersymmetry is
broken in the bulk due to the winding-shift that lifts the mass of
gravitino. The radiative corrections can then break supersymmetry
also in the open massless sector.

Before closing this section, we write for completeness the torus
partition function of this model in the Scherk-Schwarz basis \ba
{\cal T}_{{\rm W}}(R^{\rm SS}) &=& ( V_8 \bar V_8 + S_8 \bar S_8 )
\Lambda_{2m,n} + ( O_8 \bar O_8 + C_8 \bar C_8 )
\Lambda_{2m+1,n} \nonumber\\
& & - (V_8 \bar S_8 + S_8 \bar V_8)
\Lambda_{2m,n+{1\over 2}} - (O_8 \bar C_8 + C_8 \bar O_8 )
\Lambda_{2m+1,n+{1\over 2}} \, ,
\ea
where $R^{\rm SS}=2R$ and we have rescaled windings and momenta
following
\be
m \ \rightarrow 2m \ , \qquad \qquad n \ \rightarrow \frac{n}{2} \ .
\ee
\section{Brane supersymmetry breaking}
In the previous chapter we studied orbifold compactifications,
treating in particular the orbifold $T^4/\mathbb{Z}_2$. The
resulting six dimensional model, with the standard Klein-bottle
projection we made, contains O9 and O5 planes together with D9 and
D5 branes, as required by the cancellation of the total R-R and
NS-NS charges. The same Klein-bottle projection is consistent with
O-planes whose tension and charge are both reversed, requiring the
presence of antibranes in order to cancel R-R tadpoles. In this
case NS-NS tadpoles arise, and supersymmetry is broken in the open
sector. The massless closed spectrum remains unchanged.

But the $\mathbb{Z}_2$-orbifold allows another possibility. One
can reverse the sign of the twisted sector in the Klein-bottle
amplitude, changing in this way only the tension and the charge of
the $O5$-plane. The corresponding $\overline{\rm D5}$-branes,
necessary to neutralize its R-R charge, break in the open sector
supersymmetry, that instead is still preserved by the D9-branes.
This phenomenon is known as ``brane supersymmetry breaking''
\cite{bsb,dm1} and, in contrast with the Scherk-Schwarz
deformations of toroidal compactifications, for which the scale of
supersymmetry breaking is given by the compactification radius,
here supersymmetry is broken at the string scale. More precisely,
in the low-energy supergravity, it is non-linearly realized
\cite{dm1}.

Let us begin with the Klein-bottle amplitude. The first
consideration in order is that the result of the interaction of
two characters both in the untwisted sector or both in the twisted
sector has to belong to the untwisted sector, while the
interaction of a character in the twisted sector with a character
in the untwisted one has to be in the twisted sector. This means
that one has two possibilities for the Klein-bottle projection:
the first is to symmetrize both the twisted and untwisted sectors,
obtaining eq. (\ref{Klein_orb}). The second is to symmetrize the
untwisted sector while antisymmetrizing the twisted one. Thus, the
resulting Klein-bottle amplitude has only a different sign in the
twisted sector with respect to the equation (\ref{Klein_orb})
\be {\cal K} = \frac{1}{4} \left[ ( Q_o + Q_v ) ( P_m + W_n ) - 2
\times 16 ( Q_s + Q_c ){\left(\frac{\eta}{\vartheta_4}\right)}^2
\right] \ . \ee
The unoriented closed spectrum, at the massless level, can be read
from the table (\ref{tabe}), from which one has to take from the
first two rows the same states we considered in the supersymmetric
$T^4/\mathbb{Z}_2$ orbifold, while the last row, corresponding to
the twisted sector, the 16 NS-NS scalars, the 16 antiself dual R-R
2-forms, and the 16 right-handed spinors. The massless spectrum is
still organized in multiplets of $\mathcal{N}=(1,0)$, and
precisely contains the usual graviton multiplet, 17 tensor
multiplets, 16 of which from the twisted sector, and 4 hyper
multiplets.

The transverse channel is simply equal to the one in the
supersymmetric case, but for a different sign in the S-transform
of the direct twisted sector
\be \tilde{\cal K} = \frac{2^5}{4} \left[ (Q_o + Q_v) \left( v_4
W_n^{(e)} + \frac{1}{v_4} P_m^{(e)} \right) - 2 (Q_o - Q_v)
\left( \frac{2 \eta}{\vartheta_2} \right)^2 \right] \, . \ee
From the amplitude at the origin of the lattice
\be \tilde{\cal K}_0 = \frac{2^5}{4} \left[ Q_o \left( \sqrt{v_4}
- \frac{1}{\sqrt{v_4}}\right)^2 + Q_v \left( \sqrt{v_4}  +
\frac{1}{\sqrt{v_4}}\right)^2 \right] \ , \ee
one can see clearly the relative sign between $\sqrt{v_4}$ and
 $1/\sqrt{v_4}$ in the squared coefficient multiplying $Q_o$.
Such a sign led to a negative mixed term, and thus it corresponds
to a configuration with $O9_+$ and $O5_-$ planes (or the T-dual
configuration with $O9_-$ and $O5_+$ planes). The presence of the
$O5_-$ planes then requires the introduction in the model of
$\overline{\rm D 5}$-branes that absorb the R-R charge. There is a
relative sign in the R-R charge of the $D9$ branes and
$\overline{\rm D5}$ branes, so that the only difference in the
transverse annulus amplitude with respect to the supersymmetric
case (\ref{Annulus_orb}) is the sign of the R-R states in the
mixed $ND$ and $R_N R_D$ sectors. Therefore, the transverse
amplitude for the annulus is simply
\ba \tilde {\cal A} &=& \frac{2^{-5}}{4} \Biggl[ (Q_o + Q_v)
\left( N^2 v_4 W_n + \frac{D^2}{v_4}P_m \right)  + 16 \left(R_N^2
+ R_D^2 \right) (Q_s + Q_c) \left( {\eta \over
\vartheta_4}\right)^2
\nonumber\\
&& + 2 N D \, (V_4O_4+C_4C_4 - O_4V_4-S_4S_4 ) \left( {2 \eta \over \vartheta_2}\right)^2
\nonumber \\
& &  - 2 \times 4 R_N R_D \, (O_4C_4+S_4O_4 - V_4S_4-C_4V_4 )
\left( {\eta \over \vartheta_3}\right)^2
\Biggr] \, ,
\ea
that at the origin of the lattice reads
\be \tilde{\cal A}_0 = \frac{2^{-5}}{4} \left[ (V_4 O_4 - S_4 S_4)
\left( N \sqrt{v_4}  + \frac{D}{\sqrt{v_4}}\right)^2  +  (O_4 V_4
- C_4 C_4) \left( N\sqrt{v_4}  -
\frac{D}{\sqrt{v_4}}\right)^2\right]  \, . \ee
One can see from the coefficients multiplying $V_4O_4$ that the
product of the tensions is always positive, while from the
 coefficient of $-C_4C_4$ we see that the mixed term is
 negative. Thus, the partition function we wrote describes the
right configuration of branes and antibranes.

In the direct channel the Annulus amplitude is
\ba \label{bsb_direct_A} {\cal A} &=& \frac{1}{4} \left[(Q_o +
Q_v) ( N^2 P_m  + D^2 W_n ) + (R_N^2 + R_D^2) (Q_o - Q_v)
{\left(\frac{2
\eta}{\vartheta_2}\right)}^2 \right.\nonumber \\
&& \left. +
2 N D ( O_4 S_4 - C_4 O_4 + V_4 C_4 -
S_4 V_4) {\left(\frac{\eta}{\vartheta_4}\right)}^2
\right. \nonumber\\
&& \left.+ 2 R_N R_D ( - O_4 S_4 - C_4 O_4 + V_4 C_4 + S_4 V_4
){\left(\frac{ \eta}{\vartheta_3}\right)}^2 \right] \ , \ea
and finally we have to write the M\"{o}bius amplitude. First of
all, from $\tilde{\cal A}_0$  and $\tilde{\cal K}_0$, we write the
contribution to the transverse channel at the origin of the
lattice,
\ba \tilde{\cal M}_0 &=& - \frac{1}{2} \left[ \hat{V}_4 \hat{O}_4
\left( \sqrt{v_4}  - \frac{1}{\sqrt{v_4}}\right) \left( N
\sqrt{v_4}  +
\frac{D}{\sqrt{v_4}}\right) \right.\nonumber \\
&& \left.+ \hat{O}_4
\hat{V}_4 \left( \sqrt{v_4}  +
\frac{1}{\sqrt{v_4}}\right) \left( N \sqrt{v_4}  -
\frac{D}{\sqrt{v_4}}\right) \right.\nonumber \\
&& \left.-\hat{C}_4 \hat{C}_4 \left(
\sqrt{v_4}  -
\frac{1}{\sqrt{v_4}}\right) \left( N \sqrt{v_4}  -
\frac{D}{\sqrt{v_4}}\right) \right.\nonumber \\
&& \left. - \hat{S}_4 \hat{S}_4 \left( \sqrt{v_4}  +
\frac{1}{\sqrt{v_4}}\right) \left( N \sqrt{v_4}  +
\frac{D}{\sqrt{v_4}}\right)
\right] \, ,
\ea
from which one can recognize all the single exchanges between the
$O9_+$ and $O5_-$ planes and the $D9$ and $\overline{\rm{D}5}$
branes. Then, completing the reticular sums one obtains
\ba \tilde{\cal M} &=& -\frac{1}{2}\Bigl[v_4 N
 W_n^{(e)}(\hat{V}_4\hat{O}_4+\hat{O}_4\hat{V}_4
-\hat{S}_4\hat{S}_4-\hat{C}_4\hat{C}_4)\nonumber\\
&&
+\frac{1}{v_4} D P_m^{(e)}(-\hat{V}_4\hat{O}_4-\hat{O}_4\hat{V}_4
-\hat{S}_4\hat{S}_4-\hat{C}_4\hat{C}_4) \nonumber\\
&&+N(-\hat{V}_4\hat{O}_4+\hat{O}_4\hat{V}_4
-\hat{S}_4\hat{S}_4+\hat{C}_4\hat{C}_4)\left( {2\hat \eta \over \hat \vartheta_2}\right)^2\nonumber\\
&&
+D(\hat{V}_4\hat{O}_4-\hat{O}_4\hat{V}_4
-\hat{S}_4\hat{S}_4+\hat{C}_4\hat{C}_4)\left( {2\hat \eta \over \hat \vartheta_2}\right)^2\Bigr] \ ,
\ea
and after a $P$-modular transformation
\ba \label{bsb_directM} {\cal M} &=& - \frac{1}{4} \Biggl[ N P_m (
\hat{O}_4 \hat{V}_4  + \hat{V}_4 \hat{O}_4  - \hat{S}_4 \hat{S}_4
- \hat{C}_4
\hat{C}_4 ) \nonumber \\
&& -  D W_n ( \hat{O}_4
\hat{V}_4  + \hat{V}_4 \hat{O}_4  + \hat{S}_4 \hat{S}_4 + \hat{C}_4
\hat{C}_4 )  \nonumber \\
&& - N(
\hat{O}_4 \hat{V}_4 - \hat{V}_4 \hat{O}_4 - \hat{S}_4 \hat{S}_4
+ \hat{C}_4 \hat{C}_4 )\left(
{2{\hat{\eta}}\over{\hat{\vartheta}}_2}\right)^2  \nonumber \\
&& + D( \hat{O}_4
\hat{V}_4 - \hat{V}_4 \hat{O}_4 + \hat{S}_4 \hat{S}_4
- \hat{C}_4 \hat{C}_4)\left(
{2{\hat{\eta}}\over{\hat{\vartheta}}_2}\right)^2  \Biggr] \, .
\ea
The R-R tadpole conditions can be extracted from $\tilde{\cal
A}_0$, $\tilde{\cal K}_0$ and $\tilde{\cal M}_0$, and give
\be N=D=32 \ ,\qquad \qquad R_N=R_D = 0 \ , \ee
but in contrast with the supersymmetric case, the vector now also
flows in the M\"{o}bius amplitude, and thus the right
parametrization for the Chan-Paton multiplicities is
\ba
N=n_1+ n_2 \, , \qquad D=d_1+ d_2 \, , \nonumber \\
R_N=n_1- n_2 \, , \qquad R_D=d_1- d_2 \, , \ea
that leads to the solution
\be n_1=n_2=d_1=d_2=16 \ . \ee
In terms of the given parametrization, the massless unoriented
open spectrum is given by
\ba {\cal A}_0 + {\cal M}_0 &=& \frac{n_1(n_1-1) + n_2(n_2-1) +
d_1(d_1+1) +  d_2(d_2+1) }{2} \ V_4 O_4 \nonumber \\
 &&- \frac{n_1(n_1-1) + n_2(n_2-1) + d_1(d_1-1) + d_2(d_2-1) }{2} \ C_4
C_4 \nonumber \\ &&+ (n_1 n_2 + d_1 d_2 ) ( O_4 V_4 - S_4 S_4 ) + (
n_1 d_2 + n_2 d_1 ) \ O_4 S_4 \nonumber \\
&& - (n_1 d_1 + n_2 d_2 ) \ C_4 O_4 \, .
\ea
The gauge group is $[SO(16) \times SO(16) ]_9 \times  [ USp(16)
\times USp(16) ]_5$, where the first two factors refer to the D9
branes and the last to the $\overline{\rm D5}$ branes. The sector
with NN boundary conditions has a supersymmetric massless
spectrum, consisting of a gauge multiplet in the adjoint
representation of $SO(16)\times SO(16)$ and a hyper multiplet in
the $(16,16,1,1)$. The DD spectrum is clearly not supersymmetric,
because it contains vectors in the adjoint representation of the
gauge group $USp(16)\times USp(16)$, but the corresponding
fermions are in the symmetric representation $(1,1,120,1)$ and
$(1,1,1,120)$. Moreover, this sector contains four scalars and a
left-handed Weyl spinor in the $(1,1,16,16)$. The mixed ND sector
also breaks supersymmetry, and contains two scalars in the
$(1,16,16,1)$ and in $(16,1,1,16)$ and a Majorana-Weyl fermion in
the $(16,1,16,1)$ and $(1,16,1,16)$, where the Majorana condition
here is implemented by a conjugation matrix in the pseudo-real
gauge group representation.

Supersymmetry breaking on the $\overline{\rm D5}$ branes
leaves uncancelled NS-NS tadpoles, that one can read from the
transverse amplitudes
\be \left[ (N-32){\sqrt v_4}+{D+32 \over{\sqrt v_4}}\right]^2 \
V_4O_4 +\left[ (N-32){\sqrt v_4}-{D+32 \over{\sqrt v_4}}\right]^2
\ O_4V_4 \ . \ee
The coefficients of $V_4O_4$ and $O_4V_4$ are as usual are
proportional to the squared one-point functions in front of
boundaries and crosscaps. From the space-time point of view, one
can derive them from a term in the low-energy effective action of
type
\be \Delta S \sim  (N-32)\sqrt{v_4} \, \int d^6 x \, \sqrt{-g} \,
e^{-\varphi_6} + \frac{D+32}{\sqrt{v_4}} \, \int d^6 x \,
\sqrt{-g} \, e^{-\varphi_6} \ . \ee
In fact, the derivatives of $\Delta S$ with respect to the
deviation of the six-dimensional dilaton $\varphi_6$ and the
internal volume $\sqrt{v}_4$ from their background values give the
square roots of the coefficients multiplying respectively $V_4O_4$
and $O_4V_4$. The first contribution in $\Delta S$ refers to the
system of D9 branes with the corresponding O9 plane. The
supersymmetry of the D9-O9 sector ensures that not only the R-R
charge but also the total tension vanishes. Thus the R-R tadpole
condition $N=32$ eliminates this term. The second contribution
refers to the $\overline{\rm D5}$ branes with their O5$_-$ planes,
but their tensions now are summed. Therefore, this term is not
cancelled and gives origin to a positive dilaton potential, whose
presence signals that the Minkowski background is not a good
vacuum for this model.
\section{Supersymmetry breaking and magnetic deformations}
In this section we want to analyze the possibility to break
supersymmetry coupling the ends of open strings to a background
magnetic field
\cite{bachasmag,Angelantonj:2000rw,magnetic,Larosa}. The reason
why supersymmetry can be broken in this way is that modes with
different spins couple differently to a magnetic filed, thus
splitting the mass of bosons and fermions. But there is another
interesting way to see this. Let us turn on a uniform abelian
magnetic field whose vector potential is given by
$A_\mu=-\textstyle{\frac{1}{2}}F_{\mu\nu}X^\nu$. The minimal
coupling of such a field with the ends of an open string leads to
an action with new boundary terms. For instance, for a bosonic
string
\ba S &=& - \frac{1}{4 \pi \alpha'} \, \int d\tau \int_0^{\pi}
d\sigma
\partial_\alpha X \cdot \partial^\alpha X \nonumber\\
 &-&
\left. q_{\rm L} \int d\tau A_\mu \partial_\tau X^{\mu}
\right|_{\sigma =0} - \left. q_{\rm R} \int d\tau A_\mu
\partial_\tau X^{\mu} \right|_{\sigma =\pi} \ ,
\ea
and the equations of motion of such action have to be implemented
together with the boundary conditions
\ba
\partial_\sigma X^\mu - 2\pi \alpha' q_{\rm L} F^{\mu}{}_{\nu} \,
\partial_\tau X^\nu &=& 0 \, , \qquad \sigma=0 \ , \nonumber\\
\partial_\sigma X^\mu + 2\pi\alpha' q_{\rm R} F^{\mu}{}_{\nu} \,
\partial_\tau X^\nu &=& 0 \, , \qquad \sigma=\pi \ .
\ea
Now, if for example we consider a magnetic field living only in
the plane defined by the direction $X^1$, $X^2$,
$F_{12}=-F_{21}=H$, and we perform a $T$ duality along the $X^2$
direction, whose dual coordinate is now denoted with $Y^2$, the
boundary conditions become
\be
\partial_\sigma \left( X^1 -  2 \pi \alpha '
q_{\rm L} H \, Y^2 \right) = 0 \,, \qquad
\partial_\tau \left( Y^2 +  2 \pi \alpha ' q_{\rm R} H \,
X^1 \right) = 0 \,,
\label{bound_cond}
\ee
since $T$-duality interchanges Dirichlet with Neumann boundary
conditions. Therefore, a coordinate in the direction identified by
the first equation in (\ref{bound_cond}) has Neumann boundary
conditions, and so it describes a coordinate parallel to a brane
rotated by an angle $\theta_{L} = \tan^{-1} (2 \pi \alpha' q_L
H)$. On the other hand, the coordinate in the direction identified
by the second equation in (\ref{bound_cond}) satisfies only
Dirichlet boundary conditions, and so it is orthogonal to a brane
rotated by the angle $\theta_{R} = \tan^{-1} (2 \pi \alpha' q_R
H)$. Supersymmetry breaking is therefore understood in this
language, because strings terminating at two rotated branes are
stretched and their modes take mass.

At this point we discuss the canonical quantization. It is useful
to introduce the complex coordinates
\be X_{\pm} = \frac{1}{\sqrt{2}} (X^1 \pm i X^2)\,, \ee together
with their canonically conjugate momenta \be P_\mp(\tau,\sigma) =
\frac{1}{2 \pi \alpha'} \left\{ \partial_\tau  X_\mp (\tau,\sigma)
+ i X_\mp (\tau,\sigma) \ 2\pi\alpha' H \ \left[ q_L
\delta(\sigma) + q_R \delta(\pi - \sigma) \right] \right\} \, .
\ee
The solution of the equations of motion leads to different mode
expansions if the total charge $q_L+q_R$  is different or equal to
zero. In the first case, $q_L+q_R\neq 0$, $X_+(\tau,\sigma)$, $ \
X_-=X_+^\dag$, can be expanded as
\be X_+(\tau,\sigma) = x_+ + i \sqrt{2 \alpha'} \left[
\sum_{n=1}^\infty a_n \psi_n(\tau,\sigma) - \sum_{m=0}^\infty
b^\dag_m \psi_{-m}(\tau,\sigma) \right] \, , \ee where \be
\psi_n(\tau,\sigma) = \frac{1}{\sqrt{|n-z|}} \, \cos\left[ (n -
z)\sigma + \gamma \right]\, e^{-i(n-z)\tau} \, , \ee and $z$,
$\gamma$ and $\gamma'$ are defined by \be z = {1\over \pi} (\gamma
+ \gamma' ) \,, \qquad \gamma= \tan ^{-1} (2\pi\alpha'q_L H ) \, ,
\quad \gamma' =\tan^{-1} (2\pi\alpha' q_R H ) \, . \ee
One can see that in this case the frequencies of the oscillators
are shifted by $z$. The commutation relations for the independent
oscillators $a_n, a^\dag_m$ and  $b_n, b^\dag_m$ are the usual
ones, while the zero modes $x_+$, $x_-$ do not commute
\be [x_+,x_-]=\frac{1}{H(q_L+q_R)} \, , \ee
and in fact describe the usual creation and annihilation operators
for the Landau levels of a particle in a uniform magnetic field,
giving, in the small field limit, the following contribution to
the mass formula:
\be \Delta M^2 = (2n+1)(q_L+q_R)H \ . \ee
On the other hand, in the case of vanishing total charge,
$q_L=-q_R=q$, the oscillators do not feel the magnetic field,
because $z=0$, and thus their frequencies are not shifted, but
there is a new zero mode
\be X_+(\tau,\sigma) = \frac{x_+ + p_-\left[ \tau -
i2\pi\alpha'qH(\sigma - \frac{1}{2} \pi) \right]}{\sqrt{1 +
(2\pi\alpha'qH)^2}} + i \sqrt{2 \alpha'} \sum_{n=1}^\infty \left[
a_n \psi_n(\tau,\sigma) - b^\dag_n \psi_{-n}(\tau,\sigma) \right]
. \label{dip} \ee
In this case, to which we will refer as the ``dipole string'', the
Landau levels do not affect the mass formula because they would be
proportional to $(q_L+q_R)H$, that is zero.

At this point we focus our attention on the case of a constant
magnetic field defined in a compact space. For example, let us
compactify $X_1$ and $X_2$ on a torus. The motivation is that in such
a case the degeneracy of Landau levels is finite. If $2\pi R_1$ and
$2\pi R_2$ are the two sides of the fundamental cell of the torus,
the Landau degeneracy $k$ is simply given by
\be k= 2\pi R_1 R_2 q H = 2\pi \alpha' v q H \label{Land} \ee
where the second equality is due to the usual definition,
$v=R_1R_2/\alpha'$. Actually, $k$ has also another meaning. It is
the integer that enters in the Dirac quantization condition due to
the fact that a magnetic field on the torus is a monopole field
producing a non-vanishing flux. On the fundamental cell of the
torus one can define the vector potential
\be A_1=a_1 \ , \qquad \qquad A_2 = a_2 + H X_1 \ , \ee
so that $F_{12}=H$. The constants $a_{1,2}$ are related to the
Wilson lines that break the gauge group, and for our present
discussion we can set them to zero. The gauge transformation
\be A_i = A_i -i e^{-i\varphi}\dd_i e^{i\varphi} \ \qquad \qquad
i=1,2 \ee
allows to join with continuity $X_1=0$ to $X_1=2\pi R_1$, if
$\varphi$ is chosen to be
\be \varphi=2\pi R_1 H X_2 \ . \ee
Imposing the monodromy of the function $q\varphi$ one can recover
the Dirac quantization condition $qH=k/2\pi\alpha'v$, and we can
see that here the integer $k$ is the same as the one that defines
the Landau degeneracy (\ref{Land}). In the $T$-dual picture $k$
can be interpreted as the number of wrappings before the rotated
branes close, as can be understood using the definition of the
angle of rotation and the relation (\ref{Land}) with $R_2^T=
\alpha'/R_2$
\be \tan{\theta} = k\frac{{R^T_2}}{R_1} \ . \ee

After this introduction, we can analyze how the vacuum amplitudes
have to be deformed in the presence of a constant magnetic field.
For what said, we want to consider the magnetic field in a compact
space. For example, let us consider the six dimensional model
given by $\mathcal{M}_6 \times [T^2(H_1)\times
T^2(H_2)]/\mathbb{Z}_2$, with two different abelian magnetic
fields in the two $T^2$-tori. The reason why we are interested in
this peculiar orbifold, that admits D9 and D5 (or $\overline{\rm
D5}$) branes, is that for the particular choice of magnetic fields
satisfying the relation $H_1=\pm H_2$, a system of magnetized D9
branes can emulate a number of D5 (or $\overline{\rm D5}$) branes.
We can see this fact already at the low-energy effective field
theory \cite{Angelantonj:2000rw,magnetic}, where the action of a
stack of magnetized D9 branes is
\ba S_9 & = & -T_9\sum_{a=1}^{32}\int_{{\cal
M}_{10}}d^{10}Xe^{-\phi} \sqrt{-{\rm
    det}(g_{10}+2\pi\alpha'q_a F)} \nonumber\\
&& -\mu_9\sum_{p,a}\int_{{\cal M}_{10}}e^{2\pi\alpha' q_a F}\wedge C_{p+1}
\ ,
\label{BI-WZ}
\ea
and where $q_a$ labels different type of Chan-Paton charges. Here
the first contribution is the Born-Infeld action, that is known to
describe the low-energy dynamics of an open string, while the
second contribution is the Wess-Zumino term that couples the
magnetic field to different R-R forms $C_{p+1}$. $T_9$ and $\mu_9$
are respectively the tension and the R-R charge of the brane. For
a generic BPS D$p$ brane the following relation holds
\be T_p=|\mu_p|=\sqrt{\frac{\pi}{2k^2}}(2\pi\sqrt{\alpha'})^{3-p}
, \label{tens} \ee
where $k^2=8\pi G_N^{(10)}$ defines Newton's constant in $10$
dimensions. Now, compactifying $4$ dimensions on $T^2\times T^2$,
and allowing two different magnetic fields on the two tori, the
action becomes
\ba S_9 &=& -T_9 \int_{{\cal M}_{10}}d^{10}Xe^{-\phi}
\sum_{a=1}^{32}\sqrt{-g_6}\sqrt{(1+2\pi\alpha'q_a H_1^2)
(1+2\pi\alpha'q_a H_2^2)}\nonumber\\
&&-32 \ \mu_9\int_{{\cal M}_{10}} C_{10} \ - \
\mu_5 \ v_1 v_2 H_1 H_2 \sum_{a=1}^{32}(2\pi q_a)^2 \int_{{\cal M}_{6}} C_{6}
 \ ,
\ea
where we used (\ref{tens}), and $v_i=R_i^1R_i^2/\alpha'$ refers to
the two volumes of the two tori. The linear term in the expansion
of the Wess-Zumino action is not present because the generator of
the $U(1)$ abelian group is traceless, since the $U(1)$ is a
subgroup of the original $SO(32)$ gauge group for the D9 branes of
the Type I string. Then, if $H_1=\pm H_2$,  and using the Dirac
quantization condition for each of the two magnetic fields
$k_i=2\pi\alpha' v_i q H_i$, the action reduces to
\ba S_9 &=& - 32 \int_{{\cal M}_{10}} \left( d^{10}X
\sqrt{-g_6} \ e^{-\phi} \ T_9 \ + \ \mu_9 \ C_{10} \right)\nonumber\\
&&
-\sum_{a=1}^{32}\left(\frac{q_a}{q}\right)^2 \int_{{\cal M}_{6}}\left( d^6 X
\sqrt{-g_6} \ |k_1k_2| \ T_5 \ e^{-\phi} \ + \ k_1 k_2 \ \mu_5 \ C_{6} \right)
 \ ,
\ea
from which one can recognize a system of D9 branes together with
$|k_1k_2|$ D5 branes, if $k_1k_2 > 0 \ $ ($H_1=+H_2$), or
$\overline{\rm D5}$ branes, if $k_1k_2<0 \ $ ($H_1=-H_2$).

After this digression, we come back to write the amplitudes. Let
us start to modify the Klein-bottle amplitude of the
supersymmetric model $T^4/\mathbb{Z}_2$ to obtain the magnetized
orbifold $[T^2(H_1) \times T^2(H_2)]/\mathbb{Z}_2$. Denoting with
$P_i$ and $W_i$ the sums over momenta and windings for the two
tori, and decomposing the internal characters in the
representation of $SO(2)\times SO(2)$ according to
\begin{eqnarray}
Q_o (z_1 ; z_2) &=& V_4 (0) \left[ O_2 (z_1 ) O_2 (z_2  ) +
V_2 (z_1 ) V_2 (z_2  ) \right] \nonumber\\
&&- C_4 (0) \left[ S_2 (z_1 ) C_2 (z_2  ) +
C_2 (z_1 ) S_2 (z_2  ) \right] \, ,
\nonumber
\\
Q_v (z_1 ; z_2 ) &=&  O_4 (0) \left[ V_2 (z_1 ) O_2 (z_2  ) +
O_2 (z_1 ) V_2 (z_2  ) \right] \nonumber \\
&&- S_4 (0) \left[ S_2 (z_1 )
S_2 (z_2  ) + C_2 (z_1 ) C_2 (z_2  ) \right] \, ,
\nonumber
\\
Q_s (z_1 ;z_2  ) &=& O_4 (0) \left[ S_2 (z_1 ) C_2 (z_2  ) +
C_2 (z_1 ) S_2 (z_2  ) \right] \nonumber \\
&&- S_4 (0) \left[  O_2 (z_1 ) O_2 (z_2  ) +
V_2 (z_1 ) V_2 (z_2  ) \right] \, ,
\nonumber
\\
Q_c (z_1 ; z_2 )
&=&  V_4 (0) \left[ S_2 (z_1 ) S_2 (z_2  ) +
C_2 (z_1 ) C_2 (z_2  ) \right] \nonumber \\
&&- C_4 (0)
\left[  V_2 (z_1 ) O_2 (z_2  ) +
O_2 (z_1 ) V_2 (z_2 ) \right] \, ,
\end{eqnarray}
the Klein-bottle amplitude can be written
\be {\cal K} = \frac{1}{4} \Biggl\{ (Q_o + Q_v) (0;0) \left[ P_1
P_2 + W_1 W_2 \right] + 16\times 2 (Q_s + Q_c ) (0;0) \left( {\eta
\over \vartheta_4 (0)} \right)^2 \Biggr\} \, , \ee
where the arguments $z_1$ and $z_2$, that take into account the
presence of the two magnetic fields on the two internal tori, are
here set to zero since the closed sector does not couple to the
magnetic fields. The characters of the level-1 affine extension of
$O(2n)$, in the presence of a magnetic field, are expressed in
terms of the Jacobi theta-functions for non-vanishing arguments
\ba \label{so2n} O_{2n}(z) &=& \frac{1}{2 \eta^n (\tau)} \left[
\vartheta_3^n(z|\tau) + \vartheta_4^n(z|\tau)\right] \, ,
\nonumber \\
V_{2n}(z) &=&\frac{1}{2 \eta^n (\tau)} \left[
\vartheta_3^n(z|\tau)
- \vartheta_4^n(z|\tau) \right] \, , \nonumber \\
S_{2n}(z) &=& \frac{1}{2 \eta^n (\tau)} \left[
\vartheta_2^n(z|\tau) + i^{-n} \vartheta_1^n(z|\tau)\right] \,
, \nonumber \\
C_{2n}(z) &=& \frac{1}{2 \eta^n (\tau)} \left[
\vartheta_2^n(z|\tau)
- i^{-n} \vartheta_1^n(z|\tau) \right] \, .
\ea

Let us now consider the open sector with a unitary gauge group, as
in the case of the original supersymmetric $T^4/\mathbb{Z}_2$
model, and let $N_0=n+\bar n$ be the number of neutral D9 branes,
while $m$ and $\bar m$ count the number of the magnetized D9
branes with $U(1)$-charges equal to $+1$ or $-1$. In addition, we
also have the D5 branes with their multiplicity $d+\bar d$. The
annulus amplitude is obtained deforming the one of the original
$T^4/\mathbb{Z}_2$ model. First of all, we have to substitute the
original factor $(n+\bar n)$ with $(n+\bar n)+(m+\bar m)$, since
now $m+\bar m$ of the D9 branes are magnetized. Therefore, in the
annulus amplitude, apart from the same neutral strings, whose
multiplicities do not depend from $m$ and $\bar m$, now we have
also a ``dipole'' term, with multiplicity $m\bar m$, also neutral,
and charged terms with multiplicities proportional to $m$, $\bar
m$, $m^2$ and $\bar m^2$. As we already know, the oscillator
frequencies in the case of a ``dipole'' string, are not shifted by
the magnetic field, but there is a new zero mode whose
normalization in (\ref{dip}) says us that momenta have to be
quantized in units of $1/R\sqrt{1+(2\pi\alpha'
 \ H_i)^2}$. Thus, for the ``dipole'' string, we only have to
substitute $P_1P_2$ with the sum $\tilde{P}_1 \tilde{P}_2$, that
is defined with momenta $m_i/R\sqrt{1+(2\pi\alpha'
 \ H_i)^2}$.
On the other hand, the charged terms in the annulus are associated
to theta functions of non-vanishing argument, since their
frequencies are shifted. The deformed annulus amplitude is then
\begin{eqnarray}
{\cal A} &=& \frac{1}{4} \Biggl\{ (Q_o + Q_v)(0;0) \left[
(n+\bar n)^2 P_1 P_2 + (d+\bar d)^2 W_1 W_2
+ 2 m \bar{m} \tilde P_1 \tilde P_2 \right]
\nonumber
\\
&-& 2 (m+\bar m) (n + \bar{n}) (Q_o + Q_v )(z_1 \tau ; z_2 \tau
) {k_1 \eta \over
\vartheta_1 (z_1 \tau)} {k_2 \eta \over \vartheta_1 (z_2 \tau)}
\nonumber
\\
&-& ( m^2 + \bar{m}^2 ) (Q_o + Q_v ) (2 z_1 \tau ; 2 z_2 \tau )
{2 k_1 \eta \over
\vartheta_1 (2 z_1 \tau)} {2 k_2 \eta \over \vartheta_1 (2 z_2 \tau)}
\nonumber
\\
&-& \left[ (n-\bar n)^2 -2 m\bar m + (d-\bar d)^2 \right] (Q_o - Q_v ) (0;0)
\left( {2\eta \over \vartheta_2 (0)}\right)^2
\nonumber
\\
&-& 2 (m-\bar m) (n - \bar{n}) (Q_o - Q_v ) (z_1 \tau ; z_2 \tau)
{2\eta \over \vartheta_2
(z_1 \tau)} {2\eta \over \vartheta_2 (z_2 \tau)}
\nonumber
\\
&-& (m^2 + \bar{m}^2) (Q_o - Q_v ) (2z_1 \tau ; 2z_2 \tau)
{2\eta \over \vartheta_2
(2z_1 \tau)} {2\eta \over \vartheta_2 (2z_2 \tau)}
\nonumber
\\
&+& 2 (n+\bar n ) (d+\bar d) (Q_s + Q_c) (0;0) \left({\eta \over
\vartheta_4 (0)}\right)^2
\nonumber
\\
&+& 2 (m + \bar{m})(d+\bar d)(Q_s + Q_c) (z_1 \tau ; z_2 \tau)
{\eta \over \vartheta_4
(z_1 \tau )} {\eta \over \vartheta_4 (z_2 \tau )}
\nonumber
\\
&-& 2 (n-\bar n) (d - \bar d) (Q_s - Q_c )
(0;0) \left( {\eta \over \vartheta_3 (0)}\right)^2 \label{annsusy}
\\
&-& 2  (m - \bar{m})(d-\bar d) (Q_s - Q_c) (z_1 \tau ; z_2 \tau)
 {\eta \over \vartheta_3
(z_1 \tau )} {\eta \over \vartheta_3 (z_2 \tau )} \Biggr\} \, ,
\nonumber
\end{eqnarray}
while the corresponding M\"{o}bius amplitude is
\begin{eqnarray}
{\cal M} &=& -\frac{1}{4} \Biggl[
(\hat Q_o + \hat Q_v )(0;0) \left[ (n+\bar n) P_1 P_2 + (d+\bar d) W_1
W_2 \right]
\nonumber
\\
&-& ( m + \bar{m}) (\hat Q_o + \hat Q_v ) (2z_1 \tau ; 2z_2
\tau) {2 k_1
\hat\eta \over \hat \vartheta_1 (2z_1\tau)} {2 k_2
\hat\eta \over \hat \vartheta_1 (2z_2\tau)}
\nonumber
\\
&-& \left( n+ \bar n + d + \bar d \right) (\hat Q_o - \hat Q_v )(0;0) \left(
{2\hat\eta \over \hat \vartheta_2 (0)}\right)^2 \label{mobsusy}
\\
&-& (m + \bar{m}) (\hat Q_o - \hat Q_v ) (2 z_1 \tau ; 2 z_2 \tau )
{2\hat\eta \over \hat\vartheta_2 (2z_1\tau)}
{2\hat\eta \over \hat\vartheta_2 (2z_2\tau)} \Biggr] \, ,
\nonumber
\end{eqnarray}
where terms with opposite $U(1)$ charges, and thus with opposite
arguments $z_i$, have been grouped together, using the symmetry of
the Jacobi theta-functions, and both the modulus of ${\cal A}$ and
that of ${\cal M}$ are indicated with $\tau$. Moreover, we note
that strings with one or two charged ends are associated with
functions respectively of arguments $z_i$ or $2z_i$.

At this point, we begin to study the spectrum, that is generically
not supersymmetric and can develop some tachyonic modes, giving
the so called Nielsen-Olesen instabilities \cite{nole}. In fact,
in the untwisted sector, and for small magnetic fields, the mass
formula acquires a correction of the type
\be \Delta M^2 = \frac{1}{2 \pi \alpha'} \; \sum_{i=1,2}
\Bigl[  (2 n_i + 1) |2 \pi \alpha ' (q_{\rm L} + q_{\rm R}) H_i| \nonumber\\
+ 4 \pi\alpha' (q_{\rm L} + q_{\rm R}) \Sigma_i H_i \Bigr] \, ,
\ee
where the first term is the contribution of the Landau levels and
the second is the coupling of the magnetic moments of spins
$\Sigma_i$ to the magnetic fields. From the formula for $\Delta
M^2$ it is clear that, for generic values of magnetic fields, the
magnetic couplings of the internal vectors can lower the Landau
zero-level energy, thus generating tachyons. On the other hand,
the internal fermions can at most compensate it. In the twisted
sector there are no Landau levels, and while the fermionic part
$S_4O_4$ of $Q_s$ develops no tachyons, as the internal characters
are scalars, and the scalars do not have magnetic couplings, the
bosonic part $O_4C_4$ has a mass shift that again gives origin to
tachyons. An interesting thing here happens if $H_1=H_2$. In fact,
in such case all tachyonic instabilities are eliminated. Not only,
what one can show is that for $H_1=H_2$, using some Jacobi
identities, both ${\cal A}$ and ${\cal M}$ vanish identically, so
that a residual supersymmetry is present at the full string level.

We start to impose the various tadpole conditions. First of all,
we give the untwisted R-R tadpole condition, that for $C_4S_2C_2$
is \be \left[ n+\bar n + m + \bar{m} - 32 +  (2 \pi \alpha 'q)^2
H_1 H_2 (m + \bar m ) \right] \sqrt{v_1 v_2} + {1\over \sqrt{v_1
v_2}} \left[ d+\bar d - 32\right]  = 0 \, . \label{RRtad} \ee The
other untwisted R-R tadpole conditions are compatible with this
one, or vanish after the identifications $n=\bar n$, $m=\bar m$,
$d=\bar d$. The tadpole (\ref{RRtad}) is linked, in the low-energy
effective field theory, to the Wess-Zumino action we wrote in
(\ref{BI-WZ}). If we impose the Dirac quantization conditions on
the two tori, we can write (\ref{RRtad}) as
\begin{eqnarray}
& & m+\bar m + n + \bar n = 32 \, ,
\nonumber
\\
& & k_1 k_2 (m + \bar m )  + d + \bar d = 32 \, ,
\label{tadp}
\end{eqnarray}
from which one can see the phenomenon we already described at the
low-energy effective field theory level: the magnetized D9 branes
acquire the R-R charge of $|k_1k_2|$ D5 branes if $k_1k_2>0$, or
of $\overline{\rm D5}$ antibranes if $k_1k_2<0$. The untwisted
NS-NS tadpoles, apart from terms that vanish after the
identification of conjugate multiplicities, do not vanish for
general values of the magnetic fields. The contribution from
$V_4O_2O_2$,
\ba & & \left[ n+\bar n + (m + \bar m) \sqrt{\left( 1 +  (2 \pi
\alpha' q)^2 H_1^2 \right) \left( 1 + (2 \pi \alpha 'q) ^2 H_2^2
\right) }  -32 \right] \sqrt{v_1 v_2}
\nonumber \\
& & + {1\over \sqrt{v_1 v_2}} \left[ d +
\bar d - 32 \right]  \ ,
\ea
is the dilaton tadpole and is related to the derivative of the
Born-Infeld action with respect to the dilaton. There are also
tadpoles from $O_4V_2O_2$,
\ba & & \left[ n+\bar n + (m + \bar m) \, {1 - (2\pi\alpha' q
H_1)^2 \over \sqrt{ 1 + (2\pi\alpha' q H_1)^2 }}\, \sqrt{ 1 + (2
\pi\alpha' q H_2)^2 } -32 \right] \sqrt{v_1
v_2} \nonumber \\
& & - {1\over \sqrt{v_1 v_2}} \left[ d+\bar
d - 32 \right]  \, ,
\ea
and from $O_4O_2V_2$, that is the same but with
$H_1\leftrightarrow H_2$, related to the dependence of the
Born-Infeld action from the volumes of the two internal tori.

We would also like to stress that, in contrast with the usual
case, the coefficients of $O_4V_2O_2$ and $O_4O_2V_2$ in the
transverse channel are not perfect squares. The reason is that
really the transverse annulus has the form $\langle {\cal
T}(B)|q^{L_{0}}|B \rangle$, where ${\cal T}$ denotes the
time-reversal operation. As the magnetic field is odd under
time-reversal, it introduces signs that do not make the amplitudes
sesquilinear forms.

Both the dilaton tadpole and the tadpoles from $O_4V_2O_2$ and
$O_4O_2V_2$ vanish for $H_1=H_2$, using the Dirac quantization
condition, and imposing the R-R tadpole cancellation, as is usual
for a supersymmetric theory. The twisted R-R tadpole from
$S_4O_2O_2$ reflects the fact that all the D5 branes are at the
same fixed point,
\be 15 \left[ {\textstyle{1\over 4}} (m-\bar m + n -\bar n )
\right]^2 + \left[ {\textstyle{1\over 4}} (m-\bar m + n - \bar n )
- (d-\bar d) \right]^2 \ , \ee
and vanishes identifying the conjugate multiplicities. The
corresponding NS-NS tadpole
\be {2 \pi \alpha ' q \, (H_1 - H_2 ) \over \sqrt{ (1 + (2 \pi
\alpha' q H_1)^2 ) (1 + (2\pi \alpha' q H_2)^2 ) }} \,. \ee
vanishes if $H_1=H_2$.

Before closing this part, let us restrict our attention to the
supersymmetric case $H_1=H_2$ and let us describe the massless
spectrum. The closed sector has the same spectrum as the
undeformed original model $T^4/\mathbb{Z}_2$. For the massless
open sector we have to solve the R-R tadpole conditions
(\ref{tadp}), and we analyze the case $k_1=k_2=2$. There are a
number of different solutions giving anomaly-free spectra. Here we
report only the simplest one, $d=0$, $n=12$, $m=4$, just to give
an idea of what is happening. The open amplitudes at the massless
level are
\ba {\cal A}_0 + {\cal M}_0 & = & m \bar m \, Q_o (0) + n\bar n \,
Q_o (0) + \frac{(n^2-n) + (\bar n^2 - \bar n) }{2} \, Q_v (0)
\nonumber
\\
&& + \left( \frac{k_1 k_2}{2} + 2 \right) (\bar m n +  m \bar n ) \,  Q_v
(\zeta \tau)
\nonumber\\
&& + 2 \, (k_1 k_2 + 1 ) \, \frac{(m^2-m) + (\bar m^2 - \bar m )}{2} \, Q_v (\zeta \tau) \ .
\ea
The gauge group is $U(12)\times U(4)$, and apart from vector
multiplets in the adjoint representation, the spectrum contains
hyper multiplets in the $(66+\overline{66},1)$, in five copies of
the $(1,6+\bar 6)$, and in four copies of the $(\overline{12},4)$.
This type of models give origin to a rich and interesting
low-energy phenomenology. The peculiar feature common to all of
them is the emergence of multiple matter families. Moreover, one
can see that the gauge group is broken to a subgroup without
preserving the rank, giving a new possibility with respect to the
rank reduction by powers of $2$ induced by a quantized $B_{ab}$.
\chapter{Tadpoles in Quantum Field Theory}
\section{An introduction to the problem}
In String Theory the breaking of supersymmetry is generally
accompanied by the emergence of NS-NS tadpoles, one-point
functions for certain bosonic fields to go into the vacuum.
Whereas their R-R counterparts signal inconsistencies of the
field equations or quantum anomalies \cite{pc}, these tadpoles are
commonly regarded as mere signals of modifications of the
background. Still, for a variety of conceptual and technical
reasons, they are the key obstacle to a satisfactory picture of
supersymmetry breaking, an essential step to establish a proper
connection with Particle Physics. Their presence introduces
infrared divergences in string amplitudes: while these have long
been associated to the need for background redefinitions
\cite{fs}, it has proved essentially impossible to deal with them
in a full-fledged string setting. For one matter, in a theory of
gravity these redefinitions affect the background space time, and
the limited technology presently available for quantizing strings
in curved spaces makes it very difficult to implement them in
practice.

This chapter, based on the paper \cite{Dudas:2004nd}, is devoted
to exploring what can possibly be learnt if one insists on working
in a Minkowski background, that greatly simplifies string
amplitudes, even when tadpoles arise. This choice may appear
contradictory since, from the world-sheet viewpoint, the emergence
of tadpoles signals that the Minkowski background becomes a
``wrong vacuum''. Indeed, loop and perturbative expansions cease
in this case to be equivalent, while the leading infrared
contributions need suitable resummations. In addition, in String
Theory $NS$-$NS$ tadpoles are typically large, so that a
perturbative approach is not fully justified. While we are well
aware of these difficulties, we believe that this approach has the
advantage of making a concrete string analysis possible, if only
of qualitative value in the general case, and has the potential of
providing good insights into the nature of this crucial problem. A
major motivation for us is that the contributions to the vacuum
energy from Riemann surfaces with arbitrary numbers of boundaries,
where $NS$-$NS$ tadpoles can emerge already at the disk level,
play a key role in orientifold models. This is particularly
evident for the mechanism of brane supersymmetry breaking
\cite{sugimoto,bsb}, where the simultaneous presence of branes and
antibranes of different types, required by the simultaneous
presence of $O_+$ and $O_-$ planes, and possibly of additional
brane-antibrane systems \cite{sugimoto,bsb,aiq}, is generically
accompanied by $NS$-$NS$ tadpoles that first emerge at the disk
and projective disk level. Similar considerations apply to
non-supersymmetric intersecting brane models
\cite{bachasmag,intersecting,Larosa}\footnote{Or, equivalently,
models with internal magnetic fields.}, and the three mechanisms
mentioned above have a common feature: in all of them
supersymmetry is preserved, to lowest order, in the closed sector,
while it is broken in the open (brane) sector. However, problems
of this type are ubiquitous also in closed-string constructions
\cite{schschwstr} based on the Scherk-Schwarz mechanism
\cite{scherkschwarz}, where their emergence is only postponed to
the torus amplitude.

To give a flavor of the difficulties one faces, let us begin by
considering models where only a tadpole $\Delta^{(0)}$ for the
dilaton $\varphi$ is present. The resulting higher-genus
contributions to the vacuum energy are then plagued with infrared
(IR) divergences originating from dilaton propagators that go into
the vacuum at zero momentum, so that the leading (IR dominated)
contributions to the vacuum energy have the form
\ba \Lambda_0 &=& e^{-\varphi} \Delta^{(0)} +  \frac{1}{2} \
\Delta^{(m)}  \left( i \, {\cal D}_0^{mn}\right) \Delta^{(n)}
+ \frac{1}{2} \ e^{\varphi}  \Delta^{(m)}  \left( i \, {\cal
D}_0^{mn}\right)  \Sigma^{np}  \left( i\, {\cal
D}_0^{pq}\right) \Delta^{(q)} +
\cdots  \nonumber \\
&=&  e^{-\varphi} \ \Delta^{(0)} \ + \ \frac{1}{2}\ \Delta \ [\,
1-e^{\varphi} \left( i\, {\cal D}_0 \right) \Sigma \, ]^{-1}
\left( i\, {\cal D}_0 \right) \ \Delta \ + \cdots \ . \label{i1}
\ea
Eq.~(\ref{i1}) contains in general contributions from the dilaton
and from its massive Kaluza-Klein recurrences, implicit in its
second form, where they are taken to fill a vector $\Delta$ whose
first component is the dilaton tadpole $\Delta^{(0)}$. Moreover,
\be \langle m |{\cal D}_0 (p^2) |n \rangle \ \equiv \
 {\cal D}_0^{(mn)} (p^2) =
\delta^{mn} {\cal D}_0^{(mm)} (p^2)  \ee \label{i01} and \be
\langle m |\Sigma_0 (p^2) |n \rangle \ \equiv \ \Sigma^{(mn)}
(p^2) \label{i011} \ee
denote the sphere-level propagator of a dilaton recurrence of mass
$m$ and the matrix of two-point functions for dilaton recurrences
of masses $m$ and $n$ on the disk. They are both evaluated at zero
momentum in (\ref{i1}), where the first term is the disk
(one-boundary) contribution, the second is the cylinder
(two-boundary) contribution, the third is the genus 3/2
(three-boundary) contribution, and so on.  The resummation in the
last line of (\ref{i1}) is thus the string analogue of the more
familiar Dyson propagator resummation in Field Theory,
\be -i\; {\cal D}^{-1} (p^2) \ = \ -i\; {\cal D}_0^{-1} (p^2) -
e^{\varphi} \, \Sigma (p^2) \ , \label{i111} \ee
where in our conventions the self-energy $\Sigma (p^2)$ does not
include the string coupling in its definition. Even if the
individual terms in (\ref{i1}) are IR divergent, the resummed
expression is in principle perfectly well defined at zero
momentum, and yields
\ba \Lambda_0 &=& e^{-\varphi}\Delta^{(0)} \, - \, \frac{1}{2}\,
e^{-\varphi} \, \Delta^{(0)} \, \Sigma^{-1 \ (00)} \, \Delta^{(0)}
\, \nonumber\\ &+&  \, \frac{1}{2} \sum_{m,n \neq 0} \! \Delta^{(m)}
\left([1-e^{\varphi} \left( i\; {\cal D}_0 \right) \Sigma ]^{-1}
\left( i\; {\cal D}_0 \right) \right)^{mn} \Delta^{(n)} \ .
\label{i02} \ea
In addition, the soft dilaton theorem implies that
\be e^\varphi \, \Sigma^{(00)} = \frac{\partial}{\partial \varphi}
\ \left(e^\varphi \, \Delta^{(0)}\right) \ , \ee
so that the first two contributions cancel one another, up to a
relative factor of two. This is indeed a rather compact result,
but here we are describing for simplicity only a partial
resummation, that does not take into account higher-point
functions: a full resummation is in general far more complicated
to deal with, and therefore it is essential to identify possible
simplifications of the procedure.

A lesson we shall try to provide in this chapter, via a number of toy
examples  based on model field theories meant to shed light on
different features of the realistic string setting, is that
\emph{when a theory is expanded around a ``wrong'' vacuum, the
vacuum energy is typically driven to its value at a nearby
extremum (not necessarily a minimum)}, while the IR divergences
introduced by the tadpoles are simultaneously eliminated. In an
explicit example we also display some
wrong vacua in which higher-order tadpole insertions cancel both
in the field {\it v.e.v.} and in the vacuum energy, so that the
lowest corrections determine the full resummations. Of course,
subtle issues related to modular invariance or to its counterparts
in open-string diagrams are of crucial importance if this program
is to be properly implemented in String theory, and make the
present considerations somewhat incomplete. For this reason, we
plan to return to this key problem in a future work
\cite{bdnps}. The special
treatment reserved to the massless modes has nonetheless a clear
motivation: tadpoles act as external fields that in general lift
the massless modes, eliminating the corresponding infrared
divergences if suitable resummations are taken into account. On
the other hand, for massive modes such modifications are expected
to be less relevant, if suitably small. We present a number of
examples that are meant to illustrate this fact: small tadpoles
can at most deform slightly the massive spectrum, without any
sizable effect on the infrared behavior. The difficulty associated
with massless modes, however, is clearly spelled out in eq.
(\ref{i02}): resummations in a wrong vacuum, even within a
perturbative setting of small $g_s$, give rise to effects that are
typically large, of disk (tree) level, while the last term in
(\ref{i02}) due to massive modes is perturbatively small provided
the string coupling $e^\varphi$ satisfies the natural bound
$e^\varphi < m^2 / M_s^2$, where for the Kaluza-Klein case $m$
denotes the mass of the lowest recurrences and $M_s$ denotes the
string scale. The behavior of massless fields in simple models can
give a taste of similar difficulties that they introduce in String
Theory, and is also a familiar fact in Thermal Field Theory
\cite{tft}, where a proper treatment of IR divergences points
clearly to the distinct roles of two power-series expansions, in
coupling constants and in tadpoles. As a result, even models with
small couplings can well be out of control, and unfortunately this
is what happens in the most natural (and, in fact, in all known
perturbative) realizations of supersymmetry breaking in String
Theory.

Despite all these difficulties, at times string perturbation
theory can retain some meaning even in the presence of tadpoles.
For instance, in some cases one can identify subsets of the
physical observables that are insensitive to $NS$-$NS$ tadpoles.
There are indeed some physical quantities for which the IR effects
associated to the dilaton going into the vacuum are either absent
or are at least protected by perturbative vertices and/or by the
propagation of massive string modes. Two such examples are
threshold corrections to differences of gauge couplings for gauge
groups related by Wilson line breakings and scalar masses induced
by Wilson lines. For these quantities, the breakdown of
perturbation theory occurs at least at higher orders. There are
also models with ``small'' tadpoles. For instance, with suitable
fluxes \cite{fluxes} it is possible to concoct ``small'' tadpoles,
and one can then define a second perturbative expansion, organized
by the number of tadpole insertions, in addition to the
conventional expansion in powers of the string coupling
\cite{bdnps}.

\section{``Wrong vacua'' and the effective action}

The standard formulation of Quantum Field Theory refers implicitly
to a choice of ``vacuum'', instrumental for defining the
perturbative expansion, whose key ingredient is the generating
functional of connected Green functions $W[J]$. Let us refer for
definiteness to a scalar field theory, for which
\be e^{\frac{i}{\hbar}W[J]} \ = \ \int\,[ D\phi ] \,
e^{\,\frac{i}{\hbar}\,(S(\phi)+  \int d^{\cal D} x \, J \phi )} \
, \ee
where
\be S(\phi) \ = \ \int d^{\cal D} x \left(\; - \;
\frac{1}{2}\,\partial_\mu\phi\,\partial^\mu\phi-V(\phi)\right) \ ,
\ee
written here symbolically for a collection of scalar fields $\phi$
in ${\cal D}$ dimensions. Whereas the conventional saddle-point
technique rests on a shift
\be \phi \ = \ \varphi \ + \ \phi_0 \ee
about an extremum of the full action with the external source,
here we are actually interested in expanding around a ``wrong''
vacuum $\phi_0$, defined by
\be \left. \frac{\delta S}{\delta \phi}\right|_{\phi=\phi_0} \ = \
- \; (J + \Delta ) \ , \label{ft9} \ee
where, for simplicity, we let $\Delta$ be
a constant, field-independent quantity, to be regarded as the
classical manifestation of a tadpole. In the following
Sections, however, we shall also discuss examples where
$\Delta$ depends on $\phi_0$.

The shifted action then expands according to
\be S(\varphi + \phi_0) \ = \ S(\phi_0) \ - \  (J + \Delta) \,
\varphi \ + \ \frac{1}{2} \, \varphi \, \left( i\, {\cal D}^{-1}
\right)_{\phi_0} \, \varphi \ + \ S_I(\phi_0,\varphi) \ ,
\label{ft1} \ee where \be \left( i\, {\cal D}^{-1}\right)_{\phi_0}
\ = \ \left. {\delta^2 \ S \over \delta \phi^2}\right|_{\phi_0} \
, \label{ft2} \ee
while $S_I(\phi_0,\varphi)$ denotes the interacting part, that in
general begins with cubic terms. After the shift, the generating
functional
\be e^{\frac{i}{\hbar}W[J]}=e^{\frac{i}{\hbar}[S(\phi_0)+ J\phi_0
]}\int [ D\varphi ] \, e^{\frac{i}{\hbar}[\frac{1}{2} \varphi \; i
\mathcal{D}^{-1}(\phi_0) \; \varphi - \Delta \varphi +
S_I(\varphi,\phi_0)]}  \label{ft3} \ee
can be put in the form
\be W [J] = S(\phi_0 ) +  \phi_0 J  +  \frac{i\hbar}{2}
\tr\ln\left(i\left. \mathcal{D}^{-1} \right|_{\phi_0}\right) + W_2
[J] + {i \over 2} \Delta \mathcal{D} \Delta \ , \label{ft4} \ee
where \be e^{\frac{i}{\hbar}W_2[J]}=\frac{\int [ D\varphi ] \,
e^{\frac{i}{\hbar}[\frac{1}{2}\varphi \; i \mathcal{D}^{-1}
(\phi_0)\; \varphi - \varphi \Delta + S_I(\varphi,\phi_0)]}} {\int
[ D\varphi ] \, e^{\frac{i}{\hbar} ({1 \over 2} \varphi \; i
\mathcal{D}^{-1} \; \varphi - \varphi \Delta )}} \ .
 \label{ft5} \ee

In the standard approach, $W_2$ is computed perturbatively
\cite{cw}, expanding $\exp(\frac{iS_I}{\hbar})$ in a power series,
and contributes starting from two loops. On the contrary, if
classical tadpoles are present it also gives tree-level
contributions to the vacuum energy, but these are at least ${\cal
O}(\Delta^3)$. Defining as usual the effective action as
\be \Gamma(\bar\phi) \ = \ W[J] \ - \ J\bar\phi \ , \label{ft6}
\ee
the classical field is
\be \bar\phi=\frac{\delta W}{\delta J}= \phi_0 + \frac{\delta
\phi_0}{\delta J}\,\bigg(\; - \; \Delta+\frac{1}{2}
\Delta^2\frac{\delta}{\delta \phi_0} i \mathcal{D}(\phi_0)+
\frac{i \hbar}{2}\tr\frac{\delta}{\delta \phi_0}
\ln(i\mathcal{D}^{-1}({\bar \phi_0})) \bigg) + \cdots \ .
\label{ft12} \ee
Notice that $\bar\phi$ is no longer a small quantum correction to
the original ``wrong'' vacuum configuration $\phi_0$, and indeed
the second and third terms on the \emph{r.h.s.} of (\ref{ft12}) do
not carry any $\hbar$ factors. Considering only tree-level terms
and working to second order in the tadpole, one can solve for
$\phi_0$ in terms of $\bar\phi$, obtaining
\be \phi_0\ = \ \bar\phi \ + \ i \mathcal{D}(\bar\phi)\Delta \ + \
\frac{1}{2}\,i \mathcal{D}(\bar\phi)
\,\frac{\delta}{\delta\bar\phi}i\mathcal{D}(\bar\phi)\,\Delta^2 \
+ \ {\cal O}(\Delta^3) \ ,\ee
and substituting in the expression for $\Gamma[\bar\phi]$ then
gives
\be \Gamma[\bar\phi] \ = \ S(\bar\phi)+\left( \left. {\delta \ S
\over \delta \phi}\right|_{\bar{\phi}}+J\right)\,
\left[i\mathcal{D}(\bar\phi)\,\Delta+
\frac{1}{2}\,i\mathcal{D}(\bar\phi)\,\frac{\delta}{\delta\bar\phi}\,
i\mathcal{D}(\bar\phi)\,\Delta^2\right]+{\cal O}(\Delta^3) \ .
\label{gammawrong} \ee
One can now relate $J$ to $\,\bar\phi\,$ using eq.~(\ref{ft9}),
and the result is
\be J  = -\; \left. {\delta \ S \over \delta
\phi}\right|_{\bar{\phi}}\, -i\; \mathcal{D}^{-1}\,
(\bar\phi)(i\mathcal{D}(\bar\phi)\,\Delta) -\Delta+{\cal
O}(\Delta^2)=-\left. {\delta \ S \over \delta
\phi}\right|_{\bar{\phi}}+{\cal O}(\Delta^2) \ . \ee
Making use of this expression in (\ref{gammawrong}), all explicit
corrections depending on $\Delta$ cancel, and the tree-level
effective action reduces to the classical action:
\be \Gamma[\bar\phi] \ = \ S(\bar\phi)\ + \ {\cal O}(\Delta^3) \ .
\ee \label{wrongright}
This is precisely what one would have obtained expanding around an
extremum of the theory, but we would like to stress that this
result is here recovered expanding around a ``wrong'' vacuum. The
terms ${\cal O}(\Delta^3)$, if properly taken into account, would
also cancel against other tree-level contributions originating
from $W_2[J]$, so that the recovery of the classical vacuum energy
would appear to be exact. We shall return to this important issue
in the next Section.

\begin{figure}[h]
\begin{center}
  \resizebox{14cm}{!}{\psfig{figure=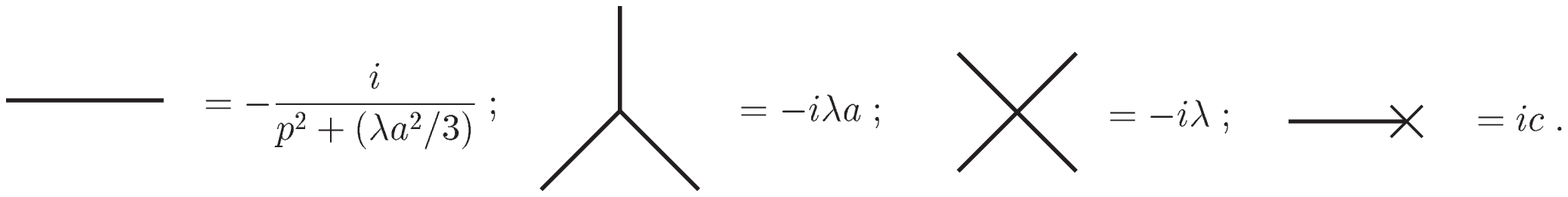,width=14cm}}
\caption{Feynman rules at $\phi_0=a$} \label{fig3}
\end{center}
\end{figure}

Let us take a closer look at the case of small tadpoles, that is
naturally amenable to a perturbative treatment. This can
illustrate a further important subtlety: the diagrams that
contribute to this series \emph{are not all 1PI}, and thus by the
usual rules \emph{should not all} contribute to $\Gamma$. For
instance, let us consider the Lagrangian
\be {\cal L} \ =\  - \, \frac{1}{2} \, (\dd_\mu\phi)^2\ -\
\frac{\lambda}{4!}(\phi^2-a^2)^2 \ + \ c\; \phi \ ,
\label{modelaction} \ee
with a Mexican-hat section potential and a driving ``magnetic
field'' represented by the tadpole $c$. The issue is to single out
the terms produced in the gaussian expansion of the path integral
of eq. (\ref{modelaction}) once the integration variable is
shifted about the ``wrong'' vacuum $\phi_0=a$, writing $\phi = a+
\chi$, so that
\be e^{\frac{i}{\hbar} \, W[J]} \ = \int [D\phi] \
e^{\frac{i}{\hbar} \int d^{\cal D} x \left( \; - \; \frac{1}{2}
\partial_\mu \chi
\partial^\mu \chi \, - \, \frac{\lambda a^2}{6} \chi^2  \, - \,
\frac{\lambda a}{3!} \chi^3 \, - \, \frac{\lambda}{4!} \chi^4 \, +
\, c(a+ \chi) \, + \, J (a + \chi) \right)} \ . \ee
Notice that, once $\chi$ is rescaled to $\hbar^{1/2} \chi$ in
order to remove all powers of $\hbar$ from the gaussian term, in
addition to the usual positive powers of $\hbar$ associated to
cubic and higher terms, a negative power $\hbar^{-1/2}$
accompanies the tadpole term in the resulting Lagrangian. As a
result, the final ${\cal O}(1/\hbar)$ contribution that
characterizes the classical vacuum energy results from infinitely
many diagrams built with the Feynman rules summarized in fig.
\ref{fig3}. The first two non-trivial contributions originate from
the three-point vertex terminating on three tadpoles, from the
four-point vertex terminating on four tadpoles and from the
exchange diagram of fig. \ref{fig4}.

\begin{figure}[h]
\begin{center}
  \resizebox{14cm}{!}{\psfig{figure=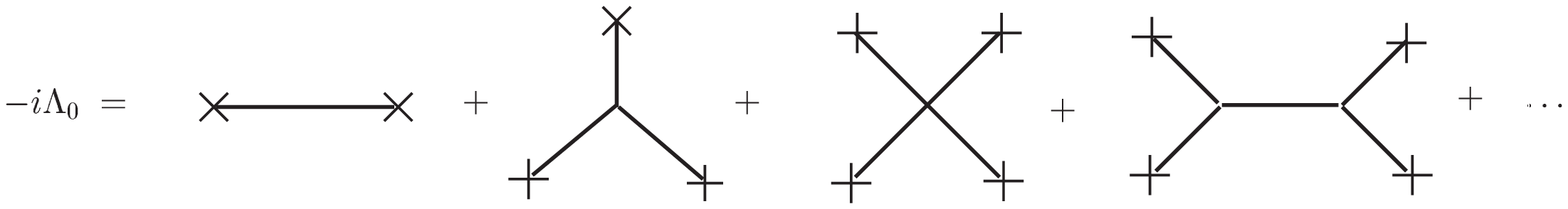,width=14cm}}
\caption{Vacuum energy} \label{fig4}
\end{center}
\end{figure}

As anticipated, tadpoles affect substantially the character of the
diagrams contributing to $\Gamma$, and in particular to the vacuum
energy, that we shall denote by $\Lambda_0$. Beginning from the
latter, let us note that, in the presence of a tadpole coupling
$c$,
\be e^{- \frac{i}{\hbar} \Lambda_0\, {\cal V}} \ = \ \int [D\phi]
\ e^{\frac{i}{\hbar} \left( S[\phi] \, + \, \int d^{\cal D} x \; c
\phi \right)} \ , \ee
where ${\cal V}$ denotes the volume of space time. Hence, the
vacuum energy is actually determined by a power series in $c$
whose coefficients are \emph{connected}, rather than 1PI,
amplitudes, since they are Green functions of W computed for a
classical value of the current determined by $c$:
\be - \, \frac{i}{\hbar} \, \Lambda_0 \, {\cal V} \  = \ \sum_n \
\frac{(i \; c )^n}{n! \; \hbar^n} \ W^{(n)}[\{p_j = 0\}] \ . \ee
A similar argument applies to the higher Green functions of
$\Gamma$: the standard Legendre transform becomes effectively in
this case
\be W[J+c] \ = \ \Gamma[\bar{\phi}] \ + \ J \bar{\phi} \ ,
\label{shiftleg} \ee
since the presence of a tadpole shifts the argument of W. However,
the \emph{l.h.s.} of (\ref{shiftleg}) contains an infinite series
of conventional connected Green functions, and after the Legendre
transform only those portions that do not depend on the tadpole
$c$ are turned into 1PI amplitudes. The end conclusion is indeed
that the contributions to $\Gamma$ that depend on the tadpoles
involve arbitrary numbers of connected, but also non 1PI,
diagrams.

The vacuum energy is a relatively simple and most important
quantity that one can deal with from this viewpoint, and its
explicit study will help to clarify the meaning of
eq.~(\ref{gammawrong}). Using eqs. (\ref{ft4}) and (\ref{ft6}) one
can indeed conclude that, at the classical level,
\be \Lambda_0 \ = \ - \, \frac{W(J=0)}{\cal V} \ = \ - \, V
(\phi_0) \ - \ \frac{i}{2} \ \Delta \, \left. {\cal
D}\right|_{p^2=0} \Delta + {\cal O} (\Delta^3) \ , \label{lambda1}
\ee
an equation that we shall try to illustrate via a number of
examples in this Chapter. The net result of this Section is that
resummations around a wrong vacuum lead nonetheless to extrema of
the effective action. However, it should be clear from the
previous derivation that the scalar propagator must be
nonsingular, or equivalently the potential must not have an
inflection point at $\phi_0$, in order that the perturbative
corrections about the original wrong vacuum be under control. A
related question is whether the resummation flow converges
generically towards minima (local or global) or can end up in a
maximum. As we shall see in detail shortly, the end point is
generally an extremum and not necessarily a local minimum.

\section{The end point of the resummation flow}

The purpose of this Section is to investigate, for some explicit
forms of the scalar potential $V (\phi)$ and for arbitrary initial
values of the scalar field $\phi_0$, the end point reached by the
system after classical tadpole resummations are performed. The
answer, that will be justified in a number of examples, is as
follows: {\it starting from a wrong vacuum $\phi_0$, the system
typically reaches a nearby extremum (be it a minimum or a maximum)
of the potential not separated from it by any inflection.} While
this is the generic behavior, we shall also run across a notable
exception to this simple rule: there exist some peculiar ``large''
flows, corresponding to special values of $\phi_0$, that can
actually reach an extremum by going past an inflection, and in
fact even crossing a barrier, but are nonetheless captured by the
low orders of the perturbative expansion!

An exponential potential is an interesting example that is free of
such inflection points, and is also of direct interest for
supersymmetry breaking in String Theory. Let us therefore begin by
considering a scalar field with the Lagrangian
\be {\cal L} \ = \ - \, \frac{1}{2} \, \partial^\mu \phi \,
\partial_\mu \phi \ - \ \alpha \ e^{b\phi} \ , \ee
where for definiteness the two coefficients $\alpha$ and $b$ are
both taken to be positive. The actual minimum is reached as $\phi
\to - \infty$, where the classical vacuum energy vanishes.

In order to recover this result from a perturbative expansion
around a generic ``wrong'' vacuum $\phi_0$, let us shift the
field, writing $\phi=\phi_0+\chi$. The Feynman rules can then be
extracted from
\be {\cal L}_{eff} = - \, \frac{1}{2}\
\partial^\mu \chi \, \partial_\mu \chi \ - \ \frac{\Delta}{b}\
e^{b\chi} \ , \ee
where $\Delta$, the one-point function in the ``wrong'' vacuum, is
defined by
\be - \ \left . \frac{\delta {\cal
L}_{eff}}{\delta\phi}\right|_{\phi_0} \ = \ \Delta \ = \ \alpha \
b \ e^{b \phi_0} \ , \ee
and the first few contributions to the classical vacuum energy are
as in fig. \ref{fig5}.
\begin{figure}[h]
\begin{center}
\resizebox{14cm}{!}{\psfig{figure=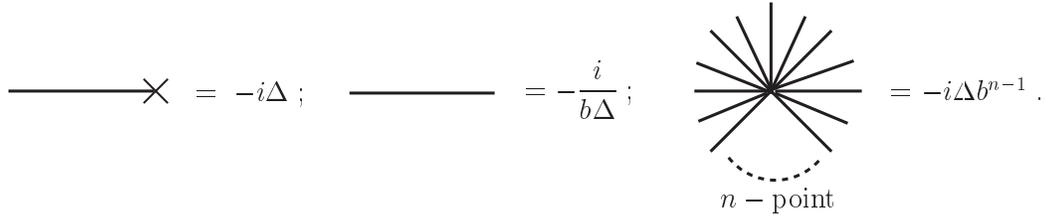,width=14cm}}
\caption{Feynman rules for the exponential potential} \label{fig5}
\end{center}
\end{figure}

It is fairly simple to compute the first few diagrams. For
instance, the two-tadpole correction to $- i V(\phi_0)$ is
\be \ \frac{1}{2} \ \frac{-i}{b\Delta} \ (-i\Delta)^2 =
-\frac{1}{2}\left(\frac{-i\Delta}{b}\right) \ , \ee while the
three-tadpole correction, still determined by a single diagram, is
\be \frac{1}{3!} \ (-i \Delta \ b^2) \ (-i \Delta)^3
\left(\frac{-i}{b \Delta}\right)^3 \ = \ - \
\frac{1}{6}\left(\frac{-i\Delta}{b}\right) \ . \ee
On the other hand, the quartic contribution is determined by two
distinct diagrams, and equals
\be \frac{1}{4!} \ (-i \Delta \ b^3) \ (-i \Delta )^4
\left(\frac{-i}{b \Delta}\right)^4 + \frac{1}{8} \ (-i \Delta \
b^2)^2 \ (-i \Delta)^4 \left(\frac{-i}{b \Delta}\right)^5 \ = \
 -\ \frac{1}{12}\left(\frac{-i\Delta}{b}\right) \ , \ee
while the quintic contribution originates from three diagrams.
Putting it all together, one obtains
\be \Lambda_0 \ = \ \frac{\Delta}{b} \
\left(1-\frac{1}{2}-\frac{1}{6}-\frac{1}{12}-\frac{1}{20}+\ldots
\right) \ = \ \frac{\Delta}{b}\left[1- \sum_{n=1}^{\infty}
\frac{1}{n(n+1)} \right] \ . \label{expwrong} \ee
The resulting pattern is clearly identifiable, and suggests in an
obvious fashion the series in (\ref{expwrong}). Notice that,
despite the absence of a small expansion parameter, in this
example the series in (\ref{expwrong}) actually \emph{converges}
to $1$, so that the correct vanishing value for the classical
vacuum energy can be exactly recovered from an arbitrary wrong
vacuum $\phi_0$.

\begin{figure}[h]
\begin{center}
\resizebox{5cm}{!}{\psfig{figure=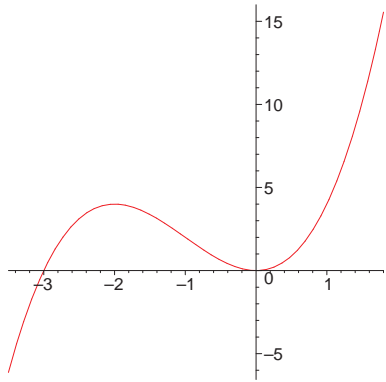,width=5cm}} \caption{A
cubic potential} \label{cubic}
\end{center}
\end{figure}

We can now turn to a more intricate example and consider the model
\be {\cal L} \ = \ - \, \frac{1}{2} (\dd_\mu\phi)^2\ -\  \frac{m^2
\phi^2}{2} \ - \ \frac{\lambda \phi^3}{6} \ , \label{ft33} \ee
the simplest setting where one can investigate the role of an
inflection. Strictly speaking this example is pathological, since
its Hamiltonian is unbounded from below, but for our purpose of
gaining some intuition on classical resummations it is nonetheless
instructive. The two extrema of the scalar potential $v_{1,2}$ and
the inflection point $v_I$ are
\be v_1 = 0 \ , \quad v_2 = - \frac{2 m^2}{\lambda} \ , \quad v_I
= - \frac{ m^2}{\lambda} \ . \label{ft14} \ee
Starting from an arbitrary initial value $\phi_0$, let us
investigate the convergence of the resummation series and the
resulting resummed value $\langle \phi \rangle$. A close look at
the diagrammatic expansion indicates that
\ba && \langle \phi \rangle \ = \ \phi_0 + \ \frac{V'}{V^{''}} \
\sum_{n=0}^{\infty} c_n \left[ \frac {\lambda V'}{(V^{''})^2}
\right]^n \ \equiv \ \phi_0 + \frac{V'}{V^{''}} \ f (x) \ ,
\label{ft15}\ea
where the actual expansion variable is
\be x \ = \ \frac {\lambda V'}{(V^{''})^2} \ , \label{xvar} \ee
to be contrasted with the naive dimensionless expansion variable
\be z = \frac{\lambda \phi_0}{m^2} \ . \ee
According to eq.~(\ref{xvar}), their relation is
\be x = \frac{z(z+2)}{2(z+1)^2} \ , \ee
whose inverse is
\be z = -1 \ \mp \ \frac{1}{\sqrt{1 - 2 x }} \ , \label{zdouble}
\ee
where the upper sign corresponds to the region $(L)$ to the left
of the inflection, while the lower sign corresponds to the region
$(R)$ to the right of the inflection. In other words: $\phi_0$,
and thus the naive variable $z$ of the problem, is actually a
double-valued function of $x$, while the actual range covered by
$x$ terminates at the inflection.

A careful evaluation of the symmetry factors of various diagrams
with variable numbers of tadpole insertions shows that
\be
 f (x) \ = \ - \, \frac{\sqrt{\pi}}{2} \sum_{n=0}^{\infty} \frac{(-1)^n
   \ 2^{n+1}}{(n+1)!\ \Gamma {(1/2-n)}} \ x^n\ ,
\label{ft166} \ee
a series that for $|x|<1/2$ converges to
\be
 f(x)  \ = \ -\, \frac{1}{x} \ + \
 \frac{\sqrt{1-2x}}{x} \ . \label{ft16}
\ee
The relation between $z$ and $x$ implies that both $\phi_0$ and
$V^{''}$ have two different expressions in terms of $x$ on the two
sides of the inflection,
\be V^{''} = \pm \frac{m^2}{\sqrt{1-2x}} \quad , \quad \phi_0 =
-\frac{m^2}{\lambda} \mp \frac{m^2}{\lambda} \frac{1}{\sqrt{1-2x}}
\ , \label{ft17} \ee
where the upper signs apply to the region $(L)$ that lies to the
left of the inflection, while the lower signs apply to the region
$(R)$ that lies to the right of the inflection. Combining
(\ref{ft16}) and (\ref{ft17}) finally yields the announced result:
\ba && \langle \phi \rangle \rightarrow 0 \quad {\rm in \
the \ (R) \ region: } \quad - \frac{m^2}{\lambda}  < \phi_0 < \infty \ , \nonumber \\
&& \langle \phi \rangle \rightarrow - \frac{2m^2}{\lambda} \quad
 {\rm for \ in \ (L) \ region:} \quad - \infty < \phi_0 < - \frac{m^2}{\lambda} \
 . \label{ft18}
\ea
The resummation clearly breaks down near the inflection point
$v_I$. In the present case, the series in (\ref{ft16}) converges
for $|x| < 1/2$, and this translates into the condition
\be \phi_0 \in \left(- \infty,-\frac{m^2}{\lambda} \left(1 +
\frac{1}{\sqrt{2}}\right)\right) \ \bigcup \
\left(-\frac{m^2}{\lambda} \left(1 -
\frac{1}{\sqrt{2}}\right),\infty\right) \
 . \label{ft19} \ee

A symmetric interval around the inflection point thus lies outside
this region, while in the asymptotic regions $\phi_0 \rightarrow
\pm \infty$ the parameter $x$ tends to $1/2$, a limiting value for
the convergence of the series (\ref{ft166}).

The vacuum energy $\Lambda_0$ is another key quantity for this
problem. Starting as before from an arbitrary initial value
$\phi_0$, standard diagrammatic methods indicate that
\be \Lambda_0 = \ V(\phi_0) + \ \frac{(V')^2}{V^{''}} \
\sum_{n=0}^{\infty} d_n \left[ \frac {\lambda V'}{(V^{''})^2}
\right]^n \ \equiv \ V(\phi_0) + \frac{(V')^2}{V^{''}} \ h (x) \ .
\label{ft20} \ee
A careful evaluation of the symmetry factors of various diagrams
with arbitrary numbers of tadpole insertions then shows that
\be
 h (x) \ = \ - \, \frac{\sqrt{\pi}}{4} \sum_{n=0}^{\infty} \frac{(-1)^n
   \ 2^{n+2}}{(n+2)!\ \Gamma {(1/2-n)}} \ x^n  \ , \label{ft21} \ee
that for $|x| <1/2$ converges to
\be h(x) \ = \ \frac{1}{3x^2} \left[1-3x - (1-2x)^{3/2} \right] \
. \ee

The two different relations between $z$ and $x$ in (\ref{zdouble})
that apply to the two regions $(L)$ and $(R)$ finally yield:
\ba && \Lambda_0 \rightarrow 0 \ , \quad {\rm in \ the \ (R) \
  region } \ - \frac{m^2}{\lambda}  < \phi_0 < \infty \ , \nonumber \\
&& \Lambda_0 \rightarrow \frac{2m^6}{3 \lambda^2} \ , \quad
 {\rm in \ the \ (L) \ region} \   -\infty < \phi_0 < - \frac{m^2}{\lambda} \
 . \label{ft22}
\ea
To reiterate, we have seen how in this model the resummations
approach nearby extrema (local minima or maxima) not separated
from the initial value $\phi_0$ by any inflection.

\begin{figure}[h]
\begin{center}
\resizebox{5cm}{!}{\psfig{figure=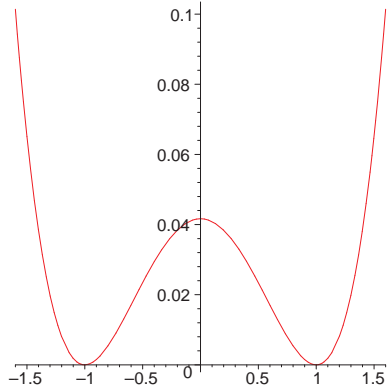,width=5cm}}
\caption{A quartic potential} \label{quartic}
\end{center}
\end{figure}

A physically more interesting example is provided by a real scalar
field described by the Lagrangian
\be {\cal L} \ = \ - \, \frac{1}{2} (\dd_\mu\phi)^2 \ - \
\frac{\lambda}{4!}\; (\phi^2-a^2)^2 \ , \label{phitothefour} \ee
whose potential has three extrema, two of which are a pair of
degenerate minima, $v_1= -a$ and $v_3=+a$, separated by a
potential wall, while the third is a local maximum at the origin.
Two inflection points are now present,
\be v^{(1)}_I = - \frac{a}{\sqrt{3}} \ , \qquad v^{(2)}_I =
\frac{a}{\sqrt{3}} \ , \ee
that also form a symmetric pair with respect to the vertical axis.

Starting from an arbitrary initial value $\phi_0$, one can again
in principle sum all the diagrams, and a closer look shows that
there are a pair of natural variables, $y_1$ and $y_2$, defined as
\be y_1 \ = \ \phi_0 \ \frac {\lambda V'}{(V^{''})^2} \qquad {\rm
and} \qquad y_2 \ =\ \frac {\lambda {V'}^2}{(V^{''})^3} \ ,
\label{ft23} \ee
that reflect the presence of the cubic and quartic vertices and
depend on the square of $\phi_0$. On the other hand, the naive
dimensionless variable to discuss the resummation flow is in this
case
\be z=\phi_0^2 / {a}^2 \ , \ee
and eqs.~(\ref{ft23}) imply that
\ba y_1 &=& 6\, \frac{z(z-1)}{(3 z -1 )^2} \ , \\
y_2 &=& 6\, \frac{z(z-1)^2}{(3 z -1 )^3} \ . \ea

The diagrammatic expansions of $\langle \phi \rangle$ and
$\Lambda_0$ are now more complicated than in the $\phi^{3}$
example. A careful evaluation of the symmetry factors of various
diagrams, however, uncovers an interesting pattern, since
\ba \langle \phi \rangle &=& \phi_0 + \frac{V'}{V^{''}} \left[-1
-\frac{1}{2}\left(y_1-\frac{y_2}{3}\right)-
\frac{1}{2}\left(y_1-\frac{y_2}{3}\right)
\left(y_1-\frac{y_2}{2}\right) \right. \nonumber
\\&-& \left. \frac{5}{8}\left(y_1-\frac{y_2}{3}\right) \left(y_1 - \frac{2 y_2}{3}\right)\left(y_1 -
\frac{2 y_2}{5}\right) \right. \nonumber \\
&-& \left. \frac{7}{8} \left(y_1 - \frac{y_2}{3}\right) \left(
y_1^3 - \frac{5}{3} y_1^2 y_2 + \frac{55}{63} y_1 y_2^2 -
\frac{55}{378} y_2^3 \right) \right. \nonumber \\
&-& \frac{21}{16}\ \left.
\left(y_1-\frac{y_2}{3}\right)\left(y_1^4-\frac{16}{7} y_1^3
y_2+\frac{13}{7} y_1^2 y_2^2 -
\frac{52}{81} y_1 y_2^3+\frac{13}{162} y_2^4\right)  + \cdots \right] \ , \nonumber \\
\Lambda_0 &=& V(\phi_0) + \frac{(V')^2}{V^{''}} \left[-\frac{1}{2}
-\frac{1}{6}\left(y_1-\frac{y_2}{4}\right)-
\frac{1}{8}\left(y_1-\frac{y_2}{3}\right)^2 \right. \nonumber \\
&-&\left.
\frac{1}{8}\left(y_1-\frac{y_2}{3}\right)^2
\left(y_1- \frac{1}{2} y_2\right) -\frac{7}{48}  \left(y_1-\frac{y_2}{3}\right)^2 \, \left(
y_1^2 - \frac{22}{21}
y_1 y_2 + \frac{11}{42} y_2^2 \right) \right. \nonumber \\
&-& \left. \frac{3}{16}\left(y_1-\frac{y_2}{3}\right)^2
\left(y_1^3-\frac{13}{8} y_1^2 y_2+ \frac{91}{108} y_1
y_2^2-\frac{91}{648} y_2^3\right) + \cdots \right] \ ,
\label{ft24} \ea
where all linear and higher-order corrections in $\langle \phi
\rangle$ and all quadratic and higher order corrections in
$\Lambda_0$ apparently disappear at the special point $y_1=y_2/3$.
Notice that this condition identifies the three extrema
$\phi_0=\pm a$ and $\phi_0=0$, but also, rather surprisingly, the
two additional points $\phi_0 = \pm \frac{a}{2}$. In all these
cases the series expansions for $\langle \phi \rangle$ and
$\Lambda_0$ apparently end after a few terms.

If one starts from a wrong vacuum {\it sufficiently close} to one
of the extrema, one can convince oneself that, in analogy with the
previous example,
\ba
&& \langle \phi \rangle \rightarrow -a \quad , \quad {\rm for
 \
  region \ 1:} \quad -\infty < \phi_0 <  v^{(1)}_I\ , \nonumber \\
&& \langle \phi \rangle \rightarrow 0 \quad , \quad
 {\rm for \ region \ 2:} \quad v^{(1)}_I  < \phi_0 < v^{(2)}_I \
 , \nonumber \\
&& \langle \phi \rangle \rightarrow +a \quad , \quad
 {\rm for \ region \ 3:} \quad v^{(2)}_I   < \phi_0 < \infty \  ,
\label{regions} \ea
but we have not arrived at a single natural expansion parameter
for this problem, an analog of the variable $x$ of the cubic
potential. In addition, while these perturbative flows follow the
pattern of the previous example, since they are separated by
inflections that act like barriers, a puzzling and amusing result
concerns the special initial points
\be
 \phi_0 = \pm \frac{a}{2} \ . \label{nonre}
\ee
In this case $y_2 = 3 y_1$ and, as we have seen, apparently all
but the first few terms in $\langle \phi \rangle$ and all but the
first few terms in $\Lambda_0$ vanish. The non-vanishing terms in
eq.~(\ref{ft24}) show explicitly that the endpoints of these
resummation flows correspond to $\langle \phi \rangle = \pm a$ for
$\phi_0 = \mp \frac{a}{2}$, and that $ \Lambda_0 = 0$, so that
these two flows apparently ``cross'' the potential barrier and
pass beyond an inflection. One might be tempted to dismiss this
phenomenon, since after all this is a case with large tadpoles
(and large values of $y_1$ and $y_2$), that is reasonably outside
the region of validity of perturbation theory and hence of the
strict range of applicability of eq. (\ref{ft24}). Still, toward
the end of this Section we shall encounter a similar
phenomenon, clearly within a perturbative setting, where the
resummation will unquestionably collapse to a few terms to land at
an extremum, and therefore it is worthwhile to pause and devote to
this issue some further thought.

Interestingly, the tadpole resummations that we are discussing
have a simple interpretation in terms of Newton's method of
tangents, a very effective iterative procedure to derive the roots
of non-linear algebraic equations. It can be simply adapted to our
case, considering the function $V'(\phi)$, whose zeroes are the
extrema of the scalar potential. The method begins with guess, a
``wrong vacuum'' $\phi_0$, and proceeds via a sequence of
iterations determined by the zeros of the sequence of straight
lines
\be y - V' (\phi^{(n)}) \ = \ V^{''} (\phi^{(n)}) \ (x-\phi^{(n)})
\ , \label{tangent} \ee
that are tangent to the curve at subsequent points, defined
recursively as
\be \phi^{(n+1)} \ = \ \phi^{(n)} \ - \  \frac{V'
(\phi^{(n)})}{V^{''} (\phi^{(n)})} \ , \ee
where $\phi^{(n)}$ denotes the $n$-th iteration of the wrong
vacuum $\phi^{(0)}= \phi_0$.

When applied to our case, restricting our attention to the first
terms the method gives
\ba \phi^{(1)} &=& \phi_0 - \frac{V'}{V^{''}} \ ,
\nonumber \\
\phi^{(2)} &=& \phi^{(1)} -
\frac{V'(\phi^{(1)})}{V^{''}(\phi^{(1)})}=
 \phi_0 + \frac{V'}{V^{''}} \left[ \, - 1 \ - \ \frac{ \frac{V'\; V^{'''}}{2 (V^{''})^2} - \frac{(V')^2 \;
V^{''''}}{6 (V^{''})^3}}{1 - \frac{V'V^{'''}}{(V^{''})^2} +
\frac{(V')^2
V^{''''}}{2 (V^{''})^3}}\, \right] \nonumber \\
&\cong & \phi_0  + \frac{V'}{V^{''}} \biggl[-1
-\frac{1}{2}(y_1-\frac{y_2}{3})- \frac{1}{2}(y_1-\frac{y_2}{3})
(y_1-\frac{y_2}{2}) \biggr] \ , \label{ft25} \ea
where $y_1$ and $y_2$ are defined in (\ref{ft23}) and, for
brevity, the arguments are omitted whenever they are equal to
$\phi_0$. Notice the precise agreement with the first four terms
in (\ref{ft24}), that imply that our tadpole resummations have a
simple interpretation in terms of successive iterations of the
solutions of the vacuum equations $V'=0$ by Newton's method.
Notice the emergence of the combination $y_1-y_2/3$ after the
first iteration: as a result, the pattern of eqs. (\ref{ft24})
continues indeed to all orders.

In view of this interpretation, the non-renormalization points
$\phi_0 = \pm a/2$ acquire a clear geometrical interpretation: in
these cases the iteration stops after the first term $\phi^{(1)}$,
since the tangent drawn at the original ``wrong'' vacuum, say at
$a/2$, happens to cross the real axis precisely at the extremum on
the other side of the barrier, at $\langle \phi \rangle = -a$.
Newton's method can also shed some light on the behavior of the
iterations, that stay on one side of the extremum or pass to the
other side according to the concavity of the potential, and on the
convergence radius of our tadpole resummations, that the second
iteration already restricts $y_1$ and $y_2$ to the region
\be \left|y_1\ - \ \frac{y_2}{2}\right| \ < \ 1 \ . \label{ft27}
\ee

However, the tangent method behaves as a sort of Dyson resummation
of the naive diagrammatic expansion, and has therefore better
convergence properties. For instance, starting near the
non-renormalization point $\phi_0 \simeq -a/2$, the first
iteration lands far away, but close to the minimum $\phi^{(1)}
\simeq a$. The second correction, that when regarded as a
resummation in (\ref{ft24}) is large, is actually small in the
tangent method, since it is proportional to $(1/16) (y_1-y_2/3)$.
It should be also clear by now that not only the points
$\phi_0=\pm a/2$, but finite intervals around them, move across
the barrier as a result of the iteration. These steps, however, do
not have a direct interpretation in terms of Feynman diagram
tadpole resummations, since (\ref{ft27}) is violated, so that the
corresponding diagrammatic expansion actually diverges. The reason
behind the relative simplicity of the cubic potential (\ref{ft33})
is easily recognized from the point of view of the tangent method:
the corresponding $V'(\phi)$ is a parabola, for which Newton's
method never leads to tangents crossing the real axis past an
inflection point.

\begin{figure}[h]
\begin{center}
\resizebox{5cm}{!}{\psfig{figure=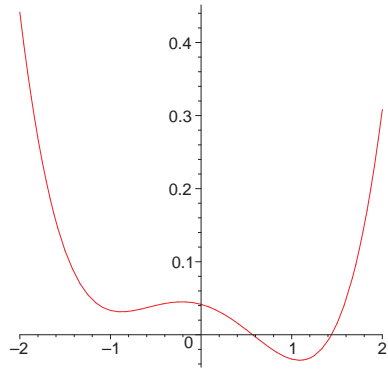,width=5cm}}
\caption{A quartic potential with a ``magnetic'' deformation}
\label{magnetic}
\end{center}
\end{figure}

There is a slight technical advantage in returning to the example
of eq. (\ref{modelaction}), since for a small magnetic field
(tadpole) $c$ one can expand the complete expressions for the
vacuum energy and the scalar {\it v.e.v.}'s in powers of the
tadpole. The expansions (\ref{ft24}) still apply, with an obvious
change in the one-point function $V'$, and their sums should
coincide, term by term, with the tadpole resummations obtained
starting from the undeformed ``wrong'' vacua $\phi_0 = \pm a,0$.

In this case the vacuum energy is given by
\be \Lambda_0 \ = \ \frac{\lambda}{4!}\, (v^2-a^2)^2 \ - \ c \; v
\ , \label{condens} \ee
where the correct vacuum value $\langle\phi\rangle = v $ is
determined by the cubic equation
\be \lambda \frac{v}{6} \; (v^2 - a^2) \ - \ c \ = \ 0 \ ,
\label{cubiceq} \ee
that can be easily solved perturbatively in the tadpole $c$, so
that if one starts around $\phi_0=a$,
\be \Lambda_0 \ = \ - \,  c \; a \ - \ \frac{3 c^2}{2 \lambda a^2}
\ + \ {\cal O}(c^3) \ . \ee
This result can be also recovered rather simply starting from the
wrong vacuum $\phi_0=a$ and making use of eq.~(\ref{lambda1}),
since in this case $\Delta=-c$ and ${\cal D}(p^2=0)=- 3i /\lambda
a^2$. However, the cubic equation (\ref{cubiceq}) can be also
solved exactly in terms of radicals, and in the small tadpole
limit its three solutions are real and can be written in the form
\be v \ = \ \frac{2a}{\sqrt{3}} \ \cos\left[ \frac{\alpha}{3} +
\frac{2 \pi k}{3} \right] \qquad (k=0,1,2) \ , \label{cubicsol}
\ee where \be \cos(\alpha) = \xi \ , \qquad \sin(\alpha) = \sqrt{1
- \xi^2} \ , \quad {\rm with} \quad \xi = \frac{9 \sqrt{3}
c}{\lambda a^3} \ . \ee
For definiteness, let us consider a tadpole $c$ that is small and
positive, so that the absolute minimum of the deformed potential
lies in the vicinity of the original minimum of the Mexican-hat
section at $v=a$ and corresponds to $k=0$. We can now describe the
fate of the resummations that start from two different wrong
vacua:

i) $\phi_0 = a$. In this case and in the small tadpole limit $\xi
<<1$ resummations in the diagrammatic language produce the first
corrections
\be
 \langle\phi\rangle = a + \frac{3 c}{\lambda a^2} -\frac{27 c^2}{2
 \lambda^2 a^5} + \cdots \ . \label{ft7}
\ee
Alternatively, this result could be obtained solving
eq.~(\ref{cubiceq}) in powers of the tadpole $c$ with the initial
value $\phi_0=a$, so that, once more, starting from a wrong vacuum
close to an extremum and resumming one can recover the correct
answer order by order in the expansion parameter. In this case
both a cubic and a quartic vertex are present, and the complete
expression for the vacuum energy, obtained substituting
(\ref{cubicsol}) in (\ref{condens}), is
\be \Lambda_0 \ = \ \frac{2\; \lambda a^4 \; \xi}{27} \ \left[
\frac{\xi}{16} \, \frac{1}{\cos^2\left( \frac{\alpha}{3} \right)}
\ - \ \cos\left( \frac{\alpha}{3} \right)  \right] \ .
\label{clvaen} \ee
This can be readily expanded in a power series in $\xi$, whose
first few terms,
\be \Lambda_0 \ = \ - a \; c \ - \ \frac{\sqrt{3}}{18} \; a \; c
\; \xi \ + \ \frac{1}{54} \; a \; c \; \xi^2 \ - \
\frac{\sqrt{3}}{243} \; a \; c \; \xi^3 \ + \ {\cal O}(\xi^4) \ .
\label{powerser} \ee match precisely the tadpole expansion
(\ref{ft24}).

ii)  $\phi_0 = 0$. In this case the first corrections obtained
resumming tadpole diagrams are
\be
 \langle\phi\rangle = - \frac{6 c}{\lambda a^2} -\frac{216 c^3}{
 \lambda^3 a^8} + \cdots \ . \label{ft8}
\ee
The same result can be obtained expanding the solution of
(\ref{cubiceq}) in powers of the tadpole $c$, starting from the
initial value $\phi_0=0$. We can now compare (\ref{cubicsol}) with
(\ref{ft24}), noting that in this case only the quartic vertex is
present, so that $y_1=0$. Since $y_2 = - 8 \xi^2/9$ for
$\phi_0=0$, the {\it v.e.v.} $\langle \phi \rangle$ contains only
odd powers of the tadpole $c$, a property that clearly holds in
(\ref{cubicsol}) as well, since $\xi \rightarrow - \xi$
corresponds to $\alpha \rightarrow \alpha + (2l+1) 3 \pi$, with
$l$ integer and $v \rightarrow -v$ in (\ref{cubicsol}). The
contributions to $\langle \phi \rangle$ are small and negative,
and therefore, starting from a wrong vacuum close to a maximum of
the theory, the resummation flow leads once more to a nearby
extremum (the local maximum slightly to the left of the origin, in
this case), rather than rolling down to the minimum corresponding
to the $k=0$ solution of (\ref{cubicsol}). The important points in
the scalar potential are again the extrema and the inflections,
precisely as we had seen in the example with a cubic potential.
Barring the peculiar behavior near the points identified by the
condition $y_1=y_2/3$, the scalar field flows in general to the
nearest extremum (minimum or maximum) of this potential, without
passing through any inflection point along the whole resummation
flow. It should be appreciated how the link with Newton's method
associates a neat geometrical interpretation to this behavior.

For $\xi = 1$ the two extrema of the unperturbed potential located
at $ \ v=-a \ $ and at $ \ v=0 \ $ coalesce with the inflection at
$-a/\sqrt{3}$. If the potential is deformed further, increasing
the value of the tadpole, the left minimum disappears and one is
left with only one real solution, corresponding to $k = 0$. The
correct parametrization for $ \ \xi>1 \ $ is $ \ \cosh(\alpha)=\xi
\ $ and $ \ \sinh(\alpha) = \sqrt{\xi^2-1} \ $, and the classical
vacuum energy is like in (\ref{clvaen}), but with $ \
\cos(\alpha/3)$ replaced by $\cosh(\alpha/3)$. In order to recover
the result (\ref{powerser}) working perturbatively in the
``wrong'' vacuum $v = a$, one should add the contributions of an
infinite series of diagrams that build a power series in $1/\xi$,
but this cannot be regarded as a tadpole resummation anymore. The
meaning of the parameter $\xi$ should by now have become apparent:
it is proportional to the product of the tadpole and the
propagator in the wrong vacuum, $V'/V''$, a natural expansion
parameter for problems of this type. Notice that the ratio
$V'/V^{''}$ is twice as large (and of opposite sign) at the origin
$\phi_0=0$ than at $\phi_0=a$. Therefore, the tadpole expansion
first breaks down around $\phi_0=0$. As we have seen, the endpoint
of the resummation flow for $\phi_0=0$ is the local maximum
corresponding to $k=2$ in (\ref{cubicsol}), that is reached for
$\xi < 1$. At $\xi=1$, however, the two extrema corresponding to
$k=1$ and $k=2$ coalesce with the inflection at $-a/\sqrt{3}$, and
hence there is no possible endpoint for the resummation flow. This
is transparent in (\ref{cubicsol}), since for $\xi \ge 1$ the
$\xi$ expansion clearly breaks down. It should therefore be clear
why, if the potential is deformed too extensively, corresponding
to $ \ \xi >1$, a perturbative expansion around the extrema $ \ v=
\pm a ,0\ $ in powers of $\xi$ is no longer possible. Another key
issue that should have emerged from this discussion is the need
for an independent, small expansion parameter when tadpoles are to
be treated perturbatively. In this example, as anticipated, the
expansion parameter $\xi$ can be simply related to the potential
according to
\be \xi \ = \ \frac{3 \sqrt{3}}{a} \ \left| \frac{V'}{V''} \right|
\ , \ee
that is indeed small if $c$ is small.

It is natural to ask about the fate of the points $\phi_0 = \pm
a/2$ of the previous example (\ref{phitothefour}) when the
magnetic field $c$ is turned on. Using the parameters $y_i$ in
(\ref{ft23}), the condition determining these special points,
$y_1=y_2/3$, is equivalent to
\be V' \ = \ 3 \ \phi_0 \ V^{''} \ . \label{ft99} \ee
It is readily seen that the solutions of this cubic equation are
precisely $\phi_0 = -v /2$, if $v$ denotes, collectively, the
three extrema solving (\ref{cubiceq}). Hence, in this case
$\langle \phi \rangle = - 2 \phi_0 = v$, confirming the
persistence of these non-renormalization points. In the present
case, however, there is a third non-renormalization point, $\phi_0
= - v_{k=2} /2$, that for small values of $\xi$ is well inside the
convergence region. This last point clearly admits an
interpretation in terms of tadpole resummations, and the possible
existence of effects of this type in String Theory raises the hope
that explicit vacuum redefinitions could be constructed in a few
steps for special values of the string coupling and of other
moduli.

\subsection{ On Newton's Tangent Method and the Quartic Potential}

It is actually simple to prove that, if one starts with $\phi_0$
in region 1, one ends up in the vacuum $<\phi> = -a$, while if one
starts in region 3 one ends up in $<\phi> = +a$, without ever
crossing the barrier. This is due to the fact that the function
$V'$ is convex in those regions, and that at the corresponding
extrema it has opposite signs. This implies that there is only one
zero, and that Newton's series converges to it. On the other hand
region 2, to which we now turn, is far richer, even though, in
some sense, it is ``non-perturbative''. Hence, our discussion will
rely directly on Newton's method, but will be only loosely related
to the conventional field theory expansion.

The inflections are ``initial points'' where the series clearly
diverges, since $V''$ vanishes there. If we enter region 2 and
move to the left of $a/ \sqrt{3}$ (of course, going to the right
of $-a/ \sqrt{3}$ would lead to very similar results), the first
iteration falls in region 1, while the successive ones lead
inevitably to $<\phi> = -a$.  Therefore, in this case Newton's
method leads to barrier crossing, and this happens until one hits
a point that we will call of the ``y-class'', that we shall denote
by $y_1$. Starting from this point, the first iteration crosses
the barrier and hits the $\phi$-axis exactly at $-a/ \sqrt{3}$.
$y_1$ solves the equation
$$
2 \sqrt{3} x^3 + 3 a x^2 - a^3 = 0 \qquad .
$$
Hence, after the first iteration, $y_1$ leads to a divergence. The
region ${\cal Y}_1= \{ y_1 < \phi_0 < a/ \sqrt{3} \}$ is therefore
the first ``barrier-crossing" region: starting there, one ends up
on the opposite side, at $<\phi> = -a$. This region, in
particular, contains the ``non renormalization point" $a/2$, that
solves the equation
$$
2 x^3 + 3 a x^2 - a^3 = 0 \qquad .
$$
Moving to the left of $y_1$, another region opens up.  This can be
seen noticing that the first iteration intersects the $\phi$-axis
to the right of $-a/ \sqrt{3}$, in such a way that the second
iteration ends inevitably inside region 3. Consequently, Newton's
method now leads to $<\phi> = +a$ after an infinite number of
steps and after crossing the barrier. Again, this happens until
one hits a point, that we will call of the ``z-class", say $z_1$,
such that the second iteration exactly intersects the $\phi$-axis
at the point $a/ \sqrt{3}$, bringing again to a divergence.  It
should be noticed that the region ${\cal Z}_1=\{ z_1 < \phi_0 <
y_1 \}$ is again of ``barrier-crossing type" (or, better, of
``bouncing type"), but the final vacuum is now $<\phi> = +a$. It
can also be thought of as the region such that, after the first
iteration, leads exactly to the opposite of ${\cal Y}_1$ region,
because the iteration of $z_1$ falls exactly, by definition, onto
$-y_1$.

The nature of the iterative procedure should now be apparent:
moving to the left of $z_1$, one meets another ``y-class" region,
${\cal Y}_2= \{ y_2 < \phi_0 < z_1 \}$, where $y_2$ now leads
exactly to $y_1$ after two iterations, and to the left inflexion
point $-a/ \sqrt{3}$ after three iterations. To the left of $y_2$,
there is the second ``z-class" region, ${\cal Z}_2 = \{ z_2 <
\phi_0 < y_2 \}$, where $z_2$ falls on $z_1$ after two iterations,
and on the inflection point $a/ \sqrt{3}$ after four iterations,
and so on. There is hence a countable infinity of ``y-class"
regions ${\cal Y}_n= \{ y_n < \phi_0 < z_{n-1} \}$, that alternate
with ``z-class" regions ${\cal Z}_n = \{ z_n < \phi_0 < y_n \}$.
These regions become smaller and smaller as $n$ increases, and
accumulate around a limiting point that delimits the region of
convergence to $<\phi>=0$. This point is characterized by the fact
that the first iteration leads to minus itself, in such a way
that, $V'$ being odd, the second iteration returns back to the
original point. This peculiar point solves the equation
$$
- x = x - \frac{x^3 - a^2 x}{3 x^2 - a^2} \qquad ,
$$
and is exactly $x = a/ \sqrt{5}$.  As a result, aside from the two
series of ${\cal Y}_n$ and ${\cal Z}_n$ regions, there is the
third region starting from which one ends at the maximum $<\phi> =
0$, namely  ${\cal O} = \{ -a/ \sqrt{5} < \phi_0 < a/
\sqrt{5} \}$.\\

\section{Branes and tadpoles}
\subsection{Codimension one}
Models whose tadpoles are confined to lower-dimensional surfaces
are of particular interest. In String Theory there are large
classes of examples of this type, including brane supersymmetry
breaking models \cite{sugimoto,bsb}, intersecting brane models
\cite{bachasmag,intersecting,Larosa} and models with internal fluxes \cite{fluxes}.
If the space transverse to the branes is large, the tadpoles are
"diluted" and there is a concrete hope that their corrections to
brane observables be small, as anticipated in \cite{bsb}. In the
codimension one case, tadpoles reflect themselves in boundary
conditions on the scalar (dilaton) field and hence on its
propagator, and as a result their effects on the Kaluza-Klein
spectrum and on brane-bulk couplings are nicely tractable.

Let us proceed by considering again simple toy models that display
the basic features of lower-dimensional tadpoles. The internal
space-time is taken to be $S^1/Z_2$, with $S^1$ a circle, and the
coordinate of the circle is denoted by $y$: in a string
realization its two endpoints $y = 0$ and $y = \pi R $ would be
the two fixed points of the orientifold operation $\Omega' =
\Omega \ \Pi_y $, with $\Pi_y$ the parity in $y$. We also let the
scalar field interact with a boundary gauge field, so that
\be S = \int d^4 {\bf x} \int_0^{\pi R} d y \biggl\{ \
-\frac{1}{2} ({\partial \phi})^2 \ - \ \left(T \ \phi \ + \
\frac{m}{2} \ \phi^2 \ - \ \phi \ \tr(F^2)\right) \ \delta (y -
\pi R) \ \biggr\} \ . \label{l9} \ee
The Lagrangian of this toy model describes a free massless scalar
field living in the bulk, but with a tadpole and a mass-like term
localized at one end of the interval $[0,\pi R]$. In String
Theory, both the mass-like parameter $m$ and the tadpole $T$ in
the examples we shall discuss would be perturbative in the string
coupling constant $g_s$. Any non-analytic IR behavior associated
with the possible emergence of $1/m$ terms would thus signal a
breakdown of perturbation theory, according to the discussion
presented in the first Section of this Chapter. Notice that, for dimensional
reasons, the mass term is proportional to $m$, rather than to
$m^2$ as is usually the case for bulk masses.

The starting point is the Kaluza-Klein expansion
\be
 \phi ({\bf x},y) \ = \ \phi_c (y) + \sum_k \chi_k (y) \ \phi_k ({\bf x}) \
 , \label{l3}
\ee
where $\phi_c (y)$ is the classical field and the $\phi_k ({\bf
x})$ are higher Kaluza-Klein modes. The classical field $ \ \phi_c
\ = \ - \ T/m \ $ solves the simple differential equation
\be \phi_c^{''} \ = \ 0 \  \ee
in the internal space, with the boundary conditions
\ba
&& \phi_c^{'} \ = \ 0 \  \qquad {\rm at} \ y = 0 \ , \nonumber\\
&& \phi_c^{'} \ = \ - \ T \ - \ m \ \phi_c  \ \qquad {\rm at} \ y
= \pi R \ , \label{l10} \ea
while the Kaluza-Klein modes satisfy in the internal space the
equations
\be \chi_k^{''} \ + \ M_k^2 \ \chi_k \ = \ 0 \ , \ee
with the boundary conditions
\ba &&  \chi_k^{'} \ = \ 0 \quad{\rm at} \quad y \ = \ 0 \ , \nonumber \\
&&  \chi_k^{'} \ = \ - \ m \ \chi_k  \quad {\rm at} \quad {\rm at}
\ y \ = \ \pi R \ . \label{l11} \ea
The corresponding solutions are then
\be \chi_k(y) \ = \ A_k \ \cos{(M_k y)} \ , \label{l12} \ee
where the masses $M_k$ of the Kaluza-Klein modes are determined by
the eigenvalue equation
\be M_k\tan{(M_k\pi R)} \ = \ m \ . \label{l13} \ee

The classical vacuum energy can be computed directly working in
the ``right'' vacuum. To this end, one ignores the Kaluza-Klein
fluctuations and evaluates the classical action in the correct
vacuum, as determined by the zero mode, with the end result that
\be \Lambda_0 \ = \ - \ \frac{T^2}{2m} \ . \label{l14} \ee
One can similarly compute in the ``right'' vacuum the gauge
coupling, obtaining
\be \frac{1}{g^2} \ = \ - \ \phi_c(y=\pi R) \ = \ \frac{T}{m} \ .
\label{l16} \ee

It is amusing and instructive to recover these results expanding
$\phi$ around the ``wrong'' vacuum corresponding to vanishing
values for both $T$ and $m$. The Kaluza-Klein expansion is
determined in this case by the Fourier decomposition
\be \phi ({\bf x}, y) = \frac{1}{\sqrt{\pi R}} \
\sum_{k=0}^{\infty}  b_k \cos \left(\frac{k y}{R}\right) \
\phi^{(k)} ({\bf x}) \ , \ \label{l17} \ee
where $b_0=1$ and $b_k= \sqrt{2}$ for $k \neq 0$, that turns the
action into
\be S_{KK} \ = \ \int d^4 {\bf x} \left\{\, -\,
\frac{1}{2}\sum_{k,l\geq 0} \phi_k\; \mathcal{M}^2_{k,l}\; \phi_l
\ - \ T\sum_{k\geq 0} \frac{b_k(-)^k}{\sqrt{\pi R}}\,
\phi_k\right\} \ . \label{l32} \ee
Here we are ignoring the kinetic term, since the vacuum energy is
determined by the zero momentum propagator, while the mass matrix
is
\be {\mathcal M}^2_{kl} \ = \ \frac{k^2}{R^2}\ \delta_{k,l} \ + \
\frac{m}{\pi R} \ b_k b_l \ (-)^{k+l} \ . \label{l33} \ee

The eigenvalues of this infinite dimensional matrix can be
computed explicitly using the techniques in \cite{ddg}. It is
actually a nice exercise to show that the characteristic equation
defining the eigenvalues of (\ref{l33}) is precisely (\ref{l13}),
and consequently that the eigenvectors of (\ref{l33}) are the
fields $\chi_k$ defined in (\ref{l12}). In fact, multiplying
(\ref{l33}) by normalized eigenfunctions $\Psi^{\lambda}_k$ gives
\be \Psi^{\lambda}_k \ = \ (-)^k \ \frac{m b_k}{\pi R} \ \frac{
\sum_{l}b_l(-)^l\Psi^{\lambda}_l}{\lambda^2 - \frac{k^2}{R^2}} \ ,
\label{l35} \ee
so that
\be \langle k|\lambda\rangle \ = \ \Psi_k^{\lambda} \ = \
\mathcal{N}_{\lambda} \frac{b_k(-)^k}{\frac{k^2}{R^2}-\lambda^2} \
. \label{l37} \ee
 and therefore
\be \sum_{k=0}^{\infty} \frac{b_k^2}{\lambda^2 - \frac{k^2}{R^2}}
\ = \ \frac{\pi R}{m} \ . \label{l36} \ee
The sum can be related to a well-known representation of
trigonometric functions \cite{ww},
\be {\rm cotg}(\lambda \pi R) \ = \ \frac{\lambda}{\pi R} \
\sum_{k=0}^\infty \frac{b_k^2}{\lambda^2 - \frac{k^2}{R^2}} \ ,
\ee
and hence the eigenvalues of (\ref{l33}) coincide with those of
(\ref{l13}). In order to compute the vacuum energy, one needs in
addition the k-component of the eigenvector $|\lambda\rangle$,
that can be read from (\ref{l35}). The normalization constant
$\mathcal{N}_{\lambda}$ in (\ref{l37}) is then determined by the
condition
\be 1 \ = \ \langle \lambda|\lambda\rangle \ = \
\mathcal{N}_{\lambda}^2\
\sum_{k=0}\frac{b_k^2}{\left(\frac{k^2}{R^2}-\lambda^2\right)^2} \
= \ \frac{\mathcal{N}_{\lambda}^2}{2\lambda} \ \frac{d}{d \lambda}
\ \sum_{k=0}\frac{b_k^2}{\frac{k^2}{R^2}-\lambda^2} \ ,
\label{l38} \ee
that using again eq. (\ref{l13}) can be put in the form
\be \mathcal{N}^2_\lambda \ = \ \frac{2m^2\lambda^2}{\pi^2 R^2} \
\frac{1}{\lambda^2+\alpha^2} \ , \label{l39} \ee
with
\be \alpha^2 = \frac{m}{\pi R} \left( 1 + \pi R \; m \right) \ .
\label{alpha2} \ee

Notice that in the limit $R \, m <<1$, that in a string context,
where $m$ would be proportional to the string coupling, would
correspond to the small coupling limit \cite{Sduality}, the physical
masses in (\ref{l13}) are approximately determined by the
solutions of the linearized eigenvalue equation, so that
\ba
M_0^2 &\cong& \frac{m}{\pi R} \ , \nonumber \\
M_k^2 &\cong& \frac{k^2}{R^2} + 2 \frac{m}{\pi R} \ . \label{l039}
\ea

One can now recover the classical vacuum energy using eq.
(\ref{lambda1}),
\be \Lambda_0 \ = \ -\frac{i}{2}\, \sum_{k,l}\Delta^{(k)}(y_1 =
\pi R) \, \langle k |{\cal D} ({\bf 0}~; y_1=\pi R,y_2=\pi R) |l
\rangle \, \Delta^{(l)}l(y_2 = \pi R) \ , \label{l45} \ee
that after inserting complete sets of eigenstates becomes
\be \Lambda_0 \ = \ -\, \frac{T^2}{2 \pi R}\sum_{\lambda, k,l} \
b_k b_l(-)^{k+l} \ \langle k|\Psi_\lambda\rangle \
\frac{1}{\lambda^2} \ \langle \Psi_\lambda|l\rangle \ = \
-\frac{T^2}{2 \pi R}\sum_{\lambda} \
\frac{\mathcal{N}_\lambda^2}{\lambda^2}
 \ \left(\sum_{k=0} \frac{b_k^2}{\frac{k^2}{R^2}-\lambda^2}\right)^2 \ ,
\label{l40} \ee
or, equivalently, using eq. (\ref{l36})
\be \Lambda_0 \ = \ -\, \frac{T^2}{\pi R} \
\sum_{\lambda}\frac{1}{\lambda^2+\alpha^2} \ . \label{l41} \ee

The sum over the eigenvalues in (\ref{l41}) can be finally
computed by a Sommerfeld-Watson transformation, turning it into a
Cauchy integral according to
\be \sum_{\lambda}\frac{1}{\lambda^2+\alpha^2} \ = \ \frac{1}{2} \
\oint\frac{d z}{2\pi i} \ \frac{1}{z^2+\alpha^2} \ \frac{(1+m\pi
R)\sin(\pi R z)+\pi R z\cos(\pi R z)}{z\sin(\pi R z)-m\cos(\pi R
z)} \ . \label{l42} \ee
The path of integration encircles the real axis, but can be
deformed to contain only the two poles at $ z= \pm i\alpha $. The
sum of the corresponding residues reproduces again (\ref{l14}),
since
\be \frac{1}{2\alpha} \ \frac{(1+m\pi R)\sinh(\pi R \alpha)+\pi R
\alpha\cosh(\pi R \alpha)} {\alpha\sinh(\pi R \alpha)+m\cosh(\pi R
\alpha)} \ = \ \frac{\pi R}{2 m} \ , \label{l43} \ee
where we used the definition of $\alpha$ in (\ref{alpha2}), even
though the computation was now effected starting from a wrong
vacuum.

It is useful to sort out the contributions to the vacuum energy
coming from the zero mode $\Lambda_0^{(0)}$ and from the massive
modes $\Lambda_0^{(m)}$. In a perturbative expansion using eq.
(\ref{l039}), one finds
\ba \Lambda_0^{(0)} &\cong&  - \ \frac{T^2}{2m} \ + \
\frac{T^2}{4}\ \pi R \ ,
\nonumber \\
\Lambda_0^{(m)} &\cong& - \ \frac{T^2}{4} \ \pi R \ . \label{l47}
\ea

Notice that the correct result (\ref{l14}) for the classical
vacuum energy is completely determined by the zero mode
contribution, while to leading order the massive modes simply
compensate the perturbation introduced by the tadpole, that in
String Theory would be interpreted, by open-closed duality, as the
one-loop gauge contribution to the vacuum energy. In a similar
fashion, in the wrong vacuum the gauge coupling can be read simply
from the amplitude with two external background gauge fields going
into a dilaton tadpole. In this case there are no other
corrections with internal gauge lines, since we are only
considering a background gauge field. The result for the gauge
couplings is then
\be \frac{1}{g^2} \ = \ \frac{T}{\pi R}\sum_{\lambda, k,l} \ b_k
b_l(-)^{k+l} \ \langle k|\Psi_\lambda\rangle \ \frac{1}{\lambda^2}
\ \langle \Psi_\lambda|l\rangle \ = \ \frac{T}{\pi
R}\sum_{\lambda} \ \frac{\mathcal{N}_\lambda^2}{\lambda^2}
 \ \left(\sum_{k=0} \frac{b_k^2}{\frac{k^2}{R^2}-\lambda^2}\right)^2 \ ,
\ee
so that using eq.~(\ref{l39}) for $ \ \mathcal{N}^2_\lambda \ $
and performing the sum as above one can again recover the correct
answer, displayed in (\ref{l16}).

\subsection{Higher codimension}

Antoniadis and Bachas argued that in orientifold models the
quantum corrections to brane observables \cite{ab} have a
negligible dependence on the moduli of the transverse space for
codimension larger than two. This result is due to the rapid
falloff of the Green function in the transverse space, but rests
crucially on the condition that the global $NS$-$NS$ tadpole
conditions be fulfilled. In this Subsection we would like to
generalize the analysis to models with $NS$-$NS$ tadpoles,
investigating in particular the sensitivity to scalar tadpoles of
the quantum corrections to brane observables. To this end, let us
begin by generalizing to higher codimension the example of the
previous Subsection, with
\be S = \int d^4 {\bf x} \int_0^{\pi R} d^n y \left\{ \
-\frac{1}{2} ({\partial \phi})^2 \ - \ \left(T \ \phi \ + \
\frac{m^2}{2} \ \phi^2\right) \ \delta^{(n)} (y)  \right\} \ .
\label{h1} \ee
The correct vacuum and the correct classical vacuum energy in this
example are clearly
\be
 \phi_c \ = \ - \, \frac{T}{m^2} \ , \qquad
\Lambda_0 \ = \ - \ \frac{T^2}{2m^2} \ . \label{h2} \ee
For simplicity, we are considering a symmetric compact space of
volume $V_n \equiv (\pi R)^n$, so that the Kaluza-Klein expansion
in the wrong vacuum is
\be \phi ({\bf x}, y) = \frac{1}{\sqrt{V_n}} \ \sum_{{\bf k}}
\prod_{i=1}^n \left[ b_{ k_i} \cos \left(\frac{k_i y_i}{R}\right)
\right] \ \phi^{({\bf k})} ({\bf x}) \ . \ \label{h3} \ee
After the expansion, the action reads
\be S_{KK} \ = \ \int d^4 {\bf x} \left( -\frac{1}{2}\sum_{{\bf
k},{\bf l}} \phi_{\bf k} \, \mathcal{M}^2_{{\bf k},{\bf l}}\,
\phi_{\bf l} \ - \ T\sum_{{\bf k}} \frac{b_{\bf k}}{\sqrt{V_n}}\
\phi_{\bf k} \right) \ , \label{h4} \ee
where, as in the previous Subsection, we neglected the space-time
kinetic term, that does not contribute, and where the mass matrix
is
\be {\mathcal M}^2_{{\bf k} {\bf l}} \ = \ \frac{{\bf
    k}^2}{R^2}\ \delta_{{\bf k},{\bf l}} \ + \ \frac{m^2}{V_n} \ b_{\bf
  k} b_{\bf l} \ . \label{h5}
\ee
In this case the physical Kaluza-Klein spectrum is determined by
the eigenvalues $\lambda$ of the mass matrix (\ref{h5}), and hence
is governed by the solutions of the ``gap equation''
\be 1 \ = \ \frac{m^2}{V_n} \sum_{\bf k} \frac{b_{\bf
k}^2}{\lambda^2 - {{\bf k}^2 \over
    R^2}} \ . \label{h6}
\ee

We thus face a typical problem for Field Theory in all cases of
higher codimension, the emergence of ultraviolet divergences in
sums over bulk Kaluza-Klein states. In String Theory these
divergences are generically cut off\footnote{The real situation is
actually more subtle. These divergences are infrared divergences
from the dual, gauge theory point of view, and are not regulated
by String Theory \cite{dm_1}. However, this subtlety does not
affect the basic results of this Section.} at the string scale
$|{\bf k}| < R M_s$, and in the following we shall adopt this
cutoff procedure in all UV dominated sums. In the small tadpole
limit $R m << 1$, approximate solutions to the eigenvalue equation
can be obtained, to lowest order, inserting the Kaluza-Klein
expansion (\ref{h3}) in the action, while the first correction to
the masses of the lightest modes can be obtained integrating out,
via tree-level diagrams, the heavy Kaluza-Klein states. In doing
this, one finds that to first order the physical masses are given
by
\ba M_0^2 &=& \frac{m^2}{V_n} \ \left(1 + c \, \frac{m^2}{M_s^2} +
\cdots \right) \ ,
\nonumber \\
M_k^2 &=& \frac{k^2}{R^2} + 2 \, \frac{m^2}{V_n} + \cdots \ .
\label{h7} \ea
The would-be zero mode thus acquires a small mass that, as in the
codimension-one example, signals a breakdown of perturbation
theory, whereas the corrections to the higher Kaluza-Klein masses
are very small and irrelevant for any practical purposes. We would
like to stress that the correct classical vacuum energy (\ref{h2})
is precisely reproduced, in the wrong vacuum, by the
boundary-to-boundary propagation of the single lightest mode,
since
\be \Lambda_0 \ = \ - \, \frac{1}{2} \ \frac{T}{\sqrt{V_n}} \,
\left(\frac{m^2}{V_n}\right)^{-1} \frac{T}{\sqrt{V_n}} \ ,
\label{h8} \ee
while the breaking of string perturbation theory is again manifest
in the nonanalytic behavior as $m \to 0$, so that the contribution
(\ref{h8}) is actually classical. On the other hand, as expected,
the massive modes give contributions that would not spoil
perturbation theory and that, by open-closed duality, in String
Theory could be interpreted as brane quantum corrections to the
vacuum energy. This conclusion is valid for any brane observable,
and for instance can be explicitly checked in this example for the
gauge couplings. This strongly suggests that for quantities like
differences of gauge couplings for different gauge group factors
that, to lowest order, do not directly feel the dilaton zero mode,
quantum corrections should decouple from the moduli of the
transverse space, as advocated in \cite{ab}. The main effect of
the tadpoles is then to renormalize the tree-level (disk) value,
while the resulting quantum corrections decouple as in their
absence.

\section{On the inclusion of gravity}

The inclusion of gravity, that in the Einstein frame enters the
low-energy effective field theory of strings via
\be S \ = \ \frac{1}{2 k^2} \, \int d^{\cal D} x \; \sqrt{-g}
\left( R - \frac{1}{2} (\partial \varphi)^2 \right) \ ,
\label{gravity} \ee
presents further subtleties. First, one is dealing with a gauge
theory, and the dilaton tadpole
\be \delta S \ = \ - T  \int d^{\cal D} x\ \sqrt{-g} \ e^{b
\varphi} \ , \label{gravity2} \ee
when developed in a power series around the wrong Minkowski vacuum
according to
\be g_{\mu\nu}\ = \ \eta_{\mu\nu} \ + \ 2 \;k\; h_{\mu\nu} \ ,
\qquad \varphi = \varphi_0 \ + \ \sqrt{2} \; k \; \phi \ , \ee
appears to destroy the gauge symmetry. For instance, up to
quadratic order it results in tadpoles, masses and mixings between
dilaton and graviton, since
\be T \sqrt{-g} \ e^{b \varphi} \ = \ T e^{b \varphi_0} \left[ 1 +
k h + b \phi - k^2 \left( h_{\mu\nu} h^{\mu\nu} - \frac{1}{2} h^2
\right) + k b \phi h + \frac{b^2}{2} \phi^2 + \ldots \right] \ ,
\ee
where $h$ denotes the trace of $h_{\mu\nu}$. If these terms were
treated directly to define the graviton propagator, no gauge
fixing would seem to be needed. On the other hand, since the fully
non-linear theory does possess the gauge symmetry, one should
rather insist and gauge fix the Lagrangian as in the absence of
the tadpole. Even when this is done, however, the resulting
propagators present a further peculiarity, that is already seen
ignoring the dilaton: the mass-like term for the graviton is not
of Fierz-Pauli type, so that no van Dam-Veltman-Zakharov
discontinuity \cite{vdvz} is present and a ghost propagates.
Finally, the mass term is in fact tachyonic for positive tension,
the case of direct relevance for brane supersymmetry breaking, a
feature that can be regarded as a further indication of the
instability of the Minkowski vacuum.

All these problems notwithstanding, in the spirit of this work it
is reasonable to explore some of these features referring to a toy
model, that allows to cast the problem in a perturbative setting.
This is obtained coupling the linearized Einstein theory with a
scalar field, adding to the Lagrangian (\ref{gravity})
\be \delta {\cal L} \ = \  - \frac{\lambda}{4!} (\phi^2 - a^2)^2 +
\frac{m^2}{2} (h_{\mu\nu} h^{\mu\nu} - h^2) + (\phi^2 - a^2) (h^2
+ b h) \ . \ee

This model embodies a couple of amusing features: in the correct
vacuum $\langle \phi \rangle = a$, the graviton mass is of
Fierz-Pauli type and describes five degrees of freedom in ${\cal
D}=4$, the vacuum energy vanishes, and no mixing is present
between graviton and dilaton. On the other hand, in the wrong
vacuum $\langle \phi \rangle = a(1+ \epsilon)$, the expected
${\cal O}(\epsilon)$ tadpoles are accompanied not only by a vacuum
energy
\be \Lambda_0 \ = \  \frac{\lambda a^4}{6} \, \epsilon^2   \ ,
\label{wrongvacen} \ee
but also by a mixing between $h$ and $\phi$ and by an ${\cal
O}(\epsilon)$ modification of the graviton mass, so that to
quadratic order
\be \delta {\cal L} \to   - \frac{\lambda}{4!} (4 \epsilon^2 a^4 +
8 a^3 \epsilon \phi + 4 a^2 \phi^2 ) + \frac{m^2}{2} (h_{\mu\nu}
h^{\mu\nu} - h^2) + 2 \epsilon a^2 (b h + h^2) + 2 a b \phi h \ .
\ee
Hence, in this model the innocent-looking displacement to the
wrong vacuum actually affects the degrees of freedom described by
the gravity field, since the perturbed mass term is no more of
Fierz-Pauli type. It is instructive to compute the first
contributions to the vacuum energy starting from the wrong vacuum.
To this end, one only needs the propagators for the tensor and
scalar modes at zero momentum to lowest order in $\epsilon$,
\ba && \langle h_{\mu\nu} h_{\rho\sigma} \rangle_{k=0}\ = \
\frac{i}{2 m^2} \left[ \eta_{\mu\rho} \eta_{\nu\sigma} +
\eta_{\mu\sigma} \eta_{\nu\rho} - \frac{4 a b - \frac{\lambda a
m^2}{3b}}{2 a b {\cal D} + \frac{\lambda a m^2(1-{\cal D})}{6b}}
\eta_{\mu\nu}\eta_{\rho\sigma} \right]
\ , \nonumber \\
&& \langle h_{\mu\nu} \varphi \rangle_{k=0} \ =\ \frac{i\,
\eta_{\mu\nu}}{2 a b {\cal D} +
\frac{\lambda a m^2(1-{\cal D})}{6b}} \ , \nonumber \\
&& \langle \varphi \varphi \rangle_{k=0} = - \frac{(1-{\cal
D})m^2}{2 a b} \, \frac{i}{2 a b {\cal D} + \frac{\lambda a
m^2(1-{\cal D})}{6b}} \ . \ea

There are three ${\cal O}(\epsilon^2)$ corrections to the vacuum
energy,
\ba && {\rm a: \quad} \frac{\lambda b \epsilon^2 a^4 {\cal D}}{3}
\ \frac{1}{2 b {\cal D} +
\frac{\lambda m^2}{6b}(1-{\cal D})} \ , \\
&& {\rm b: \quad} - \ \frac{2}{3} \lambda b \epsilon^2 a^4 {\cal
D} \
\frac{1}{2 b {\cal D} + \frac{\lambda m^2}{6b}(1-{\cal D})} \ , \\
&& {\rm c: \quad} - \ \frac{1}{36 b} \lambda^2 \epsilon^2 a^4
(1-{\cal D}) m^2 \ \frac{1}{2 b {\cal D} + \frac{\lambda
m^2}{6b}(1-{\cal D})} \ , \ea
coming from tensor-tensor, tensor-scalar and scalar-scalar
exchanges, according to eq. (\ref{lambda1}), and their sum is seen
by inspection to cancel the contribution from the initial wrong
choice of vacuum. Of course, there are also infinitely many
contributions that must cancel, order by order in $\epsilon$, and
we have verified explicitly that this is indeed the case to ${\cal
O}(\epsilon^3)$. The lesson, once more, is that starting from a
wrong vacuum for which the natural expansion parameter $\left|
V'/V'' \right|$ is small, one can recover nicely the correct
vacuum energy, even if there is a ghost field in the gravity
sector, as is the case in String Theory after the emergence of a
dilaton tadpole if one insists on quantizing the theory in the
wrong Minkowski vacuum.
\chapter{Tadpoles in String Theory}
\section{Evidence for a new link between string vacua}
We have already stressed that supersymmetry breaking in String
Theory is generally expected to destabilize the Minkowski vacuum
\cite{fs}, curving the background space-time. Since the
quantization of strings in curved backgrounds is a notoriously
difficult problem, it should not come as a surprise that little
progress has been made on the issue over the years. There are some
selected instances, however, where something can be said, and we
would like to begin this Chapter by discussing a notable example
to this effect.

Classical solutions of the low-energy effective action are a
natural starting point in the search for vacuum redefinitions, and
their indications can be even of quantitative value whenever the
typical curvature scales of the problem are well larger than the
string scale and the string coupling is small throughout the
resulting space time. If the configurations thus identified have
an explicit string realization, one can do even better, since the
key problem of vacuum redefinitions can then be explored at the
full string level. Our starting point are some intriguingly simple
classical configurations found in \cite{dmt}. As we shall see,
these solutions allow one to control to some extent vacuum
redefinitions at the string level in an interesting case, a
circumstance of clear interest to gain new insights into String
Theory.

Let us therefore consider the type-I' string theory, the T-dual
version of the type-I theory, with $D8/O8$ brane/orientifold
systems that we shall describe shortly, where for simplicity all
branes are placed at the end points $y=0$ and $y = \pi R$ of the
interval $S^1/Z_2$, the fixed points of the orientifold operation.
Let us also denote by $T_0$ ($q_0$) the tension (R-R charge) of
the $D8/O8$ collection at the origin, and by $T_1$ ($q_1$) the
tension (R-R charge) of the $D8/O8$ collection at the other
endpoint $y = \pi R$. The low-energy effective action for this
system then reads
\ba S &=& {1 \over 2{\kappa}^2} \int d^{10} x \, \sqrt {-G} \left[
e^{-2 \varphi} ( R + 4 ({\partial \varphi})^2)- {1 \over 2 \times
10 !} F_{10}^2
\right] \nonumber \\
&-& \int_{y=0} d^9 x \left(T_0 \ \sqrt{-\gamma} \ e^{-\varphi} \ +
q_0 \ A_9\right) - \int_{y= \pi R} d^9 x  \left(T_1 \sqrt{-\gamma}
\ e^{-\varphi} \ +q_1 A_9 \right) \ , \label{t1b} \nonumber\\
\ea
where we have included all lowest-order contributions. If
supersymmetry is broken, it was shown in \cite{dmt} that no
classical solutions exist that depend only on the transverse $y$
coordinate, a result to be contrasted with the well-known
supersymmetric case discussed by Polchinski and Witten in
\cite{Sduality}, where such solutions played an important role in
identifying the meaning of local tadpole cancellation. It is
therefore natural to inquire under what conditions warped
solutions can be found that depend on $y$ and on a single
additional spatial coordinate $z$, and to this end in the Einstein
frame one can start from the ansatz
\be ds^2=e^{2A(y,z)} \eta_{\mu\nu}dx^\mu dx^\nu \ + \ e^{2B(y,z)}
(dz^2+dy^2) \ . \ee

If the functions $A$ and $B$ and the dilaton $\varphi$ are allowed
to depend on $y$ and on $z$, the boundary conditions at the two
endpoints $0$ and $\pi R$ of the interval imply the two
inequalities \cite{dmt}
\be T_0^2 \leq q_0^2  \ , \quad  T_1^2 \leq q_1^2  \ ,
\label{condi2} \ee
necessary but not sufficient in general to guarantee that a
solution exist. As shown in \cite{dmt}, the actual solution
depends on two parameters, $\lambda$ and $\omega$, that can be
related to the boundary conditions at $y=0$ and $y = \pi R$
according to
\be \cos (\omega)= - T_0  / |q_0| \ , \quad \cos (\pi \lambda R +
\omega)=  T_1  / |q_1| \ , \label{t23} \ee
and reads
\ba e^{24 A} &=& e^{b_0+5 \varphi_0/4} \biggl[G_0 + {{3 {\kappa}^2
|q_0|} \over {2 \lambda}} \, e^{
\lambda z} \sin (\lambda |y|+\omega) \biggr] \ , \nonumber \\
e^{24 B} &=&   e^{24 \lambda z +25 b_0+5 \varphi_0/4} \biggl[G_0+
{{3 {\kappa}^2 |q_0|} \over {2 \lambda}} \, e^{
\lambda z } \sin (\lambda |y|+\omega) \biggr] \ , \nonumber \\
e^{\Phi} &=& e^{-5 b_0/6 - \varphi_0/24} \biggl[G_0 + {{3
{\kappa}^2 |q_0|} \over {2 \lambda}} \, e^{ \lambda z } \sin
(\lambda |y|+\omega) \biggr]^{-{5 \over 6}}
 \ , \label{t24}
\ea
where $b_0$, $\varphi_0$ and $G_0$ are integration constants. The
$z$ coordinate is noncompact, and as a result the effective Planck
mass is infinite in this background. There are singularities for
$z \to \pm \infty$ and, depending on the sign of $G_0$ and on the
numerical values of $\lambda$ and $\omega$, the solution may
develop additional singularities at a finite distance from the
origin in the $(y,z)$ plane.

This solution can be actually related to the supersymmetric
solution of \cite{Sduality}. Indeed, the conformal change of coordinates
\be Y \ = \ {1 \over \lambda} \ e^{\lambda z} \, \sin (\lambda y +
\omega) \ , \quad Z \ = \ {1 \over \lambda} \ e^{\lambda z} \,
\cos (\lambda y + \omega)  \ , \label{t025} \ee
or, more concisely
\be Z+ i~Y \ = \ {e^{i \omega} \over \lambda} \ e^{\lambda (z + i~
y)} \ , \label{map} \ee
maps the strip in the $(y,z)$ plane between the two $O$ planes
into a wedge in the $(Y,Z)$ plane and yields for $y >0$ the
space-time metric
\be ds^2 = \left[ G_0+ \frac{3{\kappa}^2 |q_0|}{2}\, Y \right]^{1
\over 12} \, \biggl( \eta_{\mu \nu} \, dx^{\mu} \, dx^{\nu} +dY^2+
dZ^2 \biggr) \ . \label{t026} \ee
\begin{figure}[h]
\begin{center}
\resizebox{6cm}{!}{\psfig{figure=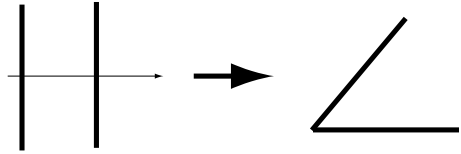,width=6cm}}
\caption{Eq.~(\ref{map}) maps a strip in the $(y,z)$ plane to a
wedge in the $(Y,Z)$ plane} \label{conformal}
\end{center}
\end{figure}

Notice that (\ref{t026}) is the metric derived by Polchinski and
Witten \cite{Sduality} in the supersymmetric case, but for one notable
difference: here the $Y$ direction is not compact. On the other
hand, in the new coordinate system $(Y,Z)$ the periodicity under
$y \to y+2 \pi R$ reflects itself in the orbifold identification
\be  Z+ i\, Y \ \to \ e^{2\pi i \lambda R} \ (Z+ i\, Y) \ ,
\label{t027} \ee
a two-dimensional rotation ${\cal R}_{\theta}$ in the $(Y,Z)$
plane by an angle $\theta = 2 \pi \lambda R$. In addition, the
orientifold identification $y\rightarrow -y$ maps into a parity
$\Pi_Y$ times a rotation ${\cal R}_{2 \omega}$ by an angle $2
\omega$, so that the new $\Omega$ projection is
\be \Omega' \ = \ \Omega \ \Pi_Y \ {\cal R}_{2 \omega} \ ,
\label{t028} \ee
where $\Omega$ denotes the conventional world-sheet parity. Notice
that both the metric and the dilaton in (\ref{t24}) depend
effectively on the real part of an analytic function, and thus
generally on a pair of real variables, aside from the case of
\cite{Sduality}, where the function is a linear one, so that one of the
real variables actually disappears. This simple observation
explains the special role of the single-variable solution of
\cite{Sduality} in this context.

As sketched in fig.~\ref{conformal}, the exponential mapping turns
the region delimited by the two parallel fixed lines of the
orientifold operations in the $(y,z)$ plane into a wedge in the
$(Y,Z)$ plane, delimited by the two lines
\ba
\Omega'~: \  Y &=& \tan \omega \ Z \ , \nonumber \\
\Omega'~ {\cal R}_{2 \pi \lambda R} : \  Y &=& \tan (\pi \lambda R
+ \omega ) \ Z \ , \label{t029} \ea
so that the orientifolds and the branes at $y=0$ form an angle
$\theta_0 = \ \omega $ with the $Z$ axis, while those at $y=\pi R$
form an angle $ \theta_1 = \ \pi\lambda R + \omega $. Notice that
the orbifold identification (\ref{t027}) implies that in general
the two-dimensional $(Y,Z)$ plane contains singularities. In order
to avoid subtleties of this type, in what follows we restrict our
attention to a case where this complication is absent.

The example we have in mind is a variant of the M-theory breaking
model of \cite{ads1}. Its oriented closed part is related by a
T-duality to a Scherk-Schwarz deformation of the toroidally
compactified IIB spectrum of \cite{ads1}, described by
\ba \mathcal{T} &=& ( |V_8|^2+|S_8|^2) \Lambda_{m,2n}  + (
|O_8|^2+|C_8|^2) \Lambda_{m,2n+1}
\nonumber\\
&-& ( V_8\bar{S}_8+S_8\bar{V}_8 ) \Lambda_{m+1/2,2n} - (
O_8\bar{C}_8+C_8\bar{O}_8 ) \Lambda_{m+1/2,2n+1} \ . \label{e3}
\ea
Here the $\Lambda$'s are toroidal lattice sums, while the
orientifold operation is based on $\Omega'=\Omega \Pi_1 $, with
$\Pi_1$ the inversion along the circle, corresponding to the
Klein-bottle amplitude
\be \mathcal{K} \ = \ \frac{1}{2} \, \left \{ (V_8-S_8) W_{2n} \
+\ (O_8-C_8) W_{2n+1} \right\} \ , \label{e4} \ee
where the $W$'s are winding sums, and introduces an $O8_{+}$ plane
at $y=0$ and an $\overline{O8}_{+}$ plane at $y=\pi R$. For
consistency, these demand that no net R-R charge be introduced,
a condition met by N pairs of D8-$\overline{D8}$ branes, where the
choice $N=16$ is singled out by the connection with M-theory
\cite{ads1}. A simple extension of the arguments in \cite{bd}
shows that the unoriented closed spectrum described by (\ref{e3})
and (\ref{e4}) precisely interpolates between the type I string in
the $R \rightarrow \infty$ limit and the type 0B orientifold with
the tachyonic orientifold projection of \cite{bs}, to be
contrasted with the non-tachyonic $0'B$ projection of \cite{susy95, c0b,bfl1},
in the $R \rightarrow 0$ limit.

In order to obtain a classical configuration as in (\ref{t24}),
without tachyons in the open sector, one can put the $N$
$\overline{D8}$ branes on top of the $O8_+$ planes and the $N$
$D8$ branes on top of the $\overline{O8}_{+}$ planes. This
configuration differs from the one emphasized in \cite{ads1} and
related to the phenomenon of ``brane supersymmetry'', with the
$D8$ on top of the $O8_+$ and the $\overline{D8}$ on top of the
$\overline{O8}_{+}$, by an overall interchange of the positions of
branes and anti-branes. This has an important physical effect:
whereas in the model of \cite{ads1} both the NS-NS tadpoles
and the R-R charges are {\it locally} saturated at the two
endpoints, in this case there is a local unbalance of charges and
tensions that results in an overall attraction between the
endpoints driving the orientifold system toward a vanishing value
for the radius $R$.

The resulting open string amplitudes\footnote{This model was
briefly mentioned in \cite{ads1} and was further analyzed in
\cite{ablau}.}
\ba \mathcal{A} &=& \frac{N^2 +M^2}{2} \ (V_8-S_8) W_{n} + N M
(O_8-C_8) W_{n+1/2}
\ , \nonumber \\
\mathcal{M} &=&- \frac{N+M}{2} \ \hat{V}_8 W_{n} - \frac{N+M}{2}
\hat{S}_8 (-1)^n W_n \  , \label{e5} \ea
describe matter charged with respect to an $ SO(N) \times SO(M)$
gauge group, where $N=M$ on account of the R-R tadpole
conditions, with nine-dimensional massless Majorana fermions in
the symmetric representations ($N(N+1)/2$,1) and (1,$M(M+1)/2$)
and massive fermions in the bi-fundamental representation ($N,M$).
Notice that tachyons appear for small values of $R$. This spectrum
should be contrasted with the one of \cite{ads1} exhibiting
``brane supersymmetry'', where the massless fermions are in
antisymmetric representations. It is a simple exercise to evaluate
tensions and R-R charges at the two ends of the interval:
\ba
&& T_0\ = \ (N-16) T_8 \ , \qquad T_1 \ = \ (N-16) T_8 \ , \nonumber \\
&& q_0\ = \ - \, (N+16) T_8 \ , \qquad q_1 \ = \ (N+16) T_8 \ .
\label{e6} \ea
These translate into corresponding values for the parameters
$\lambda$ and $\omega$ of the classical solution in (\ref{t23}),
that in this case are
\be \lambda = \frac{1}{R} \ , \quad \cos \omega =
\frac{16-N}{16+N} \ . \label{e7} \ee

Hence, in the new coordinate system (\ref{t025}) the orbifold
operation (\ref{t027}) becomes a $2 \pi$ rotation, and can thus be
related to the fermion parity $(-1)^F$, while the orientifold
operation $\Omega'$ combines a world-sheet parity with a rotation.
Notice also that $\lambda \pi R = \pi$, and therefore the $O_+$
and $\overline{O}_+$ planes are actually juxtaposed in the $(Y,Z)$
plane along the real axis $Y=0$, forming somehow a bound state
with vanishing total R-R charge. To be precise, the $O_+$ plane
lies along the half-line $Y=0,$ $Z>0,$ while the ${\bar O}_+$
plane lies along the complementary half-line $Y=0,$ $Z<0.$ The end
result is that in the $(Y,Z)$ plane one is describing the $0B$ (or
$IIB/(-1)^F$) string, subject to the orientifold projection
$\Omega' =\Omega \Pi_Y $, where $Y,$ as already stressed, is here
a {\it noncompact} coordinate. By our previous arguments, all this
is somehow equivalent to the type IIB orientifold compactified on
a circle that we started with. In more physical terms, the
attraction between the sets of $D(\overline{D})$ branes and
$O(\overline{O})$ planes at the ends of the interval drives them
to collapse into suitable systems of $D$-branes and $O$-planes
carrying no net R-R charge, that should be captured by the
static solutions of the effective action (\ref{t1b}), and these
suggest a relation to the $0B$ theory. In this respect, a
potentially singular fate of space time opens the way to a
sensible string vacuum. We would like to stress, however, that the
picture supplied by the classical solution (\ref{t24}) is
incomplete, since the origin $Y=Z=0$ is actually the site of a
singularity. Indeed, the resulting $O$-plane system has no global
R-R charge, but has nonetheless a dipole structure: its $Z<0$
portion carries a positive charge, while its $Z>0$ portion carries
a negative charge. While we are not able to provide more stringent
arguments, it is reasonable to expect that the condensation of the
open and closed-string tachyons emerging in the $R \to 0$ limit
can drive a natural redistribution of the dipole charges between
the two sides, with the end result of turning the juxtaposed $O$
and $\overline{O}$ into a charge-free type-$O$ orientifold plane.
If this were the case, not only the resulting geometry of the
bulk, but also the $D/O$ systems, would become those of the $0B$
string.

In order to provide further evidence for this, let us look more
closely at the type 0B orientifold we identified, using the
original ten-dimensional construction of \cite{bs}. The 0B torus
amplitude is \cite{dhsw}
\ba \mathcal{T} &=& (|O_8|^2+ |V_8|^2+|S_8|^2 +|C_8|^2)   \ ,
\label{e8} \ea
while the orientifold operation includes the parity
$\Omega'=\Omega \Pi_Y $, so that the Klein-bottle amplitude
\be \mathcal{K} = \frac{1}{2} \ \left ( O_8 + V_8- S_8 - C_8
\right) \ \label{e9} \ee
introduces an $O8$ plane at $Y=0$, without R-R charge and with a
tension that precisely matches that of the type-IIB
$O8$-$\overline{O8}$ bound state. In general the type-0
orientifold planes, being bound states of IIB orientifold planes,
have in fact twice their tension. In the present case, the parity
$\Pi_Y$ along a noncompact coordinate sends one of the orientifold
planes to infinity, with the net result of halving the total
tension seen in the $(Y,Z)$ plane.

One can also add to this system two different types of
brane-antibrane pairs, and the open-string amplitudes read
\cite{bs}
\ba \mathcal{A} &=& \frac{n_o^2 +n_v^2 + n_s^2 +n_c^2}{2} \ V_8 +
(n_o n_v + n_s n_c) \ O_8 \nonumber \\
&-& (n_s n_v + n_c n_o) \ S_8  - (n_s n_o + n_c n_v) \ C_8
 \ , \nonumber \\
\mathcal{M} &=& - \ \frac{n_v + n_o + n_s + n_c}{2} \ \hat{V}_8  \
, \ \label{e10} \ea
while the corresponding R-R tadpole conditions are
\be n_o \ = \ n_v = \ N \ , \quad n_s \ = \ n_c = \ M \ .
\label{e11} \ee

The gauge group of this type-0 orientifold, $SO(N)^2 \times
SO(M)^2$, becomes remarkably similar to that of the type-II
orientifold we started from, provided only branes of one type are
present, together with the corresponding antibranes, a
configuration determined setting for instance $M=0$. The resulting
spectrum is then purely bosonic, and the precise statement is
that, in the $R \rightarrow 0$ limit, the expected endpoint of the
collapse, the spectrum of the type-II orientifold should match the
purely bosonic spectrum of this type-0 orientifold, as was the
case for their closed sectors. Actually, for the geometry of the
$D/O$ configurations this was not totally evident, and the same is
true for the open spectrum, due to an apparent mismatch in the
fermionic content, but we would like to argue again that a proper
account of tachyon condensation does justice to the equivalence.

The open-string tachyon $T_{ai}$ of the type-II orientifold is
valued in the bi-fundamental, and therefore carries a pair of
indices in the fundamental of the $SO(N) \times SO(N)$ gauge
group. In the $R \rightarrow 0$ limit, {\it all} its Kaluza-Klein
excitations acquire a negative mass squared. These tachyons will
naturally condense, with $< T_{ai}
> = T(y) \ \delta_{ai}$, where $T(y)$ denotes the tachyonic kink
profile, breaking the gauge group to its diagonal $SO(N)$
subgroup, so that, after symmetry breaking and level by level, the
fermions will fall in the representations
\ba
&& C ^{(k+1/2)} ~: \frac{N (N-1)}{2} + \frac{N (N+1)}{2} \ , \nonumber \\
&& S ^{(2k)} ~: \frac{N (N+1)}{2} \ , \qquad   S ^{(2k+1)} ~:
\frac{N (N-1)}{2} \ . \label{e12} \ea
In the $R \rightarrow 0$ limit the appropriate description of
tachyon condensation is in the T-dual picture, and after a
T-duality the interactions within the open sector must respect
Kaluza-Klein number conservation. Therefore the Yukawa
interactions, that before symmetry breaking are of the type
\ba && S_{(ij)}^{(2k)} C_{ja}^{t , (k+1/2)}  T_{ai}^{(k-1/2)} \ ,
\qquad
S_{(ab)}^{(2k)} C_{bi}^{(k+1/2)}  T_{ia}^{t , (k-1/2)} \ , \nonumber \\
&& S_{[ij]}^{(2k+1)} C_{ja}^{t , (k+1/2)}  T_{ai}^{(k+1/2)} \ ,
\qquad S_{[ab]}^{(2k+1)} C_{bi}^{(k+1/2)}  T_{ia}^{t , (k+1/2)} \
, \label{e13} \ea
will give rise to the mass terms $S_{(ij)} C_{(ji)}$ , $S_{[ij]}
C_{[ji]}$. The conclusion is that the final low-lying open spectra
are bosonic on both sides and actually match precisely.

A more direct argument for the equivalence we are proposing would
follow from a natural extension of Sen's description of tachyon
condensation \cite{sen}. As we have already stressed, the $O8$ and
$\overline{O8}$ attract one another and drive the orientifold to a
collapse. In the T-dual picture, the $O9$ and $\overline{O9}$
condense into a non-BPS orientifold plane in one lower dimension,
that in the $R \rightarrow 0$ limit becomes the type-0 orientifold
plane that we have described above. This type of phenomenon can
plausibly be related to the closed-string tachyon non-trivial
profile in this model, in a similar fashion to what happens for
the open-string tachyon kink profile in $D$-$\overline{D}$
systems. At the same time, after T-duality the $D9$ and
$\overline{D9}$ branes decay into non-BPS $D8$ branes via the
appropriate tachyon kink profile. Due to the new $(-1)^F$
operation, these new non-BPS type-II branes match directly the
non-BPS type-0 branes discussed in \cite{bs,dms}, since the
$(-1)^F$ operation removes the unwanted additional fermions. Let
us stress that String Theory can resolve in this fashion the
potential singularity associated to an apparent collapse of space-time:
after tachyon condensation, the $O$-$\overline{O}$
attraction can give birth to a well defined type-0 vacuum.

In this example one is confronted with the ideal situation in
which a vacuum redefinition can be analyzed to some extent in
String Theory. In general, however, a string treatment in such
detail is not possible, and it is therefore worthwhile to take a
closer look, on the basis of the intuition gathered from Field
Theory, at how the conventional perturbative string setting can be
adapted to systems in need of vacuum redefinitions, and especially
at what it can teach us about the generic features of the
redefinitions. We intend to return to this issue in a future
publication \cite{bdnps}.
\section{Threshold corrections in open strings and $NS$-$NS$ tadpoles}
While $NS$-$NS$ tadpoles ask for classical resummations that are
very difficult to perform systematically, it is often possible to
identify physical observables for which resummations are needed
only at higher orders of perturbation theory. This happens
whenever, in the appropriate limit of moduli space (infinite tube
length, in the case of disk tadpoles), massless exchanges cannot
be attached to the sources. Examples of quantities of such type are provided by the
quantum corrections to gauge couplings, commonly known as
threshold corrections \cite{Kaplunovsky:1992vs}.
If the tree-level gauge coupling is $1/g^2$,
the one-loop threshold corrections $\Delta$ are defined as
\be
\left. \frac{4\pi^2}{g^2}\right|_{\rm one-loop} \ = \
\left. \frac{4\pi^2}{g^2}\right|_{\rm tree-level} \ + \ \Delta(\mu,\Phi_i) \ ,
\ee
where $\Delta$ depends on the energy scale $\mu$ and on the moduli $\Phi_i$.

The study of quantum corrections to gauge couplings is very
important from a phenomenological point of view. From Quantum
Field Theory, we know that coupling constants run with energy due
to the loop contributions of charged particles. Therefore, it is
necessary to take into account the radiative corrections predicted
by String Theory, in order to achieve a more realistic matching
with the low-energy world, and indeed the study of threshold
corrections is closely related to the issue of unification in
supersymmetric theories. Moreover, as we said, threshold
corrections depend on the moduli of the model in consideration,
and the knowledge of this dependence is important for the issue of
moduli stabilization after supersymmetry breaking.

Let us now discuss some properties of threshold corrections. From
string computations one can see that $\Delta$ has the following
general structure \be \Delta \ = \ \frac{1}{4} \int_0^\infty
\frac{d t}{t} {\cal B}(t) \ . \ee This integral is typically
divergent in the (open) infrared limit, due to the massless
charged particles circulating in the loop. Moreover, thinking of
the model we want to study as compactified to four dimensions, for
$t\to \infty \ $ $\Delta$ has to reproduce the correct logarithmic
divergence of the four-dimensional low-energy effective field
four-dimensional low-energy effective field theory. Therefore we expect that, in such a
converge to the usual one-loop $\beta$-function coefficient of the
four-dimensional low energy gauge theory under consideration \be
\lim_{t\rightarrow\infty} {\cal B}(t) \ = \ b \ . \ee where $b$ is
given by the usual formula \be \label{indice} b \ = \
-\frac{11}{3}C_2(adj)+\frac{2}{3}\sum_R T(R) +\frac{1}{3}\sum_r
T(r) \ , \ee with $C_2(adj)$ the quadratic Casimir of the adjoint
representation, and $T(R)$ and $T(r)$ the Dynkin indices of the
representation $R$ for the four-dimensional Weyl fermions and of
the representation $r$ for the four-dimensional complex scalars
\footnote{If we denote with $T^a$ the generators of a group, the
quadratic Casimir $C_2(R)$ and the Dynkin index $T(R)$ of a
representation $R$ are defined respectively by
$\sum_a\left(T^a(R)T^a(R)\right)^i_j=C_2(R)\delta^i_j$ and ${\rm
tr}\left(T^a(R)T^b(R)\right)=T(R)\delta^{ab}$. Moreover, if $d(R)$
is the dimension of the representation $R$, the following relation
$d(R)C_2(R)=d({\rm adj})T(R)$ implies that $C_2({\rm adj})=T({\rm
adj})$.}.

On the other hand, the (open) ultraviolet limit, corresponding to the infrared
$\ell \to \infty$ limit in the transverse closed channel, in all supersymmetric cases is finite
since the divergences arising from the propagation of massless closed states going into the
vacuum are eliminated by tadpole cancellations.

In this Section we are interested in the ultraviolet
limit of one-loop threshold corrections for models with
supersymmetry breaking and NS-NS tadpoles.
We will consider at first the Sugimoto model compactified on $T^6$. Then
we will proceed to analyze the orbifold $T^4/{\mathbb Z}_2\times T^2$,
both in the case of brane supersymmetry breaking and in the supersymmetric case, but with
the addition of a brane-antibrane system that breaks
supersymmetry. Finally, we will analyze the $0^\prime B$ model compactified on a torus $T^6$.
In all these models, that have parallel branes and are free of closed tachyons,
we will show that the one-loop threshold
corrections are ultraviolet finite, despite the presence of uncancelled
NS-NS tadpoles.
One can understand this finiteness noting that in the $l
\rightarrow \infty$ limit the string amplitudes acquire a
field-theory interpretation in terms of dilaton and graviton
exchanges between $Dp$-branes and $Op$-planes. For parallel
localized sources, the relevant terms in the effective Lagrangian
are
\ba S &=& \frac{1}{2 k_{10}^2} \int d^{10} x \sqrt {-G} \biggl \{ R -
\frac{1}{2} (\partial \varphi)^2 - \frac{1}{2 (p+2)~!} e^{(5-p-2)
\varphi /2}
F_{p+2}^2 \biggr \} \nonumber \\
&-& \int_{{\bf y} = {\bf y_i}} d^{p+1} \xi  \biggl \{
\sqrt{-\gamma} \left[ T_p  e^{(p-3) \varphi /4} + e^{(p-7) \varphi
/4} \tr F_{\mu \nu}^2 \right] + q C^{(p+1)} \biggr \} \ ,
\label{p5} \ea
where ${\bf \xi}$ are brane world-volume coordinates, $q = \pm 1$
distinguishes between branes or $O$-planes and antibranes or
$\overline{O}$-planes, $G$ is the 10-dimensional metric, $\gamma$
is the induced metric and $C^{(p+1)}$ denotes a $R$-$R$ form that
couples to the branes. The result for the one-loop corrections to
gauge couplings, obtained using (\ref{p5}) while treating for
simplicity the K-K momenta as a continuum, is proportional to
\be
T_p \int \frac{d^{9-p} k}{(2 \pi)^{9-p}} \biggl \{ - T ^{\mu
\nu} \, \langle h_{\mu \nu} h_{\rho \sigma}\rangle \, \eta^{\rho
\sigma} \ + \ \tr F^2 \langle \, \varphi \varphi \rangle\,
\frac{(p-3)(p-7)}{16} \biggr \} \ , \label{p6}
\ee
where
\be
\langle h_{\mu \nu} h_{\rho \sigma}\rangle \ = \ \frac{1}{2k^2}
\left( \eta_{\mu \rho} \eta_{\nu \sigma}+ \eta_{\mu \sigma}
\eta_{\nu \rho} - \frac{1}{4} \eta_{\mu \nu} \eta_{\rho
\sigma}\right)
\ee
is the ten-dimensional graviton propagator in De Donder gauge,
\be \langle \varphi \varphi \rangle \ = \ \frac{1}{k^2} \ee
is the ten-dimensional dilaton propagator, and
\be T_{\mu \nu} = \tr \left( F_{\mu \rho} F_{\nu}^{\rho} -
\frac{1}{4} \eta_{\mu
   \nu} F^2 \right)  \label{p7} \ee
is the vector energy-momentum tensor.

The nice thing to notice in (\ref{p6}) is that the dilaton and graviton
exchanges cancel precisely, source by source, in the
threshold corrections, ensuring that the result is actually
finite in spite of the presence of the dilaton and graviton tadpoles.
This fact could be explained considering that in models with parallel
branes with supersymmetry broken on the branes,
the bulk is supersymmetric and therefore, in the (closed) infrared limit, threshold corrections are essentially
given by supersymmetric expressions. actually, this type of argument applies
also the non-tachyonic type $0'B$ orientifold.
\subsection{Background field method}

The main goal of this Section is the computation of the one-loop corrections
to gauge couplings in a number of string models with supersymmetry breaking.
We will focus our attention on the ultraviolet behavior of such quantities, showing their
UV finiteness in spite of the emergence of NS-NS tadpoles.
Here, we review the background field method \cite{bf}, the
tool we shall use to extract from the one-loop open partition functions
the expressions for threshold corrections.

The starting point is the one-loop vacuum energy that
(in the Euclidean case) is provided by (\ref{imp})
\be
\Gamma \ = \ -\frac{V}{2}\int_{\epsilon}^{\infty} \frac{dt}{t} \
\int\frac{d^Dp}{(2\pi)^D} \ e^{-tp^2} \ \mathrm{Str}\left(e^{-tM^2}\right) \ ,
\ee
We now restrict the previous formula to the case
of open strings for which $M^2=\frac{1}{\alpha'}(N+a)$.
Performing the integral over momenta
\be
\int\frac{dp^D}{(2\pi)^D} \ e^{-tp^2} \ = \ \frac{1}{(4\pi t)^\frac{D}{2}} \ ,
\ee
after a rescaling of the integration variable, $t=\alpha'\pi\tau$,
we obtain the general form of the annulus amplitude
\be
\Gamma \ = \ -\frac{V_D}{2(4\alpha'\pi^2)^{\frac{D}{2}}} \
\int_{0}^{\infty} \frac{d\tau}{\tau^{\frac{D}{2}+1}} \
\mathrm{Str} \ e^{-\pi\tau \ (N+a)} \ .
\ee
The four-dimensional case is then recovered compactifying six dimensions
on a torus, for example. Thus, let us fix $D=4$ and add the sum over
internal momenta
\be
\label{somma}
P \ = \ \sum_{m} e^{-\alpha'\pi\tau \ m^{\rm
T}g^{-1}m} \ .
\ee
Of course, we have to add also the M\"{o}bius contribution, that as usual
is obtained from the annulus amplitude projecting it with the
orientifold operation $P=(1+\epsilon\Omega)/2$,
where $\Omega$ is the world-sheet parity, while $\epsilon$ is a sign.
Putting the factor $1/2$ of the orientifold projection inside the definition of
$\cal{A}$ and $\cal{M}$, and considering the energy per unit volume,
the general amplitudes for the unoriented open sector are given by
\ba
\cal{A} \ &=& \ -\frac{1}{(8\alpha'\pi^2)^2} \ \int_{0}^{\infty}
\frac{d\tau}{\tau^3} \
  \mathrm{Str} \ e^{-\pi\tau \ (N+a)} \
P \ , \nonumber\\
\cal{M} \ &=& \ -\frac{1}{(8\alpha'\pi^2)^2} \ \int_{0}^{\infty}
\frac{d\tau}{\tau^3} \
  \mathrm{Str} \ \left(e^{-\pi\tau(N+a)} \ \epsilon\Omega \ \right) \
P \ .
\ea

We now turn on a background magnetic field, for example in the
$X^1$ direction
\be F_{23} \ = \ B Q \ ,
\ee
where $X^0 \ldots X^3$ are the uncompactified dimensions
and $Q$ is a generator of the gauge group, chosen with a suitable normalization
\footnote{We recall that with the normalization ${\rm tr}Q^2=1/2$, the Yang-Mills Lagrangian
is ${\cal L}=\frac{1}{2g^2}{\rm tr} F^2 $.}.
The net effect of the magnetic field on the open amplitudes
\cite{acny} is to shift the oscillator frequencies of the complex
coordinate $X_2+iX_3$ by an amount $\epsilon$, where
\be
\pi\epsilon \ = \ \tan^{-1}(\pi q_a B)+ \tan^{-1}(\pi q_b B) \ ,
\ee
and $q_{a(b)}$ are the eigenvalues of the generator $Q$ acting on
the Chan-Paton charges at the left (right) endpoints of the
string. Moreover, we fixed $2\alpha'=1$, and we will use this
convention in all the following computations.\\
The partition functions in the presence of $B$ are simply obtained
replacing
\be
p^{\mu}p_{\mu} \qquad {\rm with} \qquad -(p_0)^2 \ + \ (p_1)^2 \ + (2n+1)\epsilon
\ + \ 2\epsilon \ \Sigma_{23}
\ee
and
\be
(\sum_{bos}-\sum_{ferm}) \ \int\frac{d^4p}{(2\pi)^4} \qquad {\rm
  with} \qquad  (\sum_{bos}-\sum_{ferm}) \frac{(q_a+q_b)B}{2\pi}
\ \sum_{n=0} \int\frac{d^2p}{(2\pi)^2} \ ,
\ee
where $n$ labels the Landau levels, $(q_a+q_b)B/2\pi$ is the
degeneracy of Landau levels per unit area, and $\Sigma$ is the spin
along the direction of the magnetic field.
Threshold corrections at one-loop are then obtained
 expanding the total one-loop vacuum energy in terms of $B$ \cite{bf}
\be \Lambda(B) \ = \left({\cal T}+{\cal K}+{\cal A}(B)+{\cal
M}(B)\right) \ = \ \Lambda_0 +\frac{1}{2}
\left(\frac{B}{2\pi}\right)^2 \Lambda_2 +o(B^4) \ , \ee
where the closed sector does not contribute to the expansion,
since one can charge only the ends of open strings. The first
term, $\Lambda_0$, is the one-loop cosmological constant and
vanishes for a supersymmetric theory, while the quadratic term in
the background field, $\Lambda_2$, gives just the one-loop
threshold corrections to gauge couplings.
\subsection{Sugimoto model}
Let us consider the Sugimoto model compactified on $T^6$.
We recall that the gauge group is  $USp(32)$ and that
the model contains $N$ $\overline{{\rm D}9}$ branes and an
O9$_-$ plane and thus the NS-NS tadpole is not cancelled.
The starting point to compute the threshold corrections is
to write the open amplitudes. First of all we consider the case $B=0$.
The annulus and M\"{o}bius amplitudes, with their normalization and
the integral and the contributions of the internal bosons explicitly displayed,
are (see eq. (\ref{Sugimoto_amplitudes}) with
$n_+=0 \ $, $ \ n_-=N=32 \ $, $ \ \epsilon_{NS}=\epsilon_{R}=+1$)
\ba
{\cal A} & = & -\frac{N^2}{(4\pi^2)^2} \int \frac{d\tau}{\tau^3} \
\frac{(V_8-S_8)}{\eta^8} \ P^{(6)}
\ ,\nonumber\\
{\cal M} & = & -\frac{N}{(4\pi^2)^2} \int \frac{d\tau}{\tau^3} \
\frac{(\hat V_8+\hat
  S_8)}{\hat\eta^8}
\ P^{(6)} \ ,
\ea
where $P^{(6)}$ is defined as in (\ref{somma}). The integration moduli, $i\tau/2$ for the annulus amplitude
and $i\tau/2+1/2$ for the M\"{o}bius strip, are understood.
Expressing the $so(8)$ characters
in terms of $\vartheta$-functions (\ref{characters}), the amplitudes read
\ba
\label{SUGIMOTOsenzaB}
{\cal A} & = & - \ \frac{N^2}{2(4\pi^2)^2} \int \frac{d\tau}{\tau^3} \
\sum_{\alpha,\beta}\eta_{\alpha\beta} \
\frac{\vartheta^4\left[{\textstyle{\alpha \atop \beta}}\right]
}{\eta^{12}} \ P^{(6)} \ , \nonumber\\
{\cal M} & = & - \ \frac{N}{2(4\pi^2)^2} \int \frac{d\tau}{\tau^3} \
\sum_{\alpha,\beta}\eta_{\alpha\beta} \ e^{2\pi i \alpha} \
\frac{\vartheta^4\left[{\textstyle{\alpha\atop\beta}}\right]
}{\eta^{12}}
\ P^{(6)} \ ,
\ea
where $\alpha$ and $\beta$ are $0,1/2$, and $\eta_{\alpha\beta}=(-)^{2\alpha+2\beta+4\alpha\beta}$.
In Appendix A we recall the definition of
the Jacobi $\vartheta$-functions and some of their properties.

At this point we turn on a background magnetic field, $F_{23}=QB$,
where $Q$ is a $U(1)$ generator in the Cartan subalgebra of the gauge
group $USp(32)$.
The magnetic filed shifts the frequency of string oscillators by
\be
\pi\epsilon = \left\{ \begin{array}{ll}
\tan^{-1}{(\pi q_a B)}+\tan^{-1}{(\pi q_b B)} \ \simeq \
\pi(q_a+q_b)B-\frac{\pi^3}{3}(q_a^3+q_b^3)B^3 \ , & \quad{\cal A}\\
& \\
2\tan^{-1}{\pi q_a B} \ \simeq  \ 2\pi q_a B -\frac{2\pi^3}{3}q_a^3 B^3
\ . & \quad {\cal M} \
\end{array} \right.
\label{epsi}
\ee
In practice, if the magnetic field acts only on two coordinates, in our case
$X^2$ and $X^3$, one has to perform the replacement
\be
\frac{\vartheta\left[{\textstyle{\alpha\atop\beta}}\right](0|\tau)}{\eta} \ \longrightarrow
\frac{\vartheta\left[{\textstyle{\alpha\atop\beta}}\right]\left(i\epsilon\tau/2|\tau\right)}{\eta}
\ ,
\ee
for the two fermionic degrees of freedom affected by $B$, and
\be
\frac{1}{\eta^2} \ \longrightarrow
\frac{\eta}{\vartheta_1(i\epsilon\tau/2)} \ ,
\ee
for the bosonic coordinates.
Then the amplitudes in the presence of $B$ read
\ba
{\cal A} & = & - \ \frac{i\pi B}{2(4\pi^2)^2} \int \frac{d\tau}{\tau^2} \
\sum_{a,b=1}^{32}(q_a+q_b) \
\sum_{\alpha,\beta}\eta_{\alpha\beta} \
\frac{\vartheta\left[{\textstyle{\alpha \atop \beta}}\right]
\left(\frac{i\epsilon\tau}{2}\right)}{\vartheta_1\left(\frac{i\epsilon\tau}{2}\right)} \
\frac{\vartheta^3\left[{\textstyle{\alpha \atop \beta}}\right]}
{\eta^9} \ P^{(6)} \ , \nonumber\\
{\cal M} & = & - \ \frac{i\pi B}{(4\pi^2)^2} \int \frac{d\tau}{\tau^2} \
\sum_{a=1}^{32}q_a \
\sum_{\alpha,\beta}\eta_{\alpha\beta} \ e^{2\pi i \alpha} \
\frac{\vartheta\left[{\textstyle{\alpha\atop\beta}}\right]
\left(\frac{i\epsilon\tau}{2}\right)}{\vartheta_1\left(\frac{i\epsilon\tau}{2}\right)} \
\frac{\vartheta^3\left[{\textstyle{\alpha\atop\beta}}\right]}
{\eta^9} \ P^{(6)} \ ,
\ea
and one can verify that in the small magnetic field limit these
expressions reduce to the ones in (\ref{SUGIMOTOsenzaB}).

Since we are interested in the open ultraviolet limit, it is convenient
to pass in the transverse channel where such a limit corresponds
to the infrared $\ell\rightarrow\infty$ limit.
For the annulus amplitude the transformation is $t=\tau/2$ and then
$S: t \rightarrow 1/\ell$. We get a factor $2^{-3}$ from the lattice sum
and a factor $2^{-1}$ from the integral measure, that together with
the factor $1/2$ from the normalization of ${\cal A}$ reconstruct the
right power $2^{-5}$.
For the M\"{o}bius strip one has to perform a $P$-modular transformation,
$i\tau/2+1/2 \rightarrow i/2t+1/2$ and then make the substitution
$\ell=t/2$. From the measure we gain a factor $2$, and
defining
\be
W^{(6)} \ = \ \sum_{n} \ e^{-\pi\ell \ n^{\rm T}g n / 2 \alpha'}
\ ,
\ee
and
\be
\label{normvol}
v^{(d)} \ = \ \sqrt{\frac{{\rm det} \ g}{\alpha'^{ \ d}}} \ ,
\ee
the amplitudes in the transverse channel are given by
\ba
\label{SUGIMOTOtransverse}
\tilde{{\cal A}} & = & - \ \frac{\pi B \ v^{(6)} 2^{-5}}{(4\pi^2)^2} \int d\ell \
\sum_{a,b=1}^{32}(q_a+q_b) \
\sum_{\alpha,\beta}\eta_{\alpha\beta} \
\frac{\vartheta\left[{\textstyle{\alpha \atop \beta}}\right]
(\epsilon)}{\vartheta_1(\epsilon)} \
\frac{\vartheta^3\left[{\textstyle{\alpha \atop \beta}}\right]}
{\eta^9} \ W^{(6)} \ , \nonumber\\
\tilde{{\cal M}} & = & - \ \frac{2\pi B v^{(6)}}{(4\pi^2)^2} \int d\ell \
\sum_{a=1}^{32}q_a \
\sum_{\alpha,\beta}\eta_{\alpha\beta} \ e^{2\pi i \alpha} \
\frac{\vartheta\left[{\textstyle{\alpha\atop\beta}}\right]
\left(\frac{\epsilon}{2}\right)}{\vartheta_1\left(\frac{\epsilon}{2}\right)} \
\frac{\vartheta^3\left[{\textstyle{\alpha\atop\beta}}\right]}
{\eta^9} \ W^{(6)}_{e} \ ,
\ea
where the integrands are computed for $\tau=i\ell$ for the annulus and
for $\tau=i\ell+1/2$ for the M\"{o}bius, and where as usual in the transverse
M\"{o}bius only even windings propagate.
Moreover, we used the fact that $\vartheta_{2,3,4}(z)$ are even
functions of $z$, while $\vartheta_1(z)$ is odd.
At this point one can use the identity (\ref{A.13})
\be
\vartheta_3(z)\vartheta_3^3-\vartheta_4(z)\vartheta_4^3-\vartheta_2(z)\vartheta_2^3
\ = \ 2\vartheta_1^4(z/2) \ ,
\ee
obtaining for the transverse amplitudes
\ba
\label{SUGIMOTOconB}
\tilde{{\cal A}} & = & - \ \frac{2\pi B \ v^{(6)} 2^{-5}}{(4\pi^2)^2} \int d\ell \
\sum_{a,b=1}^{32}(q_a+q_b) \
\frac{\vartheta_1^4(\epsilon/2)}
{\vartheta_1(\epsilon)\eta^9} \ W^{(6)} \ , \nonumber\\
\tilde{{\cal M}} & = & - \ \frac{4\pi B v^{(6)}}{(4\pi^2)^2} \int d\ell \
\sum_{a=1}^{32}q_a \
\frac{\vartheta_2(\epsilon/2)\vartheta_2^3+\vartheta_1^4(\epsilon/4)}
{\vartheta_1(\epsilon/2)\eta^9} \ W^{(6)}_{e} \ .
\ea
The contribution proportional to $\vartheta_1^4$ is what one would obtain
in the supersymmetric case from the character $V_8-S_8$.
Since $\mathcal{N}=1$ in $D=10$ corresponds to $\mathcal{N}=4$ in $D=4$,
one expects no threshold corrections from such a term.
And in fact, since for small $B$
\be
\vartheta_1(z) \ \simeq \ 2\pi z \eta^3 \ ,
\ee
the term proportional to $\vartheta_1^4$ in the annulus and in the M\"{o}bius amplitudes
starts at the quartic order in $B$, giving no
corrections to gauge coupling.
On the other hand, the $\vartheta_2$ term in the
M\"{o}bius strip contributes to threshold corrections.
Expanding the $\vartheta$-functions for a small magnetic field
\ba
&&\vartheta_2(\epsilon/2) \ \simeq \ \vartheta_2 +
\frac{\epsilon^2}{8}\vartheta_2^{\prime\prime} \ , \nonumber\\
&&\vartheta_1(\epsilon/2) \ \simeq \
\frac{\epsilon}{2}\vartheta_1^{\prime}+
\frac{\epsilon^3}{48}\vartheta_1^{\prime\prime\prime} \ =
\ \pi\epsilon \ \eta^3 +
\frac{\epsilon^3}{48}\vartheta_1^{\prime\prime\prime} \ ,
\ea
and recalling that for the M\"{o}bius amplitude
\be
\pi\epsilon \ \simeq \ 2\pi B q_a -\frac{2\pi^3}{3}q_a^3 B^3 \ ,
\ee
we obtain at the second order in the magnetic field
\ba
\tilde{\cal M} &=& - \ \frac{2v^{(6)}}{(4\pi^2)^2} \ N \ \int d\ell
\frac{\vartheta_2^4}{\eta^{12}} \ W^{(6)}_{e} \nonumber\\
&& - \frac{v^{(6)}B^2 \ {\rm tr} Q^2}{(4\pi^2)^2}  \int d \ell \
\frac{\vartheta_2^4}{\eta^{12}} \
\left(\frac{2\pi^2}{3}+\frac{\vartheta_2^{\prime\prime}}{\vartheta_2}-
\frac{\vartheta_1^{\prime\prime\prime}}{6\pi\eta^3}\right)
W^{(6)}_{e} + o(B^4) \ ,
\ea
where $N=32$ and ${\rm tr }Q^2=\sum_{a=1}^{32}q_a^2$, where
the zeroth order is the contribution due to the uncancelled NS-NS tadpole.

Finally, from the second order terms one extracts the one-loop threshold corrections
to the gauge coupling, that is
\be
\label{SUGIMOTOthreshold}
\Lambda_2 \ = \ - \ \frac{v^{(6)} \ {\rm tr} Q^2}{2\pi^2}  \int d \ell \
\frac{\vartheta_2^4}{\eta^{12}} \
\left(\frac{2\pi^2}{3}+\frac{\vartheta_2^{\prime\prime}}{\vartheta_2}-
\frac{\vartheta_1^{\prime\prime\prime}}{6\pi\eta^3}\right)
W^{(6)}_{e} \ .
\ee
The good infrared behavior of the last expression can be checked
considering that the term proportional to $\vartheta_2(\epsilon/2)\vartheta_2^3$ in the expression
for the M\"{o}bius transverse amplitude (\ref{SUGIMOTOconB})
is independent of $B$ in the $\ell \rightarrow\infty$ limit
\be
\label{IRgood}
\lim_{\ell\rightarrow\infty}
\pi q_a B \
\frac{\vartheta_2(\epsilon/2) \ \vartheta_2^3}{\vartheta_1(\epsilon/2)
  \ \eta^9}
 \ = \ \frac{8\pi q_a B}{\tan{(\pi\epsilon/2)}} \ = \ 8 ,
\ee
where in the last equality we used the expression for $\epsilon$
given in (\ref{epsi}).

One can also check that the open infrared limit of threshold
corrections reconstruct just the right one-loop $\beta$-function
coefficient of the low-energy effective field theory corresponding
to the model. To study the behavior of (\ref{SUGIMOTOthreshold})
in such a limit it is convenient to consider the open channel.
Threshold corrections from the open channel come only from the
M\"{o}bius strip since the annulus is the same as in the
supersymmetric theory, and in particular from a term \be
\frac{-i\pi B}{(4\pi^2)^2} \ \int \frac{d\tau}{\tau^2} \
\sum_{a=1}^{32}q_a
 \ \frac{2\vartheta_2(i\epsilon\tau/2)\vartheta_2^3}{\vartheta_1(i\epsilon\tau/2)\eta^9} \ P^{(6)} \ ,
 \ee
that in the $\tau \to \infty$ limit and to the second order in the magnetic field behaves as
\be
- \ \frac{1}{3\pi^2} B^2 {\rm tr}Q^2 \ \int\frac{d\tau}{\tau} \ + \ {\rm IR \ finite \ terms} \ .
\ee
Hence, the leading open infrared limit of the threshold corrections, with the
normalization ${\rm tr_{fund}} Q^2 = 1/2$ of the generator $Q$ in the
fundamental representation, are given by
\be
\Lambda_{2} = \frac{1}{4} \ \left(-\frac{16}{3}\right) \ \int \frac{d\tau}{\tau} \ ,
\ee
where $b=-16/3$ should be the one-loop $\beta$-function coefficient given by
the usual four-dimensional formula \ref{indice}.
In our case the model at the massless field theory level contains a
four-dimensional gauge vector together with 3 complex scalars in the adjoint representation
 and 4 four-dimensional Weyl fermions in the antisymmetric representation, and thus one recovers
\be
b=\ -\frac{11}{3}\times 17+\frac{2}{3}\times4\times 15 +\frac{1}{3}\times 3\times 17 = -\frac{16}{3} \ ,
\ee
where for $USp(N)$, with the normalization chosen, $T({\rm fund.})=1/2$, $ \ T({\rm adj})=C_2({\rm adj})=N/2+1 \ $
and $T({\rm antisym.})=N/2-1$.
\subsection{Brane supersymmetry breaking}
In this Subsection we will compute the threshold corrections for the
$T^4/{\mathbb Z}_2$ brane
supersymmetry breaking model \cite{bsb} compactified on a further $T^2$.
This model contains D9 branes and $\overline{\rm D5}$ antibranes
together with  O9$_+$ and O5$_-$ orientifold planes,
and thus the NS-NS tadpoles on the $\overline{\rm D5}$ antibranes
are not cancelled. The gauge group is $[SO(16)\times
  SO(16)]_9\times[USp(16)\times USp(16)]_5$.

The partition functions, that one can write starting from eqs (\ref{bsb_direct_A})
and (\ref{bsb_directM}) are
\ba
{\cal A} &=& -\ \frac{1}{2(4\pi^2)^2} \ \int \frac{d\tau}{\tau^3} \
\left[(Q_o + Q_v) \left( N^2 \frac{P^{(4)}}{\eta^4}  + D^2
  \frac{W^{(4)}}{\eta^4}\right)\right.
\nonumber\\
&&\left.+ (R_N^2 + R_D^2) (Q_o - Q_v) {\left(\frac{2
\eta}{\vartheta_2}\right)}^2 \right.\nonumber\\
&& \left. +
2 N D ( O_4 S_4 - C_4 O_4 + V_4 C_4 -
S_4 V_4) {\left(\frac{\eta}{\vartheta_4}\right)}^2
\right. \nonumber\\
&& \left.+ 2 R_N R_D ( - O_4 S_4 - C_4 O_4 + V_4 C_4 +
S_4 V_4 ){\left(\frac{
\eta}{\vartheta_3}\right)}^2 \right] \ \frac{P^{(2)}}{\eta^2} \
\frac{1}{\eta^2} \ , \nonumber\\
{\cal M} &=& \frac{1}{2(4\pi^2)^2} \ \int \frac{d\tau}{\tau^3} \Biggl[
  N \frac{P^{(4)}}{\hat\eta^4}
( \hat{O}_4 \hat{V}_4  + \hat{V}_4 \hat{O}_4  - \hat{S}_4 \hat{S}_4 - \hat{C}_4
\hat{C}_4 ) \nonumber \\
&& -  D \frac{W^{(4)}}{\hat\eta^4} ( \hat{O}_4
\hat{V}_4  + \hat{V}_4 \hat{O}_4  + \hat{S}_4 \hat{S}_4 + \hat{C}_4
\hat{C}_4 )  \nonumber \\
&& - N(
\hat{O}_4 \hat{V}_4 - \hat{V}_4 \hat{O}_4 - \hat{S}_4 \hat{S}_4
+ \hat{C}_4 \hat{C}_4 )\left(
{2{\hat{\eta}}\over{\hat{\vartheta}}_2}\right)^2  \nonumber \\
&& + D( \hat{O}_4
\hat{V}_4 - \hat{V}_4 \hat{O}_4 + \hat{S}_4 \hat{S}_4
- \hat{C}_4 \hat{C}_4)\left(
{2{\hat{\eta}}\over{\hat{\vartheta}}_2}\right)^2  \Biggr] \
\frac{P^{(2)}}{\hat\eta^2}\frac{1}{\hat\eta^2} \ ,
\ea
where we wrote explicitly the integrals with their right normalization
and the contributions of the two transverse bosons and of the internal bosons.
The moduli, $i\tau/2$ for the annulus amplitude, and $i\tau/2+1/2$ for the M\"{o}bius strip,
are understood. Moreover we recall that
\ba
N=n_1+ n_2 \, , \qquad D=d_1+ d_2 \, , \nonumber \\
R_N=n_1- n_2 \, , \qquad R_D=d_1- d_2 \, ,
\ea
and that
\be
N=D=32 \ , \qquad \qquad R_N=R_D=0 \ .
\ee
The lattice sums $P^{(4)}$ and $W^{(4)}$ on one hand, and $P^{(2)}$ on
the other hand, refer respectively to the $T^4$ torus of the orbifold
compactification and to the additional $T^2$ torus.

At this point we magnetize the $\overline{\rm D5}$ antibranes and, as usual, we turn on a
magnetic field, $F_{23}=BQ$, along the complex direction $X^2+iX^3$, where $Q$ corresponds to a
$U(1)$ subgroup of one of the two gauge factors $USp(16)$, normalized to ${\rm tr}_{fund}Q^2=1/2$.
Thinking of such $U(1)$ factor as embedded in $SO(32)$, a good choice for it could be
$Q_{32\times 32}={\rm diag} (1/2,-1/2,0\cdots 0)$ if $Q\subset USp(d_1)\subset SO(32)$,
or $Q_{32\times 32}={\rm diag} (0\cdots 0,1/2,-1/2 )$ if $Q\subset USp(d_2)\subset SO(32)$.

Then, since only the $\overline{\rm D5}$ antibranes are magnetized, let
us consider only those terms of the amplitudes that can couple to $B$,
let us say with an obvious notation ${\cal A}_{\bar{5}\bar{5}+\bar{5}9}$
and ${\cal M}_{\overline{5}}$.
Notice that the term proportional to $Q_o+Q_v$ in the annulus amplitude is the
usual supersymmetric combination of $\vartheta$-functions (it is like $V_8-S_8$)
\be
 {\cal A}^{({\cal N}=4)} \ = \
 - \ \frac{1}{4(4\pi^2)^2} \ \int\frac{d\tau}{\tau^3} \ D^2 W^{(4)} P^{(2)}
\sum_{\alpha\beta}\eta_{\alpha,\beta}\frac{\vartheta^4\left[{\textstyle{\alpha\atop\beta}}\right]}
{\eta^{12}}
\ee
and thus, in the presence of a magnetic field $B$, it would start at
the order $B^4$, exactly as we already saw in the case of the Sugimoto model.
This term in fact would give, after two $T$
dualities along the directions of $T^2$ that transform the $\overline{\rm
  D5}$ antibranes into $\overline{\rm D3}$ antibranes, the threshold
corrections to gauge couplings (actually vanishing) of the four-dimensional ${\mathcal N}=4$ super
Yang-Mills theory.

Using eqs. (\ref{so4}) and (\ref{so2n}) to express the characters
in terms of $\vartheta$-functions, the amplitudes read
\ba
{\cal A}_{\bar{5}\bar{5}+\bar{5}9}&=& {\cal
  A}^{({\cal N}=4)} \ - \
\frac{1}{4(4\pi^2)^2} \ \int\frac{d\tau}{\tau^3} \
\left( 4R_D^2 \ \frac{\vartheta_3^2 \ \vartheta_4^2 \ - \ \vartheta_4^2
  \ \vartheta_3^2}{\eta^6 \ \vartheta_2^2} \right.\nonumber\\
&& \left. + \ 2ND \ \frac{\vartheta_3^2 \ \vartheta_2^2 \ - \ \vartheta_2^2 \
  \vartheta_3^2}{\eta^6 \ \vartheta_4^2}\right.
\left. - 2R_NR_D \ \frac{\vartheta_4^2 \ \vartheta_2^2 \ + \ \vartheta_2^2 \
  \vartheta_4^2}{\eta^6 \ \vartheta_3^2}
\right) \ P^{(2)}\ , \nonumber\\
{\cal M}_{\bar{5}} & = &
- \ \frac{1}{4(4\pi^2)^2} \ \int\frac{d\tau}{\tau^3} \
\Biggl(
D W^{(4)} P^{(2)} \
\sum_{\alpha\beta}\eta_{\alpha,\beta} \ e^{2\pi i\alpha} \
\frac{\vartheta^4\left[{\textstyle{\alpha\atop\beta}}\right]}
{\eta^{12}}  \nonumber\\
&&+ \ 4DP^{(2)} \ \frac{\vartheta_3^2 \ \vartheta_4^2 \ - \ \vartheta_4^2 \
  \vartheta_3^2}{\eta^6 \ \vartheta_2^2} \Biggr) \ ,
\ea
where we used $\vartheta_1(0|\tau)=0$ without losing any of the terms that can couple to $B$,
since, when we will turn on the magnetic
field, in each term containing $\vartheta_1$-functions there will be
at least one $\vartheta_1$ evaluated at $z=0$.
Notice that the first contribution to ${\cal M}$, apart from a factor $1/2$ due to the orbifold
projection, is the same that we found in the case of Sugimoto model, and so it will
contribute to the threshold corrections with the same term
we already wrote in eq. (\ref{SUGIMOTOthreshold}) (apart from the additional factor $1/2$
originating from the orbifold projection).

After the turning on of the magnetic field, whose action on the oscillator frequencies is given by
\be
\pi\epsilon = \left\{ \begin{array}{ll}
\tan^{-1}{(\pi q_a B)}+\tan^{-1}{(\pi q_b B)} \ \simeq \
\pi(q_a+q_b)B-\frac{\pi^3}{3}(q_a^3+q_b^3)B^3 \ , & \quad{\cal
  A}_{{\bar 5} {\bar 5}}\\
& \\
\tan^{-1}{(\pi q_a B)}  \ \simeq \ \pi q_aB-\frac{\pi^3}{3}q_a^3 B^3 \
, & \quad {\cal A}_{\bar{5}9}\\
& \\
2\tan^{-1}{\pi q_a B} \ \simeq  \ 2\pi q_a B -\frac{2\pi^3}{3}q_a^3 B^3
\ , & \quad {\cal M}_{\bar{5}}
\end{array} \right. \
\label{BSBepsi}
\ee
the open amplitudes read
\ba
\label{MobBSB}
&&{\cal A}_{\bar{5}\bar{5}+\bar{5}9} \ = \ {\cal
  A}^{({\cal N}=4)}(B) \nonumber\\
&&- \ \frac{i\pi B}{(4\pi^2)^2}\int\frac{d\tau}{\tau^2} \
\sum_{a,b=1}^{32} (q_a+q_b)\hat{R}_{aa}\hat{R}_{bb} \
\frac{\vartheta_3(i\epsilon\tau/2)\vartheta_3\vartheta_4^2 \ - \
  \vartheta_4(i\epsilon\tau/2)\vartheta_4\vartheta_3^2}{\eta^3 \ \vartheta_1(i\epsilon\tau/2)
  \vartheta_2^2}
\ P^{(2)}\nonumber\\
&& - \ \frac{i\pi B 16}{(4\pi^2)^2}\int\frac{d\tau}{\tau^2} \
\sum_{a=1}^{32}q_a \
\frac{\vartheta_3(i\epsilon\tau/2)\vartheta_3\vartheta_2^2
-\vartheta_2(i\epsilon\tau/2)\vartheta_2\vartheta_3^2}
{\eta^3 \ \vartheta_1(i\epsilon\tau/2) \vartheta_4^2}
 \ P^{(2)} \ , \nonumber\\
&& {\cal M}_{\bar{5}} \ = \
- \ \frac{i\pi B}{2(4\pi^2)^2} \int \frac{d\tau}{\tau^2} \
\sum_{a=1}^{32}q_a \
\sum_{\alpha,\beta}\eta_{\alpha\beta} \ e^{2\pi i \alpha} \
\frac{\vartheta\left[{\textstyle{\alpha\atop\beta}}\right]
\left(\frac{i\epsilon\tau}{2}\right)}{\vartheta_1\left(\frac{i\epsilon\tau}{2}\right)} \
\frac{\vartheta^3\left[{\textstyle{\alpha\atop\beta}}\right]}
{\eta^9} \ W^{(4)}P^{(2)} \ \nonumber\\
&& - \ \frac{ 2 i\pi B}{(4\pi^2)^2}\int\frac{d\tau}{\tau^2} \
\sum_{a=1}^{32}q_a \
\frac{\vartheta_3(i\epsilon\tau/2)\vartheta_3\vartheta_4^2
-\vartheta_4(i\epsilon\tau/2)\vartheta_4\vartheta_3^2}
{\eta^3 \ \vartheta_1(i\epsilon\tau/2) \vartheta_2^2}
 \ P^{(2)} \ ,
\ea
where we used the fact that $R_N$ is identically zero and we fixed $N=32$.
The matrix $\hat{R}$ is defined as $\hat{R}={\rm diag}(1\cdots
1,-1\cdots -1)$ with $d_1=16$ entrances equal to $+1$  and $d_2=16$
entrances equal to $-1$.

Now it is easy to perform an $S$-modular transformation for the
annulus amplitude, and a $P$ transformation for the M\"{o}bius strip, to write the
amplitudes in the transverse channel. The annulus gets a factor $1/2$
from the lattice sum and a factor $1/2$ from the integral measure,
while the M\"{o}bius gets only a factor $2$ from the integral measure.
If we denote with $v_3$ the normalized (as in \ref{normvol}) volume
of the further $T^2$, and with $v_1v_2$ the normalized volume of the
internal $T^4$ (that we think of as $T^2\times T^2$), the amplitudes in
the transverse channel are given by
\ba
\tilde{\cal A}_{\overline{5}\overline{5}+\overline{5}9} & = & \tilde{{\cal
  A}}^{({\cal N}=4)}(B) \nonumber\\
&-&  \frac{\pi Bv_3}{4(4\pi^2)^2}\int d\ell \
\sum_{a,b=1}^{32} (q_a+q_b)\hat{R}_{aa}\hat{R}_{bb} \
\frac{\vartheta_3(\epsilon)\vartheta_3\vartheta_2^2 \ - \
  \vartheta_2(\epsilon)\vartheta_2\vartheta_3^2}{\eta^3 \ \vartheta_1(\epsilon)
  \vartheta_4^2}
\ W^{(2)}\nonumber\\
&-& \frac{4 \pi B \ v_3}{(4\pi^2)^2}\int d \ell \
\sum_{a=1}^{32}q_a \
\frac{\vartheta_3(\epsilon)\vartheta_3\vartheta_4^2
-\vartheta_4(\epsilon)\vartheta_4\vartheta_3^2}
{\eta^3 \ \vartheta_1(\epsilon) \vartheta_2^2}
 \ W^{(2)} \ , \nonumber\\
\tilde{\cal M}_{\overline{5}} &=&
- \ \frac{\pi B}{(4\pi^2)^2} \frac{v_3}{v_1v_2} \ \int d\ell \
\sum_{a=1}^{32}q_a \
\sum_{\alpha,\beta}\eta_{\alpha\beta} \ e^{2\pi i \alpha} \
\frac{\vartheta\left[{\textstyle{\alpha\atop\beta}}\right]
\left(\frac{\epsilon}{2}\right)}{\vartheta_1\left(\frac{\epsilon}{2}\right)} \
\frac{\vartheta^3\left[{\textstyle{\alpha\atop\beta}}\right]}
{\eta^9} \ P^{(4)}_eW^{(2)}_e \ \nonumber\\
&-& \frac{ 4\pi B \ v_3}{(4\pi^2)^2}\int d\ell \
\sum_{a=1}^{32}q_a \
\frac{\vartheta_4(\epsilon/2)\vartheta_4\vartheta_3^2
-\vartheta_3(\epsilon/2)\vartheta_3\vartheta_4^2}
{\eta^3 \ \vartheta_1(\epsilon/2) \vartheta_2^2}
 \ W^{(2)}_e \ .
\ea
At this point we can derive the threshold corrections.
Using the definition of $\epsilon$ for ${\cal A}_{\bar 5 \bar 5}$ given in eq. (\ref{BSBepsi}),
$\epsilon\simeq (q_a+q_b)B$, and recalling that $\vartheta_{2,3,4}(\epsilon)$
are even functions of their arguments, so that in their expansion
there is the quadratic term in $\epsilon$ but not the linear one,
we can expand the first contribution to the annulus up to second order in the magnetic field, obtaining
\ba
&&B\sum_{a,b=1}^{32} (q_a+q_b)\hat{R}_{aa}\hat{R}_{bb} \
\frac{\vartheta_3(\epsilon)\vartheta_3\vartheta_2^2 \ - \
  \vartheta_2(\epsilon)\vartheta_2\vartheta_3^2}{\eta^3 \ \vartheta_1(\epsilon)
  \vartheta_4^2} \  \nonumber\\
 &&\simeq \ {\rm tr}\hat{R}^2 \
\frac{\vartheta_3^2\vartheta_2^2 \ - \
  \vartheta_2^2\vartheta_3^2}{2\pi\eta^6 \
  \vartheta_4^2}  \ + \ \pi B^2\sum_{a,b=1}^{32} (q_a+q_b)^2
\hat{R}_{aa}\hat{R}_{bb} \ ,
\ea
where we used the identity (\ref{A.14}) between $\vartheta$-functions
\be
\vartheta_3^{\prime\prime}\vartheta_3\vartheta_2^2-\vartheta_2^{\prime\prime}\vartheta_2\vartheta_3^2
\ = \ 4\pi^2\eta^6\vartheta_4^2 \ .
\ee
The term proportional to $({\rm tr}\hat{R})^2$ is identically zero
since $({\rm tr}\hat{R})^2=R_D^2=0$.
Also the second term is identically zero, being proportional to
$2{\rm tr}(Q^2\hat{R})R_D+2{\rm tr}(Q\hat{R})^2$, and recalling the
form of $Q$ and $\hat{R}$. Therefore this part of the annulus
gives neither tadpoles nor threshold corrections.

Performing the same type of expansion in the second contribution
to the annulus amplitude and in the second contribution to the
M\"{o}bius amplitude, that have the same structure, it is easy to
extract from them the term proportional to $B^2$ \be \label{super}
\ \frac{4\pi^2B^2 v_3}{(4\pi^2)^2} \ {\rm tr Q^2} \ \int d\ell \
(W^{(2)}-W^{(2)}_e) \ , \ee where we used the suitable definition
of $\epsilon$ given in (\ref{BSBepsi}), and the identity
(\ref{A.15}) \be
\vartheta_4^{\prime\prime}\vartheta_4\vartheta_3^2-\vartheta_3^{\prime\prime}\vartheta_3\vartheta_4^2
\ = \ 4\pi^2\eta^6\vartheta_2^2 \ . \ee The corresponding
contribution to the threshold corrections is then given by \be
\Lambda_2^{({\cal N}=2)} \ = \  2 v_3 \ {\rm tr }  Q^2 \ \int
d\ell \ \left(W^{(2)}-W^{(2)}_e\right) \ . \ee We have to stress
here that this result has the same structure as the
four-dimensional ${\cal N}=2$ supersymmetric $T^4/{\mathbb
Z}_2\times T^2$ model \cite{bf}, and this fact can be understood
if one notes that $ND$ term in the annulus amplitude and the last
term in the M\"{o}bius amplitude, from which we derived
(\ref{super}), are the same as the supersymmetric $T^4/{\mathbb
Z}_2$ orbifold, apart from a change of chirality. Moreover we
notice that only the short BPS states of ${\cal N}=2$ contributes
to $\Lambda_2^{({\cal N}=2)}$ while all the string oscillators
decouple according to \cite{Harvey:1995fq,bf}. In order to clarify
the relation with ${\cal N}=2$, let us consider the open infrared
limit of $\Lambda_2^{({\cal N}=2)}$. To do that we come back to
the direct channel and we consider the $B^2$ terms from the mixed
$ND$ sector in the annulus amplitude and from the second addend in
the M\"{o}bius strip. After using the identities (\ref{A.14}) and
(\ref{A.15}), we obtain \be \Lambda_2^{({\cal N}=2)} = {\rm tr}Q^2
\ \int \frac{d\tau}{\tau} \left(4P^{(2)} -P^{(2)}\right) \ , \ee
where the first lattice sum comes from the annulus amplitude while
the second comes from the M\"{o}bius amplitude. The integral can
be computed, and indeed this was done in \cite{bf}, but here let
us consider the $\tau \to \infty$ limit after which only the
massless Kaluza-Klein recurrences survive. Cutting-off the
integration variable to $\tau<1/\mu^2$ and fixing ${\rm
tr_{fund}}Q^2=1/2$, the leading term for $\tau \to \infty$ of
$\Lambda_2^{({\cal N}=2)}$ is \footnote{Really, what here we call
$\Lambda_2^{({\cal N}=2)}$ is half of the standard result that one
would obtain in the supersymmetric $T^4/{\mathbb Z}_2$ since in
that case the trace of $Q^2$ runs both on the fundamental and on
the anti-fundamental representations of the unitary gauge group.}
\be \Lambda_2^{({\cal N}=2)} \ = \ - \ \frac{1}{4}\times
\frac{1}{2} b^{({\cal N}=2)} \ \ln{\mu^2} \ + \ {\rm IR  \ finite
\ terms} \ , \ee where \be b^{({\cal N}=2)} = 12 \ , \ee is the
standard one-loop $\beta$-function coefficient of the
four-dimensional ${\cal N}=2$ super Yang-Mills theory with a gauge
multiplet in the adjoint representation of the gauge group
$SU(16)$, two hyper-multiplets in the 120, and 16 hyper multiplets
in the fundamental representation (this is just the open massless
content of the supersymmetric $T^4/{\mathbb Z}_2$ model
compactified to four dimensions) \be b^{({\cal N}=2)} = 2 \left[
-\ C_2({\rm adj}) \ + \ 2\times T(120) \ + \ 16\times T({\rm
fund})\right] = 12 \ . \ee

On the other hand the complete expression for the threshold corrections of this model is obtained
adding to $\Lambda_2^{({\cal N}=2)}$ the non supersymmetric contribution proportional
to $v_3/v_1v_2$ coming from the M\"{o}bius strip, whose structure was already analyzed in
the case of the Sugimoto model (see eq. (\ref{SUGIMOTOtransverse})).
Taking into account all the contributions, the result is
\ba
\label{bsb12}
\Lambda_2 & = &  2 v_3 \ {\rm tr }  Q^2 \ \int d\ell \ \left(W^{(2)}-W^{(2)}_e\right)
\nonumber\\
&& - \ \frac{{\rm tr}  Q^2}{4\pi^2} \ \frac{v_3}{v_1v_2} \ \int d \ell \
\frac{\vartheta_2^4}{\eta^{12}} \
\left(\frac{2\pi^2}{3}+\frac{\vartheta_2^{\prime\prime}}{\vartheta_2}-
\frac{\vartheta_1^{\prime\prime\prime}}{6\pi\eta^3}\right)
P^{(4)}_e W^{(2)}_e \ .
\ea
The leading infrared (in the open channel) term is simply obtained
adding to $\frac{b^{({\cal N}=2)}}{2}=6$ the non supersymmetric contribution
$b^{({\cal N}=0)}=-8/3$ that we already discussed  in the case of the Sugimoto model (here
there is an additional factor $1/2$ due to the orbifold projection).
This sum reproduces the right four-dimensional one-loop $\beta$-function coefficient of the low-energy
effective field theory, as one can check using the formula (\ref{indice})
\ba
b  &=& - \ \frac{11}{3}\times 9 \ + \ \frac{4}{3}\times\left(7+16\times\frac{1}{2}\right) \ + \ \frac{2}{3}\times\frac{1}{2}\times 16
\nonumber\\
&+& \ \frac{1}{3}\times\left(9+2\times\frac{1}{2}\times 16+\frac{1}{2}\times 16\right)=
\frac{10}{3} \ ,
\ea
where the massless four-dimensional fields charged  with respect to the magnetic field are:
from the $DD$ sector a gauge vector together with a complex scalar in the adjoint representation,
16$\times$2 complex scalars and 16$\times$2 Weyl fermions in the fundamental representation, 2 Weyl fermions
in the antisymmetric representation, while from the $ND$ sector 16 complex scalars and 16 Weyl fermions in the fundamental
representation.

Summarizing, the resulting threshold corrections (\ref{bsb12}) for the brane supersymmetry breaking model
compactified to four dimensions are given by a non supersymmetric term
that originates from the M\"{o}bius amplitude and by a supersymmetric term to which
only the ${\cal N}=2$ BPS states contribute while string oscillators decouple.
The remarkable property we want to stress is that,
in spite of the presence of NS-NS tadpoles induced by supersymmetry breaking,
the threshold corrections are ultraviolet (in the open channel) finite.
Moreover, performing two $T$-dualities along the directions of the further torus $T^2$,
that turn the winding sums $W^{(2)}$ into momentum sums $P^{(2)}$, the volume $v_3$ in the T-dual volume $1/v_3$
and the $\overline{\rm D5}$ in the $\overline{\rm D3}$, it is easy to see that,
in the limit of large internal volume transverse to the $\overline{\rm D3}$, the
non supersymmetric contribution is suppressed with respect to the supersymmetric one.
Therefore, at the one-loop level, despite the supersymmetry breaking at the string scale on the antibranes,
the threshold corrections are essentially determined by the supersymmetric contribution.
This result confirms the conjecture, made in
\cite{bsb,ab}, that in the brane
supersymmetry breaking model (and also in models with brane-antibrane pairs, as we will see in the next Subsection)
threshold corrections in codimension
larger than two are essentially given by supersymmetric
expressions, due to the supersymmetry of the bulk (closed string)
spectrum.
\subsection{Brane-antibrane systems}
In this Subsection we compute the threshold corrections for the model obtained adding
M D5-$\overline{\rm D5}$ pairs to the usual supersymmetric $T^4/Z_2$ orbifold of the Type
$I$ superstring \cite{bs,gp}.
Let us begin by recalling  the open amplitudes for the supersymmetric model (see (\ref{anellosuperorbifold})
and (\ref{MOebiusSuperOrbifold})), that here we write
with the integral measure and the bosonic degrees of freedom
\ba
{\cal A} &=& - \ \frac{1}{2(4\pi^2)^2} \ \int \frac{d\tau}{\tau^3}\Biggl[ \frac{Q_o + Q_v}{\eta^8} \left(
N^2 P^{(4)}_m + D^2 W^{(4)}_n\right)
\nonumber\\
&& +  \ \left(R_N^2 + R_D^2 \right) \frac{Q_o - Q_v}{\eta^4} \left( {2\eta \over
\vartheta_2}\right)^2
\ + \ 2 N D \, \frac{Q_s + Q_c }{\eta^4} \left( {\eta \over \vartheta_4}\right)^2 \nonumber\\
&&+ \ 2 R_N R_D \, \frac{Q_s - Q_c }{\eta^4} \left( {\eta \over \vartheta_3}\right)^2
\Biggr] \ P^{(2)}_m , \nonumber\\
{\cal M} &=& \frac{1}{2(4\pi^2)^2} \ \int \frac{d\tau}{\tau^3} \left[ \frac{\hat Q _o + \hat Q _v }
{\hat \eta^8} \left(
N P^{(4)}_m + D W^{(4)}_n \right)\right.\nonumber\\
&& \qquad\qquad\qquad\quad\qquad\left.- \left( N + D\right)
\frac{\hat Q _o - \hat Q _v }{\hat \eta^4}
\left( {2\hat\eta\over \hat\vartheta_2} \right)^2 \right] \ P^{(2)}_m ,
\ea
where
\ba
\label{parametriz}
&&N=n+\bar n \ , \qquad D=d+\bar d \ ,\nonumber\\
&&R_N=i(n-\bar n) \ , \qquad R_D=i(d-\bar d)
\ea
and $n=\bar n=d=\bar d=16$.
Moreover, we compactified two other dimensions on a further $T^2$ torus,
as can be seen from the lattice sum $P_m^{(2)}$.

Now we add the $M$ brane-antibrane pairs. We denote with $M_+=M$
the number of D5 branes and with $M_-=M$ the number of $\overline{\rm D5}$ antibranes.
Moreover, as in (\ref{parametriz}), we parameterize
$M_+=m_+ + \overline m_+$, and $M_-=m_- + \overline m_-$, with $m_\pm=\bar m_\pm$ to distinguish
branes (or antibranes) from their imagine.
If the additional $M_+$ D5 branes are
placed, together with the original 32, at a given fixed point of
the orbifold, while the $M_-=M \ $ $\overline{\rm D5}$ are placed at a
different fixed point, that for simplicity we take to be separated
only along one of the internal directions, the resulting gauge
group is $U(16)_9 \times [U(16+m_+) \times U(m_-)]_5$. The $M$ pairs
generate an NS-NS tadpole localized in six dimensions, that
would be expected to introduce UV divergences in one-loop
threshold corrections.

Let us first analyze the annulus amplitude.
The amplitude is obtained changing the RR signs in the transverse channel
of the sectors corresponding to exchanges between branes and antibranes, and reads
\ba
\label{susyANN}
{\cal A} &=& - \ \frac{1}{2(4\pi^2)^2} \ \int \frac{d\tau}{\tau^3}\Biggl[ \frac{Q_o + Q_v}{\eta^8} \left(
N^2 P^{(4)}_m + (D+M_+)^2 W^{(4)}_n + M_-^2W^{(4)}_n\right)
\nonumber\\
&& + \ 2(D+M_+)M_- \ \frac{O_4O_4+V_4V_4-S_4C_4-C_4S_4}{\eta^8} \ W_{n+1/2}^{(4)}\nonumber\\
&& +  \ \left(R_N^2 + R_{D+M_+}^2 + R_{M_-}^2\right) \ \frac{Q_o - Q_v}{\eta^4} \left( {2\eta \over
\vartheta_2}\right)^2\nonumber\\
&& + \ 2 N (D+M_+) \, \frac{Q_s + Q_c }{\eta^4} \left( {\eta \over \vartheta_4}\right)^2 \nonumber\\
&& + \ 2NM_- \ \frac{O_4S_4-C_4O_4+V_4C_4-S_4V_4}{\eta^4}\left( {2\eta \over
\vartheta_2}\right)^2\nonumber\\
&& + \ 2 R_N R_{D+M_+} \, \frac{Q_s - Q_c }{\eta^4} \left( {\eta \over \vartheta_3}\right)^2\nonumber\\
&& + \ 2R_N R_{M_-} \ \frac{-O_4S_4+V_4C_4+S_4V_4-C_4O_4}{\eta^4} \ \left( {\eta \over \vartheta_3}\right)^2
\Biggr] \ P^{(2)}_m , \nonumber\\
\ea
where the shift of the windings $W_{n+1/2}^{(4)}$ in the term proportional to $(D+M_+)M_-$
is due to the separation between D5 branes and $\overline{D5}$ antibranes.
Moreover, we see that there are no mixed terms proportional to $R_{D+M_+}R_{M_-}$ and the reason is that
such a term would come from the one proportional to $W_{n+1/2}^{(4)}$
that has no zero modes to project. The twisted sector is simply the one of the supersymmetric case
with $D$ replaced by $D+M_+$ and $R_D$ by $R_{D+M_+}$ together with the addition of mixed terms between the $N \ $
D9 branes and the $M_-\ $ $\overline{D5}$ antibranes, terms that apart from a change of chirality are
the same as those of the brane supersymmetry breaking model.

On the other hand, it is easy to write the M\"{o}bius amplitude for which the same characters
propagate both in the direct and in the transverse channel (see (\ref{MOebiusSuperOrbifold})
and (\ref{MOebiusSuperOrbifoldtran})), so that  in practice one can reverse the signs of the RR
sectors proportional to $M_-$ directly in the open channel. The amplitude reads
\ba
\label{susyMob}
{\cal M} &=& \frac{1}{2(4\pi^2)^2} \ \int \frac{d\tau}{\tau^3} \left[ \frac{\hat Q _o + \hat Q _v }
{\hat \eta^8} \left(
N P^{(4)}_m + (D+M_+) W^{(4)}_n \right)\right.\nonumber\\
&& \left.+ \ M_- \ \frac{V_4O_4+O_4V_4+C_4C_4+S_4S_4}{\hat \eta^8} \ W_n^{(4)}\right.\nonumber\\
&&- \left.\left( N + D + M_+\right)
\frac{\hat Q _o - \hat Q _v }{\hat \eta^4}
\left( {2\hat\eta\over \hat\vartheta_2} \right)^2 \right.\nonumber\\
&&\left.- M_- \ \frac{V_4O_4+C_4C_4-O_4V_4-S_4S_4}{\hat \eta^4}
 \ \left( {2\hat\eta\over \hat\vartheta_2} \right)^2
\right] \ P^{(2)}_m  \ .
\ea
Let us notice that, while the first and third lines in (\ref{susyMob}) are essentially the same as the
corresponding terms in the supersymmetric case
with $D$ replaced by $D+M_+$, the other two lines, proportional to $M_-$, describe respectively
the interactions of the $\overline{D5}$ antibranes with O5$_+$ planes, and with an $O9_+$ plane.
Hence, the fourth line is the same as the corresponding term in the brane supersymmetry breaking model, apart
from the usual change of chirality, while the second line has an opposite sign due to the fact that here there is an
O5$_+$ plane while in the brane supersymmetry breaking model there is an O5$_-$ plane.

At this point we can proceed with the background field method. Let
us choose a $U(1)$ generator $Q$ in the Cartan subalgebra of the
group $SU(16+m_+)$, with the usual normalization ${\rm
tr}_{fund}Q^2=1/2$, where the generator $Q$ has no components
along the anomalous $U(1)$ factor. Moreover in the Chan-Paton
basis in which we wrote the amplitudes (\ref{susyMob}) and
(\ref{susyANN}) the fundamental and anti-fundamental
representations of the group are disentangled \cite{gp}.

The only sectors that can couple to the magnetic field are the ones
proportional to the total number of antibranes $D+M_+$.
In fact, the terms linear in $R_N$ vanish identically since $R_N=0$.
Moreover, if we denote with $J$ the total number of D5 branes,
say $J=D+M_+=j+\bar j$, and we parameterize $R_J=R_{D+M_+}$ following $R_J=i(j-\bar j)$,
with $j=\bar j$, the $R^2_{J}$ term in the annulus amplitude is proportional to
\be
R_J^2 \ (Q_0-Q_v) \ = \ -(j^2+\bar j^2 -2j\bar j) \ (Q_o-Q_v) \ ,
\ee
and after the coupling with $B$ it becomes
\ba
&&-\biggl[ \sum_{a,b=1}^{j} (q_a+q_b) \ \hat R_{aa} \hat R_{bb}
+\sum_{\bar a, \bar b = \bar 1}^{\bar j} (\bar q_{\bar a}+\bar q_{\bar b}) \hat R_{\bar a  \bar a} \hat R_{\bar b \bar b}\nonumber\\
&& - \sum_{a=1}^{j}\sum_{\bar b = \bar 1}^{\bar j} (q_a+\bar
q_{\bar b}) \ \hat R_{aa} \hat R_{\bar b \bar b} -\sum_{\bar a =
\bar 1}^{\bar j}\sum_{b=1}^{j} (\bar q_{\bar a}+q_b) \ \hat
R_{\bar a\bar a} \hat R_{bb}\biggr](Q_o-Q_v)(B) , \ea where
$q_{a,b}$ and $\bar q_{\bar a, \bar b}$ run over the fundamental
or the anti-fundamental representations of the unitary gauge group
and the matrices $\hat R$ are $\hat R_{aa}=\hat
R_{bb}=\mathbf{1}_{j\times j}$ and $\hat R_{\bar a \bar a}=\hat
R_{\bar b \bar b}=\mathbf{1}_{\bar j \times \bar j}$. Moreover, we
left implicit the dependence of $Q_o-Q_v$ on $(q_a+q_b)$. Then,
expanding for small magnetic field the character $Q_o-Q_v$, the
zeroth-order terms reconstruct the term with $R_J^2$, that is
identically zero, while the quadratic order cancel thanks to the
minus signs of the mixed $j\bar j$ terms.

The last consideration to make is about the terms in the amplitudes that are proportional to the character
$Q_o+Q_v$ that is the supersymmetric one, and thus will give neither tadpole terms
nor threshold corrections (we already saw that such a term would start at the quartic order in $B$).

The only terms that contribute to threshold corrections are the one proportional to $N(D+M_+)$
from the annulus, and the one proportional to $D+M_+$ from the M\"{o}bius, that are identical,
apart from a change of chirality, to the ones already discussed in the previous Subsection.
Therefore, their contribution to threshold corrections is the same as the first term in
(\ref{bsb12}).
Moreover, there is another contribution coming from the $(D+M_+)M_-$ term in the annulus, that is
\ba
{\cal A}_{(D+M_+)M_-} &=& - \ \frac{\pi i B M_-}{2(4\pi^2)^2}\int \frac{d \tau}{\tau^2}
\sum_{a=1}^{D+M_+} q_a  \nonumber\\
&\times&\frac{\vartheta_3\left(\frac{i\epsilon\tau}{2}\right) \vartheta_3^3
+\vartheta_4\left(\frac{i\epsilon\tau}{2}\right)
\vartheta_4^3 - \vartheta_2\left(\frac{i\epsilon\tau}{2}\right)\vartheta_2^3}
{\eta^9 \ \vartheta_1\left(\frac{i\epsilon\tau}{2}\right)} \ W_{n+1/2}^{(4)}P^{(2)}_m ,
\ea
or in the transverse channel
\ba
\tilde{{\cal A}}_{(D+M_+)M_-} &=& - \frac{2^{-5} \pi B M_-}{(4\pi^2)^2}\frac{v_3}{v_1v_2}\int d\ell
\sum_{a=1}^{D+M_+} q_a \nonumber\\
&\times& \frac{\vartheta_3(\epsilon) \vartheta_3^3
+\vartheta_2(\epsilon) \vartheta_2^3 -
\vartheta_4(\epsilon)\vartheta_4^3} {\eta^9 \
\vartheta_1(\epsilon)} \ (-)^mP_m^{(4)}W^{(2)}_n . \ea The sum
over the charges runs both in the fundamental and in the
anti-fundamental representation of $SU(16+m_+)$. At this point we
can extract the threshold corrections expanding in $B$. Using the
identity (\ref{A.13}), the numerator of the integral can be
written \be \vartheta_3(\epsilon) \vartheta_3^3
+\vartheta_2(\epsilon) \vartheta_2^3 -
\vartheta_4(\epsilon)\vartheta_4^3 \ = \
2\vartheta_1^4(\epsilon/2) +2\vartheta_2(\epsilon)\vartheta_2^3 \
, \ee and one can recognize the usual term proportional to
$\vartheta_1^4(\epsilon)$ that, as we already know, starts at the
fourth order in $B$. The remaining $\vartheta_2(\epsilon)$ term,
with $\pi\epsilon\simeq \pi\epsilon q_aB-\frac{\pi^3}{3}q_a^3B^3$,
 is of the same kind we found from the M\"{o}bius
in the previous Subsection and can be treated in the same way.
Therefore the complete result for threshold corrections is
\ba
\label{bsb11}
&&\Lambda_2  =  \ 2 v_3 \ {\rm tr } \left(Q^2+\bar Q^2\right) \ \int d\ell \ \left(W^{(2)}-W^{(2)}_e\right)
\nonumber\\
&& - \ \frac{{\rm tr}  \left(Q^2+\bar Q^2\right) \ M}{2^7\pi^2} \ \frac{v_3}{v_1v_2} \ \int d \ell \
\frac{\vartheta_2^4}{\eta^{12}} \
\left(\frac{2\pi^2}{3}+\frac{\vartheta_2^{\prime\prime}}{\vartheta_2}-
\frac{\vartheta_1^{\prime\prime\prime}}{6\pi\eta^3}\right)
\ (-)^m P^{(4)}_{2m} \ W^{(2)}_{2m} \ ,
\nonumber\\
\ea
The non supersymmetric contribution originates from the annulus and reflects the interaction
between  branes and antibranes located at different orbifold fixed points. This term is
finite both in the ultraviolet and in the infrared limit (in the open channel).
The ultraviolet finiteness (infrared in the closed channel) is ensured by the same
argument of the previous examples, see eq. (\ref{IRgood}), while the infrared finiteness is
guaranteed by the separation between the branes and antibranes.
The first contribution is the usual supersymmetric one \cite{bf}
already met in the previous example.
Therefore, like in the brane supersymmetry breaking model, in a model
with brane-antibrane pairs, in spite of the presence of NS-NS tadpoles, the result
is ultraviolet (in the open channel) finite.
Moreover, in the limit of large internal volume transverse
to the branes, the non supersymmetric contribution is suppressed
with respect to the other one \cite{bsb} and the result is
essentially dominated by the supersymmetric part.

\subsection{Type 0$^\prime$B}
Let us finally consider the Type $0'B$ model \cite{susy95, c0b,bfl1}
whose open partition functions (\ref{tildeA3}) and (\ref{tildeM3})
(with $n=0$ in order to eliminate the open tachyon from the spectrum) are
\ba
\tilde{{\cal A}} & = & \frac{2^{-6}}{2} \ \left[(m+\bar m)^2 \ (V_8-C_8)
 \ - \ (m-\bar m)^2 \ (O_8-S_8)\right] \ ,\nonumber\\
 \tilde{\cal M} &=& (m+\bar m) \hat C_8 \ ,
\ea
and the gauge group is $SU(32)$, with $m =\bar m= 32$.
It is clear that from the annulus amplitude one can receive no
contribution to threshold corrections, since one term vanishes
identically after the effective identification of $m$ and $\bar
m$, while the other one is the usual supersymmetric character that
would start at the quartic order in $B$. The only possible
contribution to threshold corrections is given by the M\"{o}bius
that in the presence of a magnetic filed $B$ reads \be \tilde{\cal
M}\ = \ - \ \frac{2\pi B v^{(6)}}{(4\pi)^2} \ \int d\ell \sum q_a
\ \frac{\vartheta_2(\epsilon/2) \
\vartheta_2^3}{\vartheta_1(\epsilon/2) \ \eta^9} \ W^{(6)}_e \ ,
\ee where we compactified six dimensions on a torus and the $\sum
q_a$ runs over both the fundamental and the anti-fundamental
representations of the gauge group. We already analyzed this kind
of expression in the previous Subsections. In particular, apart
from a factor $1/2$, it is equal to the first addend in the
M\"{o}bius expression of eq. (\ref{SUGIMOTOconB}), where there is
a further term the starts at the quartic order in the magnetic
field. Therefore the threshold corrections are still the same \be
\Lambda_2 \ = \ - \ \frac{v^{(6)} \ {\rm tr} \left(Q^2+\bar
Q^2\right)}{4\pi^2}  \int d \ell \ \frac{\vartheta_2^4}{\eta^{12}}
\
\left(\frac{2\pi^2}{3}+\frac{\vartheta_2^{\prime\prime}}{\vartheta_2}-
\frac{\vartheta_1^{\prime\prime\prime}}{6\pi\eta^3}\right)
W^{(6)}_{e} \ . \ee The result is again ultraviolet finite while
in the infrared limit it gives the usual logarithmic divergence
with a one-loop $\beta$ coefficient equal to \be b \ = \ - \
\frac{11}{3}\times 32  \ +  \ 2\times \frac{2}{3}\times 4 \times
15 \ +3\times \frac{1}{3} \times 32 \ = \ -\frac{16}{3} \ , \ee
where we used the fact that the massless spectrum in four
dimensions contains a gauge vector together with three complex
scalars in the adjoint representation of $SU(32)$ and 4 Weyl
fermions in the antisymmetric representation, together with four
other Weyl fermions in its conjugate representation. For $SU(N)$
with the normalization fixed by ${\rm tr}Q^2=1/2\ $, $T({\rm
fund})=1/2$, the quadratic Casimir of the adjoint is $C_2({\rm
adj})=N$, while $T({\rm antisym})=N/2-1$.
\chapter*{Conclusions \markboth{Conclusions}{Conclusions}}
\addcontentsline{toc}{chapter}{Conclusions}
In this Thesis
we approached at first the problem of tadpoles in Field Theory, where
in a number of toy models we tried to ask what happens if one quantizes the theory around
a point that is not a real vacuum. We focused our attention on
the classical vacuum energy, a quantity of relevance in String Theory.
What we learnt from the case of a quartic scalar potential is that,
starting from an arbitrary initial value of the field,
classical tadpole resummations typically drive the classical vacuum energy to an extremum,
not necessarily a minimum. Of course the resummation does not have to
touch any inflection point where the resummation breaks down
and further subtleties can be present, since at times the procedure can lead to oscillations.
Moreover, we found some special initial ``non-renormalization'' points for which all higher
order corrections cancel, so that the flow is determined by only a few steps.
We gave an interpretation of the flow, and in
particular of such special points, in terms of Newton's tangent method.
But the convergence of tadpole resummations is a complicated issue, and
in general one has to make sure to start well within the convergence domain.
This is the case for example of a quartic potential deformed by
a magnetic field, where we saw that for a
magnetic field that is too large, there are some regions where
the tadpole expansion makes no sense
and perturbation theory breaks down.

We then analyzed the procedure for a string-inspired
toy model with tadpoles localized on lower dimensional $D$-branes, performing
explicitly the resummation.

We also tried to turn on gravity, coupling it to a scalar
field. Of course the computation here was more complicated, due to
the nature of gravity, but actually in this case another subtlety emerged.
Quantizing a theory of gravity with a tadpole term around a
Minkowski background gives a mass to the graviton that is not of
Pauli-Fierz type, so that a ghost propagates. Despite this new
complication, the tadpole resummation proved to work directly.

The resummation program is very difficult to carry out in String Theory,
where the higher order tadpole corrections correspond to amplitudes of higher
and higher genus. One could think to stop the resummation at the first
orders but generally tadpoles are large.
Moreover, the resummation in String Theory could be of practical use
if the endpoint of the resummation is a stable vacuum of the theory.
In that case the existence of the ``non-renormalization'' points
in some string models would be of a great practical value.

In String Theory at first we analyzed a model in which
the vacuum redefinition was performed at the full string level,
without the need for a resummation. In particular we found that the correct vacuum of
a Type II orientifold with local R-R tadpoles, is related to a Type 0 orientifold.
Notice that this case is not really of the kind we search
because NS-NS tadpoles are cancelled, but nevertheless is a very
interesting first example that shows  how the space configuration
of a model can be interpreted, at the full string level, as the right vacuum,
or the collapsing point, of an instable vacuum of another model.

Other quantities that we computed are the one-loop threshold corrections to gauge
couplings, showing that in a number of models with supersymmetry breaking
and parallel branes the results are essentially given by the combination of
a supersymmetric contribution and a non supersymmetric contribution that actually,
in the large internal volume limit, is suppressed with respect to the other one.
Moreover, we saw that, in spite of the NS-NS tadpoles, the one-loop threshold
corrections are (open) ultraviolet finite, and we understood this finiteness
in terms of a cancellation of the closed massless states propagating in the
bulk. There is no motivation why this cancellation should occur also at the
higher orders and indeed we are very interested in performing such an explicit computation at
genus $3/2$, that is left for a future work \cite{bdnps}.
From a filed theory analysis, what we expect to happen
in this case is that tadpole resummations lead to a breakdown of perturbation theory,
and in particular that the genus $3/2$ is of the same order as the disk contribution.

The dilaton-graviton cancellation we described in models
with parallel branes does not apply to the case of
intersecting branes. In fact the couplings of bulk fields to
a brane depend from their spin and, since the $D$-branes (and
$O$-planes) are not parallel in this models, the cancellation
can no more work. And  indeed, as observed in \cite{ls},
the one-loop threshold corrections are not UV finite in this type of models,
but what can remain finite are the differences of gauge
couplings for gauge groups related by Wilson lines,
quantities that are of direct relevance for the issue of unification.
We plan to analyze this class of models in relation with the tadpole problem
in a future work \cite{bdnps}.

\appendix
\chapter{$\vartheta$-functions}
The $\vartheta$-functions are defined through the infinite sums
\be
\label{A.1}
\vartheta \left[{\textstyle {\alpha \atop \beta}} \right] (z|\tau) =
\sum_{n\in {\mathbb Z}} \
q^{\frac{1}{2} (n+ \alpha)^2} \ e^{2 \pi i (n + \alpha)(z+\beta)} \ ,
\ee
where $ \ \alpha, \beta = 0, \ \frac{1}{2} \ $.
Equivalently,
the Jacobi $\vartheta$-functions can be defined in terms of infinite products
\ba
\label{A.2}
\vartheta \left[ {\textstyle {\alpha \atop \beta}} \right] (z|\tau) &=&
e^{2 i \pi \alpha (z+\beta)} \ q^{\alpha^2/2} \prod_{n=1}^\infty \
( 1 - q^n)\prod_{n=1}^\infty \ (1 + q^{n + \alpha - 1/2} e^{2 i \pi (z+\beta)} )\nonumber\\
&& \times \prod_{n=1}^\infty (1 + q^{n - \alpha - 1/2} e^{-2 i \pi (z+\beta)} ) \ ,
\ea
so that, in particular
\ba
\label{A.3}
&&\vartheta\left[{\textstyle{0\atop
      0}}\right](z|\tau) \ = \ \vartheta_3(z|\tau) \ = \
\prod_{m=1}^\infty (1-q^m)(1+e^{2\pi iz}\ q^{m-1/2})(1+e^{-2\pi i z} \
q^{m-1/2})
\ ,\nonumber\\
&&\vartheta\left[{\textstyle{0\atop
      1/2}}\right](z|\tau) \ = \ \vartheta_4(z|\tau) \ = \
\prod_{m=1}^\infty (1-q^m)(1-e^{2\pi iz}\ q^{m-1/2})(1-e^{-2\pi i z} \
q^{m-1/2})\ , \nonumber\\
&&\vartheta\left[{\textstyle{1/2\atop
      0}}\right](z|\tau) \ = \ \vartheta_2(z|\tau) \ = \
2q^{1/8}\cos(\pi z)\nonumber\\
&&\qquad\qquad\times\prod_{m=1}^\infty (1-q^m)(1+e^{2\pi iz}\ q^m)(1+e^{-2\pi i z} \
q^m) \ , \nonumber\\
&&\vartheta\left[{\textstyle{1/2\atop
      1/2}}\right](z|\tau) \ = \ -\vartheta_1(z|\tau) \ = \
-2q^{1/8}\sin(\pi z) \nonumber\\
&&\qquad\qquad\times\prod_{m=1}^\infty (1-q^m)(1-e^{2\pi iz}\ q^m)(1-e^{-2\pi i z} \
q^m) \ .
\ea
The modular transformations under $T$ and $S$ are expressed in a
compact notation respectively by
\be
\label{A.4}
\vartheta \left[ {\textstyle {\alpha \atop \beta}}
 \right] (z|\tau+1)
= e^{-i \pi \alpha
(\alpha -1)} \vartheta \left[ {\textstyle {\alpha \atop \beta +\alpha - 1/2}}
\right]
(z|\tau) \ ,
\ee
and
\label{A.5}
\be
\vartheta \left[ {\textstyle {\alpha \atop \beta}} \right]
\left(\frac{z}{\tau}\right|\left.-\frac{1}{\tau}\right) =
(-i \tau)^{1/2} \ e^{2 i \pi \alpha \beta + i \pi z^2/\tau} \ \vartheta
\left[{\textstyle {\beta \atop -\alpha}}\right] (z|\tau) \ .
\ee
More explicitly the modular transformations are
\ba
\label{A.6}
&&\vartheta_3(z|\tau+1) \ = \ \vartheta_4(z|\tau+1) \ , \nonumber\\
&&\vartheta_4(z|\tau+1) \ = \ \vartheta_3(z|\tau+1) \ , \nonumber\\
&&\vartheta_2(z|\tau+1) \ = \ q^{1/8} \ \vartheta_2(z|\tau+1) \ , \nonumber\\
&&\vartheta_1(z|\tau+1) \ = \ q^{1/8} \vartheta_1(z|\tau+1) \ ,
\ea
and
\ba
\label{A.7}
&&\vartheta_3(z/\tau|-1/\tau) \ = \ \sqrt{-i\tau} \ e^{\pi i z^2 /\tau}
\ \vartheta_3(z|\tau) \ , \nonumber\\
&&\vartheta_4(z/\tau|-1/\tau) \ = \ \sqrt{-i\tau} \ e^{\pi i z^2 /\tau}
\ \vartheta_2(z|\tau) \ , \nonumber\\
&&\vartheta_2(z/\tau|-1/\tau) \ = \ \sqrt{-i\tau} \ e^{\pi i z^2 /\tau}
\ \vartheta_4(z|\tau) \ , \nonumber\\
&&\vartheta_1(z/\tau|-1/\tau) \ = \ -i \sqrt{-i\tau} \ e^{\pi i z^2 /\tau}
\ \vartheta_1(z|\tau) \ . \nonumber\\
\ea
The $\vartheta$-functions satisfy the identity
\be
\label{A.8}
\vartheta_3^4-\vartheta_4^4-\vartheta_2^4 \ = \ 0 \ ,
\ee
known as the \textit{aequatio identica satis abstrusa} of Jacobi.\\
Moreover, while
\be
\label{A.9}
\vartheta(0|\tau) \ = \ 0 \ ,
\ee
the first derivative of $\vartheta_1$ at zero is
\be
\label{A.10}
\vartheta_1^\prime(0|\tau) \ = \ 2\pi \eta^3 \ ,
\ee
where the Dedekind $\eta$-function is defined by
\be
\label{A.11}
\eta(\tau) \ = \ q^{1/24} \ \prod_{n=1}^\infty(1-q^n) \ ,
\ee
and the modular transformations of \ref{A.11} are
\ba
\label{A.12}
\eta(\tau+1) \ = \ e^{i\pi /12} \ \eta(\tau) \ , \nonumber\\
\eta(-1/\tau) \ = \ \sqrt{-i\tau} \ \eta(\tau) \ .
\ea
A very useful identity is \cite{ww}
\be
\label{A.13}
\vartheta_3(z)\vartheta_3^3-\vartheta_4(z)\vartheta_4^3-\vartheta_2(z)\vartheta_2^3
\ = \ 2\vartheta_1^4(z/2) \ ,
\ee
together with the identities computed at $z=0$ \cite{ww}
\be
\label{A.14}
\vartheta_3^{\prime\prime}\vartheta_3\vartheta_2^2-\vartheta_2^{\prime\prime}\vartheta_2\vartheta_3^2
\ = \ 4\pi^2\eta^6\vartheta_4^2 \ ,
\ee
and
\be
\label{A.15}
\vartheta_4^{\prime\prime}\vartheta_4\vartheta_3^2-\vartheta_3^{\prime\prime}\vartheta_3\vartheta_4^2
\ = \ 4\pi^2\eta^6\vartheta_2^2 \ .
\ee
\addcontentsline{toc}{chapter}{\numberline {}{Bibliography}}

\end{document}